\documentclass[twocolumn,prd,amsmath,amssymb,nofootinbib,superscriptaddress,floatfix]{revtex4-2}
\usepackage{graphicx}
\usepackage{booktabs}
\newcommand{\llmfig}{ Generated with LLM assistance per Appendix~\ref{app:llm}.}
\usepackage{xcolor}
\usepackage[colorlinks=true,allcolors=blue]{hyperref}

\newcommand{\Msun}{M_\odot}
\newcommand{\Rsun}{R_\odot}

\newcommand{\vk}{w}                       \newcommand{\vbf}{\mathbf{v}}
\newcommand{\wbf}{\mathbf{w}}
\newcommand{\rbf}{\mathbf{r}}
\newcommand{\hbf}{\mathbf{h}}
\newcommand{\ebf}{\mathbf{e}}
\newcommand{\Pbound}{P_{\rm bound}}
\newcommand{\Mb}{M_{\rm b}}
\newcommand{\eps}{\varepsilon}
\newcommand{\vsys}{\mathbf{v}_{\rm sys}}

\newcommand{\ad}{\textsc{ad\_sn\_popsyn}}
\newcommand{\scorpio}{\textsc{scorpio}}
\newcommand{\cosmic}{\textsc{cosmic}}
\newcommand{\compas}{\textsc{compas}}

\providecommand\result[1]{\ifcsname #1\endcsname\csname #1\endcsname\else\textbf{??#1??}\fi}

\expandafter\providecommand\csname CatBoundTwoPct\endcsname{}\expandafter\renewcommand\csname CatBoundTwoPct\endcsname{\ensuremath{62}}
\expandafter\providecommand\csname CatMedSepOne\endcsname{}\expandafter\renewcommand\csname CatMedSepOne\endcsname{\ensuremath{135}}
\expandafter\providecommand\csname CatMedSepTwo\endcsname{}\expandafter\renewcommand\csname CatMedSepTwo\endcsname{\ensuremath{5.3}}
\expandafter\providecommand\csname ConfFrozenDetachedPct\endcsname{}\expandafter\renewcommand\csname ConfFrozenDetachedPct\endcsname{\ensuremath{97.8}}
\expandafter\providecommand\csname ConfFrozenPct\endcsname{}\expandafter\renewcommand\csname ConfFrozenPct\endcsname{\ensuremath{94.9}}
\expandafter\providecommand\csname ConfFrozenRloPct\endcsname{}\expandafter\renewcommand\csname ConfFrozenRloPct\endcsname{\ensuremath{73.7}}
\expandafter\providecommand\csname DartBoardAcceptance\endcsname{}\expandafter\renewcommand\csname DartBoardAcceptance\endcsname{\ensuremath{0.09}}
\expandafter\providecommand\csname DartBoardEvals\endcsname{}\expandafter\renewcommand\csname DartBoardEvals\endcsname{\ensuremath{9.6\times10^{4}}}
\expandafter\providecommand\csname FracBBH\endcsname{}\expandafter\renewcommand\csname FracBBH\endcsname{\ensuremath{0.29}}
\expandafter\providecommand\csname FracBNS\endcsname{}\expandafter\renewcommand\csname FracBNS\endcsname{\ensuremath{0.04}}
\expandafter\providecommand\csname FracBoundWide\endcsname{}\expandafter\renewcommand\csname FracBoundWide\endcsname{\ensuremath{0.09}}
\expandafter\providecommand\csname FracDisrOne\endcsname{}\expandafter\renewcommand\csname FracDisrOne\endcsname{\ensuremath{0.25}}
\expandafter\providecommand\csname FracDisrTwo\endcsname{}\expandafter\renewcommand\csname FracDisrTwo\endcsname{\ensuremath{0.31}}
\expandafter\providecommand\csname FracNSBH\endcsname{}\expandafter\renewcommand\csname FracNSBH\endcsname{\ensuremath{0.03}}
\expandafter\providecommand\csname FracXrb\endcsname{}\expandafter\renewcommand\csname FracXrb\endcsname{\ensuremath{0.59}}
\expandafter\providecommand\csname FrozCoalPct\endcsname{}\expandafter\renewcommand\csname FrozCoalPct\endcsname{\ensuremath{58.8}}
\expandafter\providecommand\csname FrozDisrPct\endcsname{}\expandafter\renewcommand\csname FrozDisrPct\endcsname{\ensuremath{14.8}}
\expandafter\providecommand\csname FrozFryerCoalPct\endcsname{}\expandafter\renewcommand\csname FrozFryerCoalPct\endcsname{\ensuremath{71.8}}
\expandafter\providecommand\csname FrozFryerDisrPct\endcsname{}\expandafter\renewcommand\csname FrozFryerDisrPct\endcsname{\ensuremath{4.5}}
\expandafter\providecommand\csname JEvalSeconds\endcsname{}\expandafter\renewcommand\csname JEvalSeconds\endcsname{\ensuremath{3}}
\expandafter\providecommand\csname JKickMedian\endcsname{}\expandafter\renewcommand\csname JKickMedian\endcsname{\ensuremath{5.4}}
\expandafter\providecommand\csname JKickNinety\endcsname{}\expandafter\renewcommand\csname JKickNinety\endcsname{\ensuremath{7.8}}
\expandafter\providecommand\csname JProgMass\endcsname{}\expandafter\renewcommand\csname JProgMass\endcsname{\ensuremath{1.60^{+0.14}_{-0.14}}}
\expandafter\providecommand\csname KappaC\endcsname{}\expandafter\renewcommand\csname KappaC\endcsname{\ensuremath{1.003}}
\expandafter\providecommand\csname KappaIntercept\endcsname{}\expandafter\renewcommand\csname KappaIntercept\endcsname{\ensuremath{0.013}}
\expandafter\providecommand\csname KappaScatter\endcsname{}\expandafter\renewcommand\csname KappaScatter\endcsname{\ensuremath{0.018}}
\expandafter\providecommand\csname KickMeanBoundWide\endcsname{}\expandafter\renewcommand\csname KickMeanBoundWide\endcsname{\ensuremath{19}}
\expandafter\providecommand\csname KickMeanMarginal\endcsname{}\expandafter\renewcommand\csname KickMeanMarginal\endcsname{\ensuremath{74}}
\expandafter\providecommand\csname KickMeanMerging\endcsname{}\expandafter\renewcommand\csname KickMeanMerging\endcsname{\ensuremath{15}}
\expandafter\providecommand\csname MaxClassBias\endcsname{}\expandafter\renewcommand\csname MaxClassBias\endcsname{\ensuremath{0.066}}
\expandafter\providecommand\csname MedDlogSepTwo\endcsname{}\expandafter\renewcommand\csname MedDlogSepTwo\endcsname{\ensuremath{0.019}}
\expandafter\providecommand\csname NCatalog\endcsname{}\expandafter\renewcommand\csname NCatalog\endcsname{\ensuremath{5776}}
\expandafter\providecommand\csname NFrozenValid\endcsname{}\expandafter\renewcommand\csname NFrozenValid\endcsname{\ensuremath{312}}
\expandafter\providecommand\csname NParityBound\endcsname{}\expandafter\renewcommand\csname NParityBound\endcsname{\ensuremath{110}}
\expandafter\providecommand\csname NParityEvents\endcsname{}\expandafter\renewcommand\csname NParityEvents\endcsname{\ensuremath{203}}
\expandafter\providecommand\csname NReevolve\endcsname{}\expandafter\renewcommand\csname NReevolve\endcsname{\ensuremath{400}}
\expandafter\providecommand\csname ParityBoundAgreePct\endcsname{}\expandafter\renewcommand\csname ParityBoundAgreePct\endcsname{\ensuremath{100}}
\expandafter\providecommand\csname ParityMaxSepResid\endcsname{}\expandafter\renewcommand\csname ParityMaxSepResid\endcsname{\ensuremath{2.7\times10^{-2}}}
\expandafter\providecommand\csname ParityMedEccResid\endcsname{}\expandafter\renewcommand\csname ParityMedEccResid\endcsname{\ensuremath{7\times10^{-5}}}
\expandafter\providecommand\csname ParityMedSepResid\endcsname{}\expandafter\renewcommand\csname ParityMedSepResid\endcsname{\ensuremath{1.2\times10^{-4}}}
\expandafter\providecommand\csname ReevBoundWidePct\endcsname{}\expandafter\renewcommand\csname ReevBoundWidePct\endcsname{\ensuremath{69.3}}
\expandafter\providecommand\csname ReevCoalPct\endcsname{}\expandafter\renewcommand\csname ReevCoalPct\endcsname{\ensuremath{1.0}}
\expandafter\providecommand\csname ReevDisrPct\endcsname{}\expandafter\renewcommand\csname ReevDisrPct\endcsname{\ensuremath{29.8}}
\expandafter\providecommand\csname ReevFryerBoundWidePct\endcsname{}\expandafter\renewcommand\csname ReevFryerBoundWidePct\endcsname{\ensuremath{66.75}}
\expandafter\providecommand\csname ReevFryerDisrPct\endcsname{}\expandafter\renewcommand\csname ReevFryerDisrPct\endcsname{\ensuremath{32.0}}
\expandafter\providecommand\csname ShiftDisrOnePct\endcsname{}\expandafter\renewcommand\csname ShiftDisrOnePct\endcsname{\ensuremath{-12.2}}
\expandafter\providecommand\csname ShiftDisrTwoPct\endcsname{}\expandafter\renewcommand\csname ShiftDisrTwoPct\endcsname{\ensuremath{+8.5}}
\expandafter\providecommand\csname ShiftNsbhPct\endcsname{}\expandafter\renewcommand\csname ShiftNsbhPct\endcsname{\ensuremath{+8.7}}
\expandafter\providecommand\csname TbEntrySigma\endcsname{}\expandafter\renewcommand\csname TbEntrySigma\endcsname{\ensuremath{0.025}}
\expandafter\providecommand\csname TbTierOneCoal\endcsname{}\expandafter\renewcommand\csname TbTierOneCoal\endcsname{\ensuremath{0.000}}
\expandafter\providecommand\csname TbTierOneDisr\endcsname{}\expandafter\renewcommand\csname TbTierOneDisr\endcsname{\ensuremath{0.345}}
\expandafter\providecommand\csname TbTierOneFiveCoal\endcsname{}\expandafter\renewcommand\csname TbTierOneFiveCoal\endcsname{\ensuremath{0.001}}
\expandafter\providecommand\csname TbTierOneFiveDisr\endcsname{}\expandafter\renewcommand\csname TbTierOneFiveDisr\endcsname{\ensuremath{0.363}}
\expandafter\providecommand\csname TbTierOneFiveMerge\endcsname{}\expandafter\renewcommand\csname TbTierOneFiveMerge\endcsname{\ensuremath{0.003}}
\expandafter\providecommand\csname TbTierOneFiveWide\endcsname{}\expandafter\renewcommand\csname TbTierOneFiveWide\endcsname{\ensuremath{0.633}}
\expandafter\providecommand\csname TbTierOneMerge\endcsname{}\expandafter\renewcommand\csname TbTierOneMerge\endcsname{\ensuremath{0.594}}
\expandafter\providecommand\csname TbTierOneWide\endcsname{}\expandafter\renewcommand\csname TbTierOneWide\endcsname{\ensuremath{0.061}}
\expandafter\providecommand\csname TbTierZeroCoal\endcsname{}\expandafter\renewcommand\csname TbTierZeroCoal\endcsname{\ensuremath{0.000}}
\expandafter\providecommand\csname TbTierZeroDisr\endcsname{}\expandafter\renewcommand\csname TbTierZeroDisr\endcsname{\ensuremath{0.346}}
\expandafter\providecommand\csname TbTierZeroMerge\endcsname{}\expandafter\renewcommand\csname TbTierZeroMerge\endcsname{\ensuremath{0.592}}
\expandafter\providecommand\csname TbTierZeroWide\endcsname{}\expandafter\renewcommand\csname TbTierZeroWide\endcsname{\ensuremath{0.062}}
\expandafter\providecommand\csname TbTruthCoal\endcsname{}\expandafter\renewcommand\csname TbTruthCoal\endcsname{\ensuremath{0.010}}
\expandafter\providecommand\csname TbTruthDisr\endcsname{}\expandafter\renewcommand\csname TbTruthDisr\endcsname{\ensuremath{0.298}}
\expandafter\providecommand\csname TbTruthMerge\endcsname{}\expandafter\renewcommand\csname TbTruthMerge\endcsname{\ensuremath{<10^{-3}}}
\expandafter\providecommand\csname TbTruthWide\endcsname{}\expandafter\renewcommand\csname TbTruthWide\endcsname{\ensuremath{0.693}}

\providecommand\result[1]{\ifcsname #1\endcsname\csname #1\endcsname\else\textbf{??#1??}\fi}

\expandafter\providecommand\csname AdBoxArrival\endcsname{}\expandafter\renewcommand\csname AdBoxArrival\endcsname{\ensuremath{3.2\times10^{-12}}}
\expandafter\providecommand\csname AdBoxArrivalGivenBoundPercent\endcsname{}\expandafter\renewcommand\csname AdBoxArrivalGivenBoundPercent\endcsname{\ensuremath{2.8\times10^{-11}}}
\expandafter\providecommand\csname AdBoxArrivalJcirc\endcsname{}\expandafter\renewcommand\csname AdBoxArrivalJcirc\endcsname{\ensuremath{0.14}}
\expandafter\providecommand\csname AdBoxArrivalNoDelay\endcsname{}\expandafter\renewcommand\csname AdBoxArrivalNoDelay\endcsname{\ensuremath{3.2\times10^{-12}}}
\expandafter\providecommand\csname AdBoxArrivalUngated\endcsname{}\expandafter\renewcommand\csname AdBoxArrivalUngated\endcsname{\ensuremath{9.7}}
\expandafter\providecommand\csname AdBoxBoundPercent\endcsname{}\expandafter\renewcommand\csname AdBoxBoundPercent\endcsname{\ensuremath{12}}
\expandafter\providecommand\csname AdCaseBbOffset\endcsname{}\expandafter\renewcommand\csname AdCaseBbOffset\endcsname{\ensuremath{-0.2}}
\expandafter\providecommand\csname AdCaseBbStates\endcsname{}\expandafter\renewcommand\csname AdCaseBbStates\endcsname{\ensuremath{84}}
\expandafter\providecommand\csname AdClassCArrivalErrorPercent\endcsname{}\expandafter\renewcommand\csname AdClassCArrivalErrorPercent\endcsname{\ensuremath{1.6\times10^{-11}}}
\expandafter\providecommand\csname AdClassCArrivalGivenBoundPercent\endcsname{}\expandafter\renewcommand\csname AdClassCArrivalGivenBoundPercent\endcsname{\ensuremath{1.4\times10^{-10}}}
\expandafter\providecommand\csname AdClassCArrivalPercent\endcsname{}\expandafter\renewcommand\csname AdClassCArrivalPercent\endcsname{\ensuremath{2.2\times10^{-11}}}
\expandafter\providecommand\csname AdClassCArrivalUngatedErrorPercent\endcsname{}\expandafter\renewcommand\csname AdClassCArrivalUngatedErrorPercent\endcsname{\ensuremath{0.47}}
\expandafter\providecommand\csname AdClassCArrivalUngatedPercent\endcsname{}\expandafter\renewcommand\csname AdClassCArrivalUngatedPercent\endcsname{\ensuremath{3.05}}
\expandafter\providecommand\csname AdClassCBoundPercent\endcsname{}\expandafter\renewcommand\csname AdClassCBoundPercent\endcsname{\ensuremath{16}}
\expandafter\providecommand\csname AdClassCDnsPercent\endcsname{}\expandafter\renewcommand\csname AdClassCDnsPercent\endcsname{\ensuremath{1.8\times10^{-11}}}
\expandafter\providecommand\csname AdClassCDnsUngatedPercent\endcsname{}\expandafter\renewcommand\csname AdClassCDnsUngatedPercent\endcsname{\ensuremath{2.50}}
\expandafter\providecommand\csname AdClassCKick\endcsname{}\expandafter\renewcommand\csname AdClassCKick\endcsname{\ensuremath{337}}
\expandafter\providecommand\csname AdClassWArrivalErrorPercent\endcsname{}\expandafter\renewcommand\csname AdClassWArrivalErrorPercent\endcsname{\ensuremath{0.0100}}
\expandafter\providecommand\csname AdClassWArrivalGivenBoundPercent\endcsname{}\expandafter\renewcommand\csname AdClassWArrivalGivenBoundPercent\endcsname{\ensuremath{0.37}}
\expandafter\providecommand\csname AdClassWArrivalPercent\endcsname{}\expandafter\renewcommand\csname AdClassWArrivalPercent\endcsname{\ensuremath{0.062}}
\expandafter\providecommand\csname AdClassWArrivalUngatedErrorPercent\endcsname{}\expandafter\renewcommand\csname AdClassWArrivalUngatedErrorPercent\endcsname{\ensuremath{0.0100}}
\expandafter\providecommand\csname AdClassWArrivalUngatedPercent\endcsname{}\expandafter\renewcommand\csname AdClassWArrivalUngatedPercent\endcsname{\ensuremath{0.062}}
\expandafter\providecommand\csname AdClassWBoundPercent\endcsname{}\expandafter\renewcommand\csname AdClassWBoundPercent\endcsname{\ensuremath{17}}
\expandafter\providecommand\csname AdClassWDnsPercent\endcsname{}\expandafter\renewcommand\csname AdClassWDnsPercent\endcsname{\ensuremath{0.021}}
\expandafter\providecommand\csname AdClassWDnsUngatedPercent\endcsname{}\expandafter\renewcommand\csname AdClassWDnsUngatedPercent\endcsname{\ensuremath{0.021}}
\expandafter\providecommand\csname AdClassWKick\endcsname{}\expandafter\renewcommand\csname AdClassWKick\endcsname{\ensuremath{329}}
\expandafter\providecommand\csname AdDetachedOffset\endcsname{}\expandafter\renewcommand\csname AdDetachedOffset\endcsname{\ensuremath{0.3}}
\expandafter\providecommand\csname AdWideArrival\endcsname{}\expandafter\renewcommand\csname AdWideArrival\endcsname{\ensuremath{0.01}}
\expandafter\providecommand\csname AdWideDeltaLogP\endcsname{}\expandafter\renewcommand\csname AdWideDeltaLogP\endcsname{\ensuremath{0.26}}
\expandafter\providecommand\csname AdWideDeltaLogPJcirc\endcsname{}\expandafter\renewcommand\csname AdWideDeltaLogPJcirc\endcsname{\ensuremath{0.89}}
\expandafter\providecommand\csname AdWideDeltaMass\endcsname{}\expandafter\renewcommand\csname AdWideDeltaMass\endcsname{\ensuremath{0.05}}
\expandafter\providecommand\csname AdWideLogPScatter\endcsname{}\expandafter\renewcommand\csname AdWideLogPScatter\endcsname{\ensuremath{0.50}}
\expandafter\providecommand\csname AdWideMassRank\endcsname{}\expandafter\renewcommand\csname AdWideMassRank\endcsname{\ensuremath{0.99}}
\expandafter\providecommand\csname AdWideNsMedian\endcsname{}\expandafter\renewcommand\csname AdWideNsMedian\endcsname{\ensuremath{1.42}}
\expandafter\providecommand\csname AdWidePeriodRank\endcsname{}\expandafter\renewcommand\csname AdWidePeriodRank\endcsname{\ensuremath{0.54}}
\expandafter\providecommand\csname AdWidePeriodRankJcirc\endcsname{}\expandafter\renewcommand\csname AdWidePeriodRankJcirc\endcsname{\ensuremath{0.34}}
\expandafter\providecommand\csname AdWideStates\endcsname{}\expandafter\renewcommand\csname AdWideStates\endcsname{\ensuremath{464}}
\expandafter\providecommand\csname ArchiveArrivalDoublePercent\endcsname{}\expandafter\renewcommand\csname ArchiveArrivalDoublePercent\endcsname{\ensuremath{51}}
\expandafter\providecommand\csname ArchiveArrivalFamilyPercent\endcsname{}\expandafter\renewcommand\csname ArchiveArrivalFamilyPercent\endcsname{\ensuremath{42}}
\expandafter\providecommand\csname ArchiveBaselineBirths\endcsname{}\expandafter\renewcommand\csname ArchiveBaselineBirths\endcsname{\ensuremath{2{,}000{,}000}}
\expandafter\providecommand\csname ArchiveDnsDoublePercent\endcsname{}\expandafter\renewcommand\csname ArchiveDnsDoublePercent\endcsname{\ensuremath{49}}
\expandafter\providecommand\csname ArchiveDnsFamilyPercent\endcsname{}\expandafter\renewcommand\csname ArchiveDnsFamilyPercent\endcsname{\ensuremath{32}}
\expandafter\providecommand\csname ArchiveFamilyCe\endcsname{}\expandafter\renewcommand\csname ArchiveFamilyCe\endcsname{\ensuremath{27.5}}
\expandafter\providecommand\csname ArchiveFamilyYield\endcsname{}\expandafter\renewcommand\csname ArchiveFamilyYield\endcsname{\ensuremath{0.539}}
\expandafter\providecommand\csname ArchiveJointBirths\endcsname{}\expandafter\renewcommand\csname ArchiveJointBirths\endcsname{\ensuremath{1{,}000{,}000}}
\expandafter\providecommand\csname ArchiveJointMaxPull\endcsname{}\expandafter\renewcommand\csname ArchiveJointMaxPull\endcsname{\ensuremath{1.2}}
\expandafter\providecommand\csname ArchivePartitionDraws\endcsname{}\expandafter\renewcommand\csname ArchivePartitionDraws\endcsname{\ensuremath{20{,}000{,}000}}
\expandafter\providecommand\csname ArchivePartitionEvolved\endcsname{}\expandafter\renewcommand\csname ArchivePartitionEvolved\endcsname{\ensuremath{3{,}419{,}572}}
\expandafter\providecommand\csname BoxArrivalLimit\endcsname{}\expandafter\renewcommand\csname BoxArrivalLimit\endcsname{\ensuremath{0.02}}
\expandafter\providecommand\csname BoxArrivals\endcsname{}\expandafter\renewcommand\csname BoxArrivals\endcsname{\ensuremath{0}}
\expandafter\providecommand\csname BoxBound\endcsname{}\expandafter\renewcommand\csname BoxBound\endcsname{\ensuremath{2{,}505}}
\expandafter\providecommand\csname BoxBoundPercent\endcsname{}\expandafter\renewcommand\csname BoxBoundPercent\endcsname{\ensuremath{14}}
\expandafter\providecommand\csname BoxHgCe\endcsname{}\expandafter\renewcommand\csname BoxHgCe\endcsname{\ensuremath{2{,}126}}
\expandafter\providecommand\csname BoxMerged\endcsname{}\expandafter\renewcommand\csname BoxMerged\endcsname{\ensuremath{2{,}505}}
\expandafter\providecommand\csname BoxRows\endcsname{}\expandafter\renewcommand\csname BoxRows\endcsname{\ensuremath{17{,}116}}
\expandafter\providecommand\csname ClassCAge\endcsname{}\expandafter\renewcommand\csname ClassCAge\endcsname{\ensuremath{19.7}}
\expandafter\providecommand\csname ClassCArrivalGivenBoundPercent\endcsname{}\expandafter\renewcommand\csname ClassCArrivalGivenBoundPercent\endcsname{\ensuremath{0.043}}
\expandafter\providecommand\csname ClassCArrivalPercent\endcsname{}\expandafter\renewcommand\csname ClassCArrivalPercent\endcsname{\ensuremath{0.0083}}
\expandafter\providecommand\csname ClassCArrivalSharePercent\endcsname{}\expandafter\renewcommand\csname ClassCArrivalSharePercent\endcsname{\ensuremath{0.4}}
\expandafter\providecommand\csname ClassCBoundPercent\endcsname{}\expandafter\renewcommand\csname ClassCBoundPercent\endcsname{\ensuremath{19.5}}
\expandafter\providecommand\csname ClassCCoA\endcsname{}\expandafter\renewcommand\csname ClassCCoA\endcsname{\ensuremath{1.56}}
\expandafter\providecommand\csname ClassCCoB\endcsname{}\expandafter\renewcommand\csname ClassCCoB\endcsname{\ensuremath{0.218}}
\expandafter\providecommand\csname ClassCCoResHigh\endcsname{}\expandafter\renewcommand\csname ClassCCoResHigh\endcsname{\ensuremath{2.83}}
\expandafter\providecommand\csname ClassCCoResLow\endcsname{}\expandafter\renewcommand\csname ClassCCoResLow\endcsname{\ensuremath{-1.98}}
\expandafter\providecommand\csname ClassCCompHigh\endcsname{}\expandafter\renewcommand\csname ClassCCompHigh\endcsname{\ensuremath{16.7}}
\expandafter\providecommand\csname ClassCCompLow\endcsname{}\expandafter\renewcommand\csname ClassCCompLow\endcsname{\ensuremath{2.3}}
\expandafter\providecommand\csname ClassCCoveragePercent\endcsname{}\expandafter\renewcommand\csname ClassCCoveragePercent\endcsname{\ensuremath{100}}
\expandafter\providecommand\csname ClassCDnsPercent\endcsname{}\expandafter\renewcommand\csname ClassCDnsPercent\endcsname{\ensuremath{0.0039}}
\expandafter\providecommand\csname ClassCDrawnBound\endcsname{}\expandafter\renewcommand\csname ClassCDrawnBound\endcsname{\ensuremath{36}}
\expandafter\providecommand\csname ClassCDrawnMerged\endcsname{}\expandafter\renewcommand\csname ClassCDrawnMerged\endcsname{\ensuremath{25}}
\expandafter\providecommand\csname ClassCEnclosedPercent\endcsname{}\expandafter\renewcommand\csname ClassCEnclosedPercent\endcsname{\ensuremath{90}}
\expandafter\providecommand\csname ClassCFloorHe\endcsname{}\expandafter\renewcommand\csname ClassCFloorHe\endcsname{\ensuremath{0.73}}
\expandafter\providecommand\csname ClassCFloorPercent\endcsname{}\expandafter\renewcommand\csname ClassCFloorPercent\endcsname{\ensuremath{1.5}}
\expandafter\providecommand\csname ClassCHeA\endcsname{}\expandafter\renewcommand\csname ClassCHeA\endcsname{\ensuremath{0.40}}
\expandafter\providecommand\csname ClassCHeB\endcsname{}\expandafter\renewcommand\csname ClassCHeB\endcsname{\ensuremath{0.30}}
\expandafter\providecommand\csname ClassCKick\endcsname{}\expandafter\renewcommand\csname ClassCKick\endcsname{\ensuremath{338}}
\expandafter\providecommand\csname ClassCRows\endcsname{}\expandafter\renewcommand\csname ClassCRows\endcsname{\ensuremath{929{,}808}}
\expandafter\providecommand\csname ClassCSepA\endcsname{}\expandafter\renewcommand\csname ClassCSepA\endcsname{\ensuremath{1.42}}
\expandafter\providecommand\csname ClassCSepB\endcsname{}\expandafter\renewcommand\csname ClassCSepB\endcsname{\ensuremath{0.064}}
\expandafter\providecommand\csname ClassCSepC\endcsname{}\expandafter\renewcommand\csname ClassCSepC\endcsname{\ensuremath{0.037}}
\expandafter\providecommand\csname ClassCSepResHigh\endcsname{}\expandafter\renewcommand\csname ClassCSepResHigh\endcsname{\ensuremath{0.46}}
\expandafter\providecommand\csname ClassCSepResLow\endcsname{}\expandafter\renewcommand\csname ClassCSepResLow\endcsname{\ensuremath{-1.04}}
\expandafter\providecommand\csname ClassCWeight\endcsname{}\expandafter\renewcommand\csname ClassCWeight\endcsname{\ensuremath{0.28}}
\expandafter\providecommand\csname ClassDoubleArrivalSharePercent\endcsname{}\expandafter\renewcommand\csname ClassDoubleArrivalSharePercent\endcsname{\ensuremath{51}}
\expandafter\providecommand\csname ClassWAge\endcsname{}\expandafter\renewcommand\csname ClassWAge\endcsname{\ensuremath{3.4}}
\expandafter\providecommand\csname ClassWArrivalGivenBoundPercent\endcsname{}\expandafter\renewcommand\csname ClassWArrivalGivenBoundPercent\endcsname{\ensuremath{1.2}}
\expandafter\providecommand\csname ClassWArrivalPercent\endcsname{}\expandafter\renewcommand\csname ClassWArrivalPercent\endcsname{\ensuremath{0.26}}
\expandafter\providecommand\csname ClassWArrivalSharePercent\endcsname{}\expandafter\renewcommand\csname ClassWArrivalSharePercent\endcsname{\ensuremath{42}}
\expandafter\providecommand\csname ClassWBoundPercent\endcsname{}\expandafter\renewcommand\csname ClassWBoundPercent\endcsname{\ensuremath{22.2}}
\expandafter\providecommand\csname ClassWBreakA\endcsname{}\expandafter\renewcommand\csname ClassWBreakA\endcsname{\ensuremath{20.1}}
\expandafter\providecommand\csname ClassWBreakB\endcsname{}\expandafter\renewcommand\csname ClassWBreakB\endcsname{\ensuremath{23.6}}
\expandafter\providecommand\csname ClassWCoA\endcsname{}\expandafter\renewcommand\csname ClassWCoA\endcsname{\ensuremath{0.50}}
\expandafter\providecommand\csname ClassWCoB\endcsname{}\expandafter\renewcommand\csname ClassWCoB\endcsname{\ensuremath{0.147}}
\expandafter\providecommand\csname ClassWCoResHigh\endcsname{}\expandafter\renewcommand\csname ClassWCoResHigh\endcsname{\ensuremath{1.84}}
\expandafter\providecommand\csname ClassWCoResLow\endcsname{}\expandafter\renewcommand\csname ClassWCoResLow\endcsname{\ensuremath{-1.76}}
\expandafter\providecommand\csname ClassWCompHigh\endcsname{}\expandafter\renewcommand\csname ClassWCompHigh\endcsname{\ensuremath{35.6}}
\expandafter\providecommand\csname ClassWCompLow\endcsname{}\expandafter\renewcommand\csname ClassWCompLow\endcsname{\ensuremath{11.7}}
\expandafter\providecommand\csname ClassWCoveragePercent\endcsname{}\expandafter\renewcommand\csname ClassWCoveragePercent\endcsname{\ensuremath{85}}
\expandafter\providecommand\csname ClassWDnsPercent\endcsname{}\expandafter\renewcommand\csname ClassWDnsPercent\endcsname{\ensuremath{0.10}}
\expandafter\providecommand\csname ClassWDrawnBound\endcsname{}\expandafter\renewcommand\csname ClassWDrawnBound\endcsname{\ensuremath{61}}
\expandafter\providecommand\csname ClassWDrawnMerged\endcsname{}\expandafter\renewcommand\csname ClassWDrawnMerged\endcsname{\ensuremath{59}}
\expandafter\providecommand\csname ClassWEnclosedPercent\endcsname{}\expandafter\renewcommand\csname ClassWEnclosedPercent\endcsname{\ensuremath{90}}
\expandafter\providecommand\csname ClassWFloorHe\endcsname{}\expandafter\renewcommand\csname ClassWFloorHe\endcsname{\ensuremath{0.43}}
\expandafter\providecommand\csname ClassWFloorPercent\endcsname{}\expandafter\renewcommand\csname ClassWFloorPercent\endcsname{\ensuremath{6.9}}
\expandafter\providecommand\csname ClassWHeA\endcsname{}\expandafter\renewcommand\csname ClassWHeA\endcsname{\ensuremath{0.45}}
\expandafter\providecommand\csname ClassWHeB\endcsname{}\expandafter\renewcommand\csname ClassWHeB\endcsname{\ensuremath{0.28}}
\expandafter\providecommand\csname ClassWKick\endcsname{}\expandafter\renewcommand\csname ClassWKick\endcsname{\ensuremath{328}}
\expandafter\providecommand\csname ClassWRows\endcsname{}\expandafter\renewcommand\csname ClassWRows\endcsname{\ensuremath{2{,}125{,}609}}
\expandafter\providecommand\csname ClassWSepA\endcsname{}\expandafter\renewcommand\csname ClassWSepA\endcsname{\ensuremath{2.25}}
\expandafter\providecommand\csname ClassWSepB\endcsname{}\expandafter\renewcommand\csname ClassWSepB\endcsname{\ensuremath{-0.033}}
\expandafter\providecommand\csname ClassWSepC\endcsname{}\expandafter\renewcommand\csname ClassWSepC\endcsname{\ensuremath{0.283}}
\expandafter\providecommand\csname ClassWSepResHigh\endcsname{}\expandafter\renewcommand\csname ClassWSepResHigh\endcsname{\ensuremath{0.84}}
\expandafter\providecommand\csname ClassWSepResLow\endcsname{}\expandafter\renewcommand\csname ClassWSepResLow\endcsname{\ensuremath{-0.86}}
\expandafter\providecommand\csname ClassWWeight\endcsname{}\expandafter\renewcommand\csname ClassWWeight\endcsname{\ensuremath{0.72}}
\expandafter\providecommand\csname ControlBound\endcsname{}\expandafter\renewcommand\csname ControlBound\endcsname{\ensuremath{343}}
\expandafter\providecommand\csname ControlMerged\endcsname{}\expandafter\renewcommand\csname ControlMerged\endcsname{\ensuremath{343}}
\expandafter\providecommand\csname ControlRows\endcsname{}\expandafter\renewcommand\csname ControlRows\endcsname{\ensuremath{2{,}451}}
\expandafter\providecommand\csname CosmicWideNsMedian\endcsname{}\expandafter\renewcommand\csname CosmicWideNsMedian\endcsname{\ensuremath{2.00}}
\expandafter\providecommand\csname FitBirths\endcsname{}\expandafter\renewcommand\csname FitBirths\endcsname{\ensuremath{1{,}500{,}000}}
\expandafter\providecommand\csname FitCdfLimitCe\endcsname{}\expandafter\renewcommand\csname FitCdfLimitCe\endcsname{\ensuremath{0.0435}}
\expandafter\providecommand\csname FitCdfLimitNoCe\endcsname{}\expandafter\renewcommand\csname FitCdfLimitNoCe\endcsname{\ensuremath{0.0268}}
\expandafter\providecommand\csname FitCeZero\endcsname{}\expandafter\renewcommand\csname FitCeZero\endcsname{\ensuremath{27.5}}
\expandafter\providecommand\csname FitComponents\endcsname{}\expandafter\renewcommand\csname FitComponents\endcsname{\ensuremath{36}}
\expandafter\providecommand\csname FitCorrCe\endcsname{}\expandafter\renewcommand\csname FitCorrCe\endcsname{\ensuremath{0.074}}
\expandafter\providecommand\csname FitCorrNoCe\endcsname{}\expandafter\renewcommand\csname FitCorrNoCe\endcsname{\ensuremath{0.005}}
\expandafter\providecommand\csname FitMaxCdfCe\endcsname{}\expandafter\renewcommand\csname FitMaxCdfCe\endcsname{\ensuremath{0.0242}}
\expandafter\providecommand\csname FitMaxCdfNoCe\endcsname{}\expandafter\renewcommand\csname FitMaxCdfNoCe\endcsname{\ensuremath{0.0199}}
\expandafter\providecommand\csname FitTestBirths\endcsname{}\expandafter\renewcommand\csname FitTestBirths\endcsname{\ensuremath{500{,}000}}
\expandafter\providecommand\csname FitTestRows\endcsname{}\expandafter\renewcommand\csname FitTestRows\endcsname{\ensuremath{268{,}954}}
\expandafter\providecommand\csname FitTrainRows\endcsname{}\expandafter\renewcommand\csname FitTrainRows\endcsname{\ensuremath{808{,}140}}
\expandafter\providecommand\csname GiantCeCount\endcsname{}\expandafter\renewcommand\csname GiantCeCount\endcsname{\ensuremath{71{,}653}}
\expandafter\providecommand\csname GiantCeMergedPercent\endcsname{}\expandafter\renewcommand\csname GiantCeMergedPercent\endcsname{\ensuremath{95}}
\expandafter\providecommand\csname HgCeCount\endcsname{}\expandafter\renewcommand\csname HgCeCount\endcsname{\ensuremath{91{,}590}}
\expandafter\providecommand\csname HgFinalSep\endcsname{}\expandafter\renewcommand\csname HgFinalSep\endcsname{\ensuremath{0.4}}
\expandafter\providecommand\csname HgOnsetQ\endcsname{}\expandafter\renewcommand\csname HgOnsetQ\endcsname{\ensuremath{6.7}}
\expandafter\providecommand\csname HgOnsetRadius\endcsname{}\expandafter\renewcommand\csname HgOnsetRadius\endcsname{\ensuremath{21}}
\expandafter\providecommand\csname PartitionEvolvedFraction\endcsname{}\expandafter\renewcommand\csname PartitionEvolvedFraction\endcsname{\ensuremath{0.17}}
\expandafter\providecommand\csname PartitionFloor\endcsname{}\expandafter\renewcommand\csname PartitionFloor\endcsname{\ensuremath{0.05}}
\expandafter\providecommand\csname PartitionLoss\endcsname{}\expandafter\renewcommand\csname PartitionLoss\endcsname{\ensuremath{0.5}}
\expandafter\providecommand\csname PriorFitEfficiencyErrorPercent\endcsname{}\expandafter\renewcommand\csname PriorFitEfficiencyErrorPercent\endcsname{\ensuremath{3}}
\expandafter\providecommand\csname PriorFitLogPError\endcsname{}\expandafter\renewcommand\csname PriorFitLogPError\endcsname{\ensuremath{0.02}}
\expandafter\providecommand\csname WideCompanionHigh\endcsname{}\expandafter\renewcommand\csname WideCompanionHigh\endcsname{\ensuremath{30}}
\expandafter\providecommand\csname WideCompanionLow\endcsname{}\expandafter\renewcommand\csname WideCompanionLow\endcsname{\ensuremath{14}}
\expandafter\providecommand\csname WideSepHigh\endcsname{}\expandafter\renewcommand\csname WideSepHigh\endcsname{\ensuremath{2740}}
\expandafter\providecommand\csname WideSepLow\endcsname{}\expandafter\renewcommand\csname WideSepLow\endcsname{\ensuremath{819}}

\providecommand\result[1]{\ifcsname #1\endcsname\csname #1\endcsname\else\textbf{??#1??}\fi}

\expandafter\providecommand\csname TiltBoundFlip\endcsname{}\expandafter\renewcommand\csname TiltBoundFlip\endcsname{\ensuremath{0.3389}}
\expandafter\providecommand\csname TiltBoundModerate\endcsname{}\expandafter\renewcommand\csname TiltBoundModerate\endcsname{\ensuremath{0.5083}}
\expandafter\providecommand\csname TiltBoundSmall\endcsname{}\expandafter\renewcommand\csname TiltBoundSmall\endcsname{\ensuremath{0.9806}}
\expandafter\providecommand\csname TiltBoundStrong\endcsname{}\expandafter\renewcommand\csname TiltBoundStrong\endcsname{\ensuremath{0.0258}}
\expandafter\providecommand\csname TiltCDFFlip\endcsname{}\expandafter\renewcommand\csname TiltCDFFlip\endcsname{\ensuremath{9.1\times10^{-5}}}
\expandafter\providecommand\csname TiltCDFModerate\endcsname{}\expandafter\renewcommand\csname TiltCDFModerate\endcsname{\ensuremath{7.1\times10^{-5}}}
\expandafter\providecommand\csname TiltCDFSmall\endcsname{}\expandafter\renewcommand\csname TiltCDFSmall\endcsname{\ensuremath{2.4\times10^{-5}}}
\expandafter\providecommand\csname TiltCDFStrong\endcsname{}\expandafter\renewcommand\csname TiltCDFStrong\endcsname{\ensuremath{4.6\times10^{-4}}}
\expandafter\providecommand\csname TiltMergerFlip\endcsname{}\expandafter\renewcommand\csname TiltMergerFlip\endcsname{\ensuremath{0.6806}}
\expandafter\providecommand\csname TiltMergerModerate\endcsname{}\expandafter\renewcommand\csname TiltMergerModerate\endcsname{\ensuremath{0.5423}}
\expandafter\providecommand\csname TiltMergerSmall\endcsname{}\expandafter\renewcommand\csname TiltMergerSmall\endcsname{\ensuremath{0.1806}}
\expandafter\providecommand\csname TiltMergerStrong\endcsname{}\expandafter\renewcommand\csname TiltMergerStrong\endcsname{\ensuremath{0.0000}}
\expandafter\providecommand\csname TiltResolutionFlip\endcsname{}\expandafter\renewcommand\csname TiltResolutionFlip\endcsname{\ensuremath{2.1\times10^{-4}}}
\expandafter\providecommand\csname TiltResolutionModerate\endcsname{}\expandafter\renewcommand\csname TiltResolutionModerate\endcsname{\ensuremath{1.8\times10^{-4}}}
\expandafter\providecommand\csname TiltResolutionSmall\endcsname{}\expandafter\renewcommand\csname TiltResolutionSmall\endcsname{\ensuremath{2.1\times10^{-4}}}
\expandafter\providecommand\csname TiltResolutionStrong\endcsname{}\expandafter\renewcommand\csname TiltResolutionStrong\endcsname{\ensuremath{1.1\times10^{-3}}}
\expandafter\providecommand\csname TiltRetroBoundFlip\endcsname{}\expandafter\renewcommand\csname TiltRetroBoundFlip\endcsname{\ensuremath{0.2459}}
\expandafter\providecommand\csname TiltRetroBoundModerate\endcsname{}\expandafter\renewcommand\csname TiltRetroBoundModerate\endcsname{\ensuremath{0.0000}}
\expandafter\providecommand\csname TiltRetroBoundSmall\endcsname{}\expandafter\renewcommand\csname TiltRetroBoundSmall\endcsname{\ensuremath{0.0000}}
\expandafter\providecommand\csname TiltRetroBoundStrong\endcsname{}\expandafter\renewcommand\csname TiltRetroBoundStrong\endcsname{\ensuremath{1.0000}}
\expandafter\providecommand\csname TiltRetroMergeFlip\endcsname{}\expandafter\renewcommand\csname TiltRetroMergeFlip\endcsname{\ensuremath{0.3613}}
\expandafter\providecommand\csname TiltRetroMergeModerate\endcsname{}\expandafter\renewcommand\csname TiltRetroMergeModerate\endcsname{\ensuremath{0.0000}}
\expandafter\providecommand\csname TiltRetroMergeSmall\endcsname{}\expandafter\renewcommand\csname TiltRetroMergeSmall\endcsname{\ensuremath{0.0000}}
\expandafter\providecommand\csname TiltRetroMergeStrong\endcsname{}\expandafter\renewcommand\csname TiltRetroMergeStrong\endcsname{\ensuremath{--}}
\expandafter\providecommand\csname TiltSeedFlip\endcsname{}\expandafter\renewcommand\csname TiltSeedFlip\endcsname{\ensuremath{2.9\times10^{-4}}}
\expandafter\providecommand\csname TiltSeedModerate\endcsname{}\expandafter\renewcommand\csname TiltSeedModerate\endcsname{\ensuremath{2.2\times10^{-4}}}
\expandafter\providecommand\csname TiltSeedSmall\endcsname{}\expandafter\renewcommand\csname TiltSeedSmall\endcsname{\ensuremath{1.9\times10^{-4}}}
\expandafter\providecommand\csname TiltSeedStrong\endcsname{}\expandafter\renewcommand\csname TiltSeedStrong\endcsname{\ensuremath{9.6\times10^{-4}}}
\expandafter\providecommand\csname TiltShiftFlip\endcsname{}\expandafter\renewcommand\csname TiltShiftFlip\endcsname{\ensuremath{0.1612}}
\expandafter\providecommand\csname TiltShiftModerate\endcsname{}\expandafter\renewcommand\csname TiltShiftModerate\endcsname{\ensuremath{0.0688}}
\expandafter\providecommand\csname TiltShiftSmall\endcsname{}\expandafter\renewcommand\csname TiltShiftSmall\endcsname{\ensuremath{0.1540}}
\expandafter\providecommand\csname TiltShiftStrong\endcsname{}\expandafter\renewcommand\csname TiltShiftStrong\endcsname{\ensuremath{--}}
\expandafter\providecommand\csname TiltTimeFlip\endcsname{}\expandafter\renewcommand\csname TiltTimeFlip\endcsname{\ensuremath{2.5\times10^{-6}}}
\expandafter\providecommand\csname TiltTimeModerate\endcsname{}\expandafter\renewcommand\csname TiltTimeModerate\endcsname{\ensuremath{0.0000}}
\expandafter\providecommand\csname TiltTimeSmall\endcsname{}\expandafter\renewcommand\csname TiltTimeSmall\endcsname{\ensuremath{0.0000}}
\expandafter\providecommand\csname TiltTimeStrong\endcsname{}\expandafter\renewcommand\csname TiltTimeStrong\endcsname{\ensuremath{0.0000}}
\expandafter\providecommand\csname TiltUFlip\endcsname{}\expandafter\renewcommand\csname TiltUFlip\endcsname{\ensuremath{1.2000}}
\expandafter\providecommand\csname TiltUModerate\endcsname{}\expandafter\renewcommand\csname TiltUModerate\endcsname{\ensuremath{0.8000}}
\expandafter\providecommand\csname TiltUSmall\endcsname{}\expandafter\renewcommand\csname TiltUSmall\endcsname{\ensuremath{0.3000}}
\expandafter\providecommand\csname TiltUStrong\endcsname{}\expandafter\renewcommand\csname TiltUStrong\endcsname{\ensuremath{2.2000}}

\providecommand\result[1]{\ifcsname #1\endcsname\csname #1\endcsname\else\textbf{??#1??}\fi}

\expandafter\providecommand\csname OrientCapJoint\endcsname{}\expandafter\renewcommand\csname OrientCapJoint\endcsname{\ensuremath{0.43^{+0.11}_{-0.11}}}
\expandafter\providecommand\csname OrientCapLowerJoint\endcsname{}\expandafter\renewcommand\csname OrientCapLowerJoint\endcsname{\ensuremath{0.31}}
\expandafter\providecommand\csname OrientCapLowerMerger\endcsname{}\expandafter\renewcommand\csname OrientCapLowerMerger\endcsname{\ensuremath{0.00}}
\expandafter\providecommand\csname OrientCapLowerProjected\endcsname{}\expandafter\renewcommand\csname OrientCapLowerProjected\endcsname{\ensuremath{0.22}}
\expandafter\providecommand\csname OrientCapMerger\endcsname{}\expandafter\renewcommand\csname OrientCapMerger\endcsname{\ensuremath{0.29^{+0.71}_{-0.29}}}
\expandafter\providecommand\csname OrientCapProjected\endcsname{}\expandafter\renewcommand\csname OrientCapProjected\endcsname{\ensuremath{0.42^{+0.16}_{-0.21}}}
\expandafter\providecommand\csname OrientCapUpperJoint\endcsname{}\expandafter\renewcommand\csname OrientCapUpperJoint\endcsname{\ensuremath{0.53}}
\expandafter\providecommand\csname OrientCapUpperMerger\endcsname{}\expandafter\renewcommand\csname OrientCapUpperMerger\endcsname{\ensuremath{0.99}}
\expandafter\providecommand\csname OrientCapUpperProjected\endcsname{}\expandafter\renewcommand\csname OrientCapUpperProjected\endcsname{\ensuremath{0.58}}
\expandafter\providecommand\csname OrientGradientError\endcsname{}\expandafter\renewcommand\csname OrientGradientError\endcsname{\ensuremath{6.03\times10^{-10}}}
\expandafter\providecommand\csname OrientInjectedCap\endcsname{}\expandafter\renewcommand\csname OrientInjectedCap\endcsname{\ensuremath{0.43}}
\expandafter\providecommand\csname OrientMergerMaxAngle\endcsname{}\expandafter\renewcommand\csname OrientMergerMaxAngle\endcsname{\ensuremath{38}}
\expandafter\providecommand\csname OrientMergerMaxMu\endcsname{}\expandafter\renewcommand\csname OrientMergerMaxMu\endcsname{\ensuremath{0.79}}
\expandafter\providecommand\csname OrientMergerPolarEfficiency\endcsname{}\expandafter\renewcommand\csname OrientMergerPolarEfficiency\endcsname{\ensuremath{0.168}}
\expandafter\providecommand\csname OrientNoiselessError\endcsname{}\expandafter\renewcommand\csname OrientNoiselessError\endcsname{\ensuremath{6.80\times10^{-10}}}
\expandafter\providecommand\csname OrientResponseError\endcsname{}\expandafter\renewcommand\csname OrientResponseError\endcsname{\ensuremath{1.63\times10^{-3}}}
\expandafter\providecommand\csname OrientSample\endcsname{}\expandafter\renewcommand\csname OrientSample\endcsname{\ensuremath{200}}
\expandafter\providecommand\csname OrientSecondCapJoint\endcsname{}\expandafter\renewcommand\csname OrientSecondCapJoint\endcsname{\ensuremath{0.47}}
\expandafter\providecommand\csname OrientSecondCapMerger\endcsname{}\expandafter\renewcommand\csname OrientSecondCapMerger\endcsname{\ensuremath{0.88}}
\expandafter\providecommand\csname OrientSecondCapProjected\endcsname{}\expandafter\renewcommand\csname OrientSecondCapProjected\endcsname{\ensuremath{0.69}}
\expandafter\providecommand\csname OrientSeedShift\endcsname{}\expandafter\renewcommand\csname OrientSeedShift\endcsname{\ensuremath{8.29\times10^{-4}}}

\providecommand\result[1]{\ifcsname #1\endcsname\csname #1\endcsname\else\textbf{??#1??}\fi}

\expandafter\providecommand\csname OrientPopBestEta\endcsname{}\expandafter\renewcommand\csname OrientPopBestEta\endcsname{\ensuremath{-0.045}}
\expandafter\providecommand\csname OrientPopBestKappa\endcsname{}\expandafter\renewcommand\csname OrientPopBestKappa\endcsname{\ensuremath{1.53}}
\expandafter\providecommand\csname OrientPopBestSigma\endcsname{}\expandafter\renewcommand\csname OrientPopBestSigma\endcsname{\ensuremath{241}}
\expandafter\providecommand\csname OrientPopFineJointKl\endcsname{}\expandafter\renewcommand\csname OrientPopFineJointKl\endcsname{\ensuremath{0.018}}
\expandafter\providecommand\csname OrientPopFineShapeKl\endcsname{}\expandafter\renewcommand\csname OrientPopFineShapeKl\endcsname{\ensuremath{0.050}}
\expandafter\providecommand\csname OrientPopNullCheckShift\endcsname{}\expandafter\renewcommand\csname OrientPopNullCheckShift\endcsname{\ensuremath{0.484}}
\expandafter\providecommand\csname OrientPopNullTwiceDrop\endcsname{}\expandafter\renewcommand\csname OrientPopNullTwiceDrop\endcsname{\ensuremath{3.29}}
\expandafter\providecommand\csname OrientPopRefineShift\endcsname{}\expandafter\renewcommand\csname OrientPopRefineShift\endcsname{\ensuremath{0.257}}
\expandafter\providecommand\csname OrientPopSurvivalIsotropic\endcsname{}\expandafter\renewcommand\csname OrientPopSurvivalIsotropic\endcsname{\ensuremath{0.387}}
\expandafter\providecommand\csname OrientPopSurvivalPolar\endcsname{}\expandafter\renewcommand\csname OrientPopSurvivalPolar\endcsname{\ensuremath{0.358}}

\newcommand{\PresnTwoBirths}{1080000}

\newcommand{\PresnTwoStates}{28281}
\newcommand{\PresnTwoTestStates}{452}

\newcommand{\PresnTwoESS}{5105}

\providecommand\result[1]{\ifcsname #1\endcsname\csname #1\endcsname\else\textbf{??#1??}\fi}

\expandafter\providecommand\csname ChanBinaryNatal\endcsname{}\expandafter\renewcommand\csname ChanBinaryNatal\endcsname{\ensuremath{0.62}}
\expandafter\providecommand\csname ChanBinaryNatalExclude\endcsname{}\expandafter\renewcommand\csname ChanBinaryNatalExclude\endcsname{\ensuremath{13}}
\expandafter\providecommand\csname ChanBinaryNatalTen\endcsname{}\expandafter\renewcommand\csname ChanBinaryNatalTen\endcsname{\ensuremath{0.67}}
\expandafter\providecommand\csname ChanBinaryNatalTenExclude\endcsname{}\expandafter\renewcommand\csname ChanBinaryNatalTenExclude\endcsname{\ensuremath{72}}
\expandafter\providecommand\csname ChanBinaryNatalTenWidth\endcsname{}\expandafter\renewcommand\csname ChanBinaryNatalTenWidth\endcsname{\ensuremath{0.24}}
\expandafter\providecommand\csname ChanBinaryNatalWidth\endcsname{}\expandafter\renewcommand\csname ChanBinaryNatalWidth\endcsname{\ensuremath{0.62}}
\expandafter\providecommand\csname ChanBinaryProduct\endcsname{}\expandafter\renewcommand\csname ChanBinaryProduct\endcsname{\ensuremath{0.77}}
\expandafter\providecommand\csname ChanBinaryProductExclude\endcsname{}\expandafter\renewcommand\csname ChanBinaryProductExclude\endcsname{\ensuremath{44}}
\expandafter\providecommand\csname ChanBinaryProductTen\endcsname{}\expandafter\renewcommand\csname ChanBinaryProductTen\endcsname{\ensuremath{0.83}}
\expandafter\providecommand\csname ChanBinaryProductTenExclude\endcsname{}\expandafter\renewcommand\csname ChanBinaryProductTenExclude\endcsname{\ensuremath{100}}
\expandafter\providecommand\csname ChanBinaryProductTenWidth\endcsname{}\expandafter\renewcommand\csname ChanBinaryProductTenWidth\endcsname{\ensuremath{0.16}}
\expandafter\providecommand\csname ChanBinaryProductWidth\endcsname{}\expandafter\renewcommand\csname ChanBinaryProductWidth\endcsname{\ensuremath{0.48}}
\expandafter\providecommand\csname ChanBinaryRight\endcsname{}\expandafter\renewcommand\csname ChanBinaryRight\endcsname{\ensuremath{0.50}}
\expandafter\providecommand\csname ChanBinaryRightExclude\endcsname{}\expandafter\renewcommand\csname ChanBinaryRightExclude\endcsname{\ensuremath{4}}
\expandafter\providecommand\csname ChanBinaryRightTen\endcsname{}\expandafter\renewcommand\csname ChanBinaryRightTen\endcsname{\ensuremath{0.50}}
\expandafter\providecommand\csname ChanBinaryRightTenExclude\endcsname{}\expandafter\renewcommand\csname ChanBinaryRightTenExclude\endcsname{\ensuremath{9}}
\expandafter\providecommand\csname ChanBinaryRightTenWidth\endcsname{}\expandafter\renewcommand\csname ChanBinaryRightTenWidth\endcsname{\ensuremath{0.27}}
\expandafter\providecommand\csname ChanBinaryRightWidth\endcsname{}\expandafter\renewcommand\csname ChanBinaryRightWidth\endcsname{\ensuremath{0.62}}
\expandafter\providecommand\csname ChanJointNatal\endcsname{}\expandafter\renewcommand\csname ChanJointNatal\endcsname{\ensuremath{0.10}}
\expandafter\providecommand\csname ChanJointNatalExclude\endcsname{}\expandafter\renewcommand\csname ChanJointNatalExclude\endcsname{\ensuremath{100}}
\expandafter\providecommand\csname ChanJointNatalTen\endcsname{}\expandafter\renewcommand\csname ChanJointNatalTen\endcsname{\ensuremath{0.09}}
\expandafter\providecommand\csname ChanJointNatalTenExclude\endcsname{}\expandafter\renewcommand\csname ChanJointNatalTenExclude\endcsname{\ensuremath{100}}
\expandafter\providecommand\csname ChanJointNatalTenWidth\endcsname{}\expandafter\renewcommand\csname ChanJointNatalTenWidth\endcsname{\ensuremath{0.06}}
\expandafter\providecommand\csname ChanJointNatalWidth\endcsname{}\expandafter\renewcommand\csname ChanJointNatalWidth\endcsname{\ensuremath{0.17}}
\expandafter\providecommand\csname ChanJointProduct\endcsname{}\expandafter\renewcommand\csname ChanJointProduct\endcsname{\ensuremath{0.05}}
\expandafter\providecommand\csname ChanJointProductExclude\endcsname{}\expandafter\renewcommand\csname ChanJointProductExclude\endcsname{\ensuremath{100}}
\expandafter\providecommand\csname ChanJointProductTen\endcsname{}\expandafter\renewcommand\csname ChanJointProductTen\endcsname{\ensuremath{0.03}}
\expandafter\providecommand\csname ChanJointProductTenExclude\endcsname{}\expandafter\renewcommand\csname ChanJointProductTenExclude\endcsname{\ensuremath{100}}
\expandafter\providecommand\csname ChanJointProductTenWidth\endcsname{}\expandafter\renewcommand\csname ChanJointProductTenWidth\endcsname{\ensuremath{0.05}}
\expandafter\providecommand\csname ChanJointProductWidth\endcsname{}\expandafter\renewcommand\csname ChanJointProductWidth\endcsname{\ensuremath{0.14}}
\expandafter\providecommand\csname ChanJointRight\endcsname{}\expandafter\renewcommand\csname ChanJointRight\endcsname{\ensuremath{0.47}}
\expandafter\providecommand\csname ChanJointRightExclude\endcsname{}\expandafter\renewcommand\csname ChanJointRightExclude\endcsname{\ensuremath{8}}
\expandafter\providecommand\csname ChanJointRightTen\endcsname{}\expandafter\renewcommand\csname ChanJointRightTen\endcsname{\ensuremath{0.50}}
\expandafter\providecommand\csname ChanJointRightTenExclude\endcsname{}\expandafter\renewcommand\csname ChanJointRightTenExclude\endcsname{\ensuremath{11}}
\expandafter\providecommand\csname ChanJointRightTenWidth\endcsname{}\expandafter\renewcommand\csname ChanJointRightTenWidth\endcsname{\ensuremath{0.16}}
\expandafter\providecommand\csname ChanJointRightWidth\endcsname{}\expandafter\renewcommand\csname ChanJointRightWidth\endcsname{\ensuremath{0.44}}
\expandafter\providecommand\csname ChanSingleNatal\endcsname{}\expandafter\renewcommand\csname ChanSingleNatal\endcsname{\ensuremath{0.07}}
\expandafter\providecommand\csname ChanSingleNatalExclude\endcsname{}\expandafter\renewcommand\csname ChanSingleNatalExclude\endcsname{\ensuremath{100}}
\expandafter\providecommand\csname ChanSingleNatalTen\endcsname{}\expandafter\renewcommand\csname ChanSingleNatalTen\endcsname{\ensuremath{0.06}}
\expandafter\providecommand\csname ChanSingleNatalTenExclude\endcsname{}\expandafter\renewcommand\csname ChanSingleNatalTenExclude\endcsname{\ensuremath{100}}
\expandafter\providecommand\csname ChanSingleNatalTenWidth\endcsname{}\expandafter\renewcommand\csname ChanSingleNatalTenWidth\endcsname{\ensuremath{0.05}}
\expandafter\providecommand\csname ChanSingleNatalWidth\endcsname{}\expandafter\renewcommand\csname ChanSingleNatalWidth\endcsname{\ensuremath{0.15}}
\expandafter\providecommand\csname ChanSingleProduct\endcsname{}\expandafter\renewcommand\csname ChanSingleProduct\endcsname{\ensuremath{0.02}}
\expandafter\providecommand\csname ChanSingleProductExclude\endcsname{}\expandafter\renewcommand\csname ChanSingleProductExclude\endcsname{\ensuremath{100}}
\expandafter\providecommand\csname ChanSingleProductTen\endcsname{}\expandafter\renewcommand\csname ChanSingleProductTen\endcsname{\ensuremath{0.00}}
\expandafter\providecommand\csname ChanSingleProductTenExclude\endcsname{}\expandafter\renewcommand\csname ChanSingleProductTenExclude\endcsname{\ensuremath{100}}
\expandafter\providecommand\csname ChanSingleProductTenWidth\endcsname{}\expandafter\renewcommand\csname ChanSingleProductTenWidth\endcsname{\ensuremath{0.01}}
\expandafter\providecommand\csname ChanSingleProductWidth\endcsname{}\expandafter\renewcommand\csname ChanSingleProductWidth\endcsname{\ensuremath{0.07}}
\expandafter\providecommand\csname ChanSingleRight\endcsname{}\expandafter\renewcommand\csname ChanSingleRight\endcsname{\ensuremath{0.45}}
\expandafter\providecommand\csname ChanSingleRightExclude\endcsname{}\expandafter\renewcommand\csname ChanSingleRightExclude\endcsname{\ensuremath{7}}
\expandafter\providecommand\csname ChanSingleRightTen\endcsname{}\expandafter\renewcommand\csname ChanSingleRightTen\endcsname{\ensuremath{0.50}}
\expandafter\providecommand\csname ChanSingleRightTenExclude\endcsname{}\expandafter\renewcommand\csname ChanSingleRightTenExclude\endcsname{\ensuremath{13}}
\expandafter\providecommand\csname ChanSingleRightTenWidth\endcsname{}\expandafter\renewcommand\csname ChanSingleRightTenWidth\endcsname{\ensuremath{0.20}}
\expandafter\providecommand\csname ChanSingleRightWidth\endcsname{}\expandafter\renewcommand\csname ChanSingleRightWidth\endcsname{\ensuremath{0.54}}
\expandafter\providecommand\csname DnsCatalogs\endcsname{}\expandafter\renewcommand\csname DnsCatalogs\endcsname{\ensuremath{200}}
\expandafter\providecommand\csname DnsMockNatal\endcsname{}\expandafter\renewcommand\csname DnsMockNatal\endcsname{\ensuremath{0.62}}
\expandafter\providecommand\csname DnsMockNatalExclude\endcsname{}\expandafter\renewcommand\csname DnsMockNatalExclude\endcsname{\ensuremath{10}}
\expandafter\providecommand\csname DnsMockNatalHundred\endcsname{}\expandafter\renewcommand\csname DnsMockNatalHundred\endcsname{\ensuremath{0.67}}
\expandafter\providecommand\csname DnsMockNatalHundredExclude\endcsname{}\expandafter\renewcommand\csname DnsMockNatalHundredExclude\endcsname{\ensuremath{100}}
\expandafter\providecommand\csname DnsMockNatalQuarter\endcsname{}\expandafter\renewcommand\csname DnsMockNatalQuarter\endcsname{\ensuremath{0.42}}
\expandafter\providecommand\csname DnsMockNatalQuarterExclude\endcsname{}\expandafter\renewcommand\csname DnsMockNatalQuarterExclude\endcsname{\ensuremath{14}}
\expandafter\providecommand\csname DnsMockNatalTen\endcsname{}\expandafter\renewcommand\csname DnsMockNatalTen\endcsname{\ensuremath{0.67}}
\expandafter\providecommand\csname DnsMockNatalTenExclude\endcsname{}\expandafter\renewcommand\csname DnsMockNatalTenExclude\endcsname{\ensuremath{74}}
\expandafter\providecommand\csname DnsMockNatalThreeQuarter\endcsname{}\expandafter\renewcommand\csname DnsMockNatalThreeQuarter\endcsname{\ensuremath{0.77}}
\expandafter\providecommand\csname DnsMockNatalThreeQuarterExclude\endcsname{}\expandafter\renewcommand\csname DnsMockNatalThreeQuarterExclude\endcsname{\ensuremath{2}}
\expandafter\providecommand\csname DnsMockProduct\endcsname{}\expandafter\renewcommand\csname DnsMockProduct\endcsname{\ensuremath{0.76}}
\expandafter\providecommand\csname DnsMockProductExclude\endcsname{}\expandafter\renewcommand\csname DnsMockProductExclude\endcsname{\ensuremath{42}}
\expandafter\providecommand\csname DnsMockProductHundred\endcsname{}\expandafter\renewcommand\csname DnsMockProductHundred\endcsname{\ensuremath{0.83}}
\expandafter\providecommand\csname DnsMockProductHundredExclude\endcsname{}\expandafter\renewcommand\csname DnsMockProductHundredExclude\endcsname{\ensuremath{100}}
\expandafter\providecommand\csname DnsMockProductQuarter\endcsname{}\expandafter\renewcommand\csname DnsMockProductQuarter\endcsname{\ensuremath{0.62}}
\expandafter\providecommand\csname DnsMockProductQuarterExclude\endcsname{}\expandafter\renewcommand\csname DnsMockProductQuarterExclude\endcsname{\ensuremath{62}}
\expandafter\providecommand\csname DnsMockProductTen\endcsname{}\expandafter\renewcommand\csname DnsMockProductTen\endcsname{\ensuremath{0.82}}
\expandafter\providecommand\csname DnsMockProductTenExclude\endcsname{}\expandafter\renewcommand\csname DnsMockProductTenExclude\endcsname{\ensuremath{100}}
\expandafter\providecommand\csname DnsMockProductThreeQuarter\endcsname{}\expandafter\renewcommand\csname DnsMockProductThreeQuarter\endcsname{\ensuremath{0.88}}
\expandafter\providecommand\csname DnsMockProductThreeQuarterExclude\endcsname{}\expandafter\renewcommand\csname DnsMockProductThreeQuarterExclude\endcsname{\ensuremath{0}}
\expandafter\providecommand\csname DnsMockRight\endcsname{}\expandafter\renewcommand\csname DnsMockRight\endcsname{\ensuremath{0.50}}
\expandafter\providecommand\csname DnsMockRightExclude\endcsname{}\expandafter\renewcommand\csname DnsMockRightExclude\endcsname{\ensuremath{6}}
\expandafter\providecommand\csname DnsMockRightHundred\endcsname{}\expandafter\renewcommand\csname DnsMockRightHundred\endcsname{\ensuremath{0.50}}
\expandafter\providecommand\csname DnsMockRightHundredExclude\endcsname{}\expandafter\renewcommand\csname DnsMockRightHundredExclude\endcsname{\ensuremath{10}}
\expandafter\providecommand\csname DnsMockRightQuarter\endcsname{}\expandafter\renewcommand\csname DnsMockRightQuarter\endcsname{\ensuremath{0.31}}
\expandafter\providecommand\csname DnsMockRightQuarterExclude\endcsname{}\expandafter\renewcommand\csname DnsMockRightQuarterExclude\endcsname{\ensuremath{8}}
\expandafter\providecommand\csname DnsMockRightTen\endcsname{}\expandafter\renewcommand\csname DnsMockRightTen\endcsname{\ensuremath{0.50}}
\expandafter\providecommand\csname DnsMockRightTenExclude\endcsname{}\expandafter\renewcommand\csname DnsMockRightTenExclude\endcsname{\ensuremath{11}}
\expandafter\providecommand\csname DnsMockRightThreeQuarter\endcsname{}\expandafter\renewcommand\csname DnsMockRightThreeQuarter\endcsname{\ensuremath{0.69}}
\expandafter\providecommand\csname DnsMockRightThreeQuarterExclude\endcsname{}\expandafter\renewcommand\csname DnsMockRightThreeQuarterExclude\endcsname{\ensuremath{6}}
\expandafter\providecommand\csname DnsRealAltNatal\endcsname{}\expandafter\renewcommand\csname DnsRealAltNatal\endcsname{\ensuremath{0.72}}
\expandafter\providecommand\csname DnsRealAltProduct\endcsname{}\expandafter\renewcommand\csname DnsRealAltProduct\endcsname{\ensuremath{0.90}}
\expandafter\providecommand\csname DnsRealAltRight\endcsname{}\expandafter\renewcommand\csname DnsRealAltRight\endcsname{\ensuremath{0.66}}
\expandafter\providecommand\csname DnsRealNatal\endcsname{}\expandafter\renewcommand\csname DnsRealNatal\endcsname{\ensuremath{0.67}}
\expandafter\providecommand\csname DnsRealNatalHigh\endcsname{}\expandafter\renewcommand\csname DnsRealNatalHigh\endcsname{\ensuremath{0.92}}
\expandafter\providecommand\csname DnsRealNatalLow\endcsname{}\expandafter\renewcommand\csname DnsRealNatalLow\endcsname{\ensuremath{0.33}}
\expandafter\providecommand\csname DnsRealProduct\endcsname{}\expandafter\renewcommand\csname DnsRealProduct\endcsname{\ensuremath{0.80}}
\expandafter\providecommand\csname DnsRealProductHigh\endcsname{}\expandafter\renewcommand\csname DnsRealProductHigh\endcsname{\ensuremath{0.98}}
\expandafter\providecommand\csname DnsRealProductLow\endcsname{}\expandafter\renewcommand\csname DnsRealProductLow\endcsname{\ensuremath{0.50}}
\expandafter\providecommand\csname DnsRealRight\endcsname{}\expandafter\renewcommand\csname DnsRealRight\endcsname{\ensuremath{0.56}}
\expandafter\providecommand\csname DnsRealRightHigh\endcsname{}\expandafter\renewcommand\csname DnsRealRightHigh\endcsname{\ensuremath{0.89}}
\expandafter\providecommand\csname DnsRealRightLow\endcsname{}\expandafter\renewcommand\csname DnsRealRightLow\endcsname{\ensuremath{0.24}}
\expandafter\providecommand\csname DnsSample\endcsname{}\expandafter\renewcommand\csname DnsSample\endcsname{\ensuremath{11}}
\expandafter\providecommand\csname DnsSeedShift\endcsname{}\expandafter\renewcommand\csname DnsSeedShift\endcsname{\ensuremath{0.002}}
\expandafter\providecommand\csname DnsSingleHigh\endcsname{}\expandafter\renewcommand\csname DnsSingleHigh\endcsname{\ensuremath{0.53}}
\expandafter\providecommand\csname DnsSingleLow\endcsname{}\expandafter\renewcommand\csname DnsSingleLow\endcsname{\ensuremath{0.03}}
\expandafter\providecommand\csname DnsSingles\endcsname{}\expandafter\renewcommand\csname DnsSingles\endcsname{\ensuremath{48}}
\expandafter\providecommand\csname DnsSurvivalHigh\endcsname{}\expandafter\renewcommand\csname DnsSurvivalHigh\endcsname{\ensuremath{0.47}}
\expandafter\providecommand\csname DnsSurvivalLow\endcsname{}\expandafter\renewcommand\csname DnsSurvivalLow\endcsname{\ensuremath{0.96}}
\expandafter\providecommand\csname RecoilBiasEfficiencyHigh\endcsname{}\expandafter\renewcommand\csname RecoilBiasEfficiencyHigh\endcsname{\ensuremath{0.17}}
\expandafter\providecommand\csname RecoilBiasOmittedHigh\endcsname{}\expandafter\renewcommand\csname RecoilBiasOmittedHigh\endcsname{\ensuremath{189}}
\expandafter\providecommand\csname RecoilBiasSigmasHigh\endcsname{}\expandafter\renewcommand\csname RecoilBiasSigmasHigh\endcsname{\ensuremath{4.4}}
\expandafter\providecommand\csname RecoilBiasSigmasRef\endcsname{}\expandafter\renewcommand\csname RecoilBiasSigmasRef\endcsname{\ensuremath{3}}
\expandafter\providecommand\csname RecoilBiasTrueHigh\endcsname{}\expandafter\renewcommand\csname RecoilBiasTrueHigh\endcsname{\ensuremath{300}}
\expandafter\providecommand\csname RecoilBoundPercent\endcsname{}\expandafter\renewcommand\csname RecoilBoundPercent\endcsname{\ensuremath{52}}
\expandafter\providecommand\csname RecoilCorrelationCoarse\endcsname{}\expandafter\renewcommand\csname RecoilCorrelationCoarse\endcsname{\ensuremath{0.02}}
\expandafter\providecommand\csname RecoilCorrelationJoint\endcsname{}\expandafter\renewcommand\csname RecoilCorrelationJoint\endcsname{\ensuremath{0.02}}
\expandafter\providecommand\csname RecoilCorrelationMass\endcsname{}\expandafter\renewcommand\csname RecoilCorrelationMass\endcsname{\ensuremath{0.99}}
\expandafter\providecommand\csname RecoilCorrelationRecoil\endcsname{}\expandafter\renewcommand\csname RecoilCorrelationRecoil\endcsname{\ensuremath{0.44}}
\expandafter\providecommand\csname RecoilEfficiency\endcsname{}\expandafter\renewcommand\csname RecoilEfficiency\endcsname{\ensuremath{0.516}}
\expandafter\providecommand\csname RecoilErrorImprovement\endcsname{}\expandafter\renewcommand\csname RecoilErrorImprovement\endcsname{\ensuremath{1.68}}
\expandafter\providecommand\csname RecoilFitKickOne\endcsname{}\expandafter\renewcommand\csname RecoilFitKickOne\endcsname{\ensuremath{102.3}}
\expandafter\providecommand\csname RecoilFitKickTwo\endcsname{}\expandafter\renewcommand\csname RecoilFitKickTwo\endcsname{\ensuremath{100.1}}
\expandafter\providecommand\csname RecoilFitMeanOne\endcsname{}\expandafter\renewcommand\csname RecoilFitMeanOne\endcsname{\ensuremath{1.413}}
\expandafter\providecommand\csname RecoilFitMeanTwo\endcsname{}\expandafter\renewcommand\csname RecoilFitMeanTwo\endcsname{\ensuremath{1.397}}
\expandafter\providecommand\csname RecoilGradientError\endcsname{}\expandafter\renewcommand\csname RecoilGradientError\endcsname{\ensuremath{4.24\times10^{-10}}}
\expandafter\providecommand\csname RecoilKickStdCoarse\endcsname{}\expandafter\renewcommand\csname RecoilKickStdCoarse\endcsname{\ensuremath{6.0}}
\expandafter\providecommand\csname RecoilKickStdJoint\endcsname{}\expandafter\renewcommand\csname RecoilKickStdJoint\endcsname{\ensuremath{3.6}}
\expandafter\providecommand\csname RecoilKickStdMass\endcsname{}\expandafter\renewcommand\csname RecoilKickStdMass\endcsname{\ensuremath{3186.1}}
\expandafter\providecommand\csname RecoilKickStdRecoil\endcsname{}\expandafter\renewcommand\csname RecoilKickStdRecoil\endcsname{\ensuremath{4.0}}
\expandafter\providecommand\csname RecoilMassStdCoarse\endcsname{}\expandafter\renewcommand\csname RecoilMassStdCoarse\endcsname{\ensuremath{0.008}}
\expandafter\providecommand\csname RecoilMassStdJoint\endcsname{}\expandafter\renewcommand\csname RecoilMassStdJoint\endcsname{\ensuremath{0.008}}
\expandafter\providecommand\csname RecoilMassStdMass\endcsname{}\expandafter\renewcommand\csname RecoilMassStdMass\endcsname{\ensuremath{0.047}}
\expandafter\providecommand\csname RecoilMassStdRecoil\endcsname{}\expandafter\renewcommand\csname RecoilMassStdRecoil\endcsname{\ensuremath{0.061}}
\expandafter\providecommand\csname RecoilRefineShift\endcsname{}\expandafter\renewcommand\csname RecoilRefineShift\endcsname{\ensuremath{1.06\times10^{-4}}}
\expandafter\providecommand\csname RecoilSample\endcsname{}\expandafter\renewcommand\csname RecoilSample\endcsname{\ensuremath{200}}
\expandafter\providecommand\csname RecoilVectorError\endcsname{}\expandafter\renewcommand\csname RecoilVectorError\endcsname{\ensuremath{5.46\times10^{-4}}}
\expandafter\providecommand\csname RecoilWrongKick\endcsname{}\expandafter\renewcommand\csname RecoilWrongKick\endcsname{\ensuremath{88.9}}
\expandafter\providecommand\csname RecoilWrongMean\endcsname{}\expandafter\renewcommand\csname RecoilWrongMean\endcsname{\ensuremath{1.405}}

\providecommand\result[1]{\ifcsname #1\endcsname\csname #1\endcsname\else\textbf{??#1??}\fi}

\expandafter\providecommand\csname ChanBinaryNatal\endcsname{}\expandafter\renewcommand\csname ChanBinaryNatal\endcsname{\ensuremath{0.54}}
\expandafter\providecommand\csname ChanBinaryNatalExclude\endcsname{}\expandafter\renewcommand\csname ChanBinaryNatalExclude\endcsname{\ensuremath{0}}
\expandafter\providecommand\csname ChanBinaryNatalTen\endcsname{}\expandafter\renewcommand\csname ChanBinaryNatalTen\endcsname{\ensuremath{0.54}}
\expandafter\providecommand\csname ChanBinaryNatalTenExclude\endcsname{}\expandafter\renewcommand\csname ChanBinaryNatalTenExclude\endcsname{\ensuremath{15}}
\expandafter\providecommand\csname ChanBinaryNatalTenWidth\endcsname{}\expandafter\renewcommand\csname ChanBinaryNatalTenWidth\endcsname{\ensuremath{0.39}}
\expandafter\providecommand\csname ChanBinaryNatalWidth\endcsname{}\expandafter\renewcommand\csname ChanBinaryNatalWidth\endcsname{\ensuremath{0.78}}
\expandafter\providecommand\csname ChanBinaryProduct\endcsname{}\expandafter\renewcommand\csname ChanBinaryProduct\endcsname{\ensuremath{0.68}}
\expandafter\providecommand\csname ChanBinaryProductExclude\endcsname{}\expandafter\renewcommand\csname ChanBinaryProductExclude\endcsname{\ensuremath{4}}
\expandafter\providecommand\csname ChanBinaryProductTen\endcsname{}\expandafter\renewcommand\csname ChanBinaryProductTen\endcsname{\ensuremath{0.76}}
\expandafter\providecommand\csname ChanBinaryProductTenExclude\endcsname{}\expandafter\renewcommand\csname ChanBinaryProductTenExclude\endcsname{\ensuremath{76}}
\expandafter\providecommand\csname ChanBinaryProductTenWidth\endcsname{}\expandafter\renewcommand\csname ChanBinaryProductTenWidth\endcsname{\ensuremath{0.31}}
\expandafter\providecommand\csname ChanBinaryProductWidth\endcsname{}\expandafter\renewcommand\csname ChanBinaryProductWidth\endcsname{\ensuremath{0.73}}
\expandafter\providecommand\csname ChanBinaryRight\endcsname{}\expandafter\renewcommand\csname ChanBinaryRight\endcsname{\ensuremath{0.52}}
\expandafter\providecommand\csname ChanBinaryRightExclude\endcsname{}\expandafter\renewcommand\csname ChanBinaryRightExclude\endcsname{\ensuremath{0}}
\expandafter\providecommand\csname ChanBinaryRightTen\endcsname{}\expandafter\renewcommand\csname ChanBinaryRightTen\endcsname{\ensuremath{0.50}}
\expandafter\providecommand\csname ChanBinaryRightTenExclude\endcsname{}\expandafter\renewcommand\csname ChanBinaryRightTenExclude\endcsname{\ensuremath{13}}
\expandafter\providecommand\csname ChanBinaryRightTenWidth\endcsname{}\expandafter\renewcommand\csname ChanBinaryRightTenWidth\endcsname{\ensuremath{0.39}}
\expandafter\providecommand\csname ChanBinaryRightWidth\endcsname{}\expandafter\renewcommand\csname ChanBinaryRightWidth\endcsname{\ensuremath{0.78}}
\expandafter\providecommand\csname ChanJointNatal\endcsname{}\expandafter\renewcommand\csname ChanJointNatal\endcsname{\ensuremath{0.47}}
\expandafter\providecommand\csname ChanJointNatalExclude\endcsname{}\expandafter\renewcommand\csname ChanJointNatalExclude\endcsname{\ensuremath{12}}
\expandafter\providecommand\csname ChanJointNatalTen\endcsname{}\expandafter\renewcommand\csname ChanJointNatalTen\endcsname{\ensuremath{0.47}}
\expandafter\providecommand\csname ChanJointNatalTenExclude\endcsname{}\expandafter\renewcommand\csname ChanJointNatalTenExclude\endcsname{\ensuremath{16}}
\expandafter\providecommand\csname ChanJointNatalTenWidth\endcsname{}\expandafter\renewcommand\csname ChanJointNatalTenWidth\endcsname{\ensuremath{0.14}}
\expandafter\providecommand\csname ChanJointNatalWidth\endcsname{}\expandafter\renewcommand\csname ChanJointNatalWidth\endcsname{\ensuremath{0.43}}
\expandafter\providecommand\csname ChanJointProduct\endcsname{}\expandafter\renewcommand\csname ChanJointProduct\endcsname{\ensuremath{0.39}}
\expandafter\providecommand\csname ChanJointProductExclude\endcsname{}\expandafter\renewcommand\csname ChanJointProductExclude\endcsname{\ensuremath{20}}
\expandafter\providecommand\csname ChanJointProductTen\endcsname{}\expandafter\renewcommand\csname ChanJointProductTen\endcsname{\ensuremath{0.39}}
\expandafter\providecommand\csname ChanJointProductTenExclude\endcsname{}\expandafter\renewcommand\csname ChanJointProductTenExclude\endcsname{\ensuremath{78}}
\expandafter\providecommand\csname ChanJointProductTenWidth\endcsname{}\expandafter\renewcommand\csname ChanJointProductTenWidth\endcsname{\ensuremath{0.16}}
\expandafter\providecommand\csname ChanJointProductWidth\endcsname{}\expandafter\renewcommand\csname ChanJointProductWidth\endcsname{\ensuremath{0.46}}
\expandafter\providecommand\csname ChanJointRight\endcsname{}\expandafter\renewcommand\csname ChanJointRight\endcsname{\ensuremath{0.49}}
\expandafter\providecommand\csname ChanJointRightExclude\endcsname{}\expandafter\renewcommand\csname ChanJointRightExclude\endcsname{\ensuremath{11}}
\expandafter\providecommand\csname ChanJointRightTen\endcsname{}\expandafter\renewcommand\csname ChanJointRightTen\endcsname{\ensuremath{0.50}}
\expandafter\providecommand\csname ChanJointRightTenExclude\endcsname{}\expandafter\renewcommand\csname ChanJointRightTenExclude\endcsname{\ensuremath{8}}
\expandafter\providecommand\csname ChanJointRightTenWidth\endcsname{}\expandafter\renewcommand\csname ChanJointRightTenWidth\endcsname{\ensuremath{0.15}}
\expandafter\providecommand\csname ChanJointRightWidth\endcsname{}\expandafter\renewcommand\csname ChanJointRightWidth\endcsname{\ensuremath{0.44}}
\expandafter\providecommand\csname ChanSingleNatal\endcsname{}\expandafter\renewcommand\csname ChanSingleNatal\endcsname{\ensuremath{0.44}}
\expandafter\providecommand\csname ChanSingleNatalExclude\endcsname{}\expandafter\renewcommand\csname ChanSingleNatalExclude\endcsname{\ensuremath{10}}
\expandafter\providecommand\csname ChanSingleNatalTen\endcsname{}\expandafter\renewcommand\csname ChanSingleNatalTen\endcsname{\ensuremath{0.46}}
\expandafter\providecommand\csname ChanSingleNatalTenExclude\endcsname{}\expandafter\renewcommand\csname ChanSingleNatalTenExclude\endcsname{\ensuremath{20}}
\expandafter\providecommand\csname ChanSingleNatalTenWidth\endcsname{}\expandafter\renewcommand\csname ChanSingleNatalTenWidth\endcsname{\ensuremath{0.16}}
\expandafter\providecommand\csname ChanSingleNatalWidth\endcsname{}\expandafter\renewcommand\csname ChanSingleNatalWidth\endcsname{\ensuremath{0.46}}
\expandafter\providecommand\csname ChanSingleProduct\endcsname{}\expandafter\renewcommand\csname ChanSingleProduct\endcsname{\ensuremath{0.29}}
\expandafter\providecommand\csname ChanSingleProductExclude\endcsname{}\expandafter\renewcommand\csname ChanSingleProductExclude\endcsname{\ensuremath{32}}
\expandafter\providecommand\csname ChanSingleProductTen\endcsname{}\expandafter\renewcommand\csname ChanSingleProductTen\endcsname{\ensuremath{0.30}}
\expandafter\providecommand\csname ChanSingleProductTenExclude\endcsname{}\expandafter\renewcommand\csname ChanSingleProductTenExclude\endcsname{\ensuremath{98}}
\expandafter\providecommand\csname ChanSingleProductTenWidth\endcsname{}\expandafter\renewcommand\csname ChanSingleProductTenWidth\endcsname{\ensuremath{0.18}}
\expandafter\providecommand\csname ChanSingleProductWidth\endcsname{}\expandafter\renewcommand\csname ChanSingleProductWidth\endcsname{\ensuremath{0.49}}
\expandafter\providecommand\csname ChanSingleRight\endcsname{}\expandafter\renewcommand\csname ChanSingleRight\endcsname{\ensuremath{0.47}}
\expandafter\providecommand\csname ChanSingleRightExclude\endcsname{}\expandafter\renewcommand\csname ChanSingleRightExclude\endcsname{\ensuremath{10}}
\expandafter\providecommand\csname ChanSingleRightTen\endcsname{}\expandafter\renewcommand\csname ChanSingleRightTen\endcsname{\ensuremath{0.50}}
\expandafter\providecommand\csname ChanSingleRightTenExclude\endcsname{}\expandafter\renewcommand\csname ChanSingleRightTenExclude\endcsname{\ensuremath{9}}
\expandafter\providecommand\csname ChanSingleRightTenWidth\endcsname{}\expandafter\renewcommand\csname ChanSingleRightTenWidth\endcsname{\ensuremath{0.16}}
\expandafter\providecommand\csname ChanSingleRightWidth\endcsname{}\expandafter\renewcommand\csname ChanSingleRightWidth\endcsname{\ensuremath{0.47}}
\expandafter\providecommand\csname DnsCatalogs\endcsname{}\expandafter\renewcommand\csname DnsCatalogs\endcsname{\ensuremath{200}}
\expandafter\providecommand\csname DnsMockNatal\endcsname{}\expandafter\renewcommand\csname DnsMockNatal\endcsname{\ensuremath{0.50}}
\expandafter\providecommand\csname DnsMockNatalExclude\endcsname{}\expandafter\renewcommand\csname DnsMockNatalExclude\endcsname{\ensuremath{0}}
\expandafter\providecommand\csname DnsMockNatalHundred\endcsname{}\expandafter\renewcommand\csname DnsMockNatalHundred\endcsname{\ensuremath{0.54}}
\expandafter\providecommand\csname DnsMockNatalHundredExclude\endcsname{}\expandafter\renewcommand\csname DnsMockNatalHundredExclude\endcsname{\ensuremath{30}}
\expandafter\providecommand\csname DnsMockNatalQuarter\endcsname{}\expandafter\renewcommand\csname DnsMockNatalQuarter\endcsname{\ensuremath{0.40}}
\expandafter\providecommand\csname DnsMockNatalQuarterExclude\endcsname{}\expandafter\renewcommand\csname DnsMockNatalQuarterExclude\endcsname{\ensuremath{2}}
\expandafter\providecommand\csname DnsMockNatalTen\endcsname{}\expandafter\renewcommand\csname DnsMockNatalTen\endcsname{\ensuremath{0.54}}
\expandafter\providecommand\csname DnsMockNatalTenExclude\endcsname{}\expandafter\renewcommand\csname DnsMockNatalTenExclude\endcsname{\ensuremath{10}}
\expandafter\providecommand\csname DnsMockNatalThreeQuarter\endcsname{}\expandafter\renewcommand\csname DnsMockNatalThreeQuarter\endcsname{\ensuremath{0.68}}
\expandafter\providecommand\csname DnsMockNatalThreeQuarterExclude\endcsname{}\expandafter\renewcommand\csname DnsMockNatalThreeQuarterExclude\endcsname{\ensuremath{5}}
\expandafter\providecommand\csname DnsMockProduct\endcsname{}\expandafter\renewcommand\csname DnsMockProduct\endcsname{\ensuremath{0.64}}
\expandafter\providecommand\csname DnsMockProductExclude\endcsname{}\expandafter\renewcommand\csname DnsMockProductExclude\endcsname{\ensuremath{4}}
\expandafter\providecommand\csname DnsMockProductHundred\endcsname{}\expandafter\renewcommand\csname DnsMockProductHundred\endcsname{\ensuremath{0.76}}
\expandafter\providecommand\csname DnsMockProductHundredExclude\endcsname{}\expandafter\renewcommand\csname DnsMockProductHundredExclude\endcsname{\ensuremath{100}}
\expandafter\providecommand\csname DnsMockProductQuarter\endcsname{}\expandafter\renewcommand\csname DnsMockProductQuarter\endcsname{\ensuremath{0.56}}
\expandafter\providecommand\csname DnsMockProductQuarterExclude\endcsname{}\expandafter\renewcommand\csname DnsMockProductQuarterExclude\endcsname{\ensuremath{16}}
\expandafter\providecommand\csname DnsMockProductTen\endcsname{}\expandafter\renewcommand\csname DnsMockProductTen\endcsname{\ensuremath{0.75}}
\expandafter\providecommand\csname DnsMockProductTenExclude\endcsname{}\expandafter\renewcommand\csname DnsMockProductTenExclude\endcsname{\ensuremath{77}}
\expandafter\providecommand\csname DnsMockProductThreeQuarter\endcsname{}\expandafter\renewcommand\csname DnsMockProductThreeQuarter\endcsname{\ensuremath{0.78}}
\expandafter\providecommand\csname DnsMockProductThreeQuarterExclude\endcsname{}\expandafter\renewcommand\csname DnsMockProductThreeQuarterExclude\endcsname{\ensuremath{1}}
\expandafter\providecommand\csname DnsMockRight\endcsname{}\expandafter\renewcommand\csname DnsMockRight\endcsname{\ensuremath{0.49}}
\expandafter\providecommand\csname DnsMockRightExclude\endcsname{}\expandafter\renewcommand\csname DnsMockRightExclude\endcsname{\ensuremath{0}}
\expandafter\providecommand\csname DnsMockRightHundred\endcsname{}\expandafter\renewcommand\csname DnsMockRightHundred\endcsname{\ensuremath{0.50}}
\expandafter\providecommand\csname DnsMockRightHundredExclude\endcsname{}\expandafter\renewcommand\csname DnsMockRightHundredExclude\endcsname{\ensuremath{9}}
\expandafter\providecommand\csname DnsMockRightQuarter\endcsname{}\expandafter\renewcommand\csname DnsMockRightQuarter\endcsname{\ensuremath{0.38}}
\expandafter\providecommand\csname DnsMockRightQuarterExclude\endcsname{}\expandafter\renewcommand\csname DnsMockRightQuarterExclude\endcsname{\ensuremath{1}}
\expandafter\providecommand\csname DnsMockRightTen\endcsname{}\expandafter\renewcommand\csname DnsMockRightTen\endcsname{\ensuremath{0.50}}
\expandafter\providecommand\csname DnsMockRightTenExclude\endcsname{}\expandafter\renewcommand\csname DnsMockRightTenExclude\endcsname{\ensuremath{8}}
\expandafter\providecommand\csname DnsMockRightThreeQuarter\endcsname{}\expandafter\renewcommand\csname DnsMockRightThreeQuarter\endcsname{\ensuremath{0.67}}
\expandafter\providecommand\csname DnsMockRightThreeQuarterExclude\endcsname{}\expandafter\renewcommand\csname DnsMockRightThreeQuarterExclude\endcsname{\ensuremath{6}}
\expandafter\providecommand\csname DnsRealAltNatal\endcsname{}\expandafter\renewcommand\csname DnsRealAltNatal\endcsname{\ensuremath{0.83}}
\expandafter\providecommand\csname DnsRealAltProduct\endcsname{}\expandafter\renewcommand\csname DnsRealAltProduct\endcsname{\ensuremath{0.87}}
\expandafter\providecommand\csname DnsRealAltRight\endcsname{}\expandafter\renewcommand\csname DnsRealAltRight\endcsname{\ensuremath{0.82}}
\expandafter\providecommand\csname DnsRealNatal\endcsname{}\expandafter\renewcommand\csname DnsRealNatal\endcsname{\ensuremath{0.86}}
\expandafter\providecommand\csname DnsRealNatalHigh\endcsname{}\expandafter\renewcommand\csname DnsRealNatalHigh\endcsname{\ensuremath{0.99}}
\expandafter\providecommand\csname DnsRealNatalLow\endcsname{}\expandafter\renewcommand\csname DnsRealNatalLow\endcsname{\ensuremath{0.50}}
\expandafter\providecommand\csname DnsRealProduct\endcsname{}\expandafter\renewcommand\csname DnsRealProduct\endcsname{\ensuremath{0.88}}
\expandafter\providecommand\csname DnsRealProductHigh\endcsname{}\expandafter\renewcommand\csname DnsRealProductHigh\endcsname{\ensuremath{0.99}}
\expandafter\providecommand\csname DnsRealProductLow\endcsname{}\expandafter\renewcommand\csname DnsRealProductLow\endcsname{\ensuremath{0.57}}
\expandafter\providecommand\csname DnsRealRight\endcsname{}\expandafter\renewcommand\csname DnsRealRight\endcsname{\ensuremath{0.84}}
\expandafter\providecommand\csname DnsRealRightHigh\endcsname{}\expandafter\renewcommand\csname DnsRealRightHigh\endcsname{\ensuremath{0.99}}
\expandafter\providecommand\csname DnsRealRightLow\endcsname{}\expandafter\renewcommand\csname DnsRealRightLow\endcsname{\ensuremath{0.47}}
\expandafter\providecommand\csname DnsSample\endcsname{}\expandafter\renewcommand\csname DnsSample\endcsname{\ensuremath{11}}
\expandafter\providecommand\csname DnsSeedShift\endcsname{}\expandafter\renewcommand\csname DnsSeedShift\endcsname{\ensuremath{0.001}}
\expandafter\providecommand\csname DnsSingleHigh\endcsname{}\expandafter\renewcommand\csname DnsSingleHigh\endcsname{\ensuremath{0.60}}
\expandafter\providecommand\csname DnsSingleLow\endcsname{}\expandafter\renewcommand\csname DnsSingleLow\endcsname{\ensuremath{0.51}}
\expandafter\providecommand\csname DnsSingles\endcsname{}\expandafter\renewcommand\csname DnsSingles\endcsname{\ensuremath{48}}
\expandafter\providecommand\csname DnsSurvivalHigh\endcsname{}\expandafter\renewcommand\csname DnsSurvivalHigh\endcsname{\ensuremath{0.40}}
\expandafter\providecommand\csname DnsSurvivalLow\endcsname{}\expandafter\renewcommand\csname DnsSurvivalLow\endcsname{\ensuremath{0.48}}

\providecommand\result[1]{\ifcsname #1\endcsname\csname #1\endcsname\else\textbf{??#1??}\fi}

\expandafter\providecommand\csname SurveyBoundDeviance\endcsname{}\expandafter\renewcommand\csname SurveyBoundDeviance\endcsname{\ensuremath{1.09}}
\expandafter\providecommand\csname SurveyCount\endcsname{}\expandafter\renewcommand\csname SurveyCount\endcsname{\ensuremath{1000000}}
\expandafter\providecommand\csname SurveyFastBound\endcsname{}\expandafter\renewcommand\csname SurveyFastBound\endcsname{\ensuremath{0.2765}}
\expandafter\providecommand\csname SurveyFastMerger\endcsname{}\expandafter\renewcommand\csname SurveyFastMerger\endcsname{\ensuremath{0.2529}}
\expandafter\providecommand\csname SurveyFastSelectedCount\endcsname{}\expandafter\renewcommand\csname SurveyFastSelectedCount\endcsname{\ensuremath{293316}}
\expandafter\providecommand\csname SurveyFastSelectedFraction\endcsname{}\expandafter\renewcommand\csname SurveyFastSelectedFraction\endcsname{\ensuremath{0.2933}}
\expandafter\providecommand\csname SurveyFraction\endcsname{}\expandafter\renewcommand\csname SurveyFraction\endcsname{\ensuremath{0.40}}
\expandafter\providecommand\csname SurveyGradientError\endcsname{}\expandafter\renewcommand\csname SurveyGradientError\endcsname{\ensuremath{1.04\times10^{-11}}}
\expandafter\providecommand\csname SurveyMemoryDeviance\endcsname{}\expandafter\renewcommand\csname SurveyMemoryDeviance\endcsname{\ensuremath{213.7}}
\expandafter\providecommand\csname SurveyNewbornDeviance\endcsname{}\expandafter\renewcommand\csname SurveyNewbornDeviance\endcsname{\ensuremath{1.99}}
\expandafter\providecommand\csname SurveyNewbornFraction\endcsname{}\expandafter\renewcommand\csname SurveyNewbornFraction\endcsname{\ensuremath{0.318}}
\expandafter\providecommand\csname SurveyRefineDeviance\endcsname{}\expandafter\renewcommand\csname SurveyRefineDeviance\endcsname{\ensuremath{1.21\times10^{-10}}}
\expandafter\providecommand\csname SurveyRhoFifty\endcsname{}\expandafter\renewcommand\csname SurveyRhoFifty\endcsname{\ensuremath{13761}}
\expandafter\providecommand\csname SurveyRhoHundred\endcsname{}\expandafter\renewcommand\csname SurveyRhoHundred\endcsname{\ensuremath{1664}}
\expandafter\providecommand\csname SurveyRhoThirty\endcsname{}\expandafter\renewcommand\csname SurveyRhoThirty\endcsname{\ensuremath{63940}}
\expandafter\providecommand\csname SurveyShapeDeviance\endcsname{}\expandafter\renewcommand\csname SurveyShapeDeviance\endcsname{\ensuremath{3509}}
\expandafter\providecommand\csname SurveySigmaFreeMemory\endcsname{}\expandafter\renewcommand\csname SurveySigmaFreeMemory\endcsname{\ensuremath{0.0131}}
\expandafter\providecommand\csname SurveySigmaFreeScale\endcsname{}\expandafter\renewcommand\csname SurveySigmaFreeScale\endcsname{\ensuremath{0.0035}}
\expandafter\providecommand\csname SurveySigmaKnownSpin\endcsname{}\expandafter\renewcommand\csname SurveySigmaKnownSpin\endcsname{\ensuremath{0.0023}}
\expandafter\providecommand\csname SurveyWeakBound\endcsname{}\expandafter\renewcommand\csname SurveyWeakBound\endcsname{\ensuremath{1.0000}}
\expandafter\providecommand\csname SurveyWeakMerger\endcsname{}\expandafter\renewcommand\csname SurveyWeakMerger\endcsname{\ensuremath{0.4062}}
\expandafter\providecommand\csname SurveyWrongBound\endcsname{}\expandafter\renewcommand\csname SurveyWrongBound\endcsname{\ensuremath{0.597}}
\expandafter\providecommand\csname SurveyWrongMemory\endcsname{}\expandafter\renewcommand\csname SurveyWrongMemory\endcsname{\ensuremath{0.550}}

\providecommand\result[1]{\ifcsname #1\endcsname\csname #1\endcsname\else\textbf{??#1??}\fi}

\expandafter\providecommand\csname ForwardCountChange\endcsname{}\expandafter\renewcommand\csname ForwardCountChange\endcsname{\ensuremath{-5557.22}}
\expandafter\providecommand\csname ForwardLoudCount\endcsname{}\expandafter\renewcommand\csname ForwardLoudCount\endcsname{\ensuremath{1664.11}}
\expandafter\providecommand\csname ForwardNegativeAll\endcsname{}\expandafter\renewcommand\csname ForwardNegativeAll\endcsname{\ensuremath{0.3102}}
\expandafter\providecommand\csname ForwardNegativeIntrinsic\endcsname{}\expandafter\renewcommand\csname ForwardNegativeIntrinsic\endcsname{\ensuremath{0.1118}}
\expandafter\providecommand\csname ForwardNegativeLoud\endcsname{}\expandafter\renewcommand\csname ForwardNegativeLoud\endcsname{\ensuremath{0.1349}}
\expandafter\providecommand\csname ForwardNegativeLoudSlope\endcsname{}\expandafter\renewcommand\csname ForwardNegativeLoudSlope\endcsname{\ensuremath{-0.0165}}
\expandafter\providecommand\csname ForwardRatePercent\endcsname{}\expandafter\renewcommand\csname ForwardRatePercent\endcsname{\ensuremath{-0.56}}
\expandafter\providecommand\csname ForwardRateSlope\endcsname{}\expandafter\renewcommand\csname ForwardRateSlope\endcsname{\ensuremath{-0.11}}
\expandafter\providecommand\csname ForwardSigmaAll\endcsname{}\expandafter\renewcommand\csname ForwardSigmaAll\endcsname{\ensuremath{0.0154}}
\expandafter\providecommand\csname ForwardSigmaKappaAll\endcsname{}\expandafter\renewcommand\csname ForwardSigmaKappaAll\endcsname{\ensuremath{0.0538}}
\expandafter\providecommand\csname ForwardSigmaKappaLoud\endcsname{}\expandafter\renewcommand\csname ForwardSigmaKappaLoud\endcsname{\ensuremath{0.4753}}
\expandafter\providecommand\csname ForwardSigmaLoud\endcsname{}\expandafter\renewcommand\csname ForwardSigmaLoud\endcsname{\ensuremath{0.0019}}

\providecommand\result[1]{\ifcsname #1\endcsname\csname #1\endcsname\else\textbf{??#1??}\fi}

\expandafter\providecommand\csname IntersnMatchedResidual\endcsname{}\expandafter\renewcommand\csname IntersnMatchedResidual\endcsname{\ensuremath{0.050}}
\expandafter\providecommand\csname IntersnMixedResidual\endcsname{}\expandafter\renewcommand\csname IntersnMixedResidual\endcsname{\ensuremath{0.648}}
\expandafter\providecommand\csname IntersnPostOneMedian\endcsname{}\expandafter\renewcommand\csname IntersnPostOneMedian\endcsname{\ensuremath{175.86}}
\expandafter\providecommand\csname IntersnPreTwoMedian\endcsname{}\expandafter\renewcommand\csname IntersnPreTwoMedian\endcsname{\ensuremath{5.32}}
\expandafter\providecommand\csname IntersnRealizedCE\endcsname{}\expandafter\renewcommand\csname IntersnRealizedCE\endcsname{\ensuremath{0.850}}
\expandafter\providecommand\csname IntersnRealizedCount\endcsname{}\expandafter\renewcommand\csname IntersnRealizedCount\endcsname{\ensuremath{3590}}
\expandafter\providecommand\csname IntersnRealizedInteraction\endcsname{}\expandafter\renewcommand\csname IntersnRealizedInteraction\endcsname{\ensuremath{0.922}}
\expandafter\providecommand\csname IntersnRealizedProxyResidual\endcsname{}\expandafter\renewcommand\csname IntersnRealizedProxyResidual\endcsname{\ensuremath{0.961}}
\expandafter\providecommand\csname IntersnRealizedTrackResidual\endcsname{}\expandafter\renewcommand\csname IntersnRealizedTrackResidual\endcsname{\ensuremath{0.857}}
\expandafter\providecommand\csname IntersnTimeDetached\endcsname{}\expandafter\renewcommand\csname IntersnTimeDetached\endcsname{\ensuremath{0.009}}
\expandafter\providecommand\csname IntersnTimeTransfer\endcsname{}\expandafter\renewcommand\csname IntersnTimeTransfer\endcsname{\ensuremath{0.019}}

\providecommand\result[1]{\ifcsname #1\endcsname\csname #1\endcsname\else\textbf{??#1??}\fi}

\expandafter\providecommand\csname CePortAgbLamHi\endcsname{}\expandafter\renewcommand\csname CePortAgbLamHi\endcsname{\ensuremath{23}}
\expandafter\providecommand\csname CePortAgbLamLo\endcsname{}\expandafter\renewcommand\csname CePortAgbLamLo\endcsname{\ensuremath{4}}
\expandafter\providecommand\csname CePortAgbN\endcsname{}\expandafter\renewcommand\csname CePortAgbN\endcsname{\ensuremath{7}}
\expandafter\providecommand\csname CePortAgreeDev\endcsname{}\expandafter\renewcommand\csname CePortAgreeDev\endcsname{\ensuremath{\ensuremath{3\times10^{-5}}}}
\expandafter\providecommand\csname CePortAgreeN\endcsname{}\expandafter\renewcommand\csname CePortAgreeN\endcsname{\ensuremath{160}}
\expandafter\providecommand\csname CePortN\endcsname{}\expandafter\renewcommand\csname CePortN\endcsname{\ensuremath{167}}
\expandafter\providecommand\csname CePortNsCosmic\endcsname{}\expandafter\renewcommand\csname CePortNsCosmic\endcsname{\ensuremath{1.9}}
\expandafter\providecommand\csname CePortNsOurs\endcsname{}\expandafter\renewcommand\csname CePortNsOurs\endcsname{\ensuremath{1.3}}
\expandafter\providecommand\csname CePortOnsetPeri\endcsname{}\expandafter\renewcommand\csname CePortOnsetPeri\endcsname{\ensuremath{1.03}}
\expandafter\providecommand\csname CePortSepCosmic\endcsname{}\expandafter\renewcommand\csname CePortSepCosmic\endcsname{\ensuremath{1.8}}
\expandafter\providecommand\csname CePortSepOurs\endcsname{}\expandafter\renewcommand\csname CePortSepOurs\endcsname{\ensuremath{8.1}}
\expandafter\providecommand\csname CePortSepPeri\endcsname{}\expandafter\renewcommand\csname CePortSepPeri\endcsname{\ensuremath{5.0}}

\providecommand\result[1]{\ifcsname #1\endcsname\csname #1\endcsname\else\textbf{??#1??}\fi}

\expandafter\providecommand\csname CosmicNsGainHi\endcsname{}\expandafter\renewcommand\csname CosmicNsGainHi\endcsname{\ensuremath{0.10}}
\expandafter\providecommand\csname CosmicNsGainMedian\endcsname{}\expandafter\renewcommand\csname CosmicNsGainMedian\endcsname{\ensuremath{0.01}}
\expandafter\providecommand\csname CosmicNsTwoHi\endcsname{}\expandafter\renewcommand\csname CosmicNsTwoHi\endcsname{\ensuremath{2.86}}
\expandafter\providecommand\csname CosmicNsTwoLo\endcsname{}\expandafter\renewcommand\csname CosmicNsTwoLo\endcsname{\ensuremath{1.27}}
\expandafter\providecommand\csname CosmicNsTwoMedian\endcsname{}\expandafter\renewcommand\csname CosmicNsTwoMedian\endcsname{\ensuremath{1.88}}

\providecommand\result[1]{\ifcsname #1\endcsname\csname #1\endcsname\else\textbf{??#1??}\fi}

\expandafter\providecommand\csname TwoSnNsMassBirthHi\endcsname{}\expandafter\renewcommand\csname TwoSnNsMassBirthHi\endcsname{\ensuremath{1.6}}
\expandafter\providecommand\csname TwoSnNsMassDelayedHi\endcsname{}\expandafter\renewcommand\csname TwoSnNsMassDelayedHi\endcsname{\ensuremath{2.9}}
\expandafter\providecommand\csname TwoSnNsMassHi\endcsname{}\expandafter\renewcommand\csname TwoSnNsMassHi\endcsname{\ensuremath{1.6}}
\expandafter\providecommand\csname TwoSnNsMassLo\endcsname{}\expandafter\renewcommand\csname TwoSnNsMassLo\endcsname{\ensuremath{1.1}}
\expandafter\providecommand\csname TwoSnNsMassMean\endcsname{}\expandafter\renewcommand\csname TwoSnNsMassMean\endcsname{\ensuremath{1.4}}
\expandafter\providecommand\csname TwoSnNsMassMedian\endcsname{}\expandafter\renewcommand\csname TwoSnNsMassMedian\endcsname{\ensuremath{1.4}}
\expandafter\providecommand\csname TwoSnNsMassSd\endcsname{}\expandafter\renewcommand\csname TwoSnNsMassSd\endcsname{\ensuremath{0.14}}

\providecommand\result[1]{\ifcsname #1\endcsname\csname #1\endcsname\else\textbf{??#1??}\fi}

\expandafter\providecommand\csname ClassCCompEngineBreakA\endcsname{}\expandafter\renewcommand\csname ClassCCompEngineBreakA\endcsname{\ensuremath{9.3}}
\expandafter\providecommand\csname ClassCCompEngineBreakB\endcsname{}\expandafter\renewcommand\csname ClassCCompEngineBreakB\endcsname{\ensuremath{16.0}}
\expandafter\providecommand\csname ClassCCompPanels\endcsname{}\expandafter\renewcommand\csname ClassCCompPanels\endcsname{\ensuremath{3}}
\expandafter\providecommand\csname ClassWCompEngineBreakA\endcsname{}\expandafter\renewcommand\csname ClassWCompEngineBreakA\endcsname{\ensuremath{16.0}}
\expandafter\providecommand\csname ClassWCompPanels\endcsname{}\expandafter\renewcommand\csname ClassWCompPanels\endcsname{\ensuremath{4}}
\expandafter\providecommand\csname TwoSnInjDnsFraction\endcsname{}\expandafter\renewcommand\csname TwoSnInjDnsFraction\endcsname{\ensuremath{1.6\times10^{-5}}}
\expandafter\providecommand\csname TwoSnInjPriorNodes\endcsname{}\expandafter\renewcommand\csname TwoSnInjPriorNodes\endcsname{\ensuremath{4140}}

\providecommand\result[1]{\ifcsname #1\endcsname\csname #1\endcsname\else\textbf{??#1??}\fi}

\expandafter\providecommand\csname TwoSnFitErrUpper\endcsname{}\expandafter\renewcommand\csname TwoSnFitErrUpper\endcsname{\ensuremath{2.9}}
\expandafter\providecommand\csname TwoSnFitFloorPower\endcsname{}\expandafter\renewcommand\csname TwoSnFitFloorPower\endcsname{\ensuremath{11}}
\expandafter\providecommand\csname TwoSnFitMeanErrCe\endcsname{}\expandafter\renewcommand\csname TwoSnFitMeanErrCe\endcsname{\ensuremath{0.31}}
\expandafter\providecommand\csname TwoSnFitMeanErrKick\endcsname{}\expandafter\renewcommand\csname TwoSnFitMeanErrKick\endcsname{\ensuremath{0.61}}
\expandafter\providecommand\csname TwoSnFitPriorNodes\endcsname{}\expandafter\renewcommand\csname TwoSnFitPriorNodes\endcsname{\ensuremath{147{,}456}}
\expandafter\providecommand\csname TwoSnFitReplicaErrCe\endcsname{}\expandafter\renewcommand\csname TwoSnFitReplicaErrCe\endcsname{\ensuremath{0.61}}
\expandafter\providecommand\csname TwoSnFitReplicaErrKick\endcsname{}\expandafter\renewcommand\csname TwoSnFitReplicaErrKick\endcsname{\ensuremath{1.23}}
\expandafter\providecommand\csname TwoSnFitReplicas\endcsname{}\expandafter\renewcommand\csname TwoSnFitReplicas\endcsname{\ensuremath{4}}
\expandafter\providecommand\csname TwoSnFitSobolPower\endcsname{}\expandafter\renewcommand\csname TwoSnFitSobolPower\endcsname{\ensuremath{14}}
\expandafter\providecommand\csname TwoSnMatchedOffsetCe\endcsname{}\expandafter\renewcommand\csname TwoSnMatchedOffsetCe\endcsname{\ensuremath{1.6}}
\expandafter\providecommand\csname TwoSnMatchedOffsetKick\endcsname{}\expandafter\renewcommand\csname TwoSnMatchedOffsetKick\endcsname{\ensuremath{0.2}}
\expandafter\providecommand\csname TwoSnMatchedOffsetP\endcsname{}\expandafter\renewcommand\csname TwoSnMatchedOffsetP\endcsname{\ensuremath{0.25}}
\expandafter\providecommand\csname TwoSnOrderGlMaxShift\endcsname{}\expandafter\renewcommand\csname TwoSnOrderGlMaxShift\endcsname{\ensuremath{0.7}}
\expandafter\providecommand\csname TwoSnOrderInjEffFitNodes\endcsname{}\expandafter\renewcommand\csname TwoSnOrderInjEffFitNodes\endcsname{\ensuremath{1.3\times10^{-5}}}
\expandafter\providecommand\csname TwoSnOrderInjShiftCe\endcsname{}\expandafter\renewcommand\csname TwoSnOrderInjShiftCe\endcsname{\ensuremath{0.69}}
\expandafter\providecommand\csname TwoSnOrderInjShiftKick\endcsname{}\expandafter\renewcommand\csname TwoSnOrderInjShiftKick\endcsname{\ensuremath{0.06}}
\expandafter\providecommand\csname TwoSnOrderScatterEffPercent\endcsname{}\expandafter\renewcommand\csname TwoSnOrderScatterEffPercent\endcsname{\ensuremath{24}}
\expandafter\providecommand\csname TwoSnOrderSigmaCe\endcsname{}\expandafter\renewcommand\csname TwoSnOrderSigmaCe\endcsname{\ensuremath{0.046}}
\expandafter\providecommand\csname TwoSnOrderSigmaKick\endcsname{}\expandafter\renewcommand\csname TwoSnOrderSigmaKick\endcsname{\ensuremath{0.032}}

\providecommand\result[1]{\ifcsname #1\endcsname\csname #1\endcsname\else\textbf{??#1??}\fi}

\expandafter\providecommand\csname TwoSnLocalCorr\endcsname{}\expandafter\renewcommand\csname TwoSnLocalCorr\endcsname{\ensuremath{0.20}}
\expandafter\providecommand\csname TwoSnLocalEff\endcsname{}\expandafter\renewcommand\csname TwoSnLocalEff\endcsname{\ensuremath{1.21\times10^{-5}}}
\expandafter\providecommand\csname TwoSnLocalFitCe\endcsname{}\expandafter\renewcommand\csname TwoSnLocalFitCe\endcsname{\ensuremath{1.03}}
\expandafter\providecommand\csname TwoSnLocalFitKickScale\endcsname{}\expandafter\renewcommand\csname TwoSnLocalFitKickScale\endcsname{\ensuremath{0.997}}
\expandafter\providecommand\csname TwoSnLocalMissingScoreCe\endcsname{}\expandafter\renewcommand\csname TwoSnLocalMissingScoreCe\endcsname{\ensuremath{0}}
\expandafter\providecommand\csname TwoSnLocalMissingScoreKick\endcsname{}\expandafter\renewcommand\csname TwoSnLocalMissingScoreKick\endcsname{\ensuremath{-609}}
\expandafter\providecommand\csname TwoSnLocalMissingStepCe\endcsname{}\expandafter\renewcommand\csname TwoSnLocalMissingStepCe\endcsname{\ensuremath{4}}
\expandafter\providecommand\csname TwoSnLocalMissingStepKick\endcsname{}\expandafter\renewcommand\csname TwoSnLocalMissingStepKick\endcsname{\ensuremath{19}}
\expandafter\providecommand\csname TwoSnLocalOffsetCe\endcsname{}\expandafter\renewcommand\csname TwoSnLocalOffsetCe\endcsname{\ensuremath{0.6}}
\expandafter\providecommand\csname TwoSnLocalOffsetKick\endcsname{}\expandafter\renewcommand\csname TwoSnLocalOffsetKick\endcsname{\ensuremath{0.1}}
\expandafter\providecommand\csname TwoSnLocalOffsetP\endcsname{}\expandafter\renewcommand\csname TwoSnLocalOffsetP\endcsname{\ensuremath{0.80}}
\expandafter\providecommand\csname TwoSnLocalSurvival\endcsname{}\expandafter\renewcommand\csname TwoSnLocalSurvival\endcsname{\ensuremath{0.249}}
\expandafter\providecommand\csname TwoSnLocalWidthCe\endcsname{}\expandafter\renewcommand\csname TwoSnLocalWidthCe\endcsname{\ensuremath{0.046}}
\expandafter\providecommand\csname TwoSnLocalWidthCePercent\endcsname{}\expandafter\renewcommand\csname TwoSnLocalWidthCePercent\endcsname{\ensuremath{4.6}}
\expandafter\providecommand\csname TwoSnLocalWidthKickPercent\endcsname{}\expandafter\renewcommand\csname TwoSnLocalWidthKickPercent\endcsname{\ensuremath{3.2}}

\providecommand\result[1]{\ifcsname #1\endcsname\csname #1\endcsname\else\textbf{??#1??}\fi}

\expandafter\providecommand\csname PsrCaseBbAccMyr\endcsname{}\expandafter\renewcommand\csname PsrCaseBbAccMyr\endcsname{\ensuremath{0.1}}
\expandafter\providecommand\csname PsrCaseBbEddFac\endcsname{}\expandafter\renewcommand\csname PsrCaseBbEddFac\endcsname{\ensuremath{1}}
\expandafter\providecommand\csname PsrFirstDeadPercent\endcsname{}\expandafter\renewcommand\csname PsrFirstDeadPercent\endcsname{\ensuremath{2}}
\expandafter\providecommand\csname PsrFirstGyrPercent\endcsname{}\expandafter\renewcommand\csname PsrFirstGyrPercent\endcsname{\ensuremath{98}}
\expandafter\providecommand\csname PsrFirstRecycledPercent\endcsname{}\expandafter\renewcommand\csname PsrFirstRecycledPercent\endcsname{\ensuremath{100}}
\expandafter\providecommand\csname PsrGalaxyAgeGyr\endcsname{}\expandafter\renewcommand\csname PsrGalaxyAgeGyr\endcsname{\ensuremath{10}}
\expandafter\providecommand\csname PsrYoungLifeMean\endcsname{}\expandafter\renewcommand\csname PsrYoungLifeMean\endcsname{\ensuremath{74}}
\expandafter\providecommand\csname PsrYoungLifeMedian\endcsname{}\expandafter\renewcommand\csname PsrYoungLifeMedian\endcsname{\ensuremath{38}}

\providecommand\result[1]{\ifcsname #1\endcsname\csname #1\endcsname\else\textbf{??#1??}\fi}

\expandafter\providecommand\csname TwoSnAbsoluteRefinement\endcsname{}\expandafter\renewcommand\csname TwoSnAbsoluteRefinement\endcsname{\ensuremath{1.028}}
\expandafter\providecommand\csname TwoSnBestCe\endcsname{}\expandafter\renewcommand\csname TwoSnBestCe\endcsname{\ensuremath{0.994}}
\expandafter\providecommand\csname TwoSnBestKick\endcsname{}\expandafter\renewcommand\csname TwoSnBestKick\endcsname{\ensuremath{427}}
\expandafter\providecommand\csname TwoSnCompSd\endcsname{}\expandafter\renewcommand\csname TwoSnCompSd\endcsname{\ensuremath{0.42}}
\expandafter\providecommand\csname TwoSnCoreSd\endcsname{}\expandafter\renewcommand\csname TwoSnCoreSd\endcsname{\ensuremath{0.34}}
\expandafter\providecommand\csname TwoSnCount\endcsname{}\expandafter\renewcommand\csname TwoSnCount\endcsname{\ensuremath{64}}
\expandafter\providecommand\csname TwoSnEfficiency\endcsname{}\expandafter\renewcommand\csname TwoSnEfficiency\endcsname{\ensuremath{0.749}}
\expandafter\providecommand\csname TwoSnFineFitGain\endcsname{}\expandafter\renewcommand\csname TwoSnFineFitGain\endcsname{\ensuremath{0.020}}
\expandafter\providecommand\csname TwoSnFirstSurvival\endcsname{}\expandafter\renewcommand\csname TwoSnFirstSurvival\endcsname{\ensuremath{0.779}}
\expandafter\providecommand\csname TwoSnFitGain\endcsname{}\expandafter\renewcommand\csname TwoSnFitGain\endcsname{\ensuremath{0.035}}
\expandafter\providecommand\csname TwoSnGradientError\endcsname{}\expandafter\renewcommand\csname TwoSnGradientError\endcsname{\ensuremath{3.5e-06}}
\expandafter\providecommand\csname TwoSnProfileRefinement\endcsname{}\expandafter\renewcommand\csname TwoSnProfileRefinement\endcsname{\ensuremath{0.328}}
\expandafter\providecommand\csname TwoSnRelativeRefinement\endcsname{}\expandafter\renewcommand\csname TwoSnRelativeRefinement\endcsname{\ensuremath{0.015}}
\expandafter\providecommand\csname TwoSnRepeatBestCe\endcsname{}\expandafter\renewcommand\csname TwoSnRepeatBestCe\endcsname{\ensuremath{1.116}}
\expandafter\providecommand\csname TwoSnRepeatBestKick\endcsname{}\expandafter\renewcommand\csname TwoSnRepeatBestKick\endcsname{\ensuremath{507}}
\expandafter\providecommand\csname TwoSnRepeatContrastShift\endcsname{}\expandafter\renewcommand\csname TwoSnRepeatContrastShift\endcsname{\ensuremath{0.001}}
\expandafter\providecommand\csname TwoSnRepeatFineGain\endcsname{}\expandafter\renewcommand\csname TwoSnRepeatFineGain\endcsname{\ensuremath{2.053}}
\expandafter\providecommand\csname TwoSnRepeatGain\endcsname{}\expandafter\renewcommand\csname TwoSnRepeatGain\endcsname{\ensuremath{2.052}}
\expandafter\providecommand\csname TwoSnSepSd\endcsname{}\expandafter\renewcommand\csname TwoSnSepSd\endcsname{\ensuremath{50}}

\providecommand\result[1]{\ifcsname #1\endcsname\csname #1\endcsname\else\textbf{??#1??}\fi}

\expandafter\providecommand\csname ContourCheckShift\endcsname{}\expandafter\renewcommand\csname ContourCheckShift\endcsname{\ensuremath{0.077}}
\expandafter\providecommand\csname ContourInformationResidual\endcsname{}\expandafter\renewcommand\csname ContourInformationResidual\endcsname{\ensuremath{1.107}}
\expandafter\providecommand\csname ContourInjectionDrop\endcsname{}\expandafter\renewcommand\csname ContourInjectionDrop\endcsname{\ensuremath{0.779}}
\expandafter\providecommand\csname ContourInnerCeHi\endcsname{}\expandafter\renewcommand\csname ContourInnerCeHi\endcsname{\ensuremath{1.014}}
\expandafter\providecommand\csname ContourInnerCeLo\endcsname{}\expandafter\renewcommand\csname ContourInnerCeLo\endcsname{\ensuremath{0.959}}
\expandafter\providecommand\csname ContourInnerKickHi\endcsname{}\expandafter\renewcommand\csname ContourInnerKickHi\endcsname{\ensuremath{492}}
\expandafter\providecommand\csname ContourInnerKickLo\endcsname{}\expandafter\renewcommand\csname ContourInnerKickLo\endcsname{\ensuremath{346}}
\expandafter\providecommand\csname ContourInterpolation\endcsname{}\expandafter\renewcommand\csname ContourInterpolation\endcsname{\ensuremath{0.048}}
\expandafter\providecommand\csname ContourOuterCeHi\endcsname{}\expandafter\renewcommand\csname ContourOuterCeHi\endcsname{\ensuremath{1.032}}
\expandafter\providecommand\csname ContourOuterCeLo\endcsname{}\expandafter\renewcommand\csname ContourOuterCeLo\endcsname{\ensuremath{0.944}}
\expandafter\providecommand\csname ContourOuterKickHi\endcsname{}\expandafter\renewcommand\csname ContourOuterKickHi\endcsname{\ensuremath{534}}
\expandafter\providecommand\csname ContourOuterKickLo\endcsname{}\expandafter\renewcommand\csname ContourOuterKickLo\endcsname{\ensuremath{298}}
\expandafter\providecommand\csname ContourScoreCeWidth\endcsname{}\expandafter\renewcommand\csname ContourScoreCeWidth\endcsname{\ensuremath{0.0187}}
\expandafter\providecommand\csname ContourScoreKickWidth\endcsname{}\expandafter\renewcommand\csname ContourScoreKickWidth\endcsname{\ensuremath{56.2}}
\expandafter\providecommand\csname ContourScoreWidthChange\endcsname{}\expandafter\renewcommand\csname ContourScoreWidthChange\endcsname{\ensuremath{1.3}}

\providecommand\result[1]{\ifcsname #1\endcsname\csname #1\endcsname\else\textbf{??#1??}\fi}

\expandafter\providecommand\csname SelectionContrastShift\endcsname{}\expandafter\renewcommand\csname SelectionContrastShift\endcsname{\ensuremath{0.479}}
\expandafter\providecommand\csname SelectionFineRightPenalty\endcsname{}\expandafter\renewcommand\csname SelectionFineRightPenalty\endcsname{\ensuremath{11.3}}
\expandafter\providecommand\csname SelectionFineWrongGain\endcsname{}\expandafter\renewcommand\csname SelectionFineWrongGain\endcsname{\ensuremath{148.8}}
\expandafter\providecommand\csname SelectionRightCe\endcsname{}\expandafter\renewcommand\csname SelectionRightCe\endcsname{\ensuremath{0.9849}}
\expandafter\providecommand\csname SelectionRightKick\endcsname{}\expandafter\renewcommand\csname SelectionRightKick\endcsname{\ensuremath{416.1}}
\expandafter\providecommand\csname SelectionRightPenalty\endcsname{}\expandafter\renewcommand\csname SelectionRightPenalty\endcsname{\ensuremath{11.1}}
\expandafter\providecommand\csname SelectionWrongCe\endcsname{}\expandafter\renewcommand\csname SelectionWrongCe\endcsname{\ensuremath{0.957}}
\expandafter\providecommand\csname SelectionWrongGain\endcsname{}\expandafter\renewcommand\csname SelectionWrongGain\endcsname{\ensuremath{149.2}}
\expandafter\providecommand\csname SelectionWrongKick\endcsname{}\expandafter\renewcommand\csname SelectionWrongKick\endcsname{\ensuremath{120}}

\providecommand\result[1]{\ifcsname #1\endcsname\csname #1\endcsname\else\textbf{??#1??}\fi}

\expandafter\providecommand\csname PulsarMockDnsN\endcsname{}\expandafter\renewcommand\csname PulsarMockDnsN\endcsname{\ensuremath{22}}
\expandafter\providecommand\csname PulsarMockDrpN\endcsname{}\expandafter\renewcommand\csname PulsarMockDrpN\endcsname{\ensuremath{22}}
\expandafter\providecommand\csname PulsarMockHi\endcsname{}\expandafter\renewcommand\csname PulsarMockHi\endcsname{\ensuremath{0.50}}
\expandafter\providecommand\csname PulsarMockLo\endcsname{}\expandafter\renewcommand\csname PulsarMockLo\endcsname{\ensuremath{0.27}}
\expandafter\providecommand\csname PulsarMockMCShift\endcsname{}\expandafter\renewcommand\csname PulsarMockMCShift\endcsname{\ensuremath{0.6}}
\expandafter\providecommand\csname PulsarMockMedian\endcsname{}\expandafter\renewcommand\csname PulsarMockMedian\endcsname{\ensuremath{0.38}}
\expandafter\providecommand\csname PulsarMockNuisanceHi\endcsname{}\expandafter\renewcommand\csname PulsarMockNuisanceHi\endcsname{\ensuremath{0.68}}
\expandafter\providecommand\csname PulsarMockNuisanceLo\endcsname{}\expandafter\renewcommand\csname PulsarMockNuisanceLo\endcsname{\ensuremath{0.15}}
\expandafter\providecommand\csname PulsarMockTruth\endcsname{}\expandafter\renewcommand\csname PulsarMockTruth\endcsname{\ensuremath{0.41}}
\expandafter\providecommand\csname PulsarMockWrong\endcsname{}\expandafter\renewcommand\csname PulsarMockWrong\endcsname{\ensuremath{0.51}}
\expandafter\providecommand\csname PulsarMockYoungN\endcsname{}\expandafter\renewcommand\csname PulsarMockYoungN\endcsname{\ensuremath{711}}
\expandafter\providecommand\csname PulsarMockYoungPM\endcsname{}\expandafter\renewcommand\csname PulsarMockYoungPM\endcsname{\ensuremath{46}}

\providecommand\result[1]{\ifcsname #1\endcsname\csname #1\endcsname\else\textbf{??#1??}\fi}

\expandafter\providecommand\csname JointAdError\endcsname{}\expandafter\renewcommand\csname JointAdError\endcsname{\ensuremath{5.8\times10^{-8}}}
\expandafter\providecommand\csname JointContrastShift\endcsname{}\expandafter\renewcommand\csname JointContrastShift\endcsname{\ensuremath{0.55}}
\expandafter\providecommand\csname JointDnsN\endcsname{}\expandafter\renewcommand\csname JointDnsN\endcsname{\ensuremath{20}}
\expandafter\providecommand\csname JointDrpN\endcsname{}\expandafter\renewcommand\csname JointDrpN\endcsname{\ensuremath{22}}
\expandafter\providecommand\csname JointInfoOne\endcsname{}\expandafter\renewcommand\csname JointInfoOne\endcsname{\ensuremath{0.083}}
\expandafter\providecommand\csname JointInfoTwo\endcsname{}\expandafter\renewcommand\csname JointInfoTwo\endcsname{\ensuremath{60.2}}
\expandafter\providecommand\csname JointKickOne\endcsname{}\expandafter\renewcommand\csname JointKickOne\endcsname{\ensuremath{344}}
\expandafter\providecommand\csname JointKickTwo\endcsname{}\expandafter\renewcommand\csname JointKickTwo\endcsname{\ensuremath{157}}
\expandafter\providecommand\csname JointKickTwoHi\endcsname{}\expandafter\renewcommand\csname JointKickTwoHi\endcsname{\ensuremath{196}}
\expandafter\providecommand\csname JointKickTwoLo\endcsname{}\expandafter\renewcommand\csname JointKickTwoLo\endcsname{\ensuremath{130}}
\expandafter\providecommand\csname JointProfileCheckShift\endcsname{}\expandafter\renewcommand\csname JointProfileCheckShift\endcsname{\ensuremath{0.43}}
\expandafter\providecommand\csname JointScoreSeconds\endcsname{}\expandafter\renewcommand\csname JointScoreSeconds\endcsname{\ensuremath{12.1}}
\expandafter\providecommand\csname JointStarDrop\endcsname{}\expandafter\renewcommand\csname JointStarDrop\endcsname{\ensuremath{0.23}}
\expandafter\providecommand\csname JointYoungN\endcsname{}\expandafter\renewcommand\csname JointYoungN\endcsname{\ensuremath{726}}

\providecommand\result[1]{\ifcsname #1\endcsname\csname #1\endcsname\else\textbf{??#1??}\fi}

\expandafter\providecommand\csname SharedTrailCombinedDisruptionForecast\endcsname{}\expandafter\renewcommand\csname SharedTrailCombinedDisruptionForecast\endcsname{\ensuremath{2.98}}
\expandafter\providecommand\csname SharedTrailCombinedKickForecast\endcsname{}\expandafter\renewcommand\csname SharedTrailCombinedKickForecast\endcsname{\ensuremath{7.1}}
\expandafter\providecommand\csname SharedTrailCombinedPeriodForecast\endcsname{}\expandafter\renewcommand\csname SharedTrailCombinedPeriodForecast\endcsname{\ensuremath{34.9}}
\expandafter\providecommand\csname SharedTrailCountsDisruptionForecast\endcsname{}\expandafter\renewcommand\csname SharedTrailCountsDisruptionForecast\endcsname{\ensuremath{3.36}}
\expandafter\providecommand\csname SharedTrailDnsN\endcsname{}\expandafter\renewcommand\csname SharedTrailDnsN\endcsname{\ensuremath{176}}
\expandafter\providecommand\csname SharedTrailDrpN\endcsname{}\expandafter\renewcommand\csname SharedTrailDrpN\endcsname{\ensuremath{200}}
\expandafter\providecommand\csname SharedTrailFixedKickForecast\endcsname{}\expandafter\renewcommand\csname SharedTrailFixedKickForecast\endcsname{\ensuremath{4.7}}
\expandafter\providecommand\csname SharedTrailFixedPeriodForecast\endcsname{}\expandafter\renewcommand\csname SharedTrailFixedPeriodForecast\endcsname{\ensuremath{19.7}}
\expandafter\providecommand\csname SharedTrailFreeKickForecast\endcsname{}\expandafter\renewcommand\csname SharedTrailFreeKickForecast\endcsname{\ensuremath{12.3}}
\expandafter\providecommand\csname SharedTrailFreePeriodForecast\endcsname{}\expandafter\renewcommand\csname SharedTrailFreePeriodForecast\endcsname{\ensuremath{76.4}}
\expandafter\providecommand\csname SharedTrailKickTruth\endcsname{}\expandafter\renewcommand\csname SharedTrailKickTruth\endcsname{\ensuremath{170}}
\expandafter\providecommand\csname SharedTrailPeriodTruth\endcsname{}\expandafter\renewcommand\csname SharedTrailPeriodTruth\endcsname{\ensuremath{54.8}}
\expandafter\providecommand\csname SharedTrailPropertiesDisruptionForecast\endcsname{}\expandafter\renewcommand\csname SharedTrailPropertiesDisruptionForecast\endcsname{\ensuremath{9.55}}
\expandafter\providecommand\csname SharedTrailYoungN\endcsname{}\expandafter\renewcommand\csname SharedTrailYoungN\endcsname{\ensuremath{286}}

\providecommand\result[1]{\ifcsname #1\endcsname\csname #1\endcsname\else\textbf{??#1??}\fi}

\expandafter\providecommand\csname KklGammaHulseTaylor\endcsname{}\expandafter\renewcommand\csname KklGammaHulseTaylor\endcsname{\ensuremath{0.263}}
\expandafter\providecommand\csname KklGammaSecondSystem\endcsname{}\expandafter\renewcommand\csname KklGammaSecondSystem\endcsname{\ensuremath{0.247}}
\expandafter\providecommand\csname KklLossEffectHulseTaylor\endcsname{}\expandafter\renewcommand\csname KklLossEffectHulseTaylor\endcsname{\ensuremath{2.61}}
\expandafter\providecommand\csname KklLossEffectSecondSystem\endcsname{}\expandafter\renewcommand\csname KklLossEffectSecondSystem\endcsname{\ensuremath{1.24}}

\newcommand{\NativeDNSRateMedian}{25.33}
\newcommand{\NativeDNSRateLowerNinety}{11.63}
\newcommand{\NativeDNSRateUpperNinety}{52.83}

\providecommand\result[1]{\ifcsname #1\endcsname\csname #1\endcsname\else\textbf{??#1??}\fi}

\expandafter\providecommand\csname ResolvedCountsKickHi\endcsname{}\expandafter\renewcommand\csname ResolvedCountsKickHi\endcsname{\ensuremath{198}}
\expandafter\providecommand\csname ResolvedCountsKickLo\endcsname{}\expandafter\renewcommand\csname ResolvedCountsKickLo\endcsname{\ensuremath{159}}
\expandafter\providecommand\csname ResolvedDirectMle\endcsname{}\expandafter\renewcommand\csname ResolvedDirectMle\endcsname{\ensuremath{0.049}}
\expandafter\providecommand\csname ResolvedDirectTail\endcsname{}\expandafter\renewcommand\csname ResolvedDirectTail\endcsname{\ensuremath{0.025}}
\expandafter\providecommand\csname ResolvedInjPeriod\endcsname{}\expandafter\renewcommand\csname ResolvedInjPeriod\endcsname{\ensuremath{54.8}}
\expandafter\providecommand\csname ResolvedKick\endcsname{}\expandafter\renewcommand\csname ResolvedKick\endcsname{\ensuremath{164}}
\expandafter\providecommand\csname ResolvedKickHi\endcsname{}\expandafter\renewcommand\csname ResolvedKickHi\endcsname{\ensuremath{175}}
\expandafter\providecommand\csname ResolvedKickLo\endcsname{}\expandafter\renewcommand\csname ResolvedKickLo\endcsname{\ensuremath{154}}
\expandafter\providecommand\csname ResolvedLooMax\endcsname{}\expandafter\renewcommand\csname ResolvedLooMax\endcsname{\ensuremath{0.004}}
\expandafter\providecommand\csname ResolvedPeriod\endcsname{}\expandafter\renewcommand\csname ResolvedPeriod\endcsname{\ensuremath{80}}
\expandafter\providecommand\csname ResolvedPeriodHi\endcsname{}\expandafter\renewcommand\csname ResolvedPeriodHi\endcsname{\ensuremath{101}}
\expandafter\providecommand\csname ResolvedPeriodLo\endcsname{}\expandafter\renewcommand\csname ResolvedPeriodLo\endcsname{\ensuremath{63}}
\expandafter\providecommand\csname ResolvedPropsKickHi\endcsname{}\expandafter\renewcommand\csname ResolvedPropsKickHi\endcsname{\ensuremath{189}}
\expandafter\providecommand\csname ResolvedPropsKickLo\endcsname{}\expandafter\renewcommand\csname ResolvedPropsKickLo\endcsname{\ensuremath{154}}
\expandafter\providecommand\csname ResolvedSupportTop\endcsname{}\expandafter\renewcommand\csname ResolvedSupportTop\endcsname{\ensuremath{131}}
\expandafter\providecommand\csname ResolvedTailDrop\endcsname{}\expandafter\renewcommand\csname ResolvedTailDrop\endcsname{\ensuremath{2.27}}

\providecommand\result[1]{\ifcsname #1\endcsname\csname #1\endcsname\else\textbf{??#1??}\fi}

\expandafter\providecommand\csname AngularCountsKickHi\endcsname{}\expandafter\renewcommand\csname AngularCountsKickHi\endcsname{\ensuremath{198}}
\expandafter\providecommand\csname AngularCountsKickLo\endcsname{}\expandafter\renewcommand\csname AngularCountsKickLo\endcsname{\ensuremath{158}}
\expandafter\providecommand\csname AngularInjPeriod\endcsname{}\expandafter\renewcommand\csname AngularInjPeriod\endcsname{\ensuremath{54.8}}
\expandafter\providecommand\csname AngularInterpMax\endcsname{}\expandafter\renewcommand\csname AngularInterpMax\endcsname{\ensuremath{0.028}}
\expandafter\providecommand\csname AngularKick\endcsname{}\expandafter\renewcommand\csname AngularKick\endcsname{\ensuremath{165}}
\expandafter\providecommand\csname AngularKickHi\endcsname{}\expandafter\renewcommand\csname AngularKickHi\endcsname{\ensuremath{176}}
\expandafter\providecommand\csname AngularKickLo\endcsname{}\expandafter\renewcommand\csname AngularKickLo\endcsname{\ensuremath{155}}
\expandafter\providecommand\csname AngularLooMax\endcsname{}\expandafter\renewcommand\csname AngularLooMax\endcsname{\ensuremath{0.003}}
\expandafter\providecommand\csname AngularPeriod\endcsname{}\expandafter\renewcommand\csname AngularPeriod\endcsname{\ensuremath{74}}
\expandafter\providecommand\csname AngularPeriodHi\endcsname{}\expandafter\renewcommand\csname AngularPeriodHi\endcsname{\ensuremath{97}}
\expandafter\providecommand\csname AngularPeriodLo\endcsname{}\expandafter\renewcommand\csname AngularPeriodLo\endcsname{\ensuremath{55}}
\expandafter\providecommand\csname AngularPropsKickHi\endcsname{}\expandafter\renewcommand\csname AngularPropsKickHi\endcsname{\ensuremath{187}}
\expandafter\providecommand\csname AngularPropsKickLo\endcsname{}\expandafter\renewcommand\csname AngularPropsKickLo\endcsname{\ensuremath{153}}
\expandafter\providecommand\csname AngularSupportTop\endcsname{}\expandafter\renewcommand\csname AngularSupportTop\endcsname{\ensuremath{131}}
\expandafter\providecommand\csname AngularTailDrop\endcsname{}\expandafter\renewcommand\csname AngularTailDrop\endcsname{\ensuremath{2.23}}

\providecommand\result[1]{\ifcsname #1\endcsname\csname #1\endcsname\else\textbf{??#1??}\fi}

\expandafter\providecommand\csname PbBeamHalfOne\endcsname{}\expandafter\renewcommand\csname PbBeamHalfOne\endcsname{\ensuremath{0.18}}
\expandafter\providecommand\csname PbBeamHalfTenth\endcsname{}\expandafter\renewcommand\csname PbBeamHalfTenth\endcsname{\ensuremath{0.51}}
\expandafter\providecommand\csname PbBeamHalfThree\endcsname{}\expandafter\renewcommand\csname PbBeamHalfThree\endcsname{\ensuremath{0.11}}
\expandafter\providecommand\csname PbBeamIsoOverPlasmaOne\endcsname{}\expandafter\renewcommand\csname PbBeamIsoOverPlasmaOne\endcsname{\ensuremath{1.6}}
\expandafter\providecommand\csname PbBeamIsoOverPlasmaTenth\endcsname{}\expandafter\renewcommand\csname PbBeamIsoOverPlasmaTenth\endcsname{\ensuremath{1.1}}
\expandafter\providecommand\csname PbBeamIsoOverPlasmaThree\endcsname{}\expandafter\renewcommand\csname PbBeamIsoOverPlasmaThree\endcsname{\ensuremath{2.3}}
\expandafter\providecommand\csname PbBeamOne\endcsname{}\expandafter\renewcommand\csname PbBeamOne\endcsname{\ensuremath{0.18}}
\expandafter\providecommand\csname PbBeamPlasmaCensusOne\endcsname{}\expandafter\renewcommand\csname PbBeamPlasmaCensusOne\endcsname{\ensuremath{0.11}}
\expandafter\providecommand\csname PbBeamPlasmaCensusTenth\endcsname{}\expandafter\renewcommand\csname PbBeamPlasmaCensusTenth\endcsname{\ensuremath{0.33}}
\expandafter\providecommand\csname PbBeamPlasmaCensusThree\endcsname{}\expandafter\renewcommand\csname PbBeamPlasmaCensusThree\endcsname{\ensuremath{0.05}}
\expandafter\providecommand\csname PbBeamPlasmaOne\endcsname{}\expandafter\renewcommand\csname PbBeamPlasmaOne\endcsname{\ensuremath{0.11}}
\expandafter\providecommand\csname PbBeamPlasmaTenth\endcsname{}\expandafter\renewcommand\csname PbBeamPlasmaTenth\endcsname{\ensuremath{0.32}}
\expandafter\providecommand\csname PbBeamPlasmaThree\endcsname{}\expandafter\renewcommand\csname PbBeamPlasmaThree\endcsname{\ensuremath{0.06}}
\expandafter\providecommand\csname PbBeamRecycledSixty\endcsname{}\expandafter\renewcommand\csname PbBeamRecycledSixty\endcsname{\ensuremath{0.36}}
\expandafter\providecommand\csname PbBeamRecycledTwenty\endcsname{}\expandafter\renewcommand\csname PbBeamRecycledTwenty\endcsname{\ensuremath{0.47}}
\expandafter\providecommand\csname PbBeamScatterPct\endcsname{}\expandafter\renewcommand\csname PbBeamScatterPct\endcsname{\ensuremath{5}}
\expandafter\providecommand\csname PbBeamTMOne\endcsname{}\expandafter\renewcommand\csname PbBeamTMOne\endcsname{\ensuremath{0.12}}
\expandafter\providecommand\csname PbBeamTMTenth\endcsname{}\expandafter\renewcommand\csname PbBeamTMTenth\endcsname{\ensuremath{0.39}}
\expandafter\providecommand\csname PbBeamTMThree\endcsname{}\expandafter\renewcommand\csname PbBeamTMThree\endcsname{\ensuremath{0.05}}
\expandafter\providecommand\csname PbBeamTenth\endcsname{}\expandafter\renewcommand\csname PbBeamTenth\endcsname{\ensuremath{0.36}}
\expandafter\providecommand\csname PbBeamThree\endcsname{}\expandafter\renewcommand\csname PbBeamThree\endcsname{\ensuremath{0.13}}
\expandafter\providecommand\csname PbCorrRecycledHalfSixty\endcsname{}\expandafter\renewcommand\csname PbCorrRecycledHalfSixty\endcsname{\ensuremath{3.4}}
\expandafter\providecommand\csname PbCorrRecycledHalfTwenty\endcsname{}\expandafter\renewcommand\csname PbCorrRecycledHalfTwenty\endcsname{\ensuremath{2.2}}
\expandafter\providecommand\csname PbCorrRecycledSixty\endcsname{}\expandafter\renewcommand\csname PbCorrRecycledSixty\endcsname{\ensuremath{2.8}}
\expandafter\providecommand\csname PbCorrRecycledTwenty\endcsname{}\expandafter\renewcommand\csname PbCorrRecycledTwenty\endcsname{\ensuremath{2.1}}
\expandafter\providecommand\csname PbDlnLAlign\endcsname{}\expandafter\renewcommand\csname PbDlnLAlign\endcsname{\ensuremath{7.5}}
\expandafter\providecommand\csname PbDlnLBothQ\endcsname{}\expandafter\renewcommand\csname PbDlnLBothQ\endcsname{\ensuremath{4.0}}
\expandafter\providecommand\csname PbDlnLBothQCensus\endcsname{}\expandafter\renewcommand\csname PbDlnLBothQCensus\endcsname{\ensuremath{2.1}}
\expandafter\providecommand\csname PbDlnLFASTPlasma\endcsname{}\expandafter\renewcommand\csname PbDlnLFASTPlasma\endcsname{\ensuremath{2.5}}
\expandafter\providecommand\csname PbDlnLHalfAll\endcsname{}\expandafter\renewcommand\csname PbDlnLHalfAll\endcsname{\ensuremath{36}}
\expandafter\providecommand\csname PbDlnLHalfExp\endcsname{}\expandafter\renewcommand\csname PbDlnLHalfExp\endcsname{\ensuremath{27}}
\expandafter\providecommand\csname PbDlnLHalfIso\endcsname{}\expandafter\renewcommand\csname PbDlnLHalfIso\endcsname{\ensuremath{27}}
\expandafter\providecommand\csname PbDlnLHalfPlasma\endcsname{}\expandafter\renewcommand\csname PbDlnLHalfPlasma\endcsname{\ensuremath{19}}
\expandafter\providecommand\csname PbDlnLHalfPlasmaQ\endcsname{}\expandafter\renewcommand\csname PbDlnLHalfPlasmaQ\endcsname{\ensuremath{19}}
\expandafter\providecommand\csname PbDlnLMeerKATPlasma\endcsname{}\expandafter\renewcommand\csname PbDlnLMeerKATPlasma\endcsname{\ensuremath{29}}
\expandafter\providecommand\csname PbDlnLMeerKATPlasmaExp\endcsname{}\expandafter\renewcommand\csname PbDlnLMeerKATPlasmaExp\endcsname{\ensuremath{2.2}}
\expandafter\providecommand\csname PbDlnLPlasma\endcsname{}\expandafter\renewcommand\csname PbDlnLPlasma\endcsname{\ensuremath{27}}
\expandafter\providecommand\csname PbDlnLPlasmaCensus\endcsname{}\expandafter\renewcommand\csname PbDlnLPlasmaCensus\endcsname{\ensuremath{25}}
\expandafter\providecommand\csname PbDlnLPlasmaExp\endcsname{}\expandafter\renewcommand\csname PbDlnLPlasmaExp\endcsname{\ensuremath{19}}
\expandafter\providecommand\csname PbDlnLPlasmaExpCensus\endcsname{}\expandafter\renewcommand\csname PbDlnLPlasmaExpCensus\endcsname{\ensuremath{17}}
\expandafter\providecommand\csname PbKSBothExp\endcsname{}\expandafter\renewcommand\csname PbKSBothExp\endcsname{\ensuremath{1.77}}
\expandafter\providecommand\csname PbKSBothIso\endcsname{}\expandafter\renewcommand\csname PbKSBothIso\endcsname{\ensuremath{2.03}}
\expandafter\providecommand\csname PbKSBothPlasma\endcsname{}\expandafter\renewcommand\csname PbKSBothPlasma\endcsname{\ensuremath{1.26}}
\expandafter\providecommand\csname PbKSBothPlasmaCensus\endcsname{}\expandafter\renewcommand\csname PbKSBothPlasmaCensus\endcsname{\ensuremath{1.29}}
\expandafter\providecommand\csname PbKSBothQ\endcsname{}\expandafter\renewcommand\csname PbKSBothQ\endcsname{\ensuremath{1.21}}
\expandafter\providecommand\csname PbKSFASTExp\endcsname{}\expandafter\renewcommand\csname PbKSFASTExp\endcsname{\ensuremath{1.13}}
\expandafter\providecommand\csname PbKSFASTIso\endcsname{}\expandafter\renewcommand\csname PbKSFASTIso\endcsname{\ensuremath{1.33}}
\expandafter\providecommand\csname PbKSFASTPlasma\endcsname{}\expandafter\renewcommand\csname PbKSFASTPlasma\endcsname{\ensuremath{0.97}}
\expandafter\providecommand\csname PbKSMeerKATExp\endcsname{}\expandafter\renewcommand\csname PbKSMeerKATExp\endcsname{\ensuremath{1.08}}
\expandafter\providecommand\csname PbKSMeerKATIso\endcsname{}\expandafter\renewcommand\csname PbKSMeerKATIso\endcsname{\ensuremath{1.68}}
\expandafter\providecommand\csname PbKSMeerKATPlasma\endcsname{}\expandafter\renewcommand\csname PbKSMeerKATPlasma\endcsname{\ensuremath{1.25}}
\expandafter\providecommand\csname PbMockAlignExpMean\endcsname{}\expandafter\renewcommand\csname PbMockAlignExpMean\endcsname{\ensuremath{0.401}}
\expandafter\providecommand\csname PbMockAlignExpSD\endcsname{}\expandafter\renewcommand\csname PbMockAlignExpSD\endcsname{\ensuremath{0.025}}
\expandafter\providecommand\csname PbMockAlignKSHi\endcsname{}\expandafter\renewcommand\csname PbMockAlignKSHi\endcsname{\ensuremath{0.80}}
\expandafter\providecommand\csname PbMockAlignN\endcsname{}\expandafter\renewcommand\csname PbMockAlignN\endcsname{\ensuremath{6}}
\expandafter\providecommand\csname PbMockAlignTauHiMyr\endcsname{}\expandafter\renewcommand\csname PbMockAlignTauHiMyr\endcsname{\ensuremath{437}}
\expandafter\providecommand\csname PbMockAlignTauLoMyr\endcsname{}\expandafter\renewcommand\csname PbMockAlignTauLoMyr\endcsname{\ensuremath{361}}
\expandafter\providecommand\csname PbMockAlignTauTruthMyr\endcsname{}\expandafter\renewcommand\csname PbMockAlignTauTruthMyr\endcsname{\ensuremath{398}}
\expandafter\providecommand\csname PbMockExpMean\endcsname{}\expandafter\renewcommand\csname PbMockExpMean\endcsname{\ensuremath{0.397}}
\expandafter\providecommand\csname PbMockExpSD\endcsname{}\expandafter\renewcommand\csname PbMockExpSD\endcsname{\ensuremath{0.023}}
\expandafter\providecommand\csname PbMockExpTruth\endcsname{}\expandafter\renewcommand\csname PbMockExpTruth\endcsname{\ensuremath{0.40}}
\expandafter\providecommand\csname PbMockKSHi\endcsname{}\expandafter\renewcommand\csname PbMockKSHi\endcsname{\ensuremath{1.03}}
\expandafter\providecommand\csname PbMockLnLMachine\endcsname{}\expandafter\renewcommand\csname PbMockLnLMachine\endcsname{\ensuremath{0.01}}
\expandafter\providecommand\csname PbMockN\endcsname{}\expandafter\renewcommand\csname PbMockN\endcsname{\ensuremath{6}}
\expandafter\providecommand\csname PbMockPlasmaExpHi\endcsname{}\expandafter\renewcommand\csname PbMockPlasmaExpHi\endcsname{\ensuremath{0.46}}
\expandafter\providecommand\csname PbMockPlasmaExpLo\endcsname{}\expandafter\renewcommand\csname PbMockPlasmaExpLo\endcsname{\ensuremath{0.34}}
\expandafter\providecommand\csname PbMockPlasmaExpMachine\endcsname{}\expandafter\renewcommand\csname PbMockPlasmaExpMachine\endcsname{\ensuremath{0.012}}
\expandafter\providecommand\csname PbMockPlasmaExpMean\endcsname{}\expandafter\renewcommand\csname PbMockPlasmaExpMean\endcsname{\ensuremath{0.41}}
\expandafter\providecommand\csname PbMockPlasmaExpSD\endcsname{}\expandafter\renewcommand\csname PbMockPlasmaExpSD\endcsname{\ensuremath{0.04}}
\expandafter\providecommand\csname PbMockPlasmaExpTruth\endcsname{}\expandafter\renewcommand\csname PbMockPlasmaExpTruth\endcsname{\ensuremath{0.40}}
\expandafter\providecommand\csname PbMockPlasmaKSHi\endcsname{}\expandafter\renewcommand\csname PbMockPlasmaKSHi\endcsname{\ensuremath{0.66}}
\expandafter\providecommand\csname PbMockPlasmaN\endcsname{}\expandafter\renewcommand\csname PbMockPlasmaN\endcsname{\ensuremath{9}}
\expandafter\providecommand\csname PbMockPlasmaNBound\endcsname{}\expandafter\renewcommand\csname PbMockPlasmaNBound\endcsname{\ensuremath{0}}
\expandafter\providecommand\csname PbMockPlasmaQHi\endcsname{}\expandafter\renewcommand\csname PbMockPlasmaQHi\endcsname{\ensuremath{1.11}}
\expandafter\providecommand\csname PbMockPlasmaQMachine\endcsname{}\expandafter\renewcommand\csname PbMockPlasmaQMachine\endcsname{\ensuremath{0.03}}
\expandafter\providecommand\csname PbMockPlasmaQMean\endcsname{}\expandafter\renewcommand\csname PbMockPlasmaQMean\endcsname{\ensuremath{0.87}}
\expandafter\providecommand\csname PbMockPlasmaQSD\endcsname{}\expandafter\renewcommand\csname PbMockPlasmaQSD\endcsname{\ensuremath{0.12}}
\expandafter\providecommand\csname PbMockPlasmaRhoSDPct\endcsname{}\expandafter\renewcommand\csname PbMockPlasmaRhoSDPct\endcsname{\ensuremath{3}}
\expandafter\providecommand\csname PbMockRhoBiasPct\endcsname{}\expandafter\renewcommand\csname PbMockRhoBiasPct\endcsname{\ensuremath{-2}}
\expandafter\providecommand\csname PbMockRhoMachinePct\endcsname{}\expandafter\renewcommand\csname PbMockRhoMachinePct\endcsname{\ensuremath{6}}
\expandafter\providecommand\csname PbMockRhoSDPct\endcsname{}\expandafter\renewcommand\csname PbMockRhoSDPct\endcsname{\ensuremath{4}}
\expandafter\providecommand\csname PbMockScatterMean\endcsname{}\expandafter\renewcommand\csname PbMockScatterMean\endcsname{\ensuremath{0.152}}
\expandafter\providecommand\csname PbMockScatterTruth\endcsname{}\expandafter\renewcommand\csname PbMockScatterTruth\endcsname{\ensuremath{0.156}}
\expandafter\providecommand\csname PbMockWidthScatterHi\endcsname{}\expandafter\renewcommand\csname PbMockWidthScatterHi\endcsname{\ensuremath{0.13}}
\expandafter\providecommand\csname PbMockWidthScatterLo\endcsname{}\expandafter\renewcommand\csname PbMockWidthScatterLo\endcsname{\ensuremath{0.05}}
\expandafter\providecommand\csname PbMockWidthScatterMachine\endcsname{}\expandafter\renewcommand\csname PbMockWidthScatterMachine\endcsname{\ensuremath{0.08}}
\expandafter\providecommand\csname PbNAllRecycled\endcsname{}\expandafter\renewcommand\csname PbNAllRecycled\endcsname{\ensuremath{21}}
\expandafter\providecommand\csname PbNAllRefs\endcsname{}\expandafter\renewcommand\csname PbNAllRefs\endcsname{\ensuremath{989}}
\expandafter\providecommand\csname PbNFAST\endcsname{}\expandafter\renewcommand\csname PbNFAST\endcsname{\ensuremath{339}}
\expandafter\providecommand\csname PbNMeerKAT\endcsname{}\expandafter\renewcommand\csname PbNMeerKAT\endcsname{\ensuremath{651}}
\expandafter\providecommand\csname PbNNormal\endcsname{}\expandafter\renewcommand\csname PbNNormal\endcsname{\ensuremath{878}}
\expandafter\providecommand\csname PbNNormalScat\endcsname{}\expandafter\renewcommand\csname PbNNormalScat\endcsname{\ensuremath{787}}
\expandafter\providecommand\csname PbNRecycled\endcsname{}\expandafter\renewcommand\csname PbNRecycled\endcsname{\ensuremath{15}}
\expandafter\providecommand\csname PbNRecycledScat\endcsname{}\expandafter\renewcommand\csname PbNRecycledScat\endcsname{\ensuremath{13}}
\expandafter\providecommand\csname PbOutlierShare\endcsname{}\expandafter\renewcommand\csname PbOutlierShare\endcsname{\ensuremath{0.07}}
\expandafter\providecommand\csname PbOutlierSharePlasma\endcsname{}\expandafter\renewcommand\csname PbOutlierSharePlasma\endcsname{\ensuremath{0.00}}
\expandafter\providecommand\csname PbPZeroCensusLoMs\endcsname{}\expandafter\renewcommand\csname PbPZeroCensusLoMs\endcsname{\ensuremath{3}}
\expandafter\providecommand\csname PbPZeroCensusMedianMs\endcsname{}\expandafter\renewcommand\csname PbPZeroCensusMedianMs\endcsname{\ensuremath{182}}
\expandafter\providecommand\csname PbPZeroCensusSigma\endcsname{}\expandafter\renewcommand\csname PbPZeroCensusSigma\endcsname{\ensuremath{0.35}}
\expandafter\providecommand\csname PbPZeroMedianMs\endcsname{}\expandafter\renewcommand\csname PbPZeroMedianMs\endcsname{\ensuremath{251}}
\expandafter\providecommand\csname PbPZeroSigma\endcsname{}\expandafter\renewcommand\csname PbPZeroSigma\endcsname{\ensuremath{0.7}}
\expandafter\providecommand\csname PbQBoth\endcsname{}\expandafter\renewcommand\csname PbQBoth\endcsname{\ensuremath{1.68}}
\expandafter\providecommand\csname PbQBothCensus\endcsname{}\expandafter\renewcommand\csname PbQBothCensus\endcsname{\ensuremath{1.35}}
\expandafter\providecommand\csname PbQBothMockSigma\endcsname{}\expandafter\renewcommand\csname PbQBothMockSigma\endcsname{\ensuremath{5.5}}
\expandafter\providecommand\csname PbQFAST\endcsname{}\expandafter\renewcommand\csname PbQFAST\endcsname{\ensuremath{0.86}}
\expandafter\providecommand\csname PbQMeerKAT\endcsname{}\expandafter\renewcommand\csname PbQMeerKAT\endcsname{\ensuremath{1.23}}
\expandafter\providecommand\csname PbRecycledLawDiffPct\endcsname{}\expandafter\renewcommand\csname PbRecycledLawDiffPct\endcsname{\ensuremath{5}}
\expandafter\providecommand\csname PbRhoExp\endcsname{}\expandafter\renewcommand\csname PbRhoExp\endcsname{\ensuremath{0.32}}
\expandafter\providecommand\csname PbRhoExpAlign\endcsname{}\expandafter\renewcommand\csname PbRhoExpAlign\endcsname{\ensuremath{0.32}}
\expandafter\providecommand\csname PbRhoExpAll\endcsname{}\expandafter\renewcommand\csname PbRhoExpAll\endcsname{\ensuremath{0.30}}
\expandafter\providecommand\csname PbRhoExpBothQ\endcsname{}\expandafter\renewcommand\csname PbRhoExpBothQ\endcsname{\ensuremath{0.36}}
\expandafter\providecommand\csname PbRhoExpBothQCensus\endcsname{}\expandafter\renewcommand\csname PbRhoExpBothQCensus\endcsname{\ensuremath{0.37}}
\expandafter\providecommand\csname PbRhoExpErr\endcsname{}\expandafter\renewcommand\csname PbRhoExpErr\endcsname{\ensuremath{0.02}}
\expandafter\providecommand\csname PbRhoExpFAST\endcsname{}\expandafter\renewcommand\csname PbRhoExpFAST\endcsname{\ensuremath{0.35}}
\expandafter\providecommand\csname PbRhoExpFASTPlasma\endcsname{}\expandafter\renewcommand\csname PbRhoExpFASTPlasma\endcsname{\ensuremath{0.38}}
\expandafter\providecommand\csname PbRhoExpHalfPlasmaSigma\endcsname{}\expandafter\renewcommand\csname PbRhoExpHalfPlasmaSigma\endcsname{\ensuremath{3.4}}
\expandafter\providecommand\csname PbRhoExpMeerKAT\endcsname{}\expandafter\renewcommand\csname PbRhoExpMeerKAT\endcsname{\ensuremath{0.32}}
\expandafter\providecommand\csname PbRhoExpMeerKATPlasma\endcsname{}\expandafter\renewcommand\csname PbRhoExpMeerKATPlasma\endcsname{\ensuremath{0.35}}
\expandafter\providecommand\csname PbRhoExpPlasma\endcsname{}\expandafter\renewcommand\csname PbRhoExpPlasma\endcsname{\ensuremath{0.36}}
\expandafter\providecommand\csname PbRhoExpPlasmaCensus\endcsname{}\expandafter\renewcommand\csname PbRhoExpPlasmaCensus\endcsname{\ensuremath{0.37}}
\expandafter\providecommand\csname PbRhoExpPlasmaErr\endcsname{}\expandafter\renewcommand\csname PbRhoExpPlasmaErr\endcsname{\ensuremath{0.04}}
\expandafter\providecommand\csname PbRhoExpScat\endcsname{}\expandafter\renewcommand\csname PbRhoExpScat\endcsname{\ensuremath{0.33}}
\expandafter\providecommand\csname PbRhoExpThree\endcsname{}\expandafter\renewcommand\csname PbRhoExpThree\endcsname{\ensuremath{0.325}}
\expandafter\providecommand\csname PbRhoOne\endcsname{}\expandafter\renewcommand\csname PbRhoOne\endcsname{\ensuremath{6.7}}
\expandafter\providecommand\csname PbRhoOneFAST\endcsname{}\expandafter\renewcommand\csname PbRhoOneFAST\endcsname{\ensuremath{6.6}}
\expandafter\providecommand\csname PbRhoOneMeerKAT\endcsname{}\expandafter\renewcommand\csname PbRhoOneMeerKAT\endcsname{\ensuremath{6.6}}
\expandafter\providecommand\csname PbRhoOnePlasma\endcsname{}\expandafter\renewcommand\csname PbRhoOnePlasma\endcsname{\ensuremath{6.0}}
\expandafter\providecommand\csname PbRhoOnePlasmaCensus\endcsname{}\expandafter\renewcommand\csname PbRhoOnePlasmaCensus\endcsname{\ensuremath{6.0}}
\expandafter\providecommand\csname PbScaleFit\endcsname{}\expandafter\renewcommand\csname PbScaleFit\endcsname{\ensuremath{0.84}}
\expandafter\providecommand\csname PbScaleFitAll\endcsname{}\expandafter\renewcommand\csname PbScaleFitAll\endcsname{\ensuremath{1.14}}
\expandafter\providecommand\csname PbScaleFitAllDlnLOne\endcsname{}\expandafter\renewcommand\csname PbScaleFitAllDlnLOne\endcsname{\ensuremath{0.2}}
\expandafter\providecommand\csname PbScaleFitAllHi\endcsname{}\expandafter\renewcommand\csname PbScaleFitAllHi\endcsname{\ensuremath{1.26}}
\expandafter\providecommand\csname PbScaleFitAllLo\endcsname{}\expandafter\renewcommand\csname PbScaleFitAllLo\endcsname{\ensuremath{1.00}}
\expandafter\providecommand\csname PbScaleFitDlnLOne\endcsname{}\expandafter\renewcommand\csname PbScaleFitDlnLOne\endcsname{\ensuremath{0.3}}
\expandafter\providecommand\csname PbScaleFitHi\endcsname{}\expandafter\renewcommand\csname PbScaleFitHi\endcsname{\ensuremath{1.00}}
\expandafter\providecommand\csname PbScaleFitLo\endcsname{}\expandafter\renewcommand\csname PbScaleFitLo\endcsname{\ensuremath{0.71}}
\expandafter\providecommand\csname PbScaleFitScat\endcsname{}\expandafter\renewcommand\csname PbScaleFitScat\endcsname{\ensuremath{0.63}}
\expandafter\providecommand\csname PbScaleFitScatDlnLOne\endcsname{}\expandafter\renewcommand\csname PbScaleFitScatDlnLOne\endcsname{\ensuremath{2.3}}
\expandafter\providecommand\csname PbScaleFitScatHi\endcsname{}\expandafter\renewcommand\csname PbScaleFitScatHi\endcsname{\ensuremath{0.71}}
\expandafter\providecommand\csname PbScaleFitScatLo\endcsname{}\expandafter\renewcommand\csname PbScaleFitScatLo\endcsname{\ensuremath{0.56}}
\expandafter\providecommand\csname PbScaleHalf\endcsname{}\expandafter\renewcommand\csname PbScaleHalf\endcsname{\ensuremath{0.55}}
\expandafter\providecommand\csname PbScaleHalfAll\endcsname{}\expandafter\renewcommand\csname PbScaleHalfAll\endcsname{\ensuremath{0.66}}
\expandafter\providecommand\csname PbScaleHalfAllDlnLOne\endcsname{}\expandafter\renewcommand\csname PbScaleHalfAllDlnLOne\endcsname{\ensuremath{2.7}}
\expandafter\providecommand\csname PbScaleHalfAllHi\endcsname{}\expandafter\renewcommand\csname PbScaleHalfAllHi\endcsname{\ensuremath{0.71}}
\expandafter\providecommand\csname PbScaleHalfAllLo\endcsname{}\expandafter\renewcommand\csname PbScaleHalfAllLo\endcsname{\ensuremath{0.56}}
\expandafter\providecommand\csname PbScaleHalfDlnLOne\endcsname{}\expandafter\renewcommand\csname PbScaleHalfDlnLOne\endcsname{\ensuremath{4.3}}
\expandafter\providecommand\csname PbScaleHalfHi\endcsname{}\expandafter\renewcommand\csname PbScaleHalfHi\endcsname{\ensuremath{0.63}}
\expandafter\providecommand\csname PbScaleHalfLo\endcsname{}\expandafter\renewcommand\csname PbScaleHalfLo\endcsname{\ensuremath{0.50}}
\expandafter\providecommand\csname PbScalePlasma\endcsname{}\expandafter\renewcommand\csname PbScalePlasma\endcsname{\ensuremath{0.94}}
\expandafter\providecommand\csname PbScalePlasmaAll\endcsname{}\expandafter\renewcommand\csname PbScalePlasmaAll\endcsname{\ensuremath{1.15}}
\expandafter\providecommand\csname PbScalePlasmaAllDlnLOne\endcsname{}\expandafter\renewcommand\csname PbScalePlasmaAllDlnLOne\endcsname{\ensuremath{0.4}}
\expandafter\providecommand\csname PbScalePlasmaAllHi\endcsname{}\expandafter\renewcommand\csname PbScalePlasmaAllHi\endcsname{\ensuremath{1.26}}
\expandafter\providecommand\csname PbScalePlasmaAllLo\endcsname{}\expandafter\renewcommand\csname PbScalePlasmaAllLo\endcsname{\ensuremath{1.00}}
\expandafter\providecommand\csname PbScalePlasmaCensus\endcsname{}\expandafter\renewcommand\csname PbScalePlasmaCensus\endcsname{\ensuremath{0.90}}
\expandafter\providecommand\csname PbScalePlasmaCensusDlnLOne\endcsname{}\expandafter\renewcommand\csname PbScalePlasmaCensusDlnLOne\endcsname{\ensuremath{0.1}}
\expandafter\providecommand\csname PbScalePlasmaCensusHi\endcsname{}\expandafter\renewcommand\csname PbScalePlasmaCensusHi\endcsname{\ensuremath{1.00}}
\expandafter\providecommand\csname PbScalePlasmaCensusLo\endcsname{}\expandafter\renewcommand\csname PbScalePlasmaCensusLo\endcsname{\ensuremath{0.79}}
\expandafter\providecommand\csname PbScalePlasmaDlnLOne\endcsname{}\expandafter\renewcommand\csname PbScalePlasmaDlnLOne\endcsname{\ensuremath{0.0}}
\expandafter\providecommand\csname PbScalePlasmaHalf\endcsname{}\expandafter\renewcommand\csname PbScalePlasmaHalf\endcsname{\ensuremath{0.64}}
\expandafter\providecommand\csname PbScalePlasmaHalfDlnLOne\endcsname{}\expandafter\renewcommand\csname PbScalePlasmaHalfDlnLOne\endcsname{\ensuremath{3.0}}
\expandafter\providecommand\csname PbScalePlasmaHalfHi\endcsname{}\expandafter\renewcommand\csname PbScalePlasmaHalfHi\endcsname{\ensuremath{0.71}}
\expandafter\providecommand\csname PbScalePlasmaHalfLo\endcsname{}\expandafter\renewcommand\csname PbScalePlasmaHalfLo\endcsname{\ensuremath{0.56}}
\expandafter\providecommand\csname PbScalePlasmaHi\endcsname{}\expandafter\renewcommand\csname PbScalePlasmaHi\endcsname{\ensuremath{1.12}}
\expandafter\providecommand\csname PbScalePlasmaLo\endcsname{}\expandafter\renewcommand\csname PbScalePlasmaLo\endcsname{\ensuremath{0.79}}
\expandafter\providecommand\csname PbScalePlasmaQ\endcsname{}\expandafter\renewcommand\csname PbScalePlasmaQ\endcsname{\ensuremath{0.98}}
\expandafter\providecommand\csname PbScalePlasmaQDlnLOne\endcsname{}\expandafter\renewcommand\csname PbScalePlasmaQDlnLOne\endcsname{\ensuremath{0.0}}
\expandafter\providecommand\csname PbScalePlasmaQHi\endcsname{}\expandafter\renewcommand\csname PbScalePlasmaQHi\endcsname{\ensuremath{1.12}}
\expandafter\providecommand\csname PbScalePlasmaQLo\endcsname{}\expandafter\renewcommand\csname PbScalePlasmaQLo\endcsname{\ensuremath{0.79}}
\expandafter\providecommand\csname PbScalePlasmaScat\endcsname{}\expandafter\renewcommand\csname PbScalePlasmaScat\endcsname{\ensuremath{0.71}}
\expandafter\providecommand\csname PbScalePlasmaScatDlnLOne\endcsname{}\expandafter\renewcommand\csname PbScalePlasmaScatDlnLOne\endcsname{\ensuremath{1.4}}
\expandafter\providecommand\csname PbScalePlasmaScatHi\endcsname{}\expandafter\renewcommand\csname PbScalePlasmaScatHi\endcsname{\ensuremath{0.79}}
\expandafter\providecommand\csname PbScalePlasmaScatLo\endcsname{}\expandafter\renewcommand\csname PbScalePlasmaScatLo\endcsname{\ensuremath{0.63}}
\expandafter\providecommand\csname PbTauFASTGyr\endcsname{}\expandafter\renewcommand\csname PbTauFASTGyr\endcsname{\ensuremath{2.6}}
\expandafter\providecommand\csname PbTauMeerKATMyr\endcsname{}\expandafter\renewcommand\csname PbTauMeerKATMyr\endcsname{\ensuremath{228}}
\expandafter\providecommand\csname PbTotalScatter\endcsname{}\expandafter\renewcommand\csname PbTotalScatter\endcsname{\ensuremath{0.14}}
\expandafter\providecommand\csname PbTotalScatterPlasma\endcsname{}\expandafter\renewcommand\csname PbTotalScatterPlasma\endcsname{\ensuremath{0.11}}

\providecommand{\apj}{ApJ}\providecommand{\apjl}{ApJL}\providecommand{\apjs}{ApJS}
\providecommand{\mnras}{MNRAS}\providecommand{\araa}{ARA\&A}\providecommand{\aap}{A\&A}
\providecommand{\prd}{Phys.~Rev.~D}
\providecommand{\nat}{Nature}
\providecommand{\nar}{New Astron.~Rev.}
\providecommand{\physrep}{Phys.~Rep.}

\begin{document}
\title{Selection-aware inference of supernova kicks}
\author{R.~O'Shaughnessy}
\affiliation{Center for Computational Relativity and Gravitation, Rochester Institute of Technology, Rochester, NY 14623, USA}
\date{2026-11-26}
\begin{abstract}
Observed compact objects (isolated or in binaries) used to infer natal kicks
are selected by class membership, which depends on disruption, subsequent
evolution, and discovery. Ignoring this selection can bias the inferred kicks.
To demonstrate this bias in a comprehensive inference framework, we construct a differentiable model
for key components of binary evolution, including correlated remnant-mass and kick
prescriptions as well as  survival probabilities.  As initial conditions prior to the first SN, we use a fiducial
\cosmic{} analysis.
Using conventional statistical methods to assess bias due to omitted selection factors in a likelihood, we demonstrate
this bias concretely. 
Reliable natal inference
requires a forward prediction for both the observed properties and the
probability of observing them.
Our framework provides a novel method to enable robust neutron star population inference combining heterogeneous
observations of multiple NS-pertinent observations. 
\end{abstract}
\maketitle

\section{Introduction}\label{sec:intro}
Neutron stars (NSs) are born in core-collapse
supernovae~\cite{1934PNAS...20..259B,1968Sci...162.1481S,1969Natur.221..453C,1968Natur.220..340L},
whose anticipated asymmetries~\cite{1994A&A...290..496J,1995ApJ...450..830B,1995PhR...256..117H,2013A&A...552A.126W,2008ARA&A..46..433W,2014Natur.506..339G}
impart natal
kicks~\cite{1997ApJ...483..399V,2025arXiv250901430D} with magnitudes of
$\sim100$--$1000\,\mathrm{km\,s^{-1}}$~\cite{2024ApJ...963...63B}.
The most compelling evidence for large natal kicks comes from observations of isolated
pulsars, which are known to move much more rapidly than their
progenitors~\cite{1970ApJ...160..979G,1994Natur.369..127L,HobbsKicks,2005ApJ...630L..61C,2017A&A...608A..57V}.
More broadly, observations of isolated pulsars and NS-containing binaries offer several avenues with which these natal
kicks can be reconstructed~\cite{2025arXiv250901430D,Igoshev2021Combined,Shariat2026Concordance,Disberg2026BinaryKicks}.
Many studies approach the recoil velocity using carefully-chosen subsets of the known NS population; for example,
some studies use  the velocities of isolated
pulsars~\cite{ArKicks,HobbsKicks,2006ApJ...643..332F,2017A&A...608A..57V,2020MNRAS.494.3663I,2025ApJ...989L...8D},
whereas others use NSs in binaries: their systemic motions and
orbits~\cite{2023MNRAS.521.2504O,Disberg2026BinaryKicks} and
reconstructions of individual supernovae in surviving
binaries~\cite{psr-kicks-WongWillemsKalogera2010,2015ApJ...801...32A,2017ApJ...846..170T}.
Because the kick distribution sets the predicted abundances, orbits and spin
orientations of X-ray binaries, recycled pulsars and compact-object
mergers~\cite{2020ApJ...898...71B,2021MNRAS.500.1380M,Igoshev2021Combined,2018MNRAS.477.4685B,2000ApJ...541..319K,ns-polarkicks-rateimplications-Postnov2007},
the natal recoil kick can conceivably be reconstructed from these observations as well~\cite{1995MNRAS.274..461B,psr-kicks-WongWillemsKalogera2010,2018MNRAS.477.4685B,2023MNRAS.521.2504O,Valli2025Polar}.
However, the distribution of NS natal kicks remains very
uncertain~\cite{2025arXiv250901430D}.

Different groups interpreting these different observations come to
substantially different conclusions about the kick magnitudes and
directions~\cite{HobbsKicks,2017A&A...608A..57V,2025ApJ...989L...8D,2023MNRAS.521.2504O,Valli2025Polar,Shariat2026Concordance,Disberg2026BinaryKicks}. Isolated-pulsar analyses favor large kicks but disagree on
the distribution's shape and number of
components~\cite{HobbsKicks,2017A&A...608A..57V,2025ApJ...989L...8D}.
Be-X-ray-binary orbits favor ultralow kicks and a higher-velocity component
aligned with the progenitor's spin~\cite{Valli2025Polar}, whereas wide
\textit{Gaia} neutron-star binaries favor low kicks with directions strongly
selected by survival~\cite{Shariat2026Concordance}.
Joint analyses of pulsars and surviving binaries favor low- and high-kick
components~\cite{Igoshev2021Combined,Shariat2026Concordance}, with isotropic
birth directions able to reproduce both the \textit{Gaia} and
Be-X-ray-binary orbits in a shared model with population-dependent
progenitors~\cite{Shariat2026Concordance}.

Interpreting these differences requires accounting for how each population
is selected, using established population-inference
methods~\cite{LIGO-P1600187-Farr-SelectionBiasaAndMonteCarlo,VitaleGerosaFarrTaylor2022}.
Kicks change both the measured orbit and whether the binary
survives~\cite{1983ApJ...267..322H,1995MNRAS.274..461B,1996ApJ...471..352K};
manifestly the isolated-pulsar sample will preferentially be drawn from the population receiving large kicks, while
surviving binaries will preferentially receive smaller natal kicks~\cite{2025ApJ...989L...8D,2023MNRAS.519.5893K,Shariat2026Concordance}.
Our study is motivated by two recent attempts to interpret NS observations
as constraints on natal kicks. Disberg and
Mandel~\cite{2025ApJ...989L...8D} infer isolated-pulsar velocities
from proper motions and distances, using a young, nonrecycled sample
and explicitly noting that low-kick NSs retained in binaries are absent
from it. Shariat et al.~\cite{Shariat2026Concordance} combine that published
pulsar constraint with forward supernova models for Galactic
\textit{Gaia} NS and Be--NS binaries. They incorporate survival selection
for the binary classes, but do not self-consistently reconcile observations of  isolated-pulsar velocities to account
for survival bias using the same model adopted for binaries.  

Population-synthesis studies have long quantified the dependence of compact
binary rates on uncertain evolution~\cite{2022LRR....25....1M}.  For example, early multidimensional StarTrack studies
mapped rates across seven parameters and used observed compact-binary and
supernova rates to constrain models~\cite{PSutility,PSutil2,PSconstraints,OShaughnessy2008Constraints}.
Subsequent studies varied common-envelope, kick and other prescriptions for compact
binaries and double neutron stars~\cite{2012ApJ...759...52D,2017arXiv170807885C,2018MNRAS.481.4009V,2018MNRAS.480.2011G,2018MNRAS.481.1908K,Broekgaarden2022Rates,2023MNRAS.524..426I,2024ApJ...976...24B,2026ApJ...997...52C,Venet2026BNS};
initial binary distributions introduce a distinct source of variation~\cite{deMink2015Initial}.
Model selection, Fisher-matrix forecasts and hierarchical inference use
gravitational-wave catalogs, Galactic double neutron stars and pulsars to
constrain common-envelope and kick parameters~\cite{2015ApJ...810...58S,2018MNRAS.477.4685B,gwastro-DanielW-PopsynKickPaper2017,2021ApJ...910..152Z,2021MNRAS.507.5224B,Richards2023Multiple,2024ApJ...963...80D,2025MNRAS.544L..89C}.
Because each model is expensive, these analyses interpolate or emulate
population-synthesis output across continuous
parameters~\cite{2018MNRAS.481.4685T,2019PhRvD.100h3015W,2021PhRvD.103h3021W,2023MNRAS.521.2201R,2024MNRAS.534.3506R,2025ApJ...988..189C}.
Survival can be self-consistently folded into such a Bayesian inference by
interpolating observational likelihoods over formation
models~\cite{popsyn-gwastro-STInterpFinal-Vera2023,Delfavero2025DWD} and iteratively adding
models where the posterior concentrates~\cite{popsyn-gwastro-STInterpFinal-Vera2023}.
Other analyses sample initial conditions and binary parameters through the
binary-evolution code itself, without
gradients~\cite{2018ApJS..237....1A,2023ApJ...950..181W}.
Comparisons of black-hole binaries also illustrate how evolutionary tracks and
inter-supernova transport affect inferred outcomes~\cite{Gerosa2018Spin,Guerrero2026Codes}.

In this work, we develop a toolkit for self-consistent inference on
isolated and binary neutron stars, accounting for the common physics and
selection biases associated with binary survival.   We implement the first full differentiable approximation to binary
evolution through two supernovae, starting from uncertain configurations
immediately before the first collapse. Its stellar prescriptions are drawn
from COSMIC (Compact Object Synthesis and Monte Carlo Investigation Code),
a rapid population-synthesis suite for compact binaries and their
progenitors~\cite{2020ApJ...898...71B}. COSMIC builds on the binary-star
evolution algorithm BSE~\cite{2002MNRAS.329..897H}, which combines binary
interactions with the single-star evolution formulae
SSE~\cite{2000MNRAS.315..543H}. These formulae approximate stellar properties
as functions of mass, metallicity and age; BSE adds prescriptions for mass
transfer, common-envelope evolution, tides, winds and supernovae.

Our implementation ports hydrogen-rich and helium-star SSE formulae from
COSMIC into differentiable functions and couples them to an approximate
inter-supernova channel model. The calculation follows wind mass loss,
Roche-lobe contact, circularization, common-envelope evolution, stable mass
transfer and accretion to determine the masses and orbit entering the second
collapse (Sec.~\ref{sec:configurationmethod}). Automatic differentiation
then propagates changes in the adopted kick and binary-evolution parameters
through both explosions, the intervening evolution, and the survival
normalization of the population likelihood. COSMIC supplies the prescription
baseline and stored evolutionary histories for comparison; the likelihood
evaluates our approximation rather than calling COSMIC or taking its
pre-second-supernova states as fixed inputs.

Equally critically,  our implementation also provides a differentiable model for NS selection by radio pulsar
surveys~\cite{Chunglee-nsns-1,psr-SpinFb,Kim2015DoublePulsar,Pol2019DNSRate}. This ingredient allows us to self-consistently model the full forward population and observation process,
encompassing both physical (survival) and observational (survey) selection effects.

Joint analyses of several NS populations given a remnant mass and natal kick model already
exist~\cite{Igoshev2021Combined,Disberg2026BinaryKicks,Shariat2026Concordance}.
Our work adds two key new improvements.  First, because we
follow each binary through both explosions and the evolution between
them, we account for  disruption, merger before the second collapse (and visibility as a pulsar)
all self-consistently in the same selection function. Second, automatic derivatives of that likelihood
enable us to more effectively perform inference and propagate uncertainties for the
kick and binary evolution properties.

Once in possession of a differentiable likelihood and selection model, we use conventional statistical techniques to
demonstrate the model and biases introduced by omitting factors. Specifically, we generate synthetic populations drawn
from our model; use a population (inhomogeneous Poisson) likelihood to characterize observations; find the best-fitting
population parameters; characterize uncertainties via a Fisher matrix calculation (i.e., using derivatives)~\cite{CutlerFlanagan1994,Vallisneri2008,GairAntonelliBarbieri2023}; and
estimate parameter biases using the Fisher matrix and neglected likelihood factors~\cite{CutlerVallisneri2007,VallisneriYunes2013}.    Similar techniques have been
widely used elsewhere in for example gravitational wave astrophysics.

This paper is organized as follows. 
Section~\ref{sec:method} defines that likelihood and the evolution model.
Section~\ref{sec:demo} tests kick/CE recovery from double-NS endpoints
and its sensitivity to survival normalization, then combines isolated and
binary pulsar constraints. Conditional SN2 orientation tests are in
Appendix~\ref{app:conditional_sn2_experiments}.
Section~\ref{sec:conclude} summarizes the implications and findings. Appendix~\ref{app:physics} details the inter-supernova physics.
Appendix~\ref{app:native_survey_checks} gives separate native survey-tool
validation.

\section{Methods}\label{sec:method}
A kick changes the orbit and recoil of a binary, but also whether the
binary survives and can be observed. We include both effects when comparing
a supernova prescription with a population of compact objects.

\subsection{Evolutionary trajectories: Notation and framework}
\begin{figure*}[t]
 \centering
 \includegraphics[width=\textwidth]{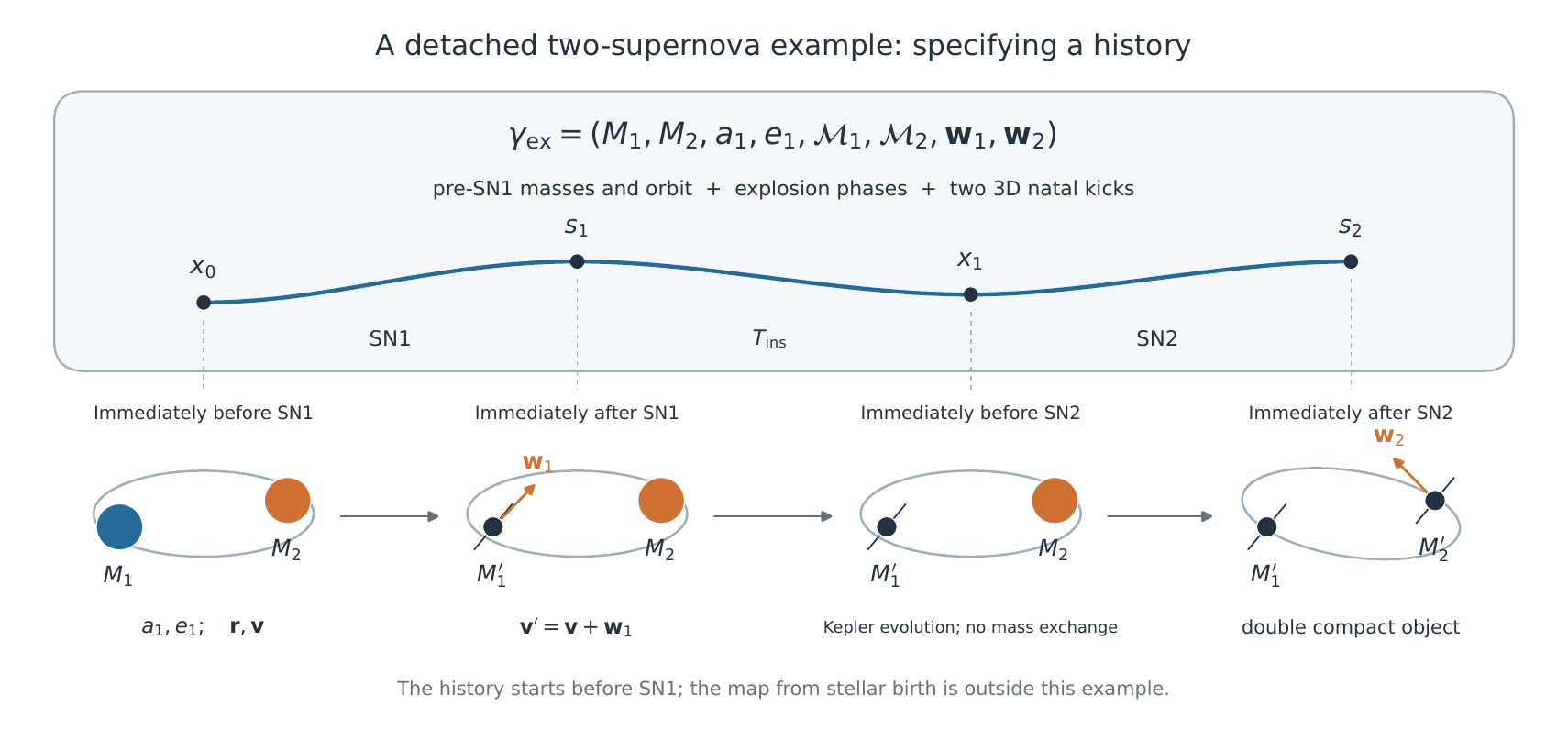}
 \caption{A concrete example of the history variables $\gamma$ for a
 detached binary with two supernovae. The contour encloses the input
 variables and one resulting path; it is schematic, with no metric or
 probability density implied. Dots mark the states immediately before and
 after each impulse, with the corresponding binaries below. Colored disks
 denote stars, small dark disks compact remnants, and arrows natal kicks;
 the kicks are three-dimensional although the sketch is planar. The path
 shown remains bound through both explosions. In this restricted example
 the inter-SN map changes orbital phase but preserves the component
 masses. The full calculation also includes disruption and interacting
 channels.\llmfig}
 \label{fig:gamma_history}
\end{figure*}

We use $\gamma$ to characterize a binary's trajectory from birth to endpoint,
reserving $\Gamma$ for the Fisher matrix (Sec.~\ref{sec:fisher}).
Figure \ref{fig:gamma_history} illustrates our notation graphically: 
we use $x_i$ and $s_i$ to
denote the pre-SN and post-SN states, respectively.

To establish notation more concretely using a simplified scenario, we outline our notation for two stars that never exchange mass, starting
immediately before the first supernova (SN1).  For known stellar structure, remnant mass, formation time, and
metallicity, and neglecting tides and winds, we can characterize a state by the progenitor orbit; the two SN kicks; and
the mean anomalies at which they occur: 
\[
 \gamma_{\rm ex}=(M_1,M_2,a_1,e_1,\mathcal M_1,\mathcal M_2,
                   \mathbf w_1,\mathbf w_2),
\]
where $M_1,M_2$ and $a_1,e_1$ are the pre-SN1 masses and orbital elements,
$\mathcal M_j$ is the mean anomaly at explosion $j$, and each natal kick
$\mathbf w_j$ has three components in a specified orbital frame.
In a violent oversimplification, in this toy model the orbital orientation is assumed fixed: no realignment due to tides
or mass transfer takes place.
We characterize the trajectory as
\begin{equation}
 x_0\xrightarrow{\mathrm{SN1}}s_1
 \xrightarrow{T_{\rm ins}}x_1\xrightarrow{\mathrm{SN2}}s_2,
 \label{eq:trajectory}
\end{equation}
where $x_0,x_1$ are the states just before the explosions and $s_1,s_2$
those just after them (Fig.~\ref{fig:gamma_history}).  In our extremely oversimplified toy model, $T_{\rm ins}$ is
point-particle evolution with unchanged component masses. SN1 replaces $M_1$ by
$M_1'=M_1-\Delta M_1$ and adds $\mathbf w_1$ to its velocity; SN2 similarly
replaces $M_2$ by $M_2'$ and imparts $\mathbf w_2$. At either event,
$\mathbf r,\mathbf v$ denote the exploding star's position and velocity
relative to its companion, so $\mathbf v'=\mathbf v+\mathbf w_j$.
Thus $\gamma$ specifies a history, while the states along its path
are derived quantities. The final state can be bound; unbound; or a merger.

Based on the above discussion, our simplified binary evolution model involves two key components: a SN model (for
remnant masses and recoil) and an inter-SN binary evolution model.  We adopt the following frameworks:
\begin{itemize}
\item \textbf{SN model}: SN are characterized by a probability of remnant masses $p(m_{ns}|x)$ and recoil kicks
  $p(w|x)$. For simplicity, we assume the remnant mass is set by the exploding star's pre-SN core structure, chiefly its CO-core and
  helium-shell masses (Appendix~\ref{sec:engine}), and that the kick
  velocity depends only on the pre-SN star's mass and potentially spin orientation.

  For simplicity, we assume all kick anisotropy
  relative to the spin orientation can be encoded in a single parameter $\kappa$ so
  $p(\cos \iota) \propto \exp \kappa \cos^2\iota$: large $\kappa$ favors polar (aligned or antialigned) kicks; large
  negative $\kappa$ favors equatorial kicks; and $\kappa=0$ favors isotropic kicks.

  The process of impulsive SN mass loss and recoil kick changes the orbit in the customary, well-determined fashion described in
  Appendix~\ref{app:sn_model}.

\item \textbf{Inter-SN model}: Between SN, a star and a NS can evolve and interact through a number of processes (eg.,
  wind mass loss, mass transfer, tides, common envelope evolution, et cetera).

  For simplicity, we adapt  a binary evolution model calibrated to COSMIC for inter-SN evolution; see Appendix~
  \ref{app:physics} for details.

\item \textbf{Pre-SN model}: Before the first SN, the progenitor binary stellar system can also evolve and interact
  through a number of processes.  For the purposes of this paper, these processes are incorporated into a prior on the
  initial state $x_o$. As with the inter-SN model, we calibrate our prior to COSMIC for pre-SN evolution.  For brevity, we treat
  binary evolution between SN and before the first SN as completely independent processes -- we specifically do not
  change our prior $p(x_o)$ when we change the binary evolution physics used during inter-SN evolution.
  Appendix~\ref{app:archive} describes the fiducial COSMIC runs from which we generate this prior.
\end{itemize}

\begin{figure*}
\centering
\includegraphics[width=.95\textwidth]{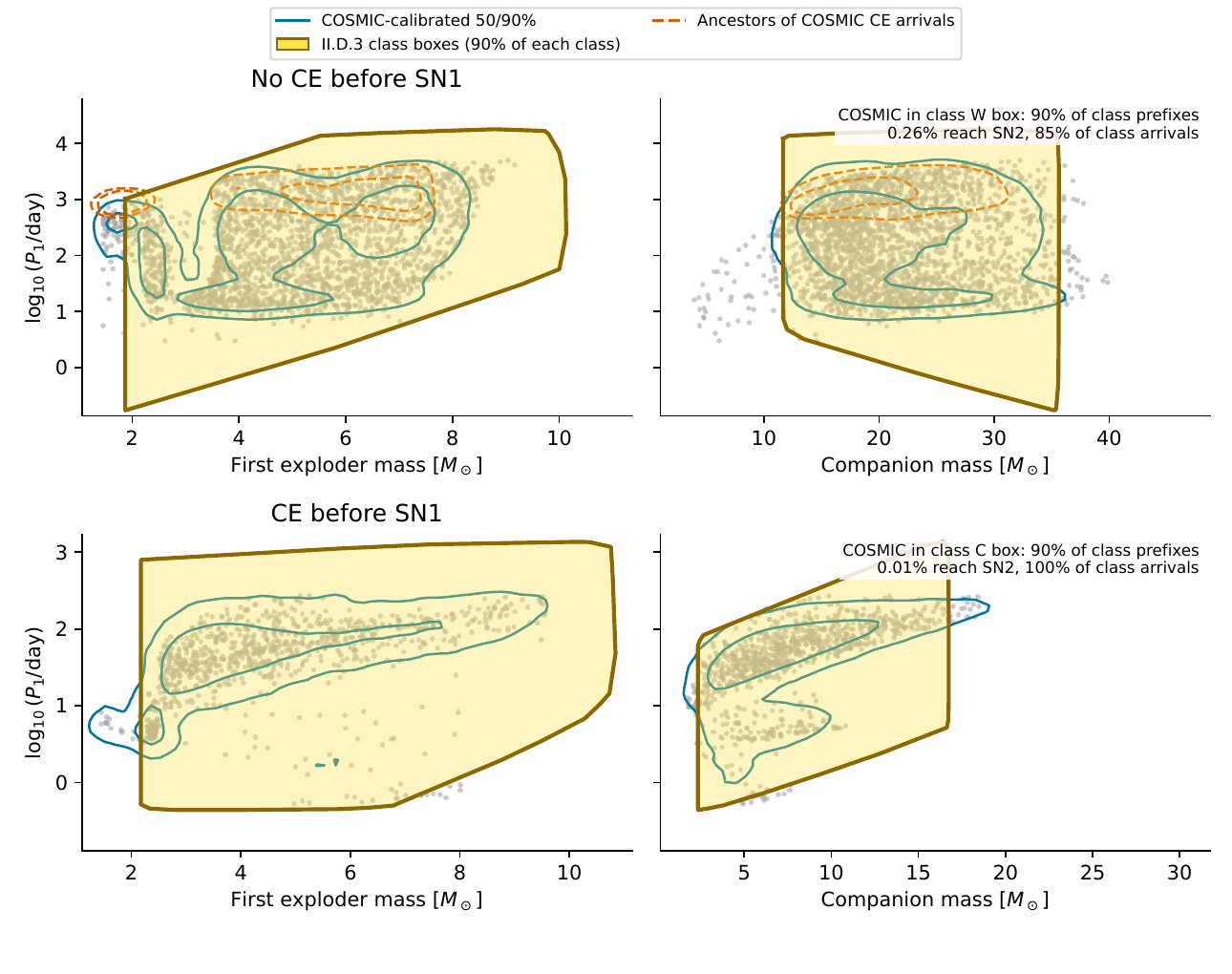}
\[
 p_\eta(x,c\mid\phi)\xrightarrow{\mathrm{SN1}(\Lambda)}s_1
 \xrightarrow{T_\phi}s_2\xrightarrow{\mathrm{SN2}(\Lambda)}
 \mathrm{descendants}\xrightarrow{p_{\rm det}}\mathrm{data}.
\]
\caption{Toy input population and the class boxes of the two-SN calculation.
Gray points are held-out \cosmic{} pre-SN1 systems, thinned for display.
Blue contours enclose approximately 50\% and 90\% of the fitted density
$p_0(x,c)$ in each projection (Appendix~\ref{app:presn_prior}).
Rows distinguish CE before SN1 from no prior CE, including stable transfer.
Yellow regions project the class boxes of the pre-SN1 prior
(Sec.~\ref{sec:two_sn_data}, Table~\ref{tab:two_sn_classes}) into mass and
Keplerian period: class W in the upper row and class C in the lower row. Each
right panel gives the share of
the class's \cosmic{} prefixes inside its box, the fraction of those that
reach SN2, and the share of the class's arrivals inside the box. Orange dashed
contours enclose the pre-SN1 states of \cosmic{} systems that reach SN2
through an inter-SN CE (Appendix~\ref{app:coverage}). In the process line, the two-SN
experiment replaces $p_\eta$ by the class boxes and varies the kick scale
and the CE efficiency in $T_\phi$.\llmfig}
\label{fig:presn_input}
\end{figure*}

We also fit a tunable toy density for the masses and periods immediately
before SN1 to a prospective \cosmic{} archive at fixed binary physics
(Fig.~\ref{fig:presn_input}). The two-SN experiment of
Sec.~\ref{sec:demo} starts instead from the class boxes of
Sec.~\ref{sec:two_sn_data}, shown in the same figure. We can therefore change the SN model without
rerunning the evolution that precedes it.
Appendix~\ref{app:presn_prior} defines the family, the fit and its validation.

\subsection{Detection models and population statistics }\label{sec:selectionmethod}
A census with well-understood selection effects enables one to reconstruct the underlying population; a good model for
the underlying population allows one to discern the mechanisms underpinning its formation. Our goal in this section is
to briefly review the essential statistical underpinnings required to perform these two steps. 

We follow the population-likelihood notation of Qazalbash, Zeeshan and
O'Shaughnessy~\cite{gwastro-mergers-zeeshan-gwkokab}: $\Lambda$ denotes the
population parameters, $\ell_i$ an individual system's likelihood, and
$\mu$ the expected number of detections; see  \cite{VitaleGerosaFarrTaylor2022} for more comprehensive discussion and references.

For population parameters $\Lambda$, the fraction of births in an element
$d\gamma$ is $p(\gamma\mid\Lambda)\,d\gamma$, normalized to unity.
An integral over $d\gamma$ includes sums over discrete explosion and
evolutionary branches. The states along the history follow from the binary
evolution.

We select a physical class $Y$, such as binaries that remain bound or merge
within a specified time. We use an indicator function $H_Y(\gamma)=1$ for histories in that class
and zero otherwise. A survey detects a system with probability
$p_{\rm det}(\gamma)$. The fraction of births that both reach $Y$ and
are detected is the expected value of $p_{\rm det}$ given $Y$:
\begin{equation}
 \epsilon_Y(\Lambda)=\int H_Y(\gamma)p_{\rm det}(\gamma)
                  p(\gamma\mid\Lambda)\,d\gamma.
 \label{eq:selected_kernel}
\end{equation}
Thus disruption, subsequent evolution, and discovery all contribute to
$\epsilon_Y$.

For the measured data $d_i$ of system $i$, let
$\ell_i(\gamma)=p(d_i,\mathrm{det}\mid\gamma)$.
This includes any relation between discovery and the reported measurements.
If discovery is determined by a cut on the data, then
$\ell_i(\gamma)=p(d_i\mid\gamma)$ for an observed system, and
\begin{equation}
 p_{\rm det}(\gamma)=\int_{\rm detected}p(d\mid\gamma)\,dd.
 \label{eq:joint_response}
\end{equation}
There is no additional detection factor in $\ell_i$ in this case.
If separate data determine discovery, one may instead write
$\ell_i=p_{\rm det}(\gamma)p(d_i\mid\gamma,\mathrm{det})$.
These expressions describe the same joint probability and prevent counting
selection twice.

Let $N_{\rm birth}$ be the expected number of progenitor binaries in the
formation history being modeled. This history includes the ages and delay
times relevant to the observed sample; it need not be confined to births
during the observing run. The expected number of detections is
$\mu(\Lambda)=N_{\rm birth}\epsilon_Y(\Lambda)$.
For independent births with Poisson counts, the log likelihood takes the
standard population form~\cite{gwastro-mergers-zeeshan-gwkokab,VitaleGerosaFarrTaylor2022},
up to a parameter-independent constant,
\begin{align}
 \ln\mathcal L&=-\mu(\Lambda)+N\ln N_{\rm birth}\nonumber\\
 &\quad+\sum_{i=1}^N\ln\!\int H_Y(\gamma)\ell_i(\gamma)
                    p(\gamma\mid\Lambda)\,d\gamma,
 \label{eq:poisson_catalog}
\end{align}
This likelihood compares both the number of detections and their measured
properties. Each integral averages one system's measurement likelihood over
the predicted binary population.
If we use only the distribution of properties at fixed observed number $N$,
the likelihood becomes~\cite{LIGO-P1600187-Farr-SelectionBiasaAndMonteCarlo,VitaleGerosaFarrTaylor2022}
\begin{equation}
 \mathcal L_N(\Lambda)=\prod_{i=1}^N
 \frac{\int H_Y(\gamma)\ell_i(\gamma)p(\gamma\mid\Lambda)\,d\gamma}
      {\epsilon_Y(\Lambda)}.
 \label{eq:catalog_likelihood}
\end{equation}
Separate
catalogs generally have different detection probabilities; a system present
in more than one catalog must be included through its joint measurements.
Unknown progenitor distributions and evolutionary channel fractions enter
$p(\gamma\mid\Lambda)$ and must be constrained along with the kick distribution.

For either explosion, let $w$ be its kick magnitude. The kick distribution
among the observed binaries is
\begin{equation}
 p(w\mid Y,\mathrm{det},\Lambda)=
 \frac{p(w\mid\Lambda)P(Y,\mathrm{det}\mid w,\Lambda)}
      {\epsilon_Y(\Lambda)}.
 \label{eq:selected_kicks}
\end{equation}
The survival and detection probability at fixed $w$ is averaged over the
other progenitor and explosion variables. A broad birth distribution can
therefore produce a low-kick binary sample. Conversely, an error in this
probability can be absorbed into the inferred birth distribution.

\subsubsection{Evolutionary conditions and observable populations}\label{sec:htransform}\label{sec:tiers}\label{sec:demo:sel}

For simplicity, we characterize merging binaries using a well-established approximation: any compact binary formed after
a burst of star formation whose
gravitational-wave merger
time is less than a Hubble time is assumed to merge.  We specifically do not perform the technically more complex task
of averaging over all past Milky Way star formation in time and metallicity; see, e.g., \cite{PSellipticals,Broekgaarden2022Rates,popsyn-gwastro-STInterpFinal-Vera2023,gwastro-PopulationReconstruct-Hierarchical-PhysicalParam1-Clusters}
for examples of this calculation. 

For a detected population, write $p_{\rm det}(\gamma)$ for the
survey detection probability and $p(d\mid\gamma,\mathrm{det})$ for the
measurement distribution among detected systems:
\begin{align}
 p(d\mid Y,\mathrm{det},\Lambda)
 &=\frac{1}{\epsilon_Y(\Lambda)}\int H_Y(\gamma)p_{\rm det}(\gamma)
 \nonumber\\
 &\quad\times p(d\mid\gamma,\mathrm{det})p(\gamma\mid\Lambda)\,d\gamma.
 \label{eq:observed}
\end{align}
This is the one-system probability used in
Eq.~\eqref{eq:catalog_likelihood}.
Evolutionary selection ($H_Y$) and survey detection ($p_{\rm det}$) are separate
factors.
We evaluate the population integrals by numerical quadrature and use AD
in \ad{} for the endpoint likelihood derivatives. The joint radio
forecast of Sec.~\ref{sec:joint_radio_results} instead evaluates a
likelihood grid using scrambled Sobol integration. \cosmic{} supplies
the pre-supernova catalog and the binary histories we compare against.

\subsubsection{An instrument-based pulsar survey model}
\label{sec:instrument_pulsar_survey}
For radio pulsars, $p_{\rm det}$ is an instrument-based survey model
following \citet{Chunglee-nsns-1,Pol2019DNSRate}. It includes Galactic
propagation, sky noise, pulse broadening and the sensitivity of five
archived surveys, so single and binary pulsars pass through one discovery
calculation. Appendix~\ref{app:native_survey_checks} specifies the model,
reproduces archived survey simulations with it, and recovers a five-class
DNS merger rate.

\subsection{Two-supernova endpoint likelihood}
\label{sec:configurationmethod}
We combine the supernova model of Appendix~\ref{app:sn_model} with the
inter-supernova evolution of Appendix~\ref{app:physics}, starting from
uncertain configurations immediately before SN1. The inferred parameters
in the kick-amplitude and common-envelope experiment are
\begin{equation}
 \theta_k=\ln s_k,
 \qquad \theta_{\rm CE}=\ln\alpha_{\rm CE}.
 \label{eq:two_sn_parameters}
\end{equation}
Here $s_k$ multiplies every natal-kick speed of \cosmic{}'s prescription
(Appendix~\ref{sec:cosmic_engine}), so $\theta_k=0$ is \cosmic{}'s kick distribution.
The kick scale acts at both explosions; $\alpha_{\rm CE}$ controls
the intervening common envelope. Other prescription coefficients remain fixed.
For a physical kick coordinate, the ordinary core-collapse component has
median $\widetilde w_{\rm CC}=e^{5.60}s_k\,\mathrm{km\,s^{-1}}$;
the reduced-kick channel shares the multiplier $s_k$.

Using the states of Eq.~\eqref{eq:trajectory}, draw the pre-SN1 state
$x_0$ from the progenitor density $p(x_0)$. At each explosion,
$p(s_1,\mathbf w_1\mid x_0,\boldsymbol\theta)$ and
$p(s_2,\mathbf w_2\mid x_1,\boldsymbol\theta)$ describe the joint
probability of the kick and post-SN state. These joint distributions retain
the engine's correlations between remnant mass and kick. Between explosions,
$p(x_1,\mathrm{SN2}\mid s_1,\boldsymbol\theta)$ describes the probability
of reaching the next collapse in state $x_1$. Its integral is the probability
of reaching SN2, including losses through disruption or coalescence.
The final-state density for bound double neutron stars is therefore
\begin{align}
 p(s_2,\mathrm{DNS}\mid\boldsymbol\theta)
 &=H_{\rm DNS}(s_2)\int p(x_0)\nonumber\\
 &\quad\times p(s_1,\mathbf w_1\mid x_0,\boldsymbol\theta)\nonumber\\
 &\quad\times p(x_1,\mathrm{SN2}\mid s_1,\boldsymbol\theta)\nonumber\\
 &\quad\times p(s_2,\mathbf w_2\mid x_1,\boldsymbol\theta)\nonumber\\
 &\quad\times dx_0\,ds_1\,dx_1\,d^3\mathbf w_1\,d^3\mathbf w_2.
 \label{eq:two_sn_weights}
\end{align}
Here $H_{\rm DNS}(s_2)$ requires a bound orbit and two NS remnants,
including the first remnant's evolution and accretion before SN2.
The conditional probabilities marginalize over uncertain stellar structure,
explosion branches and evolutionary channels; integrals include sums over
discrete alternatives. Deterministic evolution is represented by delta
functions in these transition densities. Survival enters through this
sequence and the final-state condition, with no additional survival factor.

A DNS is seen as a pulsar binary until its GW merger, or until both pulsars
cross the radio death line $B/P^2=0.17\times10^{12}\,{\rm G\,s^{-2}}$~\cite{gwastro-BhattacharyaDecayMagneticFields1992}.
We use \cosmic{}-based spin-down, magnetic-field decay and accretion
prescriptions to calculate the radio lifetimes of both remnants
(Appendix~\ref{app:physics}). The first-born
NS can be recycled before SN2; case-BB transfer occurs during the
helium star's final \PsrCaseBbAccMyr\,Myr. The second-born has the adopted birth
spin and field distributions. For constant DNS formation over a Galactic
age $T_{\rm gal}=\PsrGalaxyAgeGyr\,$Gyr, the visible fraction is
\begin{equation}
 V(s_2)=\frac{\left\langle\min\left[t_{\rm GW}(s_2),\,T_{\rm gal},\,\max(\tau_1,\tau_2)\right]\right\rangle}{T_{\rm gal}},
 \label{eq:visible_fraction}
\end{equation}
where $\tau_1,\tau_2$ are the radio lifetimes after SN2,
$t_{\rm GW}$ is the birth-orbit merger time~\cite{Peters:1964}, and the
average is over pulsar birth properties. This endpoint experiment assumes
certain discovery while a system is visible. It neglects beaming,
spin-dependent luminosity and orbital decay during visibility; it does
not use the instrument-based survey model above.

The observed DNS fraction and the likelihood for a catalog of
fixed size are
\begin{align}
 \epsilon_{\rm DNS}(\boldsymbol\theta)
 &=\int V(s_2)\,p(s_2,\mathrm{DNS}\mid\boldsymbol\theta)\,ds_2,\nonumber\\
 K_i(\boldsymbol\theta)
 &=\int p(d_i\mid s_2)\,V(s_2)\,p(s_2,\mathrm{DNS}\mid\boldsymbol\theta)\,ds_2,\nonumber\\
 \ln\mathcal L_N&=\sum_i\ln K_i-N\ln\epsilon_{\rm DNS}.
 \label{eq:two_sn_likelihood}
\end{align}
The measurement likelihood $p(d_i\mid s_2)$ uses the final period,
eccentricity and component masses, including the first remnant's mass after
accretion. The same calculation gives the progenitor posterior,
\begin{equation}
 p(x_0\mid d_i,\mathrm{DNS},\boldsymbol\theta)=
 \frac{p(x_0)\,p(d_i,\mathrm{DNS}\mid x_0,\boldsymbol\theta)}
      {K_i(\boldsymbol\theta)},
 \label{eq:configurationposterior}
\end{equation}
where the conditional likelihood in the numerator follows the same
evolution from a fixed $x_0$. Progenitors are weighted by both their prior
probability and their probability of producing the measured DNS.

Forward-mode AD differentiates both terms of
Eq.~\eqref{eq:two_sn_likelihood} through both explosions and the
intervening evolution. A conditional last-SN calculation instead holds
the entering progenitor population fixed; it cannot propagate a changed
shared prescription through SN1. Numerical checks and uncertainty
calculations are given in Appendix~\ref{app:two_sn_numerics}.

\subsection{Fisher matrix and local parameter biases}
\label{sec:fisher}
The Fisher matrix describes the local sensitivity of a likelihood to its
parameters. For the fixed-count DNS likelihood of
Eq.~\eqref{eq:two_sn_likelihood}, define the normalized density of a selected
observation as $p_i(\boldsymbol\theta)=K_i/\epsilon_{\rm DNS}$ and write
$\ell=\ln\mathcal L_N$. With $\partial_a=\partial/\partial\theta_a$, the
log-likelihood gradient is
\begin{equation}
 g_a\equiv\partial_a\ell
 =\sum_i\partial_a\ln K_i-N\partial_a\ln\epsilon_{\rm DNS}.
 \label{eq:two_sn_information}
\end{equation}
Both terms must be differentiated: changing the kick or evolution
prescription changes the observed properties and the probability of
entering the catalog.

For independent observations from the selected model, the Fisher matrix is
\begin{align}
 (\Gamma_N)_{ab}
 &=\left\langle\partial_a\ell\,\partial_b\ell\right\rangle
 =-\left\langle\partial_a\partial_b\ell\right\rangle,\nonumber\\
 \widehat\Gamma_{N,ab}
 &=\sum_i\partial_a\ln p_i\,\partial_b\ln p_i.
 \label{eq:two_sn_fisher}
\end{align}
The second line is the empirical estimate used here. This is the population
counterpart of the familiar GW expression
$\Gamma_{ab}=(\partial_a h\mid\partial_b h)$ for a waveform $h$ in
Gaussian detector noise~\cite{CutlerFlanagan1994,GairAntonelliBarbieri2023}.
In a locally Gaussian likelihood with weak prior variation, the parameter
covariance is approximately $\Gamma_N^{-1}$ and the marginalized width is
$\sigma_a\simeq\sqrt{(\Gamma_N^{-1})_{aa}}$; finite-catalog and
non-Gaussian effects can invalidate this approximation~\cite{Vallisneri2008}.

Expanding about $\boldsymbol\theta_0$, a local estimate is
$\hat{\boldsymbol\theta}\simeq\boldsymbol\theta_0+
\widehat\Gamma_N^{-1}\boldsymbol\nabla\ell(\boldsymbol\theta_0)$.
If an incorrect model changes the objective to
$\widetilde\ell=\ell+\delta\ell$, its additional displacement is, to
first order,
\begin{equation}
 \Delta\theta_a^{\rm sys}\simeq
 \sum_b(\Gamma_N^{-1})_{ab}\,
 \partial_b\delta\ell(\boldsymbol\theta_0).
 \label{eq:fisher_bias}
\end{equation}
This linear-response estimate is analogous to the systematic-error
calculation in GW waveform inference~\cite{CutlerVallisneri2007,VallisneriYunes2013}.
Here $\delta\ell=N\ln S_1$ when only the denominator is conditioned on
first-SN survival (Sec.~\ref{sec:twosnresults}). A displacement outside the
locally Gaussian region indicates a bias direction; its magnitude does not
predict the maximum of the incorrect likelihood. Numerical evaluation is
described in Appendix~\ref{app:two_sn_numerics}.

\subsection{The synthetic catalogs}
\label{sec:populationmethods}
We generate binaries immediately before their first supernova and ask what
we could learn from the final double neutron stars. The starting masses
and orbits follow two families calibrated against \cosmic{}: one has
already passed through a common envelope, the other has not. We keep
their relative weights fixed. A route in which the companion is already
a helium star at SN1 lies outside the model; it supplies about half the
pre-SN2 arrivals in the reference archive.

Both explosions use the fiducial kick prescription, with unit inter-SN
CE efficiency. A binary must survive SN1, reach SN2 without coalescing,
and emerge as a bound pair of neutron stars. We weight each pair by the
fraction of time before merger for which at least one pulsar is radio
active, assuming certain discovery during that interval. The 640 selected
systems supply noisy periods, eccentricities and masses. The fit receives
neither the individual progenitors and kicks nor the identity of the
first-born star among the measured masses.

The conditional SN2 tests instead supply a population already at the second
explosion. Their methods and orbit-only result are presented together in
Appendix~\ref{app:conditional_sn2_experiments}. The observed field-DNS
catalog is used only in the supplemental low-kick-mixture illustration,
which also supplies its synthetic measurement-error templates.

\subsubsection{Two-SN synthetic endpoints}
\label{sec:two_sn_data}
For the evolving calculation, the starting prior is a mixture of two
classes,
\begin{equation}
 \pi(x)=\sum_{k\in\{{\rm W},{\rm C}\}}\omega_k\,\pi_k(x).
 \label{eq:two_sn_prior}
\end{equation}
Class W has no CE before SN1 and class C had one. Each $\pi_k$ is uniform
in $M_{{\rm CO},1}$, $M_{2,0}$ and $\ln a_1$ over a box aligned with the
trends of its class (Appendix~\ref{app:starting_prior_parameters},
Table~\ref{tab:two_sn_classes}). The companion mass
spans a fixed range. At each companion mass the CO core lies in a band
about a line in $M_{2,0}$, cut at the collapse floor $1.44\,\Msun$, and
$\log_{10}a_1$ lies in a band about a plane in $(M_{2,0},M_{{\rm CO},1})$.
\cosmic{}'s states at the CO floor form a separate atom. The boxes enclose
\ClassWEnclosedPercent\% and \ClassCEnclosedPercent\% of the \cosmic{}
pre-SN1 states of their classes, whether or not those reach SN2
(Fig.~\ref{fig:presn_input}, Appendix~\ref{app:coverage}).
The primary is hydrogen-free, $M_{1,0}=M_{{\rm CO},1}+M_{{\rm He},1}$, with
$M_{{\rm He},1}=a_k+b_kM_{{\rm CO},1}$ fitted to \cosmic{} in each box.
The companion starts at the median \cosmic{} age of its class, and
$\omega_k$ is the share of \cosmic{} prefixes in each box. We set
$M_{{\rm CO},2}^{\rm nom}=0.15M_{2,0}+0.6\,\Msun$ and
$M_{{\rm He},2}^{\rm nom}=0.5\,\Msun$, at $Z=0.002$.
A third route supplies \ClassDoubleArrivalSharePercent\% of \cosmic{}
arrivals: the companion is already a helium star at SN1. That state lies
outside the starting family of the map and is not included.
These priors specify a bounded formation family; they are not an inferred
Galactic birth distribution. We assume circular pre-SN1 orbits.

The injected prescription has $\boldsymbol\theta=(0,0)$: \cosmic{} kicks
and Mandel--M\"uller remnants (Appendix~\ref{sec:cosmic_engine}), with
the visibility weight of Eq.~\eqref{eq:visible_fraction}.
The main experiment uses 640 selected DNS endpoints from a
fine-quadrature injection. Recovery receives only the noisy final
coordinates $y=(\log_{10}[P/\mathrm{day}],e,m_1,m_2)$, with independent
standard deviations $(0.2,0.08,0.05\,\Msun,0.05\,\Msun)$.
Component labels are marginalized:
\begin{equation}
 p(d_i\mid y)=\tfrac12\sum_{\sigma\in\{\mathrm{id},\mathrm{swap}\}}
 \mathcal N_4(d_i;P_\sigma y,\Sigma_i).
 \label{eq:two_sn_measurement}
\end{equation}
Individual progenitors and kicks are withheld. Both injection and recovery
include the visibility weight of Eq.~\eqref{eq:visible_fraction} and
coalescence before SN2. The selected count is fixed.

Injection and recovery use different integration rules. Recovery pools
\TwoSnFitReplicas{} independently scrambled Sobol rules~\cite{Sobol1967DistributionPointsCube,Owen1995RandomlyPermutedNets,Matousek1998L2DiscrepancyAnchoredBoxes,gwastro-VirtanenScipyFundamentalAlgorithms2020}.
Their scatter corresponds to estimated errors of
\TwoSnFitMeanErrKick{} and \TwoSnFitMeanErrCe{} statistical widths in
the pooled local kick and CE estimates. The injection discretization shifts the local kick estimate by
\TwoSnOrderInjShiftKick{} statistical widths, comparable to the
\result{TwoSnLocalOffsetKick}-width displacement from the injection. These checks qualify a local
recovery test, rather than a converged precision measurement; the
numerical appendix records the rules,
refinements and matched-node controls.

The matched normalization-error control uses exactly the same 640-endpoint
catalog and forward map. Let $S_1(\boldsymbol\theta)$ be the birth-prior
probability of remaining bound immediately after SN1. We replace the
correct normalizer $\epsilon_{\rm DNS}$ by
$\epsilon_{\rm DNS}/S_1$, while leaving $K_i$ unchanged. This intentionally
conditions the denominator on first-SN survival without consistently
conditioning the numerator. We evaluate the added log-likelihood
gradient at the correctly normalized maximum and check the inconsistent
likelihood directly at selected displaced points (Sec.~\ref{sec:twosnresults}).

\subsubsection{A joint radio-selected sample}
\label{sec:joint_radio_methods}
We also construct a controlled joint sample of young isolated pulsars and
surviving DNSs. Its birth distribution is a fixed, weighted mixture of
18 prospective configurations from the W/C family above.
We restrict the histories to two NS-forming collapses and require the
first remnant to remain an NS during the intervening binary evolution.

The young isolated population includes three origins. SN1 disruption
releases the newborn primary NS; the escaped companion can subsequently
form another NS as a single star, combining its ejection velocity with an
independent natal kick. A system that survives SN1 and the
intervening evolution can instead release its newly formed second NS
when SN2 disrupts the binary. We sum these three contributions before
constructing the observed young distribution. First-born NSs released at
SN2 form a separate cohort and are excluded here, as are merger remnants.
We assume all stellar systems form in binaries.

Both observed classes pass through the five-survey radio response of
Sec.~\ref{sec:instrument_pulsar_survey}. We integrate young ages from birth
to $10\,\mathrm{Myr}$ and DNS residence from SN2 to $\min(100\,\mathrm{Myr},t_{\rm merge})$.
DNS discovery uses a generic recycled pulsar and a circular-orbit
search-loss approximation, even though we evolve an eccentric orbit to
track the system state. Astrometric follow-up has a flux-dependent probability.
Successful measurements have errors $0.2\,\mathrm{mas\,yr^{-1}}$ per
proper-motion component and $\sigma_{\ln d}=0.2$. We assume synthetic DNS
period and eccentricity measurements with errors of 0.2 dex and 0.08,
respectively. These errors vastly exceed those usually achieved with
pulsar timing; we adopt them to be conservative.

Let $A_{cj}(\boldsymbol\theta)$ be the expected selected number in
measurement cell $j$ of class $c$ per birth, summed over all latent
progenitors and origins. For independent samples with a fixed count in each class,
\begin{equation}
 \ell_{\rm joint}=\sum_{c\in\{\mathrm{iso},\mathrm{DNS}\}}\sum_j
 n_{cj}\ln p_{cj},\qquad
 p_{cj}=\frac{A_{cj}}{\sum_k A_{ck}}.
 \label{eq:joint_radio_shapes}
\end{equation}
The chosen class sizes do not constrain their relative abundance.

\section{Results: measurement capability and selection bias}\label{sec:demo}
\subsection{Surviving double neutron stars}\label{sec:twosnresults}

\begin{figure*}[t]
\centering
\includegraphics[width=\textwidth]{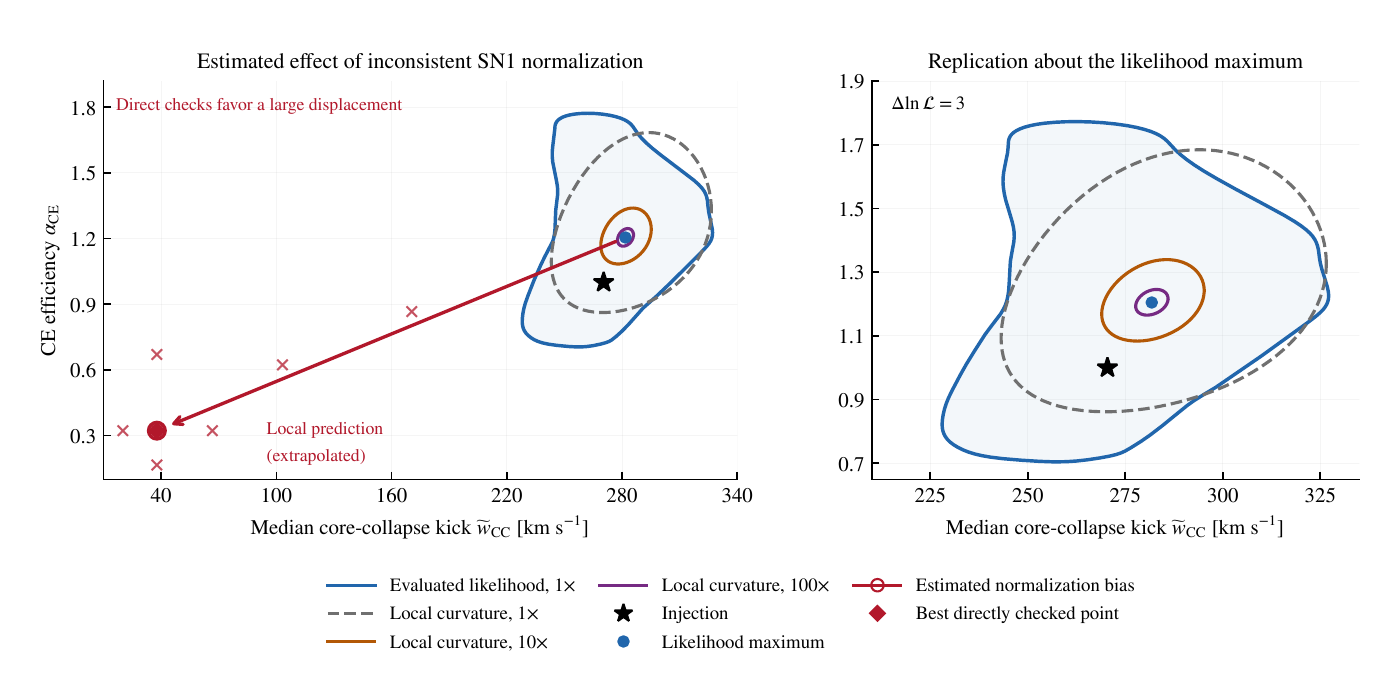}
\caption{\textbf{DNS likelihood, replication scaling and survival-normalization bias.}
One synthetic catalog of 640 radio-selected DNSs under the model of
Sec.~\ref{sec:joint_radio_methods}. Stars mark the injection; blue points
mark the interpolated likelihood maximum. The blue curve interpolates the
evaluated likelihood grid at $\Delta\ln\mathcal L=3$. Gray, orange and purple
curves use the Fisher matrix at that maximum for 1, 10 and 100
replicates of each observation, respectively, at the same likelihood drop.
Their widths in logarithmic parameters decrease as $r^{-1/2}$; the narrow contours illustrate local
scaling rather than independently resolved likelihood scans. Left: the
red circle and arrow show the extrapolated local bias from inconsistent
SN1 survival normalization. Red crosses mark additional direct likelihood
evaluations. The diamond is the best of eight directly checked points,
coincident with the local prediction, and improves the incorrect log
likelihood by 499 relative to its value at the correct maximum. These
checks do not locate the global maximum of the incorrect objective. Right: a closer view
of the correctly normalized likelihood and replication contours.\llmfig}
\label{fig:twosnlikelihood}
\end{figure*}

Figure~\ref{fig:twosnlikelihood} shows the results derived from a synthetic population of 640 identified surviving
binary neutron stars, in a synthetic Milky Way galaxy with uniform formation rate versus time. 
We generate our synthetic population using a specific kick strength and $\alpha_{CE}$ (indicated by a star).  We then
construct a two-dimensional likleihood grid in these two parameters, evaluating the observational likelihood ${\cal L}$;
due to random realization error (and quadrature error), the best-fitting parameters correspond to the blue point,
slightly offset from the injection.  Dashed gray, solid orange, and solid purple curves show Fisher isocontours, derived
to roughly correspond to  estimated credible intervals derived using this survey; a survey recovering ten times more
binary neutron stars; and a survey recovering 100 times more.  [In practice, we are simply  replicating each synthetic observation
  10 or 100 times more in our likelihood.]
Our calculation marginalizes over event rate and thus does not use the overall number of events to constrain these two
parameters, enabling the simple scaling described above. 
The likelihood contours are nearly gaussian, and well-described by the Fisher matrix approximation.

To estimate the effect of neglecting the selection bias, we use the Fisher matrix and gradient of the neglected likelihood factor,
evaluating both at the best-fitting parameters
$\hat{\boldsymbol\theta}$,
\begin{equation}
 \Delta\boldsymbol\theta\simeq\widehat\Gamma_N^{-1}\boldsymbol g,
 \label{eq:two_sn_local_step}
\end{equation}
with $\boldsymbol g=N\nabla\ln S_1$ and the Fisher matrix
$\widehat\Gamma_N$ estimated from the log-likelihood curvature.
In this expression, the bias will not scale with the number of events, but the width of the posterior
will. Specifically, in our simple data-replication method of increasing the data sample by $r$ duplicates, the  gradient and the Fisher
matrix also scale by  $r$:
\begin{equation}
 \ell_r=r\ell_1,\quad\boldsymbol g_r=r\boldsymbol g_1,\quad\widehat\Gamma_{rN}=r\widehat\Gamma_N.
 \label{eq:replicated_fisher}
\end{equation}
The local estimate is unchanged and the Fisher widths decrease as
$r^{-1/2}$.

\subsection{Joint constraints from isolated and binary pulsars}
\label{sec:joint_radio_results}
\begin{figure*}[t]
\centering
\includegraphics[width=\textwidth]{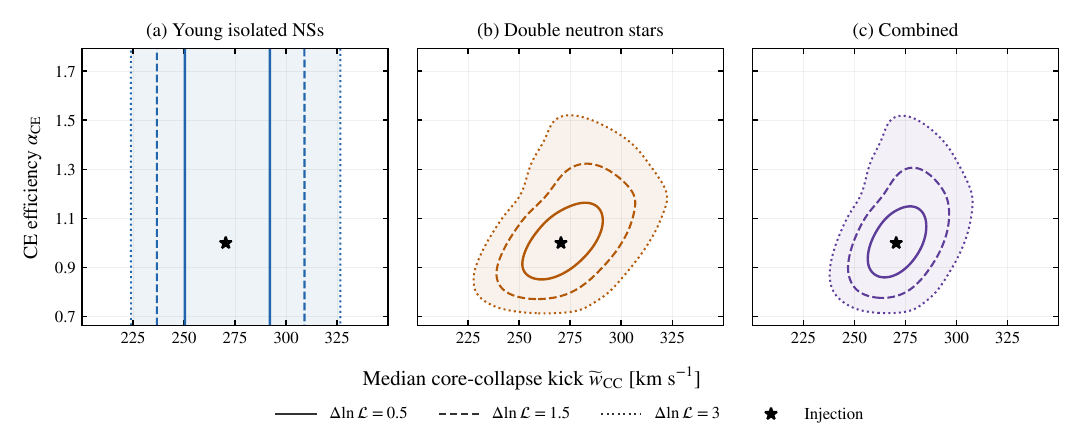}
\caption{\textbf{Joint inference from radio-selected isolated and binary NSs.}
Expected-data likelihoods for 3840 young isolated detections (about 217
with astrometry) and 640 DNS detections from the finite birth population
of Sec.~\ref{sec:joint_radio_methods}. Solid, dashed and dotted curves
show $\Delta\ln\mathcal L=0.5,1.5,3$, respectively, relative to each panel's
maximum; curves interpolate the evaluated likelihood grid. Stars mark
the injected median core-collapse kick and CE efficiency. All three
isolated origins are pooled, and each class likelihood is normalized by
its own selected intensity before the two factors are multiplied.
The isolated pulsars constrain the kick scale, while the DNS orbits
constrain the linked binary evolution.\llmfig}
\label{fig:joint_radio}
\end{figure*}

Figure~\ref{fig:joint_radio} combines two radio-selected populations under
the same kick and binary-evolution prescription. We use expected cell
counts $n_{cj}=N_c p^{\rm inj}_{cj}(\boldsymbol\theta_0)$ for $N_{\rm iso}=3840$ young isolated detections and
$N_{\rm DNS}=640$ DNS detections, generated from the same binary-evolution
assumptions as above.
These are separate, fixed-size survey designs. The DNS sample uses the
same radio response and period--eccentricity measurements as
Fig.~\ref{fig:twosnlikelihood}, using expected counts rather than one random catalog.

At the injection, 49.37\% of selected young pulsars are primary NSs
released by SN1, 50.46\% form in the later collapse of companions that
escaped at SN1, and 0.17\% are newly formed NSs released by SN2 disruption.
Thus 99.83\% arise from SN1-disrupted systems, although roughly half of
the NSs themselves are born in the companion's later explosion. The birth of either star
can therefore contribute; the likelihood receives no origin labels.
The dominant routes leave the binary before the modeled inter-SN CE,
explaining the weak CE dependence of the young likelihood at fixed
pre-SN1 population. This does not remove the earlier CE history of class C.

The isolated sample primarily constrains the kick scale.
At fixed CE efficiency, independent detections supply local information
\begin{align}
 I_{\rm iso}&=\sum_j\frac{1}{p_{{\rm iso},j}}
 \left(\frac{\partial p_{{\rm iso},j}}{\partial\ln\widetilde w_{\rm CC}}\right)^2,
 \nonumber\\
 \sigma_{\ln\widetilde w_{\rm CC}}&\simeq(N_{\rm iso}I_{\rm iso})^{-1/2}
 \simeq0.077\sqrt{\frac{3840}{N_{\rm iso}}}.
 \label{eq:isolated_kick_precision}
\end{align}
At injection, $I_{\rm iso}\simeq0.044$ per radio detection. For
$N_{\rm iso}=3840$ the local standard deviation of the median kick is
$21\,\mathrm{km\,s^{-1}}$. Only about 217 of the
3840 detections have astrometry on average, and they supply nearly all
the information. Intrinsic velocity dispersion, projection, binary recoil
and measurement noise set the coefficient; independent measurement errors
preserve the $N^{-1/2}$ scaling.

The combined expected-data likelihood reduces the CE--kick degeneracy.
Its $\Delta\ln\mathcal L=0.5$ profiled kick interval is about one-quarter
narrower than for DNSs alone.

\section{Conclusions}\label{sec:conclude}
In this work, we implement binary evolution and pulsar survey selection within the same autodifferentiable framework.
We apply this framework to demonstrate that an interpretation of binary pulsar surveys can and should account for both
survey and intrinsic selection effects, to correctly recover the NS natal kick distribution.
Within the DNS-only synthetic experiment of Sec.~\ref{sec:twosnresults},
an inconsistent survival normalization strongly favors parameters far from
the correctly normalized likelihood maximum.
The technology demonstrated here provides a more accessible method that generalizes efficiently to new binary-evolution parameters, propagating their effects through both the evolving population and its selection function. This capability extends earlier population-inference treatments of binary survival and pulsar survey selection~\cite{Belczynski2010DRP,2024arXiv231214848G,PardoAraujo2025FluxSBI}.

The joint radio example demonstrates how isolated and binary NSs constrain a shared prescription.
Our proof-of-concept investigation provides a framework that can be generalized to multi-class observations: both  other NS populations and observation
modalities  (e.g., NS in
other binaries; isolated NS; and NS found by other surveys such as Gaia) and even non-NS-containing binaries which
undergo closely related physical processes.  As a result, this work should provide a valuable starting point for future
analyses of the outcomes of isolated binary evolution.

\begin{acknowledgments}
This work used the companion remnant prescriptions~\cite{ScorpioInPrep} and pre-supernova catalog,
\cosmic{}~\cite{2020ApJ...898...71B}, and automatic differentiation for numerical
evaluation.
ROS acknowledges support from NSF PHY 2012057, 2309172 and 2206321.
ROS also acknowledges assistance from LLM agents during the implementation,
operation, and drafting of the analyses in this manuscript, as described in
Appendix~\ref{app:llm}.
\end{acknowledgments}
\appendix
\section{Binary supernova modeling}\label{app:sn_model}
\subsection{Collapse physics and orbital response}\label{sec:method:notation}\label{sec:engine}
For completeness and to establish notation, we review the well-known
response of a binary orbit to sudden mass loss and a natal
kick~\cite{1961BAN....15..265B,1983ApJ...267..322H,1995MNRAS.274..461B,1996ApJ...471..352K,2000ApJ...541..319K};
see, e.g., Refs.~\cite{2006csxs.book..623T,2014LRR....17....3P,2023pbse.book.....T}
for reviews.
At each explosion the binary state contains the exploding-star mass $M_1$,
companion mass $M_2$, semimajor axis $a$, eccentricity $e$, and stellar-core
properties. The remnant mass is $M_1'=M_1-\Delta M_1$ and its natal-kick vector
is $\mathbf w$, with magnitude $w$. We write $\Mb=M_1+M_2$,
$\Mb'=M_1'+M_2$, and $\beta=\Mb'/\Mb$. The relative position and velocity
of the exploding star are $\mathbf r$ and $\mathbf v$. Spins initially aligned
with the orbital angular momentum remain fixed during the impulse; we do not
include collapse-induced spin torques.

Let $b$ denote an explosion branch and $\xi$ a latent variable describing
core-to-core scatter. A prescription with parameters $\Lambda$ defines
\begin{equation}
 p_\Lambda(b,\xi\mid x),\qquad
 M_1'=m_\Lambda(x,b,\xi),\qquad
 w=w_\Lambda(x,b,\xi).
 \label{eq:remnantlaw}
\end{equation}
A collapse model can have successful-explosion, fallback, and
direct-collapse branches. Branch probabilities, retained mass, and kick
magnitude enter the orbital integrals together, which retains the
correlation between mass loss and recoil.

For the trajectory in Eq.~\eqref{eq:trajectory}, the probability density
$p(\gamma\mid\Lambda)$ combines progenitor properties, explosion
probabilities, core scatter, orbital phase and kick directions. A single
collapse with an already compact companion is a special case. The notation
$M_1,M_2$ in each instantaneous map refers to the exploding star and its
companion at that event; at SN2 in the detached example these masses are
$M_2,M_1'$, respectively.
\label{sec:kin}

\subsection{The instantaneous supernova--kick map}
\label{sec:kin:map}
The explosion occurs uniformly in time, so its eccentric anomaly $E$ has
measure $(1-e\cos E)\,dE/2\pi$ and sets $\rbf$ and $\vbf$~\cite{2002MNRAS.329..897H}.
A circular orbit has $r=a$ and $v=v_{\rm orb}\equiv\sqrt{G\Mb/a}$.
The kick adds to the relative velocity only,
$\vbf'=\vbf+\wbf$, and the post-supernova orbit follows from the specific energy and the
Laplace--Runge--Lenz (LRL) vector~\cite{2002ApJ...573..283P,1983ApJ...267..322H}:
\begin{equation}
  \eps'=\tfrac12 |\vbf'|^2-\frac{G\Mb'}{r},\quad
  \hbf'=\rbf\times\vbf',\quad
  \ebf'=\frac{\vbf'\times\hbf'}{G\Mb'}-\frac{\rbf}{r},
  \label{eq:lrl}
\end{equation}
with $a'=-G\Mb'/2\eps'$, $e'=|\ebf'|$, and $h'^2=G\Mb' a'(1-e'^2)$. These expressions are rational
in the inputs except for the final norm, so the orbital response and its
derivatives follow directly. The system is unbound if and only if $\eps'\ge0$.
The systemic velocity of the surviving binary
is~\cite{2002MNRAS.329..897H,1995MNRAS.274..461B}
\begin{equation}
  \vsys=\frac{M_1'}{\Mb'}\,\wbf-\frac{\Delta M_1\,M_2}{\Mb'\,\Mb}\,\vbf.
  \label{eq:vsys}
\end{equation}

\subsection{Survival probability in closed form}
\label{sec:kin:pbound}
We integrate the isotropic kick direction analytically where the endpoint allows it. At fixed phase, writing $\vartheta$ for the angle between $\wbf$ and $\vbf$ (so
$\cos\vartheta\sim U(-1,1)$), one has $|\vbf'|^2=v^2+\vk^2+2v\vk\cos\vartheta$ and hence
\begin{align}
  \Pbound(\vk\,|\,r,v)&=\tfrac12\Big[1+\mathrm{clip}_{[-1,1]}(\hat c)\Big],\nonumber\\
  \hat c&\equiv\frac{2G\Mb'/r-v^2-\vk^2}{2v\vk},
  \label{eq:pbound_phase}
\end{align}
equivalent in the circular case to Brandt \& Podsiadlowski~\cite{1995MNRAS.274..461B}.
For an eccentric orbit the phase-marginal survival probability is a single
quadrature in $E$. For spin tilts we also keep the
kick azimuth, which sets the new orbital plane.

\subsection{Orbital tilt in a single collapse}\label{sec:tiltmethod}
For a circular pre-collapse orbit, write $u=w/v_{\rm orb}$, $\zeta=\cos\vartheta$
for the kick angle relative to $\mathbf v$, and $\phi$ for its azimuth measured
from the initial angular momentum. The orbital-plane tilt
is~\cite{Dossena2026Tilts}
\begin{equation}
 z\equiv\cos\gamma=
 \frac{1+u\zeta}{\sqrt{(1+u\zeta)^2+u^2(1-\zeta^2)\cos^2\phi}}.
 \label{eq:tilt}
\end{equation}
With initially aligned spins and no impulsive spin torque, this is also the
spin--orbit tilt immediately after the collapse. An orbit density expressed
only in $(a',e')$ has discarded this orientation information, so we retain
$\mathbf h'$. At $u<1$ the orbital plane cannot reverse; at $u>1$ reversal
occurs for $\zeta<-1/u$. We write $B_1$ for survival of the collapse. In a two-collapse model both spins and the
orbital basis must be propagated through the intervening evolution and second
impulse.

\subsection{Kick orientation and survival selection}
\label{sec:orientationmethod}
Orientation-dependent survival need not imply that the natal kick-direction distribution
is measurable from surviving binaries. For a circular pre-collapse orbit,
let $\mu=\mathbf w\cdot\hat{\mathbf h}/w$ be the kick cosine relative to
the initial orbital angular momentum. This is also the spin-axis cosine
only under our initial-alignment assumption. At fixed $u=w/v_{\rm orb}$
and $\beta=\Mb'/\Mb$, a uniform azimuth gives
\begin{align}
 A(\mu;u,\beta)&=\tfrac12+
 \frac{1}{\pi}\arcsin\!\left[\mathrm{clip}_{[-1,1]}C\right],\nonumber\\
 C&=\frac{2\beta-1-u^2}{2u\sqrt{1-\mu^2}}.
 \label{eq:orientation_survival}
\end{align}
Here $A$ is the survival probability at that polar angle: the tangential
kick component is $w\sqrt{1-\mu^2}\sin\phi$, and survival requires
$1+u^2+2u\sqrt{1-\mu^2}\sin\phi<2\beta$.
At $|\mu|=1$ or $u=0$, the direct energy condition supplies the limit.
Consequently a birth density $q(\mu)$ enters a survivor sample as
\begin{equation}
 q(\mu\mid B_1)=\frac{q(\mu)A(\mu)}
 {\int_{-1}^{1}q(v)A(v)\,dv}.
 \label{eq:orientation_selected}
\end{equation}
This expression is conditional on the stated progenitor and speed;
a population calculation integrates those quantities as well.
Conditioning on the number of survivors removes the yield information.
An orientation-recovery experiment must therefore test the remaining
orbital information while allowing changes in kick magnitude and
progenitors that can mimic the survival response.

For the population recovery we use the sign-symmetric angular family
\begin{equation}
 q_\kappa(\mu)=\frac{e^{\kappa\mu^2}}
 {\int_{-1}^{1}e^{\kappa v^2}\,dv},
 \label{eq:orientation_population_law}
\end{equation}
with isotropy at $\kappa=0$ and polar preference at $\kappa>0$.
At fixed nuisance parameters, the log-likelihood derivative for a count-conditioned orbital
measurement is
\begin{equation}
 \partial_\kappa\ln p(d\mid B_1)
 =\mathbb E[\mu^2\mid d,B_1]-\mathbb E[\mu^2\mid B_1].
 \label{eq:orientation_orbit_derivative}
\end{equation}
Thus variation across measured orbits, rather than the survival fraction
alone, identifies the kick-direction distribution. We integrate over uncertain progenitor
states and kick speeds and fit their population parameters alongside
$\kappa$. We reuse the same quadrature nodes at each parameter value and
use importance weights to estimate the normalized orbital density and the
survival probability. Analytic weight derivatives and finite differences provide
an additional check independent of the trajectory AD calculation.

\subsection{Remnant and kick prescription in the two-SN experiment}\label{sec:cosmic_engine}
We use the kick prescription from the \cosmic{} comparison runs.
We change the remnant prescription so
that the model's DNS reproduce the Galactic NS masses. The \cosmic{} archive
uses the delayed prescription of Ref.~\cite{2012ApJ...749...91F}, in which
fallback onto the proto-NS grows with the CO-core mass above $3.5\,\Msun$.
Through our map and without the visibility weight, the 95th percentile of
its DNS neutron-star masses is $\TwoSnNsMassDelayedHi\,\Msun$, while
skew-normal fits to the observed DNS masses have a mean of 1.30 to
$1.41\,\Msun$ and a width of 0.13 to $0.15\,\Msun$ across the
equation-of-state models of Ref.~\cite{2025arXiv251212130A}. Those masses already include binary
evolution, so they test our channels as a whole rather than the natal map
alone. We adopt instead the prescription of
Ref.~\cite{2020MNRAS.499.3214M}, with the NS branch as implemented in
\cosmic{}. A CO core of
mass $M_{\rm CO}$ forms a black hole with probability
$(M_{\rm CO}-2\,\Msun)/(5\,\Msun)$, clipped to $[0,1]$. Otherwise the NS
baryonic mass is $1.2\,\Msun$ below $M_{\rm CO}=2\,\Msun$. It rises linearly
from 1.4 to $1.9\,\Msun$ between $M_{\rm CO}=2$ and $3\,\Msun$, then from 1.4
to $1.8\,\Msun$ between 3 and $7\,\Msun$. We keep the mean of this mass and omit
its scatter of $0.05\,\Msun$ ($0.02\,\Msun$ below $M_{\rm CO}=2\,\Msun$).
Every collapse on the black-hole branch forms a black hole, as in
Ref.~\cite{2020MNRAS.499.3214M}. \cosmic{} instead draws a baryonic mass
on that branch and, with its maximum NS mass of $3\,\Msun$, classifies the
lightest draws as neutron stars of 2.5 to $3\,\Msun$; we omit these. The baryonic mass is converted to gravitational
mass with at most $0.5\,\Msun$ lost. This prescription has no separate
electron-capture remnant.
With the visibility weight of Eq.~\eqref{eq:visible_fraction}, the
neutron stars of the selected DNS at $\boldsymbol\theta=0$ have mean
$\TwoSnNsMassMean\,\Msun$ and standard deviation $\TwoSnNsMassSd\,\Msun$,
with 5th, 50th and 95th percentiles of \TwoSnNsMassLo, \TwoSnNsMassMedian{}
and $\TwoSnNsMassHi\,\Msun$ (both components, without measurement noise).
The low end is the branch below $M_{\rm CO}=2\,\Msun$, slightly lighter than
the lightest precisely measured DNS component, $1.174\,\Msun$, in the
compilation of Ref.~\cite{2025arXiv251212130A}. Without the visibility weight
the 95th percentile is also $\TwoSnNsMassBirthHi\,\Msun$.
This prescription has no heavy tail. A neutron star is born with at most
$\simeq1.7\,\Msun$, from a baryonic mass of $1.9\,\Msun$; any heavier
neutron star in the model gained mass by accretion before SN2. The case-BB
cap keeps that gain small, so a system like GW190425, with a total mass near
$3.4\,\Msun$~\cite{2020ApJ...892L...3A}, needs two neutron stars near this
maximum.

Core-collapse kick speeds are log-normal,
$\ln(w/\mathrm{km\,s^{-1}})\sim\mathcal N(5.60,0.69^2)$, \cosmic{}'s
implementation of the fit of Ref.~\cite{2025ApJ...989L...8D}. A helium star of 1.6 to $2.25\,\Msun$
undergoes an electron-capture supernova with a Maxwellian kick of dispersion
$20\,\mathrm{km\,s^{-1}}$. These are \cosmic{}'s settings 5 for both
remnants and kicks. We replace each step in CO-core and helium-star mass by
a logistic of width $0.05\,\Msun$. The parameter $\theta_k$ of
Eq.~\eqref{eq:two_sn_parameters} scales every kick speed by $e^{\theta_k}$,
so $\theta_k=0$ is \cosmic{}'s kick distribution. Gauss--Hermite nodes in the latent
normal variable integrate the kick speed. In the \cosmic{} archive, NS masses
before SN2 equal the birth mass from the first collapse to within a median
\CosmicNsGainMedian$\,\Msun$ (\CosmicNsGainHi$\,\Msun$ at 95\%) for
helium-star--NS arrivals through an inter-SN CE, so accretion between the
explosions adds little to the mass.

The other experiments in this paper use the \scorpio{} engine, which differs
qualitatively. Its proto-NS mass follows a compactness that changes little
across these CO cores, and heavier remnants come only from a discrete
fallback branch. Its NS masses therefore cluster near two values instead of
the continuous spread of the \cosmic{} prescription. Its kick scales with the
ejecta mass and compactness, with amplitude coefficient $v_{\rm NS}$. The
two engines are interchangeable in our implementation.

\subsection{Second-collapse prescription in the two-SN experiment}
The nominal secondary core inputs are structural nuisance quantities of
this approximate evolution model. To account for shell removal before SN2,
we evaluate its explosion using
\begin{align}
 M_{{\rm CO},2}&=\min(M_{{\rm CO},2}^{\rm nom},M_{d,2}),\nonumber\\
 M_{{\rm He},2}&=\min\!\left[M_{{\rm He},2}^{\rm nom},
                     \max(M_{d,2}-M_{{\rm CO},2},0)\right].
 \label{eq:two_sn_enclosed_cores}
\end{align}
$M_{{\rm He},2}$ denotes the helium shell outside the CO core. This closure
prevents a removed shell from being counted as collapsing material; it
approximates structural response rather than predicting detailed core growth.
An exposed secondary can also be too light to undergo another collapse.
We adopt a regularized transition on the SSE core-mass scale,
\begin{align}
 u&=\mathrm{clip}_{[0,1]}\!\left[
       \frac{M_{{\rm CO},2}-1.44\,\Msun}{0.05\,\Msun}\right],\nonumber\\
 p_{{\rm CC},2}&=u^3(10-15u+6u^2).
 \label{eq:two_sn_collapse_fate}
\end{align}
Only the fraction $p_{{\rm CC},2}$ proceeds to SN2. The complement is an
absorbing non-collapse outcome and remains in the birth probability budget.
We check remnant masses against the evolved donor mass without clipping
remnants or silently discarding probability. SN2 is evaluated in a circular
orbit, which is appropriate for the interacting channel that supplies the
selected experiments below. For these non-spin endpoints we use one
birth spin/field node, $\log_{10}(P_0/{\rm s})=-0.7435$ and
$\log_{10}(B_0/{\rm G})=12.65$. The calculation carries scalar spin
evolution but does not use spin measurements or transport spin directions.

\section{Binary evolution between supernovae}\label{app:physics}
\subsection{Evolutionary channels}\label{sec:intersnphysics}
The kinematic map of Appendix~\ref{app:sn_model} is not enough on its own. The
second explosion acts on the state delivered by the first explosion and
subsequent stellar evolution. Its donor mass, core structure, orbital energy,
angular momentum, and companion mass cannot in general be varied independently.
We carry them together with the elapsed time $t$, accreted mass
$\Delta m_{\rm acc}$, and the compact companion's spin magnitude and magnetic
field. Winds, Roche-lobe overflow, common-envelope evolution, and collisions
separate post-SN1 binaries into different paths to SN2, disruption or
stellar coalescence~\cite{2002MNRAS.329..897H,2017ApJ...846..170T}.
A first-kick perturbation therefore changes both the distribution of orbits
within each channel and the probability of entering that channel.

For illustration, a fast isotropic donor wind with negligible capture obeys
$aM_{\rm b}=\mathrm{constant}$, with unchanged eccentricity. Tidal
circularization with conserved orbital angular momentum gives
$a_{\rm circ}=a(1-e^2)$. For a donor losing an envelope of mass
$M_{\rm env}$ down to core mass $M_c$, the usual common-envelope energy
prescription gives
\begin{equation}
 \frac{GM_dM_{\rm env}}{\lambda_{\rm env}R_d}=
 \alpha_{\rm CE}\left(\frac{GM_cm_a}{2a_f}
                         -\frac{GM_dm_a}{2a_i}\right).
 \label{eq:intersn_ce}
\end{equation}
Here $m_a$ is the compact accretor mass, $R_d$ the donor radius at contact,
and $\alpha_{\rm CE}$ and $\lambda_{\rm env}$ describe ejection efficiency
and envelope binding. The donor's evolutionary age sets
its remaining lifetime and when it fills its Roche lobe, and thus how much
mass the companion can accrete and how long it spins down. Contact must follow the
evolving companion's radius, mass and orbit, because it can lose mass before
first filling its Roche lobe. The spin and field prescriptions evolve only
the spin magnitude.

Let $R_k$ be the post-SN1 region following channel $k$ and
$T_k(s_1;\Lambda_{\rm bin})$ its map to the next state. Let $p(s_1\mid x_0,\Lambda)\,ds_1$ be the probability of a post-SN1 state
$s_1$ for an initial binary $x_0$. If $h_Y$ is the probability of reaching
class $Y$ after SN2, averaged over its explosion variables, then
\begin{align}
 Z_Y(x_0,\Lambda)&=\sum_k\int_{R_k(\Lambda_{\rm bin})}
 p(s_1\mid x_0,\Lambda)\nonumber\\
 &\quad\times h_Y[T_k(s_1;\Lambda_{\rm bin}),\Lambda]\,ds_1.
 \label{eq:regions}
\end{align}
Averaging this probability over the initial binaries gives the class
probability $Z_Y(\Lambda)$. Including discovery in $h_Y$ instead gives
the corresponding fraction detected.
Changing a supernova or binary-evolution parameter changes the post-SN1
state distribution, the subsequent evolution, and which binaries undergo
each interaction. Derivatives must include binaries crossing between
channels, as well as changes within a channel.
Our implementation integrates over these regions and their moving
boundaries, and returns the elapsed time, accreted mass, wind loss, spin and
field at SN2.
Table~\ref{tab:physics} lists the physics these channels carry, and a
companion paper~\cite{AdSnPopsynP3InPrep} gives their equations and
validation.
The conditional selection and orientation controls specify a pre-collapse
population or state to isolate prescription and measurement effects.
The two-SN likelihood in Sec.~\ref{sec:configurationmethod} instead
marginalizes the starting configuration and propagates the inter-SN map.

\subsection{Adopted evolution in the two-SN experiment}
We start immediately before the first collapse and evolve its surviving
binaries through the channels described above. Stellar
radii, lifetimes, contact ages and wind histories use the hydrogen-rich
and helium-star SSE formulae ported from \cosmic{}
\cite{2000MNRAS.315..543H,2002MNRAS.329..897H,2020ApJ...898...71B}.
The hydrogen-rich donor's transfer stability and common-envelope binding
energy are evaluated on its evolving track at contact. Common-envelope
survivors either remain detached or undergo stable case-BB transfer;
collisions, common-envelope mergers and accretion-induced collapse are
absorbing outcomes. The orbit, donor mass and accretor mass entering SN2
are calculated from this evolution. Recorded \cosmic{} pre-SN2 masses and
separations are not supplied to the fit.

We circularize the orbit at the post-SN1 periastron, $a(1-e)$, when the
companion fills its Roche lobe, and use the
$\alpha_{\rm CE}$ energy prescription of Eq.~\eqref{eq:intersn_ce}, with
track-dependent binding energy for the hydrogen-rich donor. As in \cosmic{},
a Hertzsprung-gap donor below the critical mass ratio $3.825$ starts its CE
at the end of the gap. We weight this delay by the stable fraction at onset,
a logistic in $\ln(q/3.825)$ with the width of the other stability gates.
Stable transfer
uses isotropic re-emission and a retained fraction $\beta_{\rm MT}=0.5$,
subject to
\begin{equation}
 \Delta m_{\rm acc}=\min\!\left[\beta_{\rm MT}\Delta M_d,
       10\dot M_{\rm Edd}(R_{\rm NS})\Delta t\right].
 \label{eq:two_sn_accretion}
\end{equation}
We use $R_{\rm NS}=9.74\,\mathrm{km}$. The segment duration follows the
nuclear or thermal clock. An accretor
crossing the adopted $2.3\,\Msun$ stability threshold is routed out of the
DNS population. Detached winds use the \cosmic{}/SSE prescriptions, with
Reimers coefficient $0.5$, helium-wind multiplier unity and Bondi--Hoyle
capture capped at the Eddington rate specified above. These are fixed
prescription choices.

\subsection{Physics inventory}
Table~\ref{tab:physics} lists the binary physics between the supernovae, as
\cosmic{} treats it~\cite{2000MNRAS.315..543H,2002MNRAS.329..897H,2020ApJ...898...71B},
and its status in our differentiable map. ``Ported'' means the same
prescription written as a smooth function of the state and the binary
parameters; ``simplified'' means a limit or closed form of it. The map has
two companion models: ``sse'' follows the single-star tracks, and ``proxy''
replaces the radius by a scaled helium-star radius and the wind by a
one-parameter envelope loss. The wind rates include those of
Refs.~\cite{1990A&A...231..134N,2001AA...369..574V}, and the
common-envelope binding energy uses the fit of
Ref.~\cite{2014A&A...563A..83C}. Remnant masses and natal kicks are not part of
this map; they come from the collapse model of Appendix~\ref{app:sn_model}. A
companion paper~\cite{AdSnPopsynP3InPrep} gives the equations, the channel
tables and the comparison with \cosmic{}.

We tested the common-envelope port against \cosmic{} on matched inputs. We
re-evolved binaries from the archive of Appendix~\ref{app:archive} with
\cosmic{} and kept the \result{CePortN} whose donor enters a common envelope
with the neutron star between the supernovae. At each recorded onset state
(the donor's initial and current mass, age, luminosity and radius) we
evaluated our track, the ported $\lambda_{\rm env}$ fit and
Eq.~\eqref{eq:intersn_ce} with \cosmic{}'s masses and separation. For the
\result{CePortAgreeN} core-helium-burning donors the radius, luminosity,
core mass, $\lambda_{\rm env}$ and final separation agree with \cosmic{} to
a fractional \result{CePortAgreeDev}. For the \result{CePortAgbN} early-AGB
donors our $\lambda_{\rm env}$ is \result{CePortAgbLamLo} to
\result{CePortAgbLamHi}\% larger.

The circularization rule sets where the envelope starts. \cosmic{} starts
it near the post-SN1 periastron: its median onset separation is
\result{CePortOnsetPeri} times $a(1-e)$. Circularizing at fixed angular
momentum gives $a(1-e^2)$, larger by $1+e$. The donor then fills its lobe
at a larger radius, where the fit gives a larger $\lambda_{\rm env}$ and the
orbit shrinks less. For class W with the \scorpio{} engine at
$(v_{\rm NS},\alpha_{\rm CE})=(400\,\mathrm{km\,s^{-1}},1)$, the median
separation before SN2 of our common-envelope survivors is
\result{CePortSepOurs}\,$R_\odot$ at fixed angular momentum and
\result{CePortSepPeri}\,$R_\odot$ at periastron; the replayed \cosmic{}
binaries give \result{CePortSepCosmic}\,$R_\odot$. We therefore circularize
at periastron. The post-SN1 states also differ:
the median neutron-star mass is \result{CePortNsOurs}\,$\Msun$ from the
\scorpio{} engine and \result{CePortNsCosmic}\,$\Msun$ in \cosmic{}, a
difference that \cosmic{}'s own prescription removes
(Appendix~\ref{sec:cosmic_engine}).

\begin{table*}
\caption{\label{tab:physics}Binary physics between the supernovae: the
\cosmic{} treatment and its status in the differentiable map.}
\begin{ruledtabular}
\begin{tabular}{p{0.16\textwidth}p{0.33\textwidth}p{0.43\textwidth}}
process & \cosmic{} & differentiable map\\
\hline
stellar structure and evolutionary age & analytic single-star tracks~\cite{2000MNRAS.315..543H} & ported for hydrogen-rich stars (types 0--6) and naked helium stars (7--9) under ``sse''; initial core masses supplied by the adopted prior, or by the catalog in stored-history comparisons\\
Roche lobe & Eggleton fit~\cite{1983ApJ...268..368E} & ported\\
stellar winds & metallicity-dependent rates for OB, LBV, giant, AGB and Wolf--Rayet phases & ported in detached phases under ``sse''; not carried for a detached core-helium-burning or AGB companion, or during overflow\\
wind accretion & Bondi--Hoyle capture~\cite{1944MNRAS.104..273B}, Eddington-limited & ported; the neutron star's spin follows the accretion torque below\\
onset of overflow & radius reaches the Roche lobe & ported as a smooth boundary on the track, with the time to onset\\
stability & critical mass ratio by donor type~\cite{1987ApJ...318..794H,2002MNRAS.329..897H} & ported, with a re-test at the donor's next type; one critical ratio under ``proxy''\\
stable mass transfer & rate integrated on thermal or nuclear time; accretion limited by the accretor and the Eddington rate & simplified: closed-form end state (donor stripped to its helium or CO core), accreted fraction $\beta$ or Eddington cap, isotropic re-emission; the rate history is not integrated\\
common envelope & $\alpha$--$\lambda$ energy formalism~\cite{1984ApJ...277..355W} with a fitted $\lambda$ & ported; $\lambda$ constant or \cosmic{}'s fit on the track; survival if core and companion fit within their lobes\\
case-BB transfer & helium-star donor overflow onto the compact object & simplified: stripping to the CO core by a smooth gate on the helium star's overflow; overflow during the helium star's last \PsrCaseBbAccMyr\,Myr; a neutron-star accretor gains at most the helium Eddington rate~\cite{2017ApJ...846..170T}\\
tides & equilibrium and dynamical tides, spin synchronization & simplified: circularization at periastron at onset; tidal spin locking in the recycling channels\\
magnetic braking & Rappaport--Verbunt--Joss form~\cite{1983ApJ...275..713R} & ported in the recycling channels\\
gravitational radiation & orbit-averaged decay of $a$ and $e$~\cite{Peters:1964} & simplified: circular decay at fixed $e$, at the orbit before case-BB overflow and then at the final orbit\\
neutron-star spin and field & accretion spin-up, dipole spin-down, field decay with time and accreted mass~\cite{psr-popsyn-AustralianDetailedPopsyn2008} & spin magnitude only, with the birth distribution on fixed quadrature nodes; braking and field decay ported; spin-up replaced by a torque toward the magnetospheric equilibrium bounded by the accreted angular momentum~\cite{gwastro-TaurisFormationMillisecondPulsars2012}\\
collisions and contact & merger of the two stars & dead-end channels\\
\end{tabular}
\end{ruledtabular}
\end{table*}

\section{\cosmic{} progenitor archive}\label{app:archive}

\subsection{Births and samples}\label{app:archive_births}
All populations in this appendix use \cosmic{} 4.1.0 at $Z=0.002$ with one
fixed configuration. Births follow the density $p_b$: circular orbits,
log-uniform primary mass $8$--$25\,M_\odot$, uniform mass ratio $0.2$--$1$,
and log-uniform period $1$--$30000$ days. Each binary evolves for
$100\,\mathrm{Myr}$. This density is an experimental proposal, not an initial
mass function or a star-formation history. Common-envelope efficiency is
$\alpha_{\rm CE}=1$, with the fitted $\lambda$ of \cosmic{}. The other
settings are \texttt{qcflag}~5, \texttt{windflag}~3, \texttt{cemergeflag}~1
and \texttt{kickflag}~5 with $\sigma=265\,\mathrm{km\,s^{-1}}$.

Each birth with a collapse yields one record: the ZAMS binary, the states
before and after each collapse, the donor type at the first transfer and CE
between collapses, and the final binary. Births without a collapse remain in
every denominator. A unit-weight baseline from $p_b$, a sample selected
toward rare classes, and a cohort targeted at pre-SN2 ancestors share this
physics. The Supplemental Material gives their weights and class yields.

\subsection{The pre-SN1 population}
\label{app:presn_prior}
To supply a correlated starting population while varying only the SN model,
we fit the distribution immediately before the first collapse. Stellar tracks,
winds, mass transfer, CE prescription and metallicity remain fixed. This is a
narrower problem than interpolation between binary-evolution prescriptions
\cite{2018MNRAS.481.4685T,2025ApJ...988..189C}; flow-based population density
estimation also has precedent in Ref.~\cite{2020arXiv200209491W}. The endpoint
experiment retains its original input distribution, and the one-SN controls use the
calibration of Appendix~\ref{app:presn2_prior}.

The family has a stripped first exploder (\cosmic{} types 7--9) with a positive
helium shell and no hydrogen envelope, a main-sequence companion, and a circular
orbit. We select on no remnant identity, kick, survival, second collapse or
final class. The label $c$ records CE before SN1; its absence includes stable
transfer. The family excludes systems whose companion has already left the
main sequence at SN1, mostly from double-core CE. They supply
\ArchiveArrivalDoublePercent\% of circular helium-star--NS arrivals, against
\ArchiveArrivalFamilyPercent\% from this family. The density therefore
describes one formation route. It is not a formation-rate model.

We fit the CO-core mass, a helium-shell scale, the companion mass and age,
the separation and the ZAMS period, in bounded coordinates. A Gaussian
mixture in each CE history defines the density, and its conditional CDFs give
an exact triangular transport. The $1.44\,\Msun$ CO-core floor is a separate
atom. The Supplemental Material compares the fit with held-out prefixes.

\begin{figure*}[t]
\centering
\includegraphics[width=.95\textwidth]{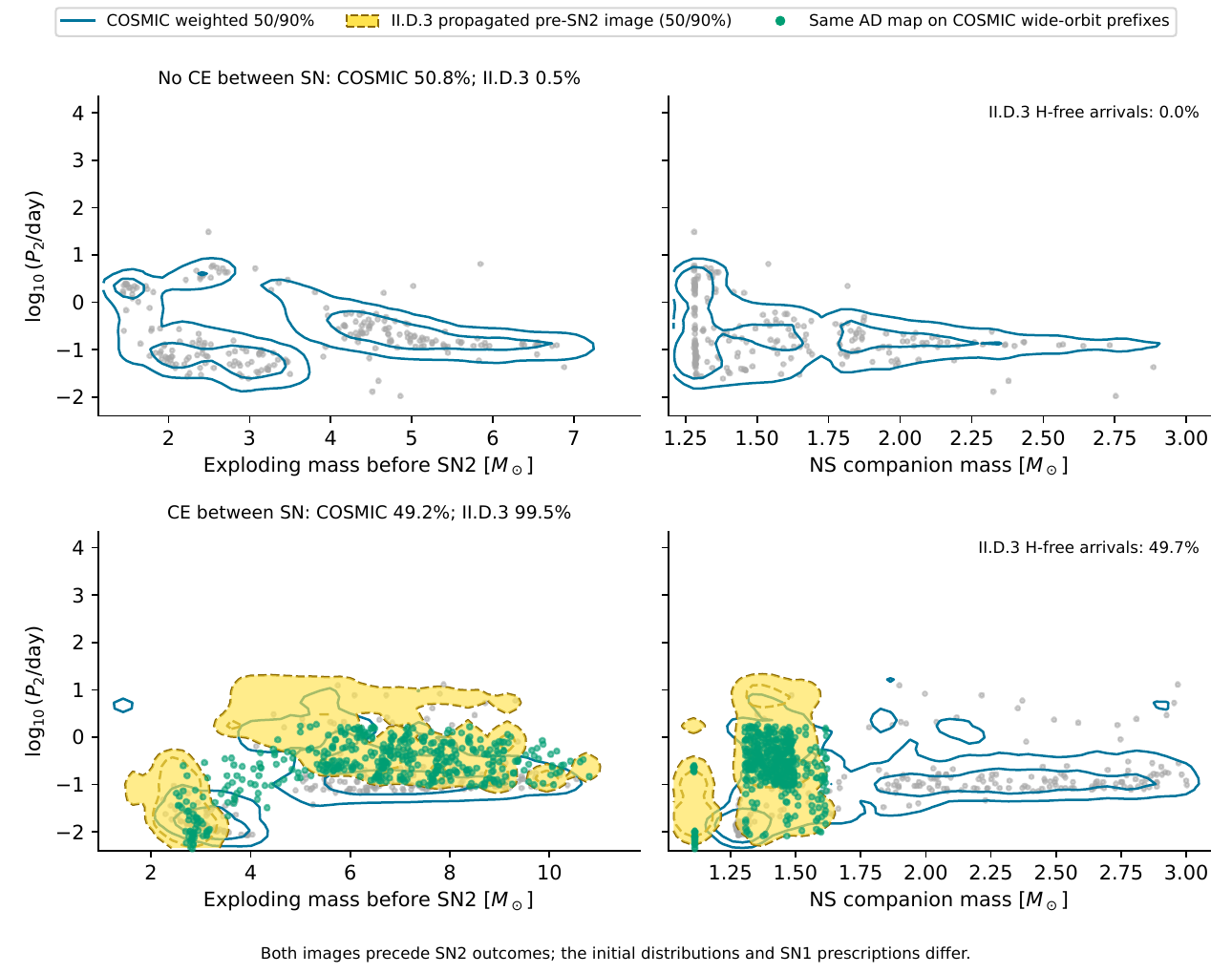}
\caption{Masses and orbital periods just before SN2. Blue contours enclose
about 50\% and 90\% of the simulated population supplied to these
one-explosion tests. Gray points show an independent \cosmic{} sample.
Rows separate binaries with and without a common envelope between explosions.
Yellow regions and dashed curves show the corresponding 50\% and 90\%
regions predicted by the separate two-explosion calculation starting from
Sec.~\ref{sec:two_sn_data}. Green points show that same evolution applied
to wide-orbit \cosmic{} progenitors. The two calculations use different
remnant-mass prescriptions. The orbit-only fit uses the blue population,
including the rare long-period tail outside the plot.\llmfig}
\label{fig:presn2_input}
\end{figure*}

The contours of Fig.~\ref{fig:presn_input} display this fit, not the data. We
draw a scrambled Sobol sample from the fitted distribution, bin each projection in a
$55\times55$ histogram, smooth with a one-bin Gaussian, and contour the levels
that enclose 50\% and 90\%. This smoothing affects the figure only. Inference
would use $p_0$ directly. The contours of Fig.~\ref{fig:presn2_input} are made
the same way from the weighted arrival rows, because that prior is empirical.

For trajectory integration we additionally restrict companions to ages
below their adopted SSE main-sequence lifetime, before applying any SN
selection. Nominal secondary CO and He layers are obtained from the fixed
isolated SSE track with no further wind loss. The existing inter-SN map
then evolves its donor and orbit and encloses those nominal cores within
the remaining donor mass. This structural closure is another tunable toy
approximation; native post-SN1 or pre-SN2 rows are not imported.

For the joint density $p_0(x,c)$ of prefix $x$ and CE history $c$, a simple
parametric period model is
\begin{equation}
 \begin{aligned}
 p_\eta(x,c)&=\frac{p_0(x,c)[P(x)/P_*]^\eta}{Z(\eta)},\\
 Z(\eta)&=\sum_c\int p_0(x,c)[P(x)/P_*]^\eta\,dx .
 \end{aligned}
 \label{eq:presn_period_tilt}
\end{equation}
Here $P$ is the pre-SN1 Keplerian period and $P_*=100$ days is an arbitrary
pivot. One global birth normalizer changes the CE mixture fraction as well as
the period distribution. SN survival and observational selection are applied
afterwards.

\subsection{Pre-SN2 arrivals for the one-explosion controls}
\label{app:presn2_prior}
The separate one-SN experiments use prospective \cosmic{} evolution under
the same fixed binary configuration and $Z=0.002$, including its fixed
native first-SN prescription, with evolution to $100\,\mathrm{Myr}$. From \PresnTwoBirths{} initial binaries we
extract bound helium-star--NS states immediately before the second
chronological event. No native second-SN kick, remnant, survival, merger or
final-DNS cut is used. We additionally require $e_2<10^{-3}$, negligible
hydrogen and exploding mass above $1.4\,\Msun$ for the circular,
nonnegative-mass-loss impulse family. This leaves \PresnTwoStates{}
calibration states; independent initial binaries supply
\PresnTwoTestStates{} comparison states. A defensive importance-sampling
proposal enriches this sparse family, and exact density ratios correct every
arrival (Supplemental Material).

The prior is the weighted joint empirical measure in
Eq.~\eqref{eq:conditional_progenitor_prior}. We retain the actual
companion masses and correlations rather than fitting independent
marginals. Each impulse experiment still sets the newborn NS mass
separately, so the new input prior does not turn these controls
into a full \cosmic{} remnant-model prediction. Changing the earlier
evolution would require recalibrating this pre-SN2 prior. Rare wide arrivals
limit calibration at strong positive period tilt: effective sample size
decreases from \PresnTwoESS{} at $\eta=0$ to about 25 at $\eta=0.5$.

\subsection{Coverage of the endpoint prior and its image}\label{app:coverage}
The archive tests whether the starting prior of Sec.~\ref{sec:two_sn_data}
lies on \cosmic{} trajectories, and whether the AD map reproduces \cosmic{}
from the same states. The class boxes of Table~\ref{tab:two_sn_classes}
enclose \ClassWCoveragePercent\% of the \cosmic{} arrivals that pass through
a wide-orbit CE between the explosions, and \ClassCCoveragePercent\% of those
with a CE before SN1.

We apply the map to the \cosmic{} prefixes of this route and compare the
\AdWideStates{} that reach SN2 with the state \cosmic{} reached from each.
The exploding mass has median difference \AdWideDeltaMass$\,M_\odot$. The
map's periods are longer by a median \AdWideDeltaLogP{} dex, with rank
correlation \AdWidePeriodRank{}. This agreement depends on circularizing at
periastron, as \cosmic{} does: at fixed angular momentum the offset is
\AdWideDeltaLogPJcirc{} dex. The median NS mass is \AdWideNsMedian$\,M_\odot$
under the map and \CosmicWideNsMedian$\,M_\odot$ in \cosmic{}, from the
remnant prescriptions at SN1 (Appendix~\ref{sec:cosmic_engine}). Survival
probabilities within the class boxes also differ between the two codes. The
boxes set the prior support, not these rates. The Supplemental Material gives
the full comparison.

\section{Numerical implementation of two-supernova uncertainty calculations}
\label{app:two_sn_numerics}
\subsection{Discretization of the population integrals}
Equations~\eqref{eq:two_sn_weights}--\eqref{eq:configurationposterior}
are evaluated by quadrature. Let $r$ index pre-SN1 configurations with
weights $\pi_r$, $a,b$ the first and second explosion atoms, $k$ an
evolutionary channel, $m$ an inter-SN node and $l$ a bound SN2 direction
node. Write $Q_{rakm}$ for the inter-SN subprobability including SN1
survival, core-collapse eligibility and other losses before SN2,
$W_{1,ra},W_{2,rakmb}$ for explosion weights, $D_{rakmbl}$ for the bound
SN2 direction weight, and $n_1,n_2$ for the engine's NS identity weights.
The quadrature contribution is
\begin{align}
 \omega_{rakmbl}&=\pi_r W_{1,ra}Q_{rakm}W_{2,rakmb}\nonumber\\
 &\quad\times D_{rakmbl}n_{1,ra}n_{2,rakmb}.
 \label{eq:two_sn_discrete_weights}
\end{align}
$Q$ includes survival of SN1 and $D$ includes survival of SN2; neither
is multiplied by another survival factor. With endpoint
$y=(\log_{10}P/{\rm day},e,m_{a,2},m_{{\rm rem},2})$, we compute
\begin{align}
 \epsilon_{\rm DNS}&\simeq\sum_{rakmbl}\omega_{rakmbl},\nonumber\\
 K_i&\simeq\sum_{rakmbl}\omega_{rakmbl}p(d_i\mid y_{rakmbl}).
 \label{eq:two_sn_discrete_likelihood}
\end{align}
For the discrete progenitor posterior, the probability assigned to node $r$
is $\pi_r K_{ir}/\sum_{r'}\pi_{r'}K_{ir'}$, where $K_{ir}$ excludes
the prior weight $\pi_r$. Each progenitor contribution is not separately
divided by its survival probability. The inter-SN distribution is retained
within each channel, rather than replaced by the channel mean.

The injection of Sec.~\ref{sec:two_sn_data} uses finer scatter and
direction nodes than the fit. With the injection's $(6,6,6)$ progenitor
rule, they raise the observed DNS fraction by
\TwoSnOrderScatterEffPercent\%, from $\TwoSnOrderInjEffFitNodes$ to
$\TwoSnInjDnsFraction$. To separate this offset from the recovery, we drew
a control catalog of 6,400 endpoints with the fit's nodes and the
$(6,6,6)$ rule, and fitted its first 640 events with that model. The fit
lies \TwoSnMatchedOffsetKick{} and \TwoSnMatchedOffsetCe{} statistical
widths from the injection, with probability \TwoSnMatchedOffsetP{}.

\subsection{Numerical evaluation of likelihood derivatives}
The Fisher matrix and likelihood derivatives are defined in
Sec.~\ref{sec:fisher}. Forward-mode AD differentiates both the endpoint
integrals $K_i$ and the common selection normalization
$\epsilon_{\rm DNS}$, propagating their response through both explosions
and the intervening evolution. The two parameter directions supply all
per-event derivatives as well as the total gradient. We check that their
sum reproduces the independently evaluated total gradient, and test the
empirical Fisher matrix against integration refinement and changes of the
reference parameters before interpreting its inverse as a local Gaussian
covariance. Further numerical diagnostics are given in the technical
supplement.

\subsection{Validation with a larger catalog}
\label{app:n6400_validation}
\begin{figure*}[t]
\centering
\includegraphics[width=\textwidth]{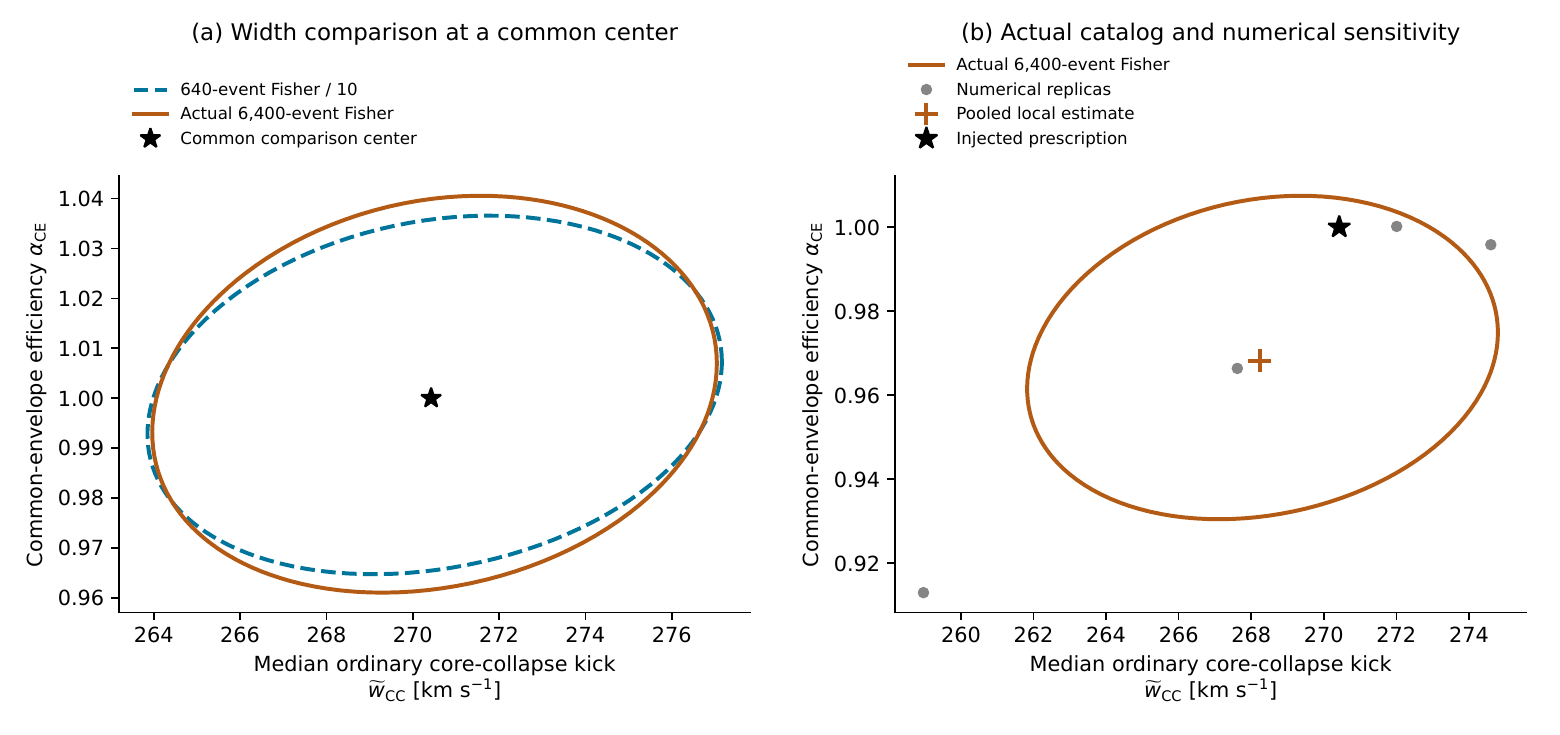}
\caption{\textbf{Larger-catalog validation of Fisher scaling.}
Left: the scaled 640-event covariance (dashed) and actual 6,400-event
covariance (solid), centered at the injection to compare their widths.
Right: the estimate from the local Fisher update and shifts from individual
Sobol integration replicas (gray). Curves are local Gaussian regions with
$\Delta\ln\mathcal L=3$, not evaluated likelihood contours.}
\label{fig:n6400_validation}
\end{figure*}

We repeated the local Fisher calculation with all 6,400 independently
injected endpoints in the matched-node control catalog.
Figure~\ref{fig:n6400_validation} compares the result with the 640-event
covariance divided by ten. The fractional widths are 0.99\% for the kick
scale and 1.62\% for the CE efficiency, respectively 1.7\% smaller and
10.7\% larger than the scaling forecast.

\subsection{Analytic estimate of the first-supernova survival gradient}
\label{app:sn1_survival_gradient}
The inconsistent normalization in Sec.~\ref{sec:twosnresults}
adds $N\ln S_1$ to the log likelihood. An analytic estimate gives the size
of its kick-scale derivative. For a circular pre-SN1 orbit let $u=w/v_{\rm orb}$ be
the kick speed in units of the relative orbital speed and $\beta$ the
fraction of total mass retained. The bound condition is
$1+u^2+2u\cos\psi<2\beta$, where $\psi$ is the angle between the kick
and the orbital velocity. Isotropic directions give the bound fraction
\begin{equation}
 B(u,\beta)=
 \operatorname{clip}_{[0,1]}\!\left[\frac12+\frac{2\beta-1}{4u}-\frac{u}{4}\right]
 \label{eq:sn1_survival_estimate}
\end{equation}
from Eq.~\eqref{eq:pbound_phase}. At fixed progenitor, remnant mass, and
unscaled kick, $u\propto s_k$ and $\partial\ln u/\partial\ln s_k=1$.
On the partially bound branch the chain rule gives
\begin{equation*}
\begin{aligned}
 \frac{\partial\ln B}{\partial\ln s_k}
 &=\frac{u}{B}\frac{\partial B}{\partial u}\\
 &=-\frac{1}{4B}\left(\frac{2\beta-1}{u}+u\right).
\end{aligned}
\end{equation*}
At $u=1$, $B=\beta/2$, so the logarithmic derivative is $-1$.

For the population, let $\langle\cdot\rangle$ denote integration over
starting binaries and remnant and kick draws, weighted by the
probability of forming a neutron star at SN1 per initial binary. The
weights retain their absolute normalization. Then $S_1=\langle B\rangle$
and
\begin{equation*}
 \frac{\partial\ln S_1}{\partial\ln s_k}
 =\frac{\left\langle\partial B/\partial\ln s_k\right\rangle}
        {\langle B\rangle}.
\end{equation*}
The average counts each progenitor's absolute contribution to the surviving
population. Thus local logarithmic slopes enter weighted by the number of
survivors they represent. A fully disrupted configuration contributes zero
away from the clipping boundary, even though its individual $\ln B$ is
undefined.
Kicks and orbital motion can be comparable in our experiment. For example,
one pre-SN1 progenitor has $v_{\rm orb}\simeq221\,\mathrm{km\,s^{-1}}$,
and the injected median core-collapse kick is $270\,\mathrm{km\,s^{-1}}$.

When kicks comparable to the orbital speed contribute appreciably
to survival, the extra log-likelihood gradient is of order $-N$ in
$\ln s_k$. At the fitted maximum the population average gives
$\partial\ln S_1/\partial\ln s_k=-0.87$, so the kick component of
$\boldsymbol g$ is $-555$ for 640 systems.

\clearpage

\section{Pulsar survey model and five-class DNS merger rate}
\label{app:native_survey_checks}
In this appendix, we describe our simplified pulsar survey model, which we use to assess pulsar survey selection
effects. To validate it, we compare against published results on binary pulsar survey selection effects.

We model radio discovery by forward simulation of surveys, following
\citet{Chunglee-nsns-1,Pol2019DNSRate} and using the propagation and
survey tools of PsrPopPy~\citep{Bates2014PsrPopPy}. Each synthetic pulsar
has a position, radio luminosity, spectrum, spin period and pulse width.
NE2001 gives its dispersion measure~\citep{psr-ElectronDensityModel-NE2001} and the
Haslam map its sky temperature~\citep{1982AAS...47....1H}. Dispersion,
scattering and instrumental sampling broaden the pulse. The broadened
width, gain, bandwidth and integration time set the radiometer
signal-to-noise ratio in each of five archived surveys, averaged over each
survey's footprint and beam offsets. Let $d_s(z)$ be the detection
probability in survey $s$ for a radio-active pulsar at state $z$ whose
beam sweeps Earth. Discovery by at least one survey has probability
\begin{equation}
 d_{\rm union}(z)=1-\prod_s[1-d_s(z)].
 \label{eq:instrument_survey_union}
\end{equation}
We treat survey coverage as independent and count a rediscovered pulsar
once. For DNSs the response also includes orbital signal loss and removal
at merger. Radio discovery is distinct from inclusion in a timing-mass
catalog, whose selection corrections are larger for orbits and spins than
for masses~\citep{Drummond2026RadioSelection}.

\begin{figure}[t]
\centering
\includegraphics[width=\columnwidth]{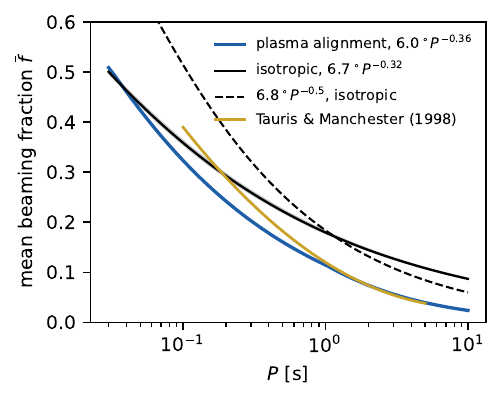}
\caption{\textbf{With plasma alignment the fitted cone reproduces the 1998
beaming fraction beyond one second.} Mean beaming fraction against period
for the fitted cone with plasma alignment (blue), for the fitted cone with
isotropic inclinations (black; the band spans opening-angle scatters from
zero to \result{PbTotalScatter}\,dex), for a $6.8^\circ P^{-1/2}$ cone
with isotropic inclinations (dashed), and for the fit of
\citet{Tauris1998Beaming} over its stated range (gold).\llmfig}
\label{fig:pulsar_beam_fraction}
\end{figure}

\textit{Beam model.} The radio module beams each pulsar as a circular cone
of half-opening angle $\rho(P)$ and averages over the magnetic inclination.
We calibrated the cone on the 10\% pulse widths of \result{PbNNormal} normal
pulsars measured near 1.3\,GHz by MeerKAT and
FAST~\citep{Posselt2021Widths,Wang2023FASTProfiles}, with a likelihood that
integrates the viewing geometry (Supplemental Material). The widths prefer
the alignment given by the torque of a plasma-filled
magnetosphere~\citep{Philippov2014Obliquity} over isotropic inclinations,
by $\Delta\ln\mathcal L=\result{PbDlnLPlasma}$ with no added parameter.
The cone is then
$\rho=\result{PbRhoOnePlasma}^\circ(P/1\,{\rm s})^{-\result{PbRhoExpPlasma}}$,
with an error of \result{PbRhoExpPlasmaErr} on the exponent from mock
samples. A second birth-period distribution changes the gain to
\result{PbDlnLPlasmaCensus} and the exponent to
\result{PbRhoExpPlasmaCensus}. The exponent agrees with the $P^{-1/3}$ lower bound drawn by
\citet{Lyne1988Shape}, and a $P^{-1/2}$ cone is lower by
\result{PbDlnLHalfPlasma}. The mean beaming fraction is
\result{PbBeamPlasmaOne} at 1\,s and \result{PbBeamPlasmaThree} at 3\,s,
against \result{PbBeamTMOne} and \result{PbBeamTMThree} from the fit of
\citet{Tauris1998Beaming} (Fig.~\ref{fig:pulsar_beam_fraction}). The same
cone with isotropic inclinations would give \result{PbBeamOne} and
\result{PbBeamThree}. For mildly recycled pulsars the widths bound the beam
scale between \result{PbScalePlasmaScat} and \result{PbScalePlasmaAll} of this
law. No result in this paper uses the cone. The two-SN endpoint experiment
omits beaming, and the five-class rate below keeps the published beaming priors.

\begin{figure}[t]
\centering
\includegraphics[width=\columnwidth]{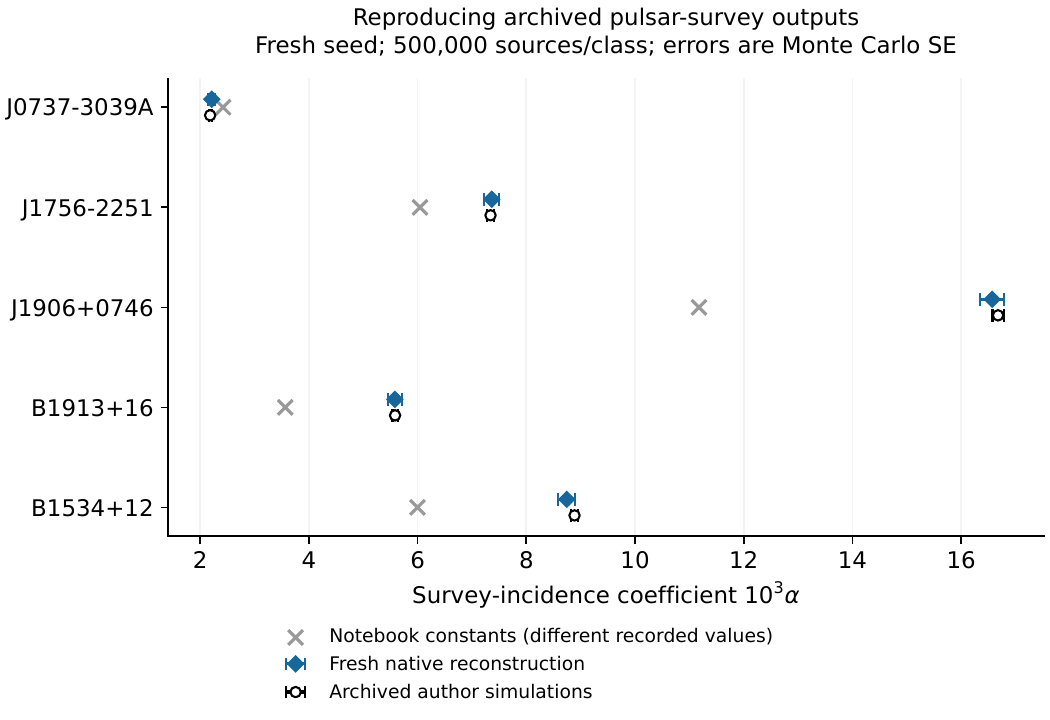}
\caption{\textbf{Reproduction of archived radio-survey simulations.}
Our calculation (blue diamonds) and the saved simulations accompanying
\citet{Pol2019DNSRate} (black circles), with Monte Carlo standard errors.
Gray crosses show values recorded separately in the authors' notebook.
The coefficient counts survey detections per active pulsar beaming toward
Earth, including rediscoveries by overlapping surveys.\llmfig}
\label{fig:native_survey_replication}
\end{figure}

\textit{Reproduction of archived simulations.} We simulated
$5\times10^5$ pulsars in each of the five DNS classes of
\citet{Pol2019DNSRate} and compared the detection coefficients with the
authors' saved simulations (Fig.~\ref{fig:native_survey_replication}).
All five agree within 0.84 combined Monte Carlo standard errors.

\begin{figure*}[t]
\centering
\includegraphics[width=.95\textwidth]{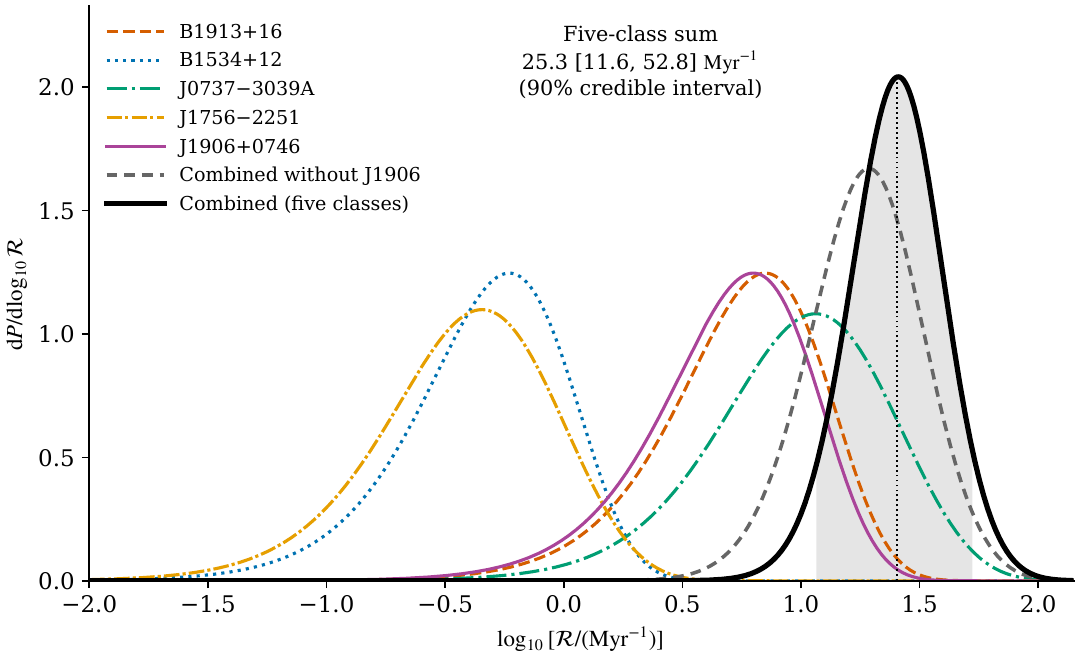}
\caption{\textbf{Merger rates represented by five historical DNS classes.}
Thin colored curves show individual class posteriors; the thick black
curve is their convolution, the posterior for the summed rate.
The vertical axis is probability per decade,
$dP/d\log_{10}\mathcal R=(\ln10)\mathcal R\,p(\mathcal R)$.
Gray shading encloses the sum's 90\% equal-tail interval, and the dotted
line marks its median. The dashed gray curve omits J1906.
The calculation uses the stated beam and lifetime prescriptions and
assumes J1906$+$0746 is a DNS.\llmfig}
\label{fig:native_dns_rate}
\end{figure*}

\textit{Five-class merger rate.} Following
\citet{Chunglee-nsns-1}, each observed binary stands for a population
of similar systems that are too faint, outside the surveyed sky, or beamed
away from Earth. We count B1913+16, B1534+12, J0737$-$3039A, J1756$-$2251
and J1906$+$0746 as one detection each, assuming J1906 has a neutron-star
companion. For class $i$, $\alpha_i$ is the fraction of simulated
radio-active, Earth-beaming pulsars detected by at least one survey,
$f_{b,i}$ is the beaming correction, and $\tau_i$ is the observable
lifetime: the pulsar's age plus its remaining radio-visible time, limited
by merger or radio death. Assuming steady formation, one Poisson detection
and a flat prior on abundance give a rate posterior at fixed beam and
lifetime~\citep{Chunglee-nsns-1}
\begin{equation}
 p(\mathcal R_i\mid\alpha_i,f_{b,i},\tau_i)
 =C_i^2\mathcal R_i e^{-C_i\mathcal R_i},
 \qquad C_i=\frac{\alpha_i\tau_i}{f_{b,i}}.
 \label{eq:native_dns_rate_gamma}
\end{equation}
We average each class over the beaming priors of
\citet{psr-SpinFb} and convolve the five class posteriors
to obtain the posterior of the summed rate
(Fig.~\ref{fig:native_dns_rate}). Its median is
\NativeDNSRateMedian$\,\mathrm{Myr}^{-1}$, with 90\% interval
$[\NativeDNSRateLowerNinety,\NativeDNSRateUpperNinety]\,\mathrm{Myr}^{-1}$.
The interval includes Poisson and beaming uncertainty but holds the
luminosity, spatial, survey and lifetime models fixed. The lifetimes
matter most: reducing J1906's remaining radio lifetime from 82 to
3\,Myr raises the median to $161.3\,\mathrm{Myr}^{-1}$.

\section{Sensitivity to the binary population and evolution model}
\label{app:compas_initial_conditions}

We test whether the normalization bias in
Fig.~\ref{fig:twosnlikelihood} depends qualitatively on the assumed
starting population. We retain the same 640-DNS catalog, stellar evolution
and radio response, and reweight each parent's absolute response before
conditioning on detection. Prospective \compas{} states set alternative
W/C formation-family proportions; within-family period tilts provide a
separate test of the initial distribution's shape.

\begin{figure*}[t]
\centering
\includegraphics[width=\textwidth]{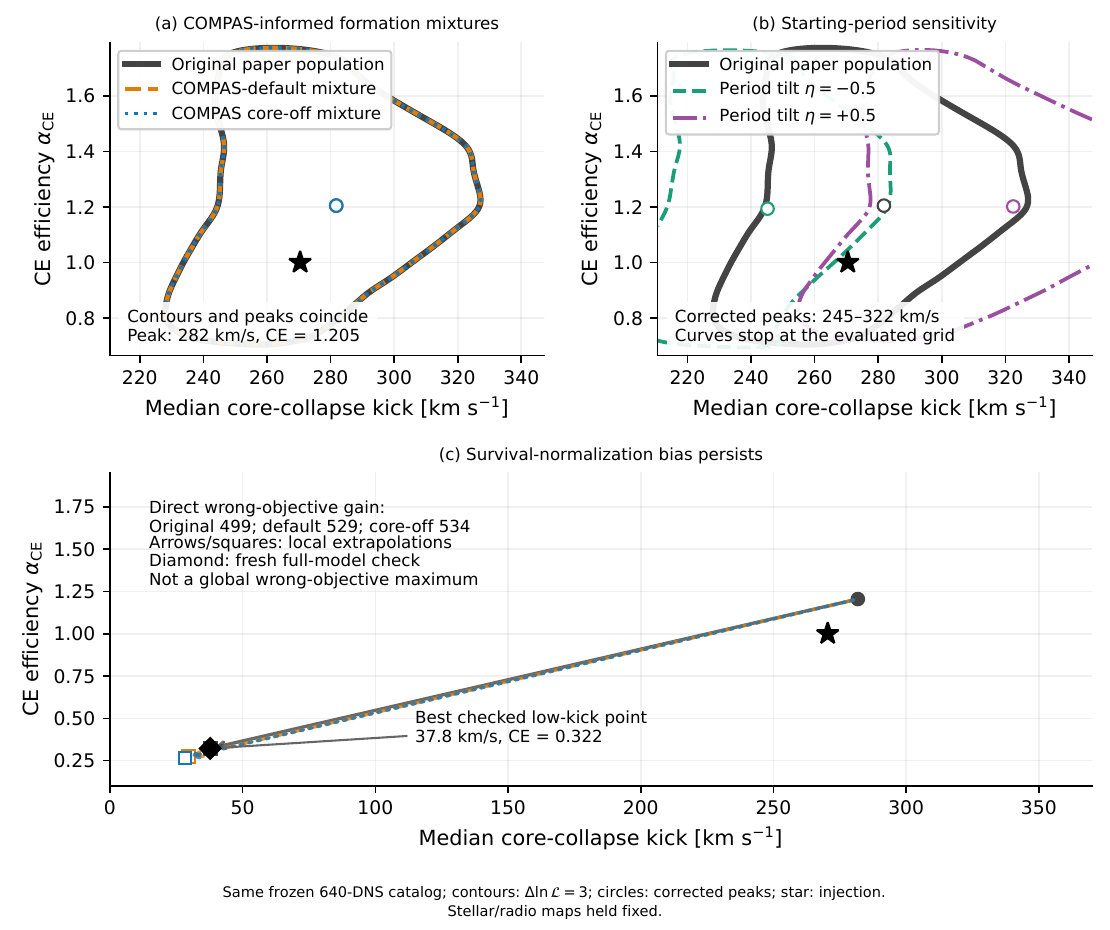}
\caption{\textbf{Initial-population sensitivity of the radio-DNS likelihood.}
The frozen 640-DNS catalog and stellar/radio maps are held fixed.
(a) Correctly normalized contours for the original and \compas{}-informed
formation mixtures coincide. (b) Pre-SN1 period tilts $P_1^{\eta}$,
$\eta=\pm0.5$, shift the corrected peak moderately. Curves show the actual
likelihood interpolant at $\Delta\ln\mathcal L=3$ from each maximum and
stop at the evaluated grid; circles mark maxima and stars the original
injection. (c) Arrows and squares show extrapolated local displacements
from omitting SN1 survival normalization. The diamond is the best directly
checked point for the original and both native-informed mixtures, at
37.8\,km\,s$^{-1}$ and CE efficiency 0.322; it improves the inconsistent
log likelihood by 499, 529 and 534, respectively (integration SE about 1.2).
These checks establish the persistent low-kick bias, without locating its
global maximum. This is a starting-measure comparison, not native
\compas{} stellar evolution.\llmfig}
\label{fig:compas_initial_conditions}
\end{figure*}

Figure~\ref{fig:compas_initial_conditions} shows the resulting likelihoods.
The native-informed mixtures leave the corrected peak and curvature
unchanged: selected DNSs overwhelmingly come from W, whose overall weight
cancels in the conditional likelihood. Period tilts move the kick peak to
245--322\,km\,s$^{-1}$, about two original local widths, while CE efficiency
stays near 1.2. The local kick/CE widths change by at most 8\%/23\%.
Fresh full-model evaluations retain the strong preference for much smaller
kicks when SN1 survival normalization is omitted. Thus these starting-population
changes affect numerical bias estimates without changing the qualitative
conclusion. This comparison changes starting weights; it does not replace
the stellar trajectories by native \compas{} evolution. The construction
and further checks are given in the Supplemental Material.

\subsection{Native binary evolution}

\begin{figure*}[t]
\centering
\includegraphics[width=\textwidth]{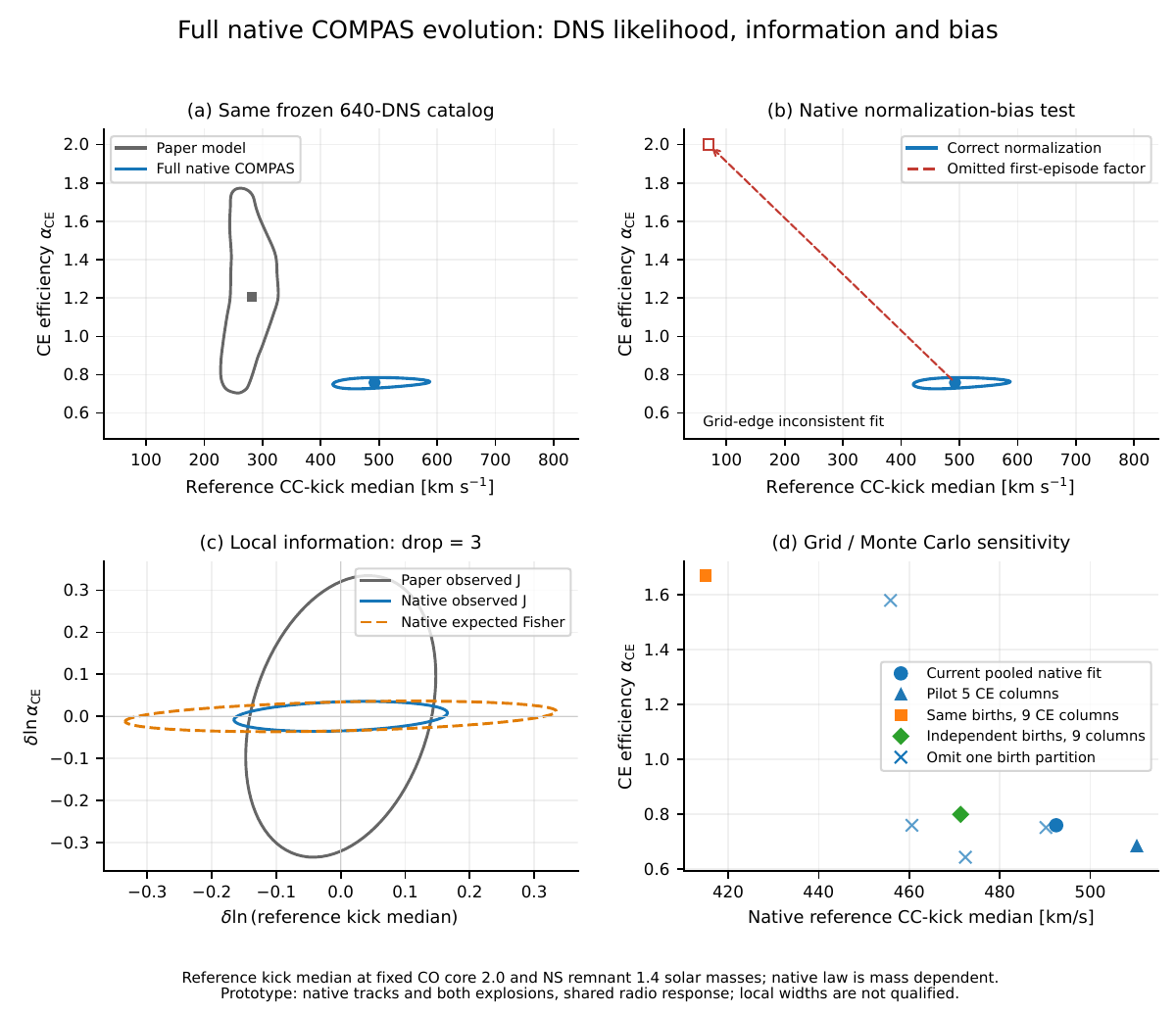}
\caption{\textbf{Sensitivity to full native \compas{} evolution.}
The same 640-DNS catalog and radio response are used with native evolution
from the initial stellar population through both explosions.
(a) Correctly normalized likelihood contours at
$\Delta\ln\mathcal L=3$ from each model's maximum.
(b) Omitting that first-collapse formation-and-survival factor shifts the fit toward
lower kicks; the arrow connects fitted points within the scanned grid,
whose CE boundary limits the inconsistent fit.
(c) Centered local observed-information ($J$) and expected-Fisher ($F$)
ellipses. (d) Grid refinement, independent births and whole-birth omissions
show sampling sensitivity exceeding the local widths, so these ellipses
are not qualified uncertainty regions. The native kick axis is the
conditional median at fixed CO-core and NS-remnant masses of
$2.0$ and $1.4\,M_\odot$, respectively, rather than a population-wide median.
Native CE efficiency acts throughout the evolution.\llmfig}
\label{fig:compas_full_native}
\end{figure*}

Figure~\ref{fig:compas_full_native} replaces the stellar evolution by native
\compas{}, retaining the frozen catalog and radio-response model. This
comparison changes both the starting population and its evolution, including
CE evolution before SN1. The corrected likelihood peak and local information
change appreciably, while omitting the probability per initial binary of
forming a bound NS system in the first collapse episode by 100 Myr still
favors substantially lower kicks. Independent birth samples and CE-grid
refinement move the native peak by more than its nominal local widths;
the numerical peak and Fisher changes therefore remain provisional.

\label{app:compas_initial_conditions:end}

\section{Use of large language models}
\label{app:llm}

The author used Claude (Opus 5/5.5, Fable 5.2) and
Codex (Sol 6/6.1, Astra 6) to assist with scientific reasoning, software
development, numerical analyses, drafting, and revision. The agents
operated the analysis and plotting tools used to produce the figures
in the article and Supplemental Material. The author directed the work,
reviewed the analyses and revisions, and takes responsibility for the
code, results, and manuscript.

The first-supernova disruption criterion follows from Newtonian orbital
energy. For isotropic kicks, the fraction of directions that leave a
binary bound is analytic at a given kick speed and mass loss. We checked
that calculation against independent direct vector impulses and checked
the orbital map against the underlying \cosmic{}~4.1 \texttt{kick.f}
routine at identical kick geometries. All 203 first-supernova events in
that comparison agreed on whether the binary remained bound.
We checked the population-averaged SN1 survival against totals from
the forward code and verified the likelihood derivatives with
finite differences. The full two-supernova likelihood was evaluated
numerically. It was checked by integration refinement, direct likelihood
evaluations at displaced parameters, and a sensitivity comparison using
native \compas{} evolution.

Separate adversarial reviews examined the code, numerical assumptions,
and interpretation of the results. For example, the native \compas{}
review checked the first-collapse classification and the zero-kick
accretion-induced-collapse branch against the C++ source and recorded
simulation events. Follow-up checks verified the corrections. The review
also documented sampling and grid sensitivity that limits the precision
we assign to the native likelihood peak. Details of the workflow and
review process are given in Ref.~\cite{whitepaper-transparent-agentic-science}.

\clearpage
\section{What can one explosion reveal if its progenitor population is known?}
\label{app:conditional_sn2_experiments}

Suppose we knew which kinds of binaries reach the second supernova (SN2)
under one prescription for the first explosion and the subsequent binary
interactions. Could the orbits left by SN2 reveal a preference for kicks
along the spin axis? Would surviving binaries and ejected neutron stars
suggest different proportions of low kicks even if they came from the same
population? We use two mock experiments to answer these questions.

\subsection{Kick directions from surviving orbits}
\label{sec:orientation_population_data}

We take a snapshot from \cosmic{} immediately before the second explosion
and retain nearly circular binaries containing a helium star and a neutron
star. We choose these binaries before learning which will survive SN2 or
what remnant the helium star will leave. Each entry supplies the two masses
and the orbital period together. The inference averages over these entries
for every observed system. It never receives the identity of that system's
progenitor. What we supply is the distribution of possible progenitors,
including correlations inherited from SN1 and the intervening evolution.
Changing the first kick or common-envelope prescription could alter those
correlations and the mix of binaries reaching SN2. Here we fix both prescriptions.

We allow one uncertainty in this population. An exponent $\eta$ changes
the relative abundance of long- and short-period progenitors. For entry
$r$ with period $P_{2,r}$ and state $x_{2,r}$, the weights are
\begin{equation}
 \begin{aligned}
 \pi_\eta(dx_2)&=\sum_r w_r(\eta)\,\delta_{x_{2,r}}(dx_2),\\
 w_r(\eta)&=\frac{b_r(P_{2,r}/{\rm day})^\eta}
                   {\sum_s b_s(P_{2,s}/{\rm day})^\eta}.
 \end{aligned}
 \label{eq:conditional_progenitor_prior}
\end{equation}
The baseline weight $b_r=p_{\rm birth}(z_r)/q_r(z_r)$ accounts for how
the original binary was sampled. We normalize once over the whole
population, including systems that went through a common envelope and
those that did not. At $\eta=0$ we recover the simulated distribution.
Positive $\eta$ gives more weight to longer orbits. All values of $\eta$
use the same set of possible progenitors (Appendix~\ref{app:presn2_prior}).

Each helium star leaves a $1.4\,\Msun$ neutron star in the direction
experiment. We draw the kick speed from a Maxwellian of dispersion
$\sigma$. The kick's azimuth is uniform. Its polar angle follows
Eq.~\eqref{eq:orientation_population_law}, with $\kappa=0$ for isotropy
and positive $\kappa$ for kicks preferentially along either pole. We assume
that the progenitor spin is aligned with the pre-explosion orbit's angular
momentum. The input catalog uses $\kappa=2$,
$\sigma=250\,\mathrm{km\,s^{-1}}$, and $\eta=0$.

We observe every bound survivor in this experiment. For each, the data
are the post-explosion orbital period and eccentricity. The independent
Gaussian errors are 0.05 in $\log_{10}(P/{\rm day})$ and 0.02 in
eccentricity. We fit $\kappa$, $\sigma$, and $\eta$ together, integrating
over each kick and progenitor. The allowed ranges are $[-5,7]$,
$[100,500]\,\mathrm{km\,s^{-1}}$ and $[-0.5,0.5]$, respectively. Three
catalogs each at 11, 100, and 640 survivors test the orbital information
at fixed sample size. A separate count experiment assumes a known number
of progenitors, so the number disrupted at SN2 also carries information.

\subsection{Low-kick fractions among survivors and ejected stars}
\label{sec:conditionalmethods}

For this test the progenitors follow the same simulation with $\eta=0$.
We assign a $1.33\,\Msun$ newborn neutron star and isotropic kicks. A
fraction $f$ of the explosions draws a low kick and the rest a high kick,
\begin{equation}
 p(w\mid f)=f\,p_L(w)+(1-f)\,p_H(w).
 \label{eq:conditional_kicklaw}
\end{equation}
The low-kick component is Maxwellian with dispersion
$30\,\mathrm{km\,s^{-1}}$. For the high-kick component,
$\ln(w/w_0)$ is Gaussian with mean 5.60 and standard deviation 0.68,
where $w_0=1\,\mathrm{km\,s^{-1}}$~\cite{2025ApJ...989L...8D}.

We make one catalog of bound binaries with post-explosion periods below
1000 days and another of neutron stars released by disruption. Wider
bound binaries enter neither. Ejected speeds are measured in the
pre-explosion center-of-mass frame and include the original orbital
motion and the effect of escaping the companion. A mock binary has noisy
orbital measurements with errors drawn from the field double-neutron-star
sample. For an ejected star we use
\begin{equation}
 \ln(v_{\rm obs}/w_0)\sim\mathcal N[\ln(v_\infty/w_0),0.2^2].
 \label{eq:ejected_noise}
\end{equation}
Both catalogs use idealized discovery. The ejected stars are observed
before any motion through the Galaxy is modeled.

To infer $f$ we ask how often a star born in each kick component reaches
its catalog. Let $\alpha_{c,k}$ be that probability for component $k$ in
catalog $c$, and let $K_{ci,k}$ be the joint density for entering the
catalog and producing measurement $d_i$. Conditioning on the size of
that catalog gives
\begin{equation}
 \mathcal L_c(f)=\prod_i
 \frac{fK_{ci,L}+(1-f)K_{ci,H}}{f\alpha_{c,L}+(1-f)\alpha_{c,H}}.
 \label{eq:conditional_channel_likelihood}
\end{equation}
If we first normalize each component within its catalog and then mix
them, the mixture weight describes the fraction of observed systems with
low kicks. Interpreting that weight as $f$ attributes survival or
disruption to the birth distribution. We compare those two fits with a third that
keeps only the numerator. The prior on $f$ is uniform. When combining
catalogs we condition on each count separately. The Supplemental Material
gives the repeated-catalog results and the field-binary illustration.

\subsection{Orbit-only test of a natal orientation preference}
\label{sec:orientation_population_results}
Can survival-selected orbits distinguish the natal kick-direction distribution from changes
in kick speed or progenitor period? The regenerated one-collapse
experiment of Sec.~\ref{sec:orientation_population_data} uses the
\cosmic{} pre-SN2 prior without supplying spin--orbit tilts or any
individual progenitor. At $\sigma=250\,\mathrm{km\,s^{-1}}$, changing
from isotropy to the injected $\kappa=2$ changes the survival probability
from $\result{OrientPopSurvivalIsotropic}$ to
$\result{OrientPopSurvivalPolar}$. A binary/disrupted count supplies only
one independent constraint on the three fitted parameters, and no yield
information remains when the survivor count is conditioned on.

\begin{figure*}[t]
\centering
\includegraphics[width=.98\textwidth]{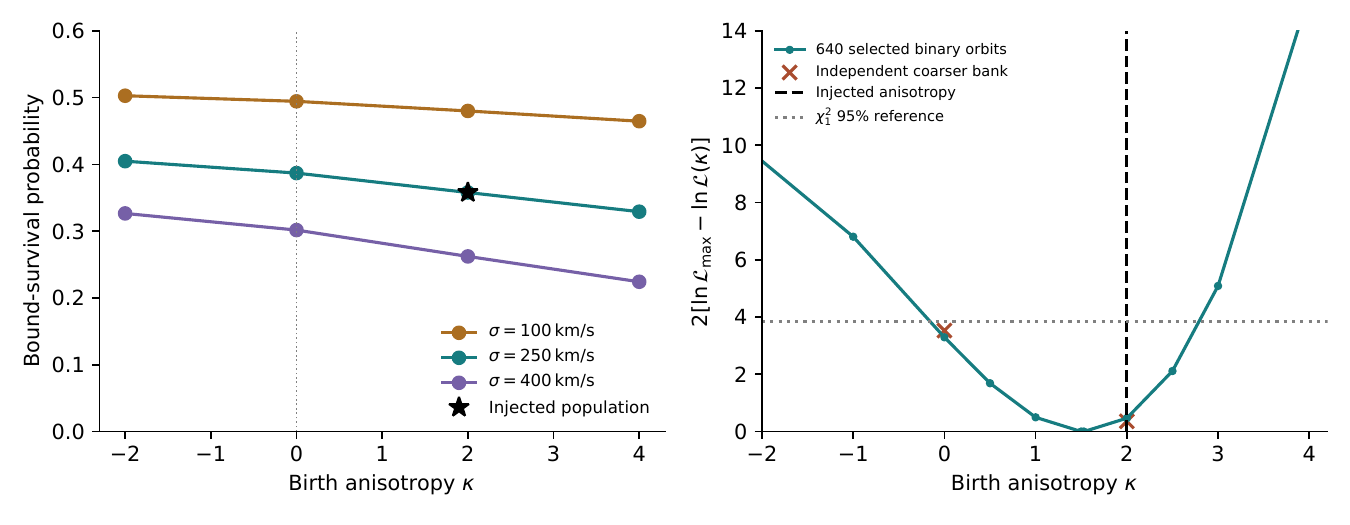}
\caption{\textbf{Survival sensitivity and orbit-only orientation profile.}
Left: survival probability under $q_\kappa(\mu)\propto e^{\kappa\mu^2}$
for the \cosmic{} pre-SN2 population of Fig.~\ref{fig:presn2_input}, at the indicated kick
dispersions and $\eta=0$. A yield change alone does not identify the
kick-direction distribution when speed and period tilt are uncertain.
Right: a count-conditioned likelihood profile for one catalog of 640
noisy surviving-binary orbits, with kick dispersion and period tilt
refitted at each $\kappa$. Crosses use the same catalog with an independent
coarser quadrature. The injected value is $\kappa=2$; no kick angle or
spin tilt is observed. Independent integration checks test the displayed
contrasts. The profile is a synthetic one-collapse demonstration with
unit discovery probability, not a calibrated confidence interval or a
Galactic constraint.\llmfig}
\label{fig:orientation_population_recovery}
\end{figure*}

Figure~\ref{fig:orientation_population_recovery} shows the first
regenerated catalog of 640 noisy survivor orbits. Fitting anisotropy,
kick dispersion and the input period tilt gives
$\kappa=\result{OrientPopBestKappa}$,
$\sigma=\result{OrientPopBestSigma}\,\mathrm{km\,s^{-1}}$ and
$\eta=\result{OrientPopBestEta}$. Isotropy is lower by
$2\Delta\ln\mathcal L=\result{OrientPopNullTwiceDrop}$ after refitting
both nuisance parameters, below the nominal one-parameter 95\% reference
of $3.84$. This reference is conditional, without coverage calibration.
None of the three 640-object catalogs exceeds it. Refits with larger independent samples change the isotropy contrast by
at most $\result{OrientPopNullCheckShift}$, and increasing the quadrature from
$2^{21}$ to $2^{23}$ nodes changes a contrast by at most
$\result{OrientPopRefineShift}$.

At $N=640$, survivor orbits alone therefore do not detect the injected
polar preference once kick dispersion and period tilt are fitted. Two of
the three catalogs favor $\kappa>0$, by less than the reference. A
detection would need larger catalogs or further observables, such as
spin--orbit tilts. Three catalogs do not calibrate coverage or
constrain a Galactic natal kick-direction distribution.

\subsection{A conditional check of low-kick mixture weights}
\label{sec:recoilresults}
Claims of reduced kicks in surviving binaries motivate a separate check
of mixture interpretation~\cite{2017ApJ...846..170T,Shariat2026Concordance}.
If the low- and high-kick components have different probabilities of
entering a catalog, the observed low-kick fraction differs from the birth
fraction $f_{\rm low}$. Survival enriches binaries in low kicks; disruption
can select the opposite mixture among released neutron stars.
A larger conditional test makes this distinction visible: across 100
mock catalogs of 11,000 surviving binaries, reading the selected fraction
as natal gives a mean of 0.549 for an injected birth fraction of 0.5.
The bias is 3.7 times the catalog-to-catalog standard deviation; accounting
for selection gives a mean of 0.502
(Appendix~\ref{app:larger_conditional_bias}). The Supplemental Material
also gives the smaller-catalog and illustrative field-DNS controls.
It holds a pre-SN2 progenitor family fixed and does not use the evolving
two-SN likelihood. This is a validation of selected-mixture interpretation;
published forward population comparisons already include
survival~\cite{Grichener2026Eccentricity,Disberg2026BinaryKicks}.
The joint two-SN experiment above addresses the additional problem of
inferring a kick scale and the CE efficiency from one parent population
when the progenitors are unknown.

\subsection{A larger sample exposes the mixture-interpretation bias}
\label{app:larger_conditional_bias}
The modest offset in the smaller catalogs reflects limited statistical
precision, rather than a disappearance of the selection effect. We repeat
the conditional low-kick test with 11,000 selected binaries in each of 100
independent catalogs, keeping the physical model, measurement errors and
injected birth fraction $f=0.5$ fixed. The sample size and random seeds
are specified before generating the catalogs.

With selection included, the inferred birth fraction averages 0.5019,
with catalog-to-catalog standard deviation 0.0132. Reading the selected
mixture weight as the birth fraction instead gives mean 0.5486 and
standard deviation 0.0131. The resulting mean bias is 3.71 times one
catalog's sampling dispersion, compared with 1.29 at 1,100 systems
(Supplemental Material, conditional low-kick-fraction recovery). These dispersions are measured
across the independent catalogs; they are not errors on the ensemble mean.
The ratio describes the systematic offset relative to statistical scatter,
rather than a guaranteed significance for each realization.

The response predicts a selected low-kick fraction of 0.5468 for equal
birth fractions. The two fitted mixture coordinates describe the same
selected density: the error is interpreting a selected fraction as natal.
An independent quadrature changes the selected-fraction estimate by at
most 0.0012 in the first ten larger catalogs, about 0.09 sampling standard
deviations. Direct Gaussian-sum likelihoods for the full first catalog
agree with the interpolation to better than 0.0002 in fitted fraction.
The calculation still supplies a fixed pre-SN2 population; it does not
extend the two-SN kick/CE recovery or measure a Galactic birth distribution.

\clearpage
\section{Parameters of the starting population}
\label{app:starting_prior_parameters}
Table~\ref{tab:two_sn_classes} specifies the two progenitor families used
in the endpoint experiment. Their construction is described in
Sec.~\ref{sec:two_sn_data}.

\begin{table}[!h]
\centering
\caption{Classes of the starting prior, Eq.~(\ref{eq:two_sn_prior}).
Masses in $\Msun$, separations $a_1$ in $\Rsun$, ages in Myr. The CO core is
$c_0+c_1M_{2,0}+r$ with $r$ in the CO band and $M_{{\rm CO},1}\ge1.44$;
$\log_{10}a_1$ is $d_0+d_MM_{2,0}+d_{\rm CO}M_{{\rm CO},1}+r'$ with $r'$ in
the separation band. The last three rows give the share of the class's
\cosmic{} prefixes inside the box, the fraction of those that reach SN2 as a
circular helium-star--NS binary, and the share of the class's arrivals
inside the box.}
\label{tab:two_sn_classes}
\begingroup\footnotesize
\begin{ruledtabular}
\setlength{\tabcolsep}{2pt}
\begin{tabular}{lcc}
 & W & C\\
\hline
CE before SN1 & no & yes\\
$M_{2,0}$ & \ClassWCompLow--\ClassWCompHigh & \ClassCCompLow--\ClassCCompHigh\\
$(c_0,c_1)$ & (\ClassWCoA, \ClassWCoB) & (\ClassCCoA, \ClassCCoB)\\
CO band & [\ClassWCoResLow, \ClassWCoResHigh] & [\ClassCCoResLow, \ClassCCoResHigh]\\
$(d_0,d_M,d_{\rm CO})$ & (\ClassWSepA, \ClassWSepB, \ClassWSepC) & (\ClassCSepA, \ClassCSepB, \ClassCSepC)\\
Separation band & [\ClassWSepResLow, \ClassWSepResHigh] & [\ClassCSepResLow, \ClassCSepResHigh]\\
CO-floor atom & \ClassWFloorPercent\% & \ClassCFloorPercent\%\\
$(a_k,b_k)$ & (\ClassWHeA, \ClassWHeB) & (\ClassCHeA, \ClassCHeB)\\
Companion age & \ClassWAge & \ClassCAge\\
$\omega_k$ & \ClassWWeight & \ClassCWeight\\
Class prefixes in box & \ClassWEnclosedPercent\% & \ClassCEnclosedPercent\%\\
\cosmic{} arrival fraction & \ClassWArrivalPercent\% & \ClassCArrivalPercent\%\\
Class arrivals in box & \ClassWCoveragePercent\% & \ClassCCoveragePercent\%\\
\end{tabular}
\end{ruledtabular}
\endgroup
\end{table}


\clearpage
\begin{thebibliography}{133}\makeatletter
\providecommand \@ifxundefined [1]{\@ifx{#1\undefined}
}\providecommand \@ifnum [1]{\ifnum #1\expandafter \@firstoftwo
 \else \expandafter \@secondoftwo
 \fi
}\providecommand \@ifx [1]{\ifx #1\expandafter \@firstoftwo
 \else \expandafter \@secondoftwo
 \fi
}\providecommand \natexlab [1]{#1}\providecommand \enquote  [1]{``#1''}\providecommand \bibnamefont  [1]{#1}\providecommand \bibfnamefont [1]{#1}\providecommand \citenamefont [1]{#1}\providecommand \href@noop [0]{\@secondoftwo}\providecommand \href [0]{\begingroup \@sanitize@url \@href}\providecommand \@href[1]{\@@startlink{#1}\@@href}\providecommand \@@href[1]{\endgroup#1\@@endlink}\providecommand \@sanitize@url [0]{\catcode `\\12\catcode `\$12\catcode
  `\&12\catcode `\#12\catcode `\^12\catcode `\_12\catcode `\%12\relax}\providecommand \@@startlink[1]{}\providecommand \@@endlink[0]{}\providecommand \url  [0]{\begingroup\@sanitize@url \@url }\providecommand \@url [1]{\endgroup\@href {#1}{\urlprefix }}\providecommand \urlprefix  [0]{URL }\providecommand \Eprint [0]{\href }\providecommand \doibase [0]{https://doi.org/}\providecommand \selectlanguage [0]{\@gobble}\providecommand \bibinfo  [0]{\@secondoftwo}\providecommand \bibfield  [0]{\@secondoftwo}\providecommand \translation [1]{[#1]}\providecommand \BibitemOpen [0]{}\providecommand \bibitemStop [0]{}\providecommand \bibitemNoStop [0]{.\EOS\space}\providecommand \EOS [0]{\spacefactor3000\relax}\providecommand \BibitemShut  [1]{\csname bibitem#1\endcsname}\let\auto@bib@innerbib\@empty
\bibitem [{\citenamefont {{Baade}}\ and\ \citenamefont
  {{Zwicky}}(1934)}]{1934PNAS...20..259B}\BibitemOpen
  \bibfield  {author} {\bibinfo {author} {\bibfnamefont {W.}~\bibnamefont
  {{Baade}}}\ and\ \bibinfo {author} {\bibfnamefont {F.}~\bibnamefont
  {{Zwicky}}},\ }\bibfield  {title} {\bibinfo {title} {{Cosmic Rays from
  Super-novae}},\ }\href {https://doi.org/10.1073/pnas.20.5.259} {\bibfield
  {journal} {\bibinfo  {journal} {Proceedings of the National Academy of
  Science}\ }\textbf {\bibinfo {volume} {20}},\ \bibinfo {pages} {259}
  (\bibinfo {year} {1934})}\BibitemShut {NoStop}\bibitem [{\citenamefont {{Staelin}}\ and\ \citenamefont
  {{Reifenstein}}(1968)}]{1968Sci...162.1481S}\BibitemOpen
  \bibfield  {author} {\bibinfo {author} {\bibfnamefont {D.~H.}\ \bibnamefont
  {{Staelin}}}\ and\ \bibinfo {author} {\bibfnamefont {E.~C.}\ \bibnamefont
  {{Reifenstein}}, \bibfnamefont {III}},\ }\bibfield  {title} {\bibinfo {title}
  {{Pulsating Radio Sources near the Crab Nebula}},\ }\href
  {https://doi.org/10.1126/science.162.3861.1481} {\bibfield  {journal}
  {\bibinfo  {journal} {Science}\ }\textbf {\bibinfo {volume} {162}},\ \bibinfo
  {pages} {1481} (\bibinfo {year} {1968})}\BibitemShut {NoStop}\bibitem [{\citenamefont {{Comella}}\ \emph {et~al.}(1969)\citenamefont
  {{Comella}}, \citenamefont {{Craft}}, \citenamefont {{Lovelace}},\ and\
  \citenamefont {{Sutton}}}]{1969Natur.221..453C}\BibitemOpen
  \bibfield  {author} {\bibinfo {author} {\bibfnamefont {J.~M.}\ \bibnamefont
  {{Comella}}}, \bibinfo {author} {\bibfnamefont {H.~D.}\ \bibnamefont
  {{Craft}}}, \bibinfo {author} {\bibfnamefont {R.~V.~E.}\ \bibnamefont
  {{Lovelace}}},\ and\ \bibinfo {author} {\bibfnamefont {J.~M.}\ \bibnamefont
  {{Sutton}}},\ }\bibfield  {title} {\bibinfo {title} {{Crab Nebula Pulsar NP
  0532}},\ }\href {https://doi.org/10.1038/221453a0} {\bibfield  {journal}
  {\bibinfo  {journal} {\nat}\ }\textbf {\bibinfo {volume} {221}},\ \bibinfo
  {pages} {453} (\bibinfo {year} {1969})}\BibitemShut {NoStop}\bibitem [{\citenamefont {{Large}}\ \emph {et~al.}(1968)\citenamefont
  {{Large}}, \citenamefont {{Vaughan}},\ and\ \citenamefont
  {{Mills}}}]{1968Natur.220..340L}\BibitemOpen
  \bibfield  {author} {\bibinfo {author} {\bibfnamefont {M.~I.}\ \bibnamefont
  {{Large}}}, \bibinfo {author} {\bibfnamefont {A.~E.}\ \bibnamefont
  {{Vaughan}}},\ and\ \bibinfo {author} {\bibfnamefont {B.~Y.}\ \bibnamefont
  {{Mills}}},\ }\bibfield  {title} {\bibinfo {title} {{A Pulsar Supernova
  Association?}},\ }\href {https://doi.org/10.1038/220340a0} {\bibfield
  {journal} {\bibinfo  {journal} {\nat}\ }\textbf {\bibinfo {volume} {220}},\
  \bibinfo {pages} {340} (\bibinfo {year} {1968})}\BibitemShut {NoStop}\bibitem [{\citenamefont {{Janka}}\ and\ \citenamefont
  {{Mueller}}(1994)}]{1994A&A...290..496J}\BibitemOpen
  \bibfield  {author} {\bibinfo {author} {\bibfnamefont {H.-T.}\ \bibnamefont
  {{Janka}}}\ and\ \bibinfo {author} {\bibfnamefont {E.}~\bibnamefont
  {{Mueller}}},\ }\bibfield  {title} {\bibinfo {title} {{Neutron star recoils
  from anisotropic supernovae.}},\ }\href@noop {} {\bibfield  {journal}
  {\bibinfo  {journal} {\aap}\ }\textbf {\bibinfo {volume} {290}},\ \bibinfo
  {pages} {496} (\bibinfo {year} {1994})}\BibitemShut {NoStop}\bibitem [{\citenamefont {{Burrows}}\ \emph {et~al.}(1995)\citenamefont
  {{Burrows}}, \citenamefont {{Hayes}},\ and\ \citenamefont
  {{Fryxell}}}]{1995ApJ...450..830B}\BibitemOpen
  \bibfield  {author} {\bibinfo {author} {\bibfnamefont {A.}~\bibnamefont
  {{Burrows}}}, \bibinfo {author} {\bibfnamefont {J.}~\bibnamefont {{Hayes}}},\
  and\ \bibinfo {author} {\bibfnamefont {B.~A.}\ \bibnamefont {{Fryxell}}},\
  }\bibfield  {title} {\bibinfo {title} {{On the Nature of Core-Collapse
  Supernova Explosions}},\ }\href {https://doi.org/10.1086/176188} {\bibfield
  {journal} {\bibinfo  {journal} {\apj}\ }\textbf {\bibinfo {volume} {450}},\
  \bibinfo {pages} {830} (\bibinfo {year} {1995})},\ \Eprint
  {https://arxiv.org/abs/astro-ph/9506061} {arXiv:astro-ph/9506061 [astro-ph]}
  \BibitemShut {NoStop}\bibitem [{\citenamefont {{Herant}}(1995)}]{1995PhR...256..117H}\BibitemOpen
  \bibfield  {author} {\bibinfo {author} {\bibfnamefont {M.}~\bibnamefont
  {{Herant}}},\ }\bibfield  {title} {\bibinfo {title} {{The convective engine
  paradigm for the supernova explosion mechanism and its consequences.}},\
  }\href {https://doi.org/10.1016/0370-1573(94)00105-C} {\bibfield  {journal}
  {\bibinfo  {journal} {\physrep}\ }\textbf {\bibinfo {volume} {256}},\
  \bibinfo {pages} {117} (\bibinfo {year} {1995})}\BibitemShut {NoStop}\bibitem [{\citenamefont {{Wongwathanarat}}\ \emph {et~al.}(2013)\citenamefont
  {{Wongwathanarat}}, \citenamefont {{Janka}},\ and\ \citenamefont
  {{M{\"u}ller}}}]{2013A&A...552A.126W}\BibitemOpen
  \bibfield  {author} {\bibinfo {author} {\bibfnamefont {A.}~\bibnamefont
  {{Wongwathanarat}}}, \bibinfo {author} {\bibfnamefont {H.-T.}\ \bibnamefont
  {{Janka}}},\ and\ \bibinfo {author} {\bibfnamefont {E.}~\bibnamefont
  {{M{\"u}ller}}},\ }\bibfield  {title} {\bibinfo {title} {{Three-dimensional
  neutrino-driven supernovae: Neutron star kicks, spins, and asymmetric
  ejection of nucleosynthesis products}},\ }\href
  {https://doi.org/10.1051/0004-6361/201220636} {\bibfield  {journal} {\bibinfo
   {journal} {\aap}\ }\textbf {\bibinfo {volume} {552}},\ \bibinfo {eid} {A126}
  (\bibinfo {year} {2013})},\ \Eprint {https://arxiv.org/abs/1210.8148}
  {arXiv:1210.8148 [astro-ph.HE]} \BibitemShut {NoStop}\bibitem [{\citenamefont {{Wang}}\ and\ \citenamefont
  {{Wheeler}}(2008)}]{2008ARA&A..46..433W}\BibitemOpen
  \bibfield  {author} {\bibinfo {author} {\bibfnamefont {L.}~\bibnamefont
  {{Wang}}}\ and\ \bibinfo {author} {\bibfnamefont {J.~C.}\ \bibnamefont
  {{Wheeler}}},\ }\bibfield  {title} {\bibinfo {title} {{Spectropolarimetry of
  supernovae.}},\ }\href
  {https://doi.org/10.1146/annurev.astro.46.060407.145139} {\bibfield
  {journal} {\bibinfo  {journal} {\araa}\ }\textbf {\bibinfo {volume} {46}},\
  \bibinfo {pages} {433} (\bibinfo {year} {2008})},\ \Eprint
  {https://arxiv.org/abs/0811.1054} {arXiv:0811.1054 [astro-ph]} \BibitemShut
  {NoStop}\bibitem [{\citenamefont {{Grefenstette}}\ \emph {et~al.}(2014)\citenamefont
  {{Grefenstette}}, \citenamefont {{Harrison}}, \citenamefont {{Boggs}},
  \citenamefont {{Reynolds}}, \citenamefont {{Fryer}}, \citenamefont
  {{Madsen}}, \citenamefont {{Wik}}, \citenamefont {{Zoglauer}}, \citenamefont
  {{Ellinger}}, \citenamefont {{Alexander}}, \citenamefont {{An}},
  \citenamefont {{Barret}}, \citenamefont {{Christensen}}, \citenamefont
  {{Craig}}, \citenamefont {{Forster}}, \citenamefont {{Giommi}}, \citenamefont
  {{Hailey}}, \citenamefont {{Hornstrup}}, \citenamefont {{Kaspi}},
  \citenamefont {{Kitaguchi}}, \citenamefont {{Koglin}}, \citenamefont {{Mao}},
  \citenamefont {{Miyasaka}}, \citenamefont {{Mori}}, \citenamefont {{Perri}},
  \citenamefont {{Pivovaroff}}, \citenamefont {{Puccetti}}, \citenamefont
  {{Rana}}, \citenamefont {{Stern}}, \citenamefont {{Westergaard}},\ and\
  \citenamefont {{Zhang}}}]{2014Natur.506..339G}\BibitemOpen
  \bibfield  {author} {\bibinfo {author} {\bibfnamefont {B.~W.}\ \bibnamefont
  {{Grefenstette}}}, \bibinfo {author} {\bibfnamefont {F.~A.}\ \bibnamefont
  {{Harrison}}}, \bibinfo {author} {\bibfnamefont {S.~E.}\ \bibnamefont
  {{Boggs}}}, \bibinfo {author} {\bibfnamefont {S.~P.}\ \bibnamefont
  {{Reynolds}}}, \bibinfo {author} {\bibfnamefont {C.~L.}\ \bibnamefont
  {{Fryer}}}, \bibinfo {author} {\bibfnamefont {K.~K.}\ \bibnamefont
  {{Madsen}}}, \bibinfo {author} {\bibfnamefont {D.~R.}\ \bibnamefont {{Wik}}},
  \bibinfo {author} {\bibfnamefont {A.}~\bibnamefont {{Zoglauer}}}, \bibinfo
  {author} {\bibfnamefont {C.~I.}\ \bibnamefont {{Ellinger}}}, \bibinfo
  {author} {\bibfnamefont {D.~M.}\ \bibnamefont {{Alexander}}}, \bibinfo
  {author} {\bibfnamefont {H.}~\bibnamefont {{An}}}, \bibinfo {author}
  {\bibfnamefont {D.}~\bibnamefont {{Barret}}}, \bibinfo {author}
  {\bibfnamefont {F.~E.}\ \bibnamefont {{Christensen}}}, \bibinfo {author}
  {\bibfnamefont {W.~W.}\ \bibnamefont {{Craig}}}, \bibinfo {author}
  {\bibfnamefont {K.}~\bibnamefont {{Forster}}}, \bibinfo {author}
  {\bibfnamefont {P.}~\bibnamefont {{Giommi}}}, \bibinfo {author}
  {\bibfnamefont {C.~J.}\ \bibnamefont {{Hailey}}}, \bibinfo {author}
  {\bibfnamefont {A.}~\bibnamefont {{Hornstrup}}}, \bibinfo {author}
  {\bibfnamefont {V.~M.}\ \bibnamefont {{Kaspi}}}, \bibinfo {author}
  {\bibfnamefont {T.}~\bibnamefont {{Kitaguchi}}}, \bibinfo {author}
  {\bibfnamefont {J.~E.}\ \bibnamefont {{Koglin}}}, \bibinfo {author}
  {\bibfnamefont {P.~H.}\ \bibnamefont {{Mao}}}, \bibinfo {author}
  {\bibfnamefont {H.}~\bibnamefont {{Miyasaka}}}, \bibinfo {author}
  {\bibfnamefont {K.}~\bibnamefont {{Mori}}}, \bibinfo {author} {\bibfnamefont
  {M.}~\bibnamefont {{Perri}}}, \bibinfo {author} {\bibfnamefont {M.~J.}\
  \bibnamefont {{Pivovaroff}}}, \bibinfo {author} {\bibfnamefont
  {S.}~\bibnamefont {{Puccetti}}}, \bibinfo {author} {\bibfnamefont
  {V.}~\bibnamefont {{Rana}}}, \bibinfo {author} {\bibfnamefont
  {D.}~\bibnamefont {{Stern}}}, \bibinfo {author} {\bibfnamefont {N.~J.}\
  \bibnamefont {{Westergaard}}},\ and\ \bibinfo {author} {\bibfnamefont
  {W.~W.}\ \bibnamefont {{Zhang}}},\ }\bibfield  {title} {\bibinfo {title}
  {{Asymmetries in core-collapse supernovae from maps of radioactive $^{44}$Ti
  in CassiopeiaA}},\ }\href {https://doi.org/10.1038/nature12997} {\bibfield
  {journal} {\bibinfo  {journal} {\nat}\ }\textbf {\bibinfo {volume} {506}},\
  \bibinfo {pages} {339} (\bibinfo {year} {2014})},\ \Eprint
  {https://arxiv.org/abs/1403.4978} {arXiv:1403.4978 [astro-ph.HE]}
  \BibitemShut {NoStop}\bibitem [{\citenamefont {{van den Heuvel}}\ and\ \citenamefont {{van
  Paradijs}}(1997)}]{1997ApJ...483..399V}\BibitemOpen
  \bibfield  {author} {\bibinfo {author} {\bibfnamefont {E.~P.~J.}\
  \bibnamefont {{van den Heuvel}}}\ and\ \bibinfo {author} {\bibfnamefont
  {J.}~\bibnamefont {{van Paradijs}}},\ }\bibfield  {title} {\bibinfo {title}
  {{Intrinsic Kicks at Birth Are Required to Explain the Observed Properties of
  Single and Binary Neutron Stars}},\ }\href {https://doi.org/10.1086/304249}
  {\bibfield  {journal} {\bibinfo  {journal} {\apj}\ }\textbf {\bibinfo
  {volume} {483}},\ \bibinfo {pages} {399} (\bibinfo {year}
  {1997})}\BibitemShut {NoStop}\bibitem [{\citenamefont {{Popov}}\ \emph {et~al.}(2025)\citenamefont
  {{Popov}}, \citenamefont {{M{\"u}ller}},\ and\ \citenamefont
  {{Mandel}}}]{2025arXiv250901430D}\BibitemOpen
  \bibfield  {author} {\bibinfo {author} {\bibfnamefont {S.}~\bibnamefont
  {{Popov}}}, \bibinfo {author} {\bibfnamefont {B.}~\bibnamefont
  {{M{\"u}ller}}},\ and\ \bibinfo {author} {\bibfnamefont {I.}~\bibnamefont
  {{Mandel}}},\ }\bibfield  {title} {\bibinfo {title} {{Natal kicks of compact
  objects}},\ }\href {https://doi.org/10.1016/j.newar.2025.101734} {\bibfield
  {journal} {\bibinfo  {journal} {\nar}\ }\textbf {\bibinfo {volume} {101}},\
  \bibinfo {eid} {101734} (\bibinfo {year} {2025})},\ \Eprint
  {https://arxiv.org/abs/2509.01430} {arXiv:2509.01430 [astro-ph.HE]}
  \BibitemShut {NoStop}\bibitem [{\citenamefont {{Burrows}}\ \emph {et~al.}(2024)\citenamefont
  {{Burrows}}, \citenamefont {{Wang}}, \citenamefont {{Vartanyan}},\ and\
  \citenamefont {{Coleman}}}]{2024ApJ...963...63B}\BibitemOpen
  \bibfield  {author} {\bibinfo {author} {\bibfnamefont {A.}~\bibnamefont
  {{Burrows}}}, \bibinfo {author} {\bibfnamefont {T.}~\bibnamefont {{Wang}}},
  \bibinfo {author} {\bibfnamefont {D.}~\bibnamefont {{Vartanyan}}},\ and\
  \bibinfo {author} {\bibfnamefont {M.~S.~B.}\ \bibnamefont {{Coleman}}},\
  }\bibfield  {title} {\bibinfo {title} {{A Theory for Neutron Star and Black
  Hole Kicks and Induced Spins}},\ }\href
  {https://doi.org/10.3847/1538-4357/ad2353} {\bibfield  {journal} {\bibinfo
  {journal} {\apj}\ }\textbf {\bibinfo {volume} {963}},\ \bibinfo {eid} {63}
  (\bibinfo {year} {2024})},\ \Eprint {https://arxiv.org/abs/2311.12109}
  {arXiv:2311.12109 [astro-ph.HE]} \BibitemShut {NoStop}\bibitem [{\citenamefont {{Gunn}}\ and\ \citenamefont
  {{Ostriker}}(1970)}]{1970ApJ...160..979G}\BibitemOpen
  \bibfield  {author} {\bibinfo {author} {\bibfnamefont {J.~E.}\ \bibnamefont
  {{Gunn}}}\ and\ \bibinfo {author} {\bibfnamefont {J.~P.}\ \bibnamefont
  {{Ostriker}}},\ }\bibfield  {title} {\bibinfo {title} {{On the Nature of
  Pulsars. III. Analysis of Observations}},\ }\href
  {https://doi.org/10.1086/150487} {\bibfield  {journal} {\bibinfo  {journal}
  {\apj}\ }\textbf {\bibinfo {volume} {160}},\ \bibinfo {pages} {979} (\bibinfo
  {year} {1970})}\BibitemShut {NoStop}\bibitem [{\citenamefont {{Lyne}}\ and\ \citenamefont
  {{Lorimer}}(1994)}]{1994Natur.369..127L}\BibitemOpen
  \bibfield  {author} {\bibinfo {author} {\bibfnamefont {A.~G.}\ \bibnamefont
  {{Lyne}}}\ and\ \bibinfo {author} {\bibfnamefont {D.~R.}\ \bibnamefont
  {{Lorimer}}},\ }\bibfield  {title} {\bibinfo {title} {{High birth velocities
  of radio pulsars}},\ }\href {https://doi.org/10.1038/369127a0} {\bibfield
  {journal} {\bibinfo  {journal} {\nat}\ }\textbf {\bibinfo {volume} {369}},\
  \bibinfo {pages} {127} (\bibinfo {year} {1994})}\BibitemShut {NoStop}\bibitem [{\citenamefont {Hobbs}\ \emph {et~al.}(2005)\citenamefont {Hobbs},
  \citenamefont {Lorimer}, \citenamefont {Lyne},\ and\ \citenamefont
  {Kramer}}]{HobbsKicks}\BibitemOpen
  \bibfield  {author} {\bibinfo {author} {\bibfnamefont {G.}~\bibnamefont
  {Hobbs}}, \bibinfo {author} {\bibfnamefont {D.~R.}\ \bibnamefont {Lorimer}},
  \bibinfo {author} {\bibfnamefont {A.~G.}\ \bibnamefont {Lyne}},\ and\
  \bibinfo {author} {\bibfnamefont {M.}~\bibnamefont {Kramer}},\ }\bibfield
  {title} {\bibinfo {title} {A statistical study of 233 pulsar proper
  motions},\ }\href {https://doi.org/10.1111/j.1365-2966.2005.09087.x}
  {\bibfield  {journal} {\bibinfo  {journal} {MNRAS}\ }\textbf {\bibinfo
  {volume} {360}},\ \bibinfo {pages} {974} (\bibinfo {year} {2005})},\ \Eprint
  {https://arxiv.org/abs/astro-ph/0504584} {arXiv:astro-ph/0504584}
  \BibitemShut {NoStop}\bibitem [{\citenamefont {{Chatterjee}}\ \emph {et~al.}(2005)\citenamefont
  {{Chatterjee}}, \citenamefont {{Vlemmings}}, \citenamefont {{Brisken}},
  \citenamefont {{Lazio}}, \citenamefont {{Cordes}}, \citenamefont {{Goss}},
  \citenamefont {{Thorsett}}, \citenamefont {{Fomalont}}, \citenamefont
  {{Lyne}},\ and\ \citenamefont {{Kramer}}}]{2005ApJ...630L..61C}\BibitemOpen
  \bibfield  {author} {\bibinfo {author} {\bibfnamefont {S.}~\bibnamefont
  {{Chatterjee}}}, \bibinfo {author} {\bibfnamefont {W.~H.~T.}\ \bibnamefont
  {{Vlemmings}}}, \bibinfo {author} {\bibfnamefont {W.~F.}\ \bibnamefont
  {{Brisken}}}, \bibinfo {author} {\bibfnamefont {T.~J.~W.}\ \bibnamefont
  {{Lazio}}}, \bibinfo {author} {\bibfnamefont {J.~M.}\ \bibnamefont
  {{Cordes}}}, \bibinfo {author} {\bibfnamefont {W.~M.}\ \bibnamefont
  {{Goss}}}, \bibinfo {author} {\bibfnamefont {S.~E.}\ \bibnamefont
  {{Thorsett}}}, \bibinfo {author} {\bibfnamefont {E.~B.}\ \bibnamefont
  {{Fomalont}}}, \bibinfo {author} {\bibfnamefont {A.~G.}\ \bibnamefont
  {{Lyne}}},\ and\ \bibinfo {author} {\bibfnamefont {M.}~\bibnamefont
  {{Kramer}}},\ }\bibfield  {title} {\bibinfo {title} {{Getting Its Kicks: A
  VLBA Parallax for the Hyperfast Pulsar B1508+55}},\ }\href
  {https://doi.org/10.1086/491701} {\bibfield  {journal} {\bibinfo  {journal}
  {\apjl}\ }\textbf {\bibinfo {volume} {630}},\ \bibinfo {pages} {L61}
  (\bibinfo {year} {2005})},\ \Eprint {https://arxiv.org/abs/astro-ph/0509031}
  {arXiv:astro-ph/0509031 [astro-ph]} \BibitemShut {NoStop}\bibitem [{\citenamefont {Verbunt}\ \emph {et~al.}(2017)\citenamefont
  {Verbunt}, \citenamefont {Igoshev},\ and\ \citenamefont
  {Cator}}]{2017A&A...608A..57V}\BibitemOpen
  \bibfield  {author} {\bibinfo {author} {\bibfnamefont {F.}~\bibnamefont
  {Verbunt}}, \bibinfo {author} {\bibfnamefont {A.}~\bibnamefont {Igoshev}},\
  and\ \bibinfo {author} {\bibfnamefont {E.}~\bibnamefont {Cator}},\ }\bibfield
   {title} {\bibinfo {title} {The observed velocity distribution of young
  pulsars},\ }\href {https://doi.org/10.1051/0004-6361/201731518} {\bibfield
  {journal} {\bibinfo  {journal} {A\&A}\ }\textbf {\bibinfo {volume} {608}},\
  \bibinfo {pages} {A57} (\bibinfo {year} {2017})},\ \Eprint
  {https://arxiv.org/abs/1708.08281} {arXiv:1708.08281} \BibitemShut {NoStop}\bibitem [{\citenamefont {Igoshev}\ \emph {et~al.}(2021)\citenamefont
  {Igoshev}, \citenamefont {Chruslinska}, \citenamefont {Dorozsmai},\ and\
  \citenamefont {Toonen}}]{Igoshev2021Combined}\BibitemOpen
  \bibfield  {author} {\bibinfo {author} {\bibfnamefont {A.~P.}\ \bibnamefont
  {Igoshev}}, \bibinfo {author} {\bibfnamefont {M.}~\bibnamefont
  {Chruslinska}}, \bibinfo {author} {\bibfnamefont {A.}~\bibnamefont
  {Dorozsmai}},\ and\ \bibinfo {author} {\bibfnamefont {S.}~\bibnamefont
  {Toonen}},\ }\bibfield  {title} {\bibinfo {title} {Combined analysis of
  neutron star natal kicks using proper motions and parallax measurements for
  radio pulsars and be x-ray binaries},\ }\href
  {https://doi.org/10.1093/mnras/stab2734} {\bibfield  {journal} {\bibinfo
  {journal} {MNRAS}\ }\textbf {\bibinfo {volume} {508}},\ \bibinfo {pages}
  {3345} (\bibinfo {year} {2021})},\ \Eprint {https://arxiv.org/abs/2109.10362}
  {arXiv:2109.10362} \BibitemShut {NoStop}\bibitem [{\citenamefont {Shariat}\ \emph {et~al.}(2026)\citenamefont
  {Shariat}, \citenamefont {El-Badry},\ and\ \citenamefont
  {Naoz}}]{Shariat2026Concordance}\BibitemOpen
  \bibfield  {author} {\bibinfo {author} {\bibfnamefont {C.}~\bibnamefont
  {Shariat}}, \bibinfo {author} {\bibfnamefont {K.}~\bibnamefont {El-Badry}},\
  and\ \bibinfo {author} {\bibfnamefont {S.}~\bibnamefont {Naoz}},\ }\bibfield
  {title} {\bibinfo {title} {A concordance model of neutron star kicks},\
  }\href {https://doi.org/10.48550/arXiv.2610.04005} {\bibfield  {journal}
  {\bibinfo  {journal} {arXiv e-prints}\ ,\ \bibinfo {eid} {arXiv:2610.04005}}
  (\bibinfo {year} {2026})},\ \Eprint {https://arxiv.org/abs/2610.04005}
  {arXiv:2610.04005} \BibitemShut {NoStop}\bibitem [{\citenamefont {Disberg}\ \emph {et~al.}(2026)\citenamefont
  {Disberg}, \citenamefont {Mandel},\ and\ \citenamefont
  {Hirai}}]{Disberg2026BinaryKicks}\BibitemOpen
  \bibfield  {author} {\bibinfo {author} {\bibfnamefont {P.}~\bibnamefont
  {Disberg}}, \bibinfo {author} {\bibfnamefont {I.}~\bibnamefont {Mandel}},\
  and\ \bibinfo {author} {\bibfnamefont {R.}~\bibnamefont {Hirai}},\ }\bibfield
   {title} {\bibinfo {title} {The kick velocities of neutron stars in binary
  systems},\ }\href {https://doi.org/10.48550/arXiv.2608.19690} {\bibfield
  {journal} {\bibinfo  {journal} {arXiv e-prints}\ ,\ \bibinfo {eid}
  {arXiv:2608.19690}} (\bibinfo {year} {2026})},\ \Eprint
  {https://arxiv.org/abs/2608.19690} {arXiv:2608.19690 [astro-ph.SR]}
  \BibitemShut {NoStop}\bibitem [{\citenamefont {{Arzoumanian}}\ \emph {et~al.}(2002)\citenamefont
  {{Arzoumanian}}, \citenamefont {{Chernoff}},\ and\ \citenamefont
  {{Cordes}}}]{ArKicks}\BibitemOpen
  \bibfield  {author} {\bibinfo {author} {\bibfnamefont {Z.}~\bibnamefont
  {{Arzoumanian}}}, \bibinfo {author} {\bibfnamefont {D.~F.}\ \bibnamefont
  {{Chernoff}}},\ and\ \bibinfo {author} {\bibfnamefont {J.~M.}\ \bibnamefont
  {{Cordes}}},\ }\bibfield  {title} {\bibinfo {title} {{The Velocity
  Distribution of Isolated Radio Pulsars}},\ }\href
  {https://doi.org/10.1086/338805} {\bibfield  {journal} {\bibinfo  {journal}
  {\apj}\ }\textbf {\bibinfo {volume} {568}},\ \bibinfo {pages} {289} (\bibinfo
  {year} {2002})},\ \Eprint {https://arxiv.org/abs/astro-ph/0106159}
  {arXiv:astro-ph/0106159 [astro-ph]} \BibitemShut {NoStop}\bibitem [{\citenamefont {{Faucher-Gigu{\`e}re}}\ and\ \citenamefont
  {{Kaspi}}(2006)}]{2006ApJ...643..332F}\BibitemOpen
  \bibfield  {author} {\bibinfo {author} {\bibfnamefont {C.-A.}\ \bibnamefont
  {{Faucher-Gigu{\`e}re}}}\ and\ \bibinfo {author} {\bibfnamefont {V.~M.}\
  \bibnamefont {{Kaspi}}},\ }\bibfield  {title} {\bibinfo {title} {{Birth and
  Evolution of Isolated Radio Pulsars}},\ }\href
  {https://doi.org/10.1086/501516} {\bibfield  {journal} {\bibinfo  {journal}
  {\apj}\ }\textbf {\bibinfo {volume} {643}},\ \bibinfo {pages} {332} (\bibinfo
  {year} {2006})},\ \Eprint {https://arxiv.org/abs/astro-ph/0512585}
  {arXiv:astro-ph/0512585 [astro-ph]} \BibitemShut {NoStop}\bibitem [{\citenamefont {{Igoshev}}(2020)}]{2020MNRAS.494.3663I}\BibitemOpen
  \bibfield  {author} {\bibinfo {author} {\bibfnamefont {A.~P.}\ \bibnamefont
  {{Igoshev}}},\ }\bibfield  {title} {\bibinfo {title} {{The observed velocity
  distribution of young pulsars - II. Analysis of complete
  PSR{\ensuremath{\pi}}}},\ }\href {https://doi.org/10.1093/mnras/staa958}
  {\bibfield  {journal} {\bibinfo  {journal} {\mnras}\ }\textbf {\bibinfo
  {volume} {494}},\ \bibinfo {pages} {3663} (\bibinfo {year} {2020})},\ \Eprint
  {https://arxiv.org/abs/2002.01367} {arXiv:2002.01367 [astro-ph.HE]}
  \BibitemShut {NoStop}\bibitem [{\citenamefont {{Disberg}}\ and\ \citenamefont
  {{Mandel}}(2025)}]{2025ApJ...989L...8D}\BibitemOpen
  \bibfield  {author} {\bibinfo {author} {\bibfnamefont {P.}~\bibnamefont
  {{Disberg}}}\ and\ \bibinfo {author} {\bibfnamefont {I.}~\bibnamefont
  {{Mandel}}},\ }\bibfield  {title} {\bibinfo {title} {{The Kick Velocity
  Distribution of Isolated Neutron Stars}},\ }\href
  {https://doi.org/10.3847/2041-8213/adf286} {\bibfield  {journal} {\bibinfo
  {journal} {\apjl}\ }\textbf {\bibinfo {volume} {989}},\ \bibinfo {pages} {L8}
  (\bibinfo {year} {2025})},\ \Eprint {https://arxiv.org/abs/2505.22102}
  {arXiv:2505.22102 [astro-ph.HE]} \BibitemShut {NoStop}\bibitem [{\citenamefont {{O'Doherty}}\ \emph {et~al.}(2023)\citenamefont
  {{O'Doherty}}, \citenamefont {{Bahramian}}, \citenamefont {{Miller-Jones}},
  \citenamefont {{Goodwin}}, \citenamefont {{Mandel}}, \citenamefont
  {{Willcox}}, \citenamefont {{Atri}},\ and\ \citenamefont
  {{Strader}}}]{2023MNRAS.521.2504O}\BibitemOpen
  \bibfield  {author} {\bibinfo {author} {\bibfnamefont {T.~N.}\ \bibnamefont
  {{O'Doherty}}}, \bibinfo {author} {\bibfnamefont {A.}~\bibnamefont
  {{Bahramian}}}, \bibinfo {author} {\bibfnamefont {J.~C.~A.}\ \bibnamefont
  {{Miller-Jones}}}, \bibinfo {author} {\bibfnamefont {A.~J.}\ \bibnamefont
  {{Goodwin}}}, \bibinfo {author} {\bibfnamefont {I.}~\bibnamefont {{Mandel}}},
  \bibinfo {author} {\bibfnamefont {R.}~\bibnamefont {{Willcox}}}, \bibinfo
  {author} {\bibfnamefont {P.}~\bibnamefont {{Atri}}},\ and\ \bibinfo {author}
  {\bibfnamefont {J.}~\bibnamefont {{Strader}}},\ }\bibfield  {title} {\bibinfo
  {title} {An observationally-derived kick distribution for neutron stars in
  binary systems},\ }\href {https://doi.org/10.1093/mnras/stad680} {\bibfield
  {journal} {\bibinfo  {journal} {\mnras}\ }\textbf {\bibinfo {volume} {521}},\
  \bibinfo {pages} {2504} (\bibinfo {year} {2023})},\ \Eprint
  {https://arxiv.org/abs/2303.01059} {arXiv:2303.01059 [astro-ph.HE]}
  \BibitemShut {NoStop}\bibitem [{\citenamefont {{Wong}}\ \emph {et~al.}(2010)\citenamefont {{Wong}},
  \citenamefont {{Willems}},\ and\ \citenamefont
  {{Kalogera}}}]{psr-kicks-WongWillemsKalogera2010}\BibitemOpen
  \bibfield  {author} {\bibinfo {author} {\bibfnamefont {T.-W.}\ \bibnamefont
  {{Wong}}}, \bibinfo {author} {\bibfnamefont {B.}~\bibnamefont {{Willems}}},\
  and\ \bibinfo {author} {\bibfnamefont {V.}~\bibnamefont {{Kalogera}}},\
  }\bibfield  {title} {\bibinfo {title} {{Constraints on Natal Kicks in
  Galactic Double Neutron Star Systems}},\ }\href
  {https://doi.org/10.1088/0004-637X/721/2/1689} {\bibfield  {journal}
  {\bibinfo  {journal} {\apj}\ }\textbf {\bibinfo {volume} {721}},\ \bibinfo
  {pages} {1689} (\bibinfo {year} {2010})},\ \Eprint
  {https://arxiv.org/abs/1008.2397} {arXiv:1008.2397 [astro-ph.GA]}
  \BibitemShut {NoStop}\bibitem [{\citenamefont {{Andrews}}\ \emph {et~al.}(2015)\citenamefont
  {{Andrews}}, \citenamefont {{Farr}}, \citenamefont {{Kalogera}},\ and\
  \citenamefont {{Willems}}}]{2015ApJ...801...32A}\BibitemOpen
  \bibfield  {author} {\bibinfo {author} {\bibfnamefont {J.~J.}\ \bibnamefont
  {{Andrews}}}, \bibinfo {author} {\bibfnamefont {W.~M.}\ \bibnamefont
  {{Farr}}}, \bibinfo {author} {\bibfnamefont {V.}~\bibnamefont {{Kalogera}}},\
  and\ \bibinfo {author} {\bibfnamefont {B.}~\bibnamefont {{Willems}}},\
  }\bibfield  {title} {\bibinfo {title} {Evolutionary channels for the
  formation of double neutron stars},\ }\href
  {https://doi.org/10.1088/0004-637X/801/1/32} {\bibfield  {journal} {\bibinfo
  {journal} {\apj}\ }\textbf {\bibinfo {volume} {801}},\ \bibinfo {eid} {32}
  (\bibinfo {year} {2015})},\ \Eprint {https://arxiv.org/abs/1410.6797}
  {arXiv:1410.6797 [astro-ph.SR]} \BibitemShut {NoStop}\bibitem [{\citenamefont {Tauris}\ \emph {et~al.}(2017)\citenamefont {Tauris},
  \citenamefont {Kramer}, \citenamefont {Freire}, \citenamefont {Wex},
  \citenamefont {Janka}, \citenamefont {Langer}, \citenamefont {Podsiadlowski},
  \citenamefont {Bozzo}, \citenamefont {Chaty}, \citenamefont {Kruckow},
  \citenamefont {van~den Heuvel}, \citenamefont {Antoniadis}, \citenamefont
  {Breton},\ and\ \citenamefont {Champion}}]{2017ApJ...846..170T}\BibitemOpen
  \bibfield  {author} {\bibinfo {author} {\bibfnamefont {T.~M.}\ \bibnamefont
  {Tauris}}, \bibinfo {author} {\bibfnamefont {M.}~\bibnamefont {Kramer}},
  \bibinfo {author} {\bibfnamefont {P.~C.~C.}\ \bibnamefont {Freire}}, \bibinfo
  {author} {\bibfnamefont {N.}~\bibnamefont {Wex}}, \bibinfo {author}
  {\bibfnamefont {H.-T.}\ \bibnamefont {Janka}}, \bibinfo {author}
  {\bibfnamefont {N.}~\bibnamefont {Langer}}, \bibinfo {author} {\bibfnamefont
  {P.}~\bibnamefont {Podsiadlowski}}, \bibinfo {author} {\bibfnamefont
  {E.}~\bibnamefont {Bozzo}}, \bibinfo {author} {\bibfnamefont
  {S.}~\bibnamefont {Chaty}}, \bibinfo {author} {\bibfnamefont {M.~U.}\
  \bibnamefont {Kruckow}}, \bibinfo {author} {\bibfnamefont {E.~P.~J.}\
  \bibnamefont {van~den Heuvel}}, \bibinfo {author} {\bibfnamefont
  {J.}~\bibnamefont {Antoniadis}}, \bibinfo {author} {\bibfnamefont {R.~P.}\
  \bibnamefont {Breton}},\ and\ \bibinfo {author} {\bibfnamefont {D.~J.}\
  \bibnamefont {Champion}},\ }\bibfield  {title} {\bibinfo {title} {Formation
  of double neutron star systems},\ }\href
  {https://doi.org/10.3847/1538-4357/aa7e89} {\bibfield  {journal} {\bibinfo
  {journal} {\apj}\ }\textbf {\bibinfo {volume} {846}},\ \bibinfo {pages} {170}
  (\bibinfo {year} {2017})},\ \Eprint {https://arxiv.org/abs/1706.09438}
  {arXiv:1706.09438} \BibitemShut {NoStop}\bibitem [{\citenamefont {{Breivik}}\ \emph {et~al.}(2020)\citenamefont
  {{Breivik}}, \citenamefont {{Coughlin}}, \citenamefont {{Zevin}},
  \citenamefont {{Rodriguez}}, \citenamefont {{Kremer}}, \citenamefont {{Ye}},
  \citenamefont {{Andrews}}, \citenamefont {{Kurkowski}}, \citenamefont
  {{Digman}}, \citenamefont {{Larson}},\ and\ \citenamefont
  {{Rasio}}}]{2020ApJ...898...71B}\BibitemOpen
  \bibfield  {author} {\bibinfo {author} {\bibfnamefont {K.}~\bibnamefont
  {{Breivik}}}, \bibinfo {author} {\bibfnamefont {S.}~\bibnamefont
  {{Coughlin}}}, \bibinfo {author} {\bibfnamefont {M.}~\bibnamefont {{Zevin}}},
  \bibinfo {author} {\bibfnamefont {C.~L.}\ \bibnamefont {{Rodriguez}}},
  \bibinfo {author} {\bibfnamefont {K.}~\bibnamefont {{Kremer}}}, \bibinfo
  {author} {\bibfnamefont {C.~S.}\ \bibnamefont {{Ye}}}, \bibinfo {author}
  {\bibfnamefont {J.~J.}\ \bibnamefont {{Andrews}}}, \bibinfo {author}
  {\bibfnamefont {M.}~\bibnamefont {{Kurkowski}}}, \bibinfo {author}
  {\bibfnamefont {M.~C.}\ \bibnamefont {{Digman}}}, \bibinfo {author}
  {\bibfnamefont {S.~L.}\ \bibnamefont {{Larson}}},\ and\ \bibinfo {author}
  {\bibfnamefont {F.~A.}\ \bibnamefont {{Rasio}}},\ }\bibfield  {title}
  {\bibinfo {title} {{COSMIC Variance in Binary Population Synthesis}},\ }\href
  {https://doi.org/10.3847/1538-4357/ab9d85} {\bibfield  {journal} {\bibinfo
  {journal} {\apj}\ }\textbf {\bibinfo {volume} {898}},\ \bibinfo {eid} {71}
  (\bibinfo {year} {2020})},\ \Eprint {https://arxiv.org/abs/1911.00903}
  {arXiv:1911.00903 [astro-ph.HE]} \BibitemShut {NoStop}\bibitem [{\citenamefont {{Mandel}}\ \emph {et~al.}(2021)\citenamefont
  {{Mandel}}, \citenamefont {{M{\"u}ller}}, \citenamefont {{Riley}},
  \citenamefont {{de Mink}}, \citenamefont {{Vigna-G{\'o}mez}},\ and\
  \citenamefont {{Chattopadhyay}}}]{2021MNRAS.500.1380M}\BibitemOpen
  \bibfield  {author} {\bibinfo {author} {\bibfnamefont {I.}~\bibnamefont
  {{Mandel}}}, \bibinfo {author} {\bibfnamefont {B.}~\bibnamefont
  {{M{\"u}ller}}}, \bibinfo {author} {\bibfnamefont {J.}~\bibnamefont
  {{Riley}}}, \bibinfo {author} {\bibfnamefont {S.~E.}\ \bibnamefont {{de
  Mink}}}, \bibinfo {author} {\bibfnamefont {A.}~\bibnamefont
  {{Vigna-G{\'o}mez}}},\ and\ \bibinfo {author} {\bibfnamefont
  {D.}~\bibnamefont {{Chattopadhyay}}},\ }\bibfield  {title} {\bibinfo {title}
  {{Binary population synthesis with probabilistic remnant mass and kick
  prescriptions}},\ }\href {https://doi.org/10.1093/mnras/staa3390} {\bibfield
  {journal} {\bibinfo  {journal} {\mnras}\ }\textbf {\bibinfo {volume} {500}},\
  \bibinfo {pages} {1380} (\bibinfo {year} {2021})},\ \Eprint
  {https://arxiv.org/abs/2007.03890} {arXiv:2007.03890 [astro-ph.HE]}
  \BibitemShut {NoStop}\bibitem [{\citenamefont {{Barrett}}\ \emph {et~al.}(2018)\citenamefont
  {{Barrett}}, \citenamefont {{Gaebel}}, \citenamefont {{Neijssel}},
  \citenamefont {{Vigna-G{\'o}mez}}, \citenamefont {{Stevenson}}, \citenamefont
  {{Berry}}, \citenamefont {{Farr}},\ and\ \citenamefont
  {{Mandel}}}]{2018MNRAS.477.4685B}\BibitemOpen
  \bibfield  {author} {\bibinfo {author} {\bibfnamefont {J.~W.}\ \bibnamefont
  {{Barrett}}}, \bibinfo {author} {\bibfnamefont {S.~M.}\ \bibnamefont
  {{Gaebel}}}, \bibinfo {author} {\bibfnamefont {C.~J.}\ \bibnamefont
  {{Neijssel}}}, \bibinfo {author} {\bibfnamefont {A.}~\bibnamefont
  {{Vigna-G{\'o}mez}}}, \bibinfo {author} {\bibfnamefont {S.}~\bibnamefont
  {{Stevenson}}}, \bibinfo {author} {\bibfnamefont {C.~P.~L.}\ \bibnamefont
  {{Berry}}}, \bibinfo {author} {\bibfnamefont {W.~M.}\ \bibnamefont
  {{Farr}}},\ and\ \bibinfo {author} {\bibfnamefont {I.}~\bibnamefont
  {{Mandel}}},\ }\bibfield  {title} {\bibinfo {title} {{Accuracy of inference
  on the physics of binary evolution from gravitational-wave observations}},\
  }\href {https://doi.org/10.1093/mnras/sty908} {\bibfield  {journal} {\bibinfo
   {journal} {\mnras}\ }\textbf {\bibinfo {volume} {477}},\ \bibinfo {pages}
  {4685} (\bibinfo {year} {2018})},\ \Eprint {https://arxiv.org/abs/1711.06287}
  {arXiv:1711.06287 [astro-ph.HE]} \BibitemShut {NoStop}\bibitem [{\citenamefont {{Kalogera}}(2000)}]{2000ApJ...541..319K}\BibitemOpen
  \bibfield  {author} {\bibinfo {author} {\bibfnamefont {V.}~\bibnamefont
  {{Kalogera}}},\ }\bibfield  {title} {\bibinfo {title} {{Spin-Orbit
  Misalignment in Close Binaries with Two Compact Objects}},\ }\href
  {https://doi.org/10.1086/309400} {\bibfield  {journal} {\bibinfo  {journal}
  {\apj}\ }\textbf {\bibinfo {volume} {541}},\ \bibinfo {pages} {319} (\bibinfo
  {year} {2000})},\ \Eprint {https://arxiv.org/abs/astro-ph/9911417}
  {arXiv:astro-ph/9911417 [astro-ph]} \BibitemShut {NoStop}\bibitem [{\citenamefont {{Postnov}}\ and\ \citenamefont
  {{Kuranov}}(2008)}]{ns-polarkicks-rateimplications-Postnov2007}\BibitemOpen
  \bibfield  {author} {\bibinfo {author} {\bibfnamefont {K.~A.}\ \bibnamefont
  {{Postnov}}}\ and\ \bibinfo {author} {\bibfnamefont {A.~G.}\ \bibnamefont
  {{Kuranov}}},\ }\bibfield  {title} {\bibinfo {title} {{Neutron star spin-kick
  velocity correlation effect on binary neutron star coalescence rates and
  spin-orbit misalignment of the components}},\ }\href
  {https://doi.org/10.1111/j.1365-2966.2007.12635.x} {\bibfield  {journal}
  {\bibinfo  {journal} {\mnras}\ }\textbf {\bibinfo {volume} {384}},\ \bibinfo
  {pages} {1393} (\bibinfo {year} {2008})},\ \Eprint
  {https://arxiv.org/abs/0710.4465} {arXiv:0710.4465 [astro-ph]} \BibitemShut
  {NoStop}\bibitem [{\citenamefont {{Brandt}}\ and\ \citenamefont
  {{Podsiadlowski}}(1995)}]{1995MNRAS.274..461B}\BibitemOpen
  \bibfield  {author} {\bibinfo {author} {\bibfnamefont {N.}~\bibnamefont
  {{Brandt}}}\ and\ \bibinfo {author} {\bibfnamefont {P.}~\bibnamefont
  {{Podsiadlowski}}},\ }\bibfield  {title} {\bibinfo {title} {{The effects of
  high-velocity supernova kicks on the orbital properties and sky distributions
  of neutron-star binaries}},\ }\href {https://doi.org/10.1093/mnras/274.2.461}
  {\bibfield  {journal} {\bibinfo  {journal} {\mnras}\ }\textbf {\bibinfo
  {volume} {274}},\ \bibinfo {pages} {461} (\bibinfo {year} {1995})},\ \Eprint
  {https://arxiv.org/abs/astro-ph/9412023} {arXiv:astro-ph/9412023}
  \BibitemShut {NoStop}\bibitem [{\citenamefont {Valli}\ \emph {et~al.}(2025)\citenamefont {Valli},
  \citenamefont {de~Mink}, \citenamefont {Justham}, \citenamefont {Callister},
  \citenamefont {Johnston}, \citenamefont {Kresse}, \citenamefont {Langer},
  \citenamefont {Rubio}, \citenamefont {Vigna-G{\'o}mez},\ and\ \citenamefont
  {Wang}}]{Valli2025Polar}\BibitemOpen
  \bibfield  {author} {\bibinfo {author} {\bibfnamefont {R.}~\bibnamefont
  {Valli}}, \bibinfo {author} {\bibfnamefont {S.~E.}\ \bibnamefont {de~Mink}},
  \bibinfo {author} {\bibfnamefont {S.}~\bibnamefont {Justham}}, \bibinfo
  {author} {\bibfnamefont {T.}~\bibnamefont {Callister}}, \bibinfo {author}
  {\bibfnamefont {C.}~\bibnamefont {Johnston}}, \bibinfo {author}
  {\bibfnamefont {D.}~\bibnamefont {Kresse}}, \bibinfo {author} {\bibfnamefont
  {N.}~\bibnamefont {Langer}}, \bibinfo {author} {\bibfnamefont {A.~C.}\
  \bibnamefont {Rubio}}, \bibinfo {author} {\bibfnamefont {A.}~\bibnamefont
  {Vigna-G{\'o}mez}},\ and\ \bibinfo {author} {\bibfnamefont {C.}~\bibnamefont
  {Wang}},\ }\bibfield  {title} {\bibinfo {title} {Evidence of polar and
  ultralow supernova kicks from the orbits of be x-ray binaries},\ }\href
  {https://doi.org/10.48550/arXiv.2505.08857} {\bibfield  {journal} {\bibinfo
  {journal} {arXiv e-prints}\ ,\ \bibinfo {eid} {arXiv:2505.08857}} (\bibinfo
  {year} {2025})},\ \Eprint {https://arxiv.org/abs/2505.08857}
  {arXiv:2505.08857} \BibitemShut {NoStop}\bibitem [{\citenamefont {{Mandel}}\ \emph {et~al.}(2019)\citenamefont
  {{Mandel}}, \citenamefont {{Farr}},\ and\ \citenamefont
  {{Gair}}}]{LIGO-P1600187-Farr-SelectionBiasaAndMonteCarlo}\BibitemOpen
  \bibfield  {author} {\bibinfo {author} {\bibfnamefont {I.}~\bibnamefont
  {{Mandel}}}, \bibinfo {author} {\bibfnamefont {W.~M.}\ \bibnamefont
  {{Farr}}},\ and\ \bibinfo {author} {\bibfnamefont {J.~R.}\ \bibnamefont
  {{Gair}}},\ }\bibfield  {title} {\bibinfo {title} {{Extracting distribution
  parameters from multiple uncertain observations with selection biases}},\
  }\href {https://doi.org/10.1093/mnras/stz896} {\bibfield  {journal} {\bibinfo
   {journal} {\mnras}\ }\textbf {\bibinfo {volume} {486}},\ \bibinfo {pages}
  {1086} (\bibinfo {year} {2019})},\ \Eprint {https://arxiv.org/abs/1809.02063}
  {arXiv:1809.02063 [physics.data-an]} \BibitemShut {NoStop}\bibitem [{\citenamefont {Vitale}\ \emph {et~al.}(2022)\citenamefont {Vitale},
  \citenamefont {Gerosa}, \citenamefont {Farr},\ and\ \citenamefont
  {Taylor}}]{VitaleGerosaFarrTaylor2022}\BibitemOpen
  \bibfield  {author} {\bibinfo {author} {\bibfnamefont {S.}~\bibnamefont
  {Vitale}}, \bibinfo {author} {\bibfnamefont {D.}~\bibnamefont {Gerosa}},
  \bibinfo {author} {\bibfnamefont {W.~M.}\ \bibnamefont {Farr}},\ and\
  \bibinfo {author} {\bibfnamefont {S.~R.}\ \bibnamefont {Taylor}},\ }\bibfield
   {title} {\bibinfo {title} {Inferring the properties of a population of
  compact binaries in presence of selection effects},\ }in\ \href
  {https://doi.org/10.1007/978-981-15-4702-7_45-1} {\emph {\bibinfo {booktitle}
  {Handbook of Gravitational Wave Astronomy}}}\ (\bibinfo  {publisher}
  {Springer},\ \bibinfo {address} {Singapore},\ \bibinfo {year} {2022})\
  p.~\bibinfo {pages} {45},\ \Eprint {https://arxiv.org/abs/2007.05579}
  {arXiv:2007.05579 [astro-ph.IM]} \BibitemShut {NoStop}\bibitem [{\citenamefont {{Hills}}(1983)}]{1983ApJ...267..322H}\BibitemOpen
  \bibfield  {author} {\bibinfo {author} {\bibfnamefont {J.~G.}\ \bibnamefont
  {{Hills}}},\ }\bibfield  {title} {\bibinfo {title} {{The effects of sudden
  mass loss and a random kick velocity produced in a supernova explosion on the
  dynamics of a binary star of arbitrary orbital eccentricity. Applications to
  X-ray binaries and to the binarypulsars.}},\ }\href
  {https://doi.org/10.1086/160871} {\bibfield  {journal} {\bibinfo  {journal}
  {\apj}\ }\textbf {\bibinfo {volume} {267}},\ \bibinfo {pages} {322} (\bibinfo
  {year} {1983})}\BibitemShut {NoStop}\bibitem [{\citenamefont {{Kalogera}}(1996)}]{1996ApJ...471..352K}\BibitemOpen
  \bibfield  {author} {\bibinfo {author} {\bibfnamefont {V.}~\bibnamefont
  {{Kalogera}}},\ }\bibfield  {title} {\bibinfo {title} {{Orbital
  Characteristics of Binary Systems after Asymmetric Supernova Explosions}},\
  }\href {https://doi.org/10.1086/177974} {\bibfield  {journal} {\bibinfo
  {journal} {\apj}\ }\textbf {\bibinfo {volume} {471}},\ \bibinfo {pages} {352}
  (\bibinfo {year} {1996})},\ \Eprint {https://arxiv.org/abs/astro-ph/9605186}
  {arXiv:astro-ph/9605186} \BibitemShut {NoStop}\bibitem [{\citenamefont {Kapil}\ \emph {et~al.}(2023)\citenamefont {Kapil},
  \citenamefont {Mandel}, \citenamefont {Berti},\ and\ \citenamefont
  {M{\"u}ller}}]{2023MNRAS.519.5893K}\BibitemOpen
  \bibfield  {author} {\bibinfo {author} {\bibfnamefont {V.}~\bibnamefont
  {Kapil}}, \bibinfo {author} {\bibfnamefont {I.}~\bibnamefont {Mandel}},
  \bibinfo {author} {\bibfnamefont {E.}~\bibnamefont {Berti}},\ and\ \bibinfo
  {author} {\bibfnamefont {B.}~\bibnamefont {M{\"u}ller}},\ }\bibfield  {title}
  {\bibinfo {title} {{Calibration of neutron star natal kick velocities to
  isolated pulsar observations}},\ }\href
  {https://doi.org/10.1093/mnras/stad019} {\bibfield  {journal} {\bibinfo
  {journal} {\mnras}\ }\textbf {\bibinfo {volume} {519}},\ \bibinfo {pages}
  {5893} (\bibinfo {year} {2023})},\ \Eprint {https://arxiv.org/abs/2209.09252}
  {arXiv:2209.09252 [astro-ph.HE]} \BibitemShut {NoStop}\bibitem [{\citenamefont {{Mandel}}\ and\ \citenamefont
  {{Broekgaarden}}(2022)}]{2022LRR....25....1M}\BibitemOpen
  \bibfield  {author} {\bibinfo {author} {\bibfnamefont {I.}~\bibnamefont
  {{Mandel}}}\ and\ \bibinfo {author} {\bibfnamefont {F.~S.}\ \bibnamefont
  {{Broekgaarden}}},\ }\bibfield  {title} {\bibinfo {title} {{Rates of compact
  object coalescences}},\ }\href {https://doi.org/10.1007/s41114-021-00034-3}
  {\bibfield  {journal} {\bibinfo  {journal} {Living Reviews in Relativity}\
  }\textbf {\bibinfo {volume} {25}},\ \bibinfo {eid} {1} (\bibinfo {year}
  {2022})},\ \Eprint {https://arxiv.org/abs/2107.14239} {arXiv:2107.14239
  [astro-ph.HE]} \BibitemShut {NoStop}\bibitem [{\citenamefont {O'Shaughnessy}\ \emph
  {et~al.}(2005{\natexlab{a}})\citenamefont {O'Shaughnessy}, \citenamefont
  {Kalogera},\ and\ \citenamefont {Belczynski}}]{PSutility}\BibitemOpen
  \bibfield  {author} {\bibinfo {author} {\bibfnamefont {R.}~\bibnamefont
  {O'Shaughnessy}}, \bibinfo {author} {\bibfnamefont {V.}~\bibnamefont
  {Kalogera}},\ and\ \bibinfo {author} {\bibfnamefont {K.}~\bibnamefont
  {Belczynski}},\ }\bibfield  {title} {\bibinfo {title} {Mapping inspiral rates
  on population synthesis parameters},\ }\href {https://doi.org/10.1086/426903}
  {\bibfield  {journal} {\bibinfo  {journal} {Astrophysical Journal}\ }\textbf
  {\bibinfo {volume} {620}},\ \bibinfo {pages} {385} (\bibinfo {year}
  {2005}{\natexlab{a}})},\ \Eprint {https://arxiv.org/abs/astro-ph/0408387}
  {arXiv:astro-ph/0408387} \BibitemShut {NoStop}\bibitem [{\citenamefont {O'Shaughnessy}\ \emph {et~al.}(2007)\citenamefont
  {O'Shaughnessy}, \citenamefont {Kalogera},\ and\ \citenamefont
  {Belczynski}}]{PSutil2}\BibitemOpen
  \bibfield  {author} {\bibinfo {author} {\bibfnamefont {R.}~\bibnamefont
  {O'Shaughnessy}}, \bibinfo {author} {\bibfnamefont {V.}~\bibnamefont
  {Kalogera}},\ and\ \bibinfo {author} {\bibfnamefont {K.}~\bibnamefont
  {Belczynski}},\ }\bibfield  {title} {\bibinfo {title} {Mapping population
  synthesis event rates on model parameters ii: Convergence and accuracy of
  multidimensional fits},\ }\href {https://doi.org/10.1086/520756} {\bibfield
  {journal} {\bibinfo  {journal} {Astrophysical Journal}\ }\textbf {\bibinfo
  {volume} {667}},\ \bibinfo {pages} {1048} (\bibinfo {year} {2007})},\ \Eprint
  {https://arxiv.org/abs/astro-ph/0609465} {arXiv:astro-ph/0609465}
  \BibitemShut {NoStop}\bibitem [{\citenamefont {O'Shaughnessy}\ \emph
  {et~al.}(2005{\natexlab{b}})\citenamefont {O'Shaughnessy}, \citenamefont
  {Kim}, \citenamefont {Fragos}, \citenamefont {Kalogera},\ and\ \citenamefont
  {Belczynski}}]{PSconstraints}\BibitemOpen
  \bibfield  {author} {\bibinfo {author} {\bibfnamefont {R.}~\bibnamefont
  {O'Shaughnessy}}, \bibinfo {author} {\bibfnamefont {C.}~\bibnamefont {Kim}},
  \bibinfo {author} {\bibfnamefont {T.}~\bibnamefont {Fragos}}, \bibinfo
  {author} {\bibfnamefont {V.}~\bibnamefont {Kalogera}},\ and\ \bibinfo
  {author} {\bibfnamefont {K.}~\bibnamefont {Belczynski}},\ }\bibfield  {title}
  {\bibinfo {title} {Constraining population synthesis models via the binary
  neutron star population},\ }\href {https://doi.org/10.1086/468180} {\bibfield
   {journal} {\bibinfo  {journal} {Astrophysical Journal}\ }\textbf {\bibinfo
  {volume} {633}},\ \bibinfo {pages} {1076} (\bibinfo {year}
  {2005}{\natexlab{b}})},\ \Eprint {https://arxiv.org/abs/astro-ph/0504479}
  {arXiv:astro-ph/0504479} \BibitemShut {NoStop}\bibitem [{\citenamefont {O'Shaughnessy}\ \emph {et~al.}(2008)\citenamefont
  {O'Shaughnessy}, \citenamefont {Kim}, \citenamefont {Kalogera},\ and\
  \citenamefont {Belczynski}}]{OShaughnessy2008Constraints}\BibitemOpen
  \bibfield  {author} {\bibinfo {author} {\bibfnamefont {R.}~\bibnamefont
  {O'Shaughnessy}}, \bibinfo {author} {\bibfnamefont {C.}~\bibnamefont {Kim}},
  \bibinfo {author} {\bibfnamefont {V.}~\bibnamefont {Kalogera}},\ and\
  \bibinfo {author} {\bibfnamefont {K.}~\bibnamefont {Belczynski}},\ }\bibfield
   {title} {\bibinfo {title} {{Constraining Population Synthesis Models via
  Empirical Binary Compact Object Merger and Supernova Rates}},\ }\href
  {https://doi.org/10.1086/523620} {\bibfield  {journal} {\bibinfo  {journal}
  {Astrophysical Journal}\ }\textbf {\bibinfo {volume} {672}},\ \bibinfo
  {pages} {479} (\bibinfo {year} {2008})},\ \Eprint
  {https://arxiv.org/abs/astro-ph/0610076} {arXiv:astro-ph/0610076}
  \BibitemShut {NoStop}\bibitem [{\citenamefont {{Dominik}}\ \emph {et~al.}(2012)\citenamefont
  {{Dominik}}, \citenamefont {{Belczynski}}, \citenamefont {{Fryer}},
  \citenamefont {{Holz}}, \citenamefont {{Berti}}, \citenamefont {{Bulik}},
  \citenamefont {{Mandel}},\ and\ \citenamefont
  {{O'Shaughnessy}}}]{2012ApJ...759...52D}\BibitemOpen
  \bibfield  {author} {\bibinfo {author} {\bibfnamefont {M.}~\bibnamefont
  {{Dominik}}}, \bibinfo {author} {\bibfnamefont {K.}~\bibnamefont
  {{Belczynski}}}, \bibinfo {author} {\bibfnamefont {C.}~\bibnamefont
  {{Fryer}}}, \bibinfo {author} {\bibfnamefont {D.~E.}\ \bibnamefont {{Holz}}},
  \bibinfo {author} {\bibfnamefont {E.}~\bibnamefont {{Berti}}}, \bibinfo
  {author} {\bibfnamefont {T.}~\bibnamefont {{Bulik}}}, \bibinfo {author}
  {\bibfnamefont {I.}~\bibnamefont {{Mandel}}},\ and\ \bibinfo {author}
  {\bibfnamefont {R.}~\bibnamefont {{O'Shaughnessy}}},\ }\bibfield  {title}
  {\bibinfo {title} {{Double Compact Objects. I. The Significance of the Common
  Envelope on Merger Rates}},\ }\href
  {https://doi.org/10.1088/0004-637X/759/1/52} {\bibfield  {journal} {\bibinfo
  {journal} {\apj}\ }\textbf {\bibinfo {volume} {759}},\ \bibinfo {eid} {52}
  (\bibinfo {year} {2012})},\ \Eprint {https://arxiv.org/abs/1202.4901}
  {arXiv:1202.4901 [astro-ph.HE]} \BibitemShut {NoStop}\bibitem [{\citenamefont {{Chruslinska}}\ \emph {et~al.}(2018)\citenamefont
  {{Chruslinska}}, \citenamefont {{Belczynski}}, \citenamefont {{Klencki}},\
  and\ \citenamefont {{Benacquista}}}]{2017arXiv170807885C}\BibitemOpen
  \bibfield  {author} {\bibinfo {author} {\bibfnamefont {M.}~\bibnamefont
  {{Chruslinska}}}, \bibinfo {author} {\bibfnamefont {K.}~\bibnamefont
  {{Belczynski}}}, \bibinfo {author} {\bibfnamefont {J.}~\bibnamefont
  {{Klencki}}},\ and\ \bibinfo {author} {\bibfnamefont {M.}~\bibnamefont
  {{Benacquista}}},\ }\bibfield  {title} {\bibinfo {title} {{Double neutron
  stars: merger rates revisited}},\ }\href
  {https://doi.org/10.1093/mnras/stx2923} {\bibfield  {journal} {\bibinfo
  {journal} {\mnras}\ }\textbf {\bibinfo {volume} {474}},\ \bibinfo {pages}
  {2937} (\bibinfo {year} {2018})},\ \Eprint {https://arxiv.org/abs/1708.07885}
  {arXiv:1708.07885 [astro-ph.HE]} \BibitemShut {NoStop}\bibitem [{\citenamefont {{Vigna-G{\'o}mez}}\ \emph {et~al.}(2018)\citenamefont
  {{Vigna-G{\'o}mez}}, \citenamefont {{Neijssel}}, \citenamefont {{Stevenson}},
  \citenamefont {{Barrett}}, \citenamefont {{Belczynski}}, \citenamefont
  {{Justham}}, \citenamefont {{de Mink}}, \citenamefont {{M{\"u}ller}},
  \citenamefont {{Podsiadlowski}}, \citenamefont {{Renzo}}, \citenamefont
  {{Sz{\'e}csi}},\ and\ \citenamefont {{Mandel}}}]{2018MNRAS.481.4009V}\BibitemOpen
  \bibfield  {author} {\bibinfo {author} {\bibfnamefont {A.}~\bibnamefont
  {{Vigna-G{\'o}mez}}}, \bibinfo {author} {\bibfnamefont {C.~J.}\ \bibnamefont
  {{Neijssel}}}, \bibinfo {author} {\bibfnamefont {S.}~\bibnamefont
  {{Stevenson}}}, \bibinfo {author} {\bibfnamefont {J.~W.}\ \bibnamefont
  {{Barrett}}}, \bibinfo {author} {\bibfnamefont {K.}~\bibnamefont
  {{Belczynski}}}, \bibinfo {author} {\bibfnamefont {S.}~\bibnamefont
  {{Justham}}}, \bibinfo {author} {\bibfnamefont {S.~E.}\ \bibnamefont {{de
  Mink}}}, \bibinfo {author} {\bibfnamefont {B.}~\bibnamefont {{M{\"u}ller}}},
  \bibinfo {author} {\bibfnamefont {P.}~\bibnamefont {{Podsiadlowski}}},
  \bibinfo {author} {\bibfnamefont {M.}~\bibnamefont {{Renzo}}}, \bibinfo
  {author} {\bibfnamefont {D.}~\bibnamefont {{Sz{\'e}csi}}},\ and\ \bibinfo
  {author} {\bibfnamefont {I.}~\bibnamefont {{Mandel}}},\ }\bibfield  {title}
  {\bibinfo {title} {{On the formation history of Galactic double neutron
  stars}},\ }\href {https://doi.org/10.1093/mnras/sty2463} {\bibfield
  {journal} {\bibinfo  {journal} {\mnras}\ }\textbf {\bibinfo {volume} {481}},\
  \bibinfo {pages} {4009} (\bibinfo {year} {2018})},\ \Eprint
  {https://arxiv.org/abs/1805.07974} {arXiv:1805.07974 [astro-ph.SR]}
  \BibitemShut {NoStop}\bibitem [{\citenamefont {{Giacobbo}}\ and\ \citenamefont
  {{Mapelli}}(2018)}]{2018MNRAS.480.2011G}\BibitemOpen
  \bibfield  {author} {\bibinfo {author} {\bibfnamefont {N.}~\bibnamefont
  {{Giacobbo}}}\ and\ \bibinfo {author} {\bibfnamefont {M.}~\bibnamefont
  {{Mapelli}}},\ }\bibfield  {title} {\bibinfo {title} {{The progenitors of
  compact-object binaries: impact of metallicity, common envelope and natal
  kicks}},\ }\href {https://doi.org/10.1093/mnras/sty1999} {\bibfield
  {journal} {\bibinfo  {journal} {\mnras}\ }\textbf {\bibinfo {volume} {480}},\
  \bibinfo {pages} {2011} (\bibinfo {year} {2018})},\ \Eprint
  {https://arxiv.org/abs/1806.00001} {arXiv:1806.00001 [astro-ph.HE]}
  \BibitemShut {NoStop}\bibitem [{\citenamefont {{Kruckow}}\ \emph {et~al.}(2018)\citenamefont
  {{Kruckow}}, \citenamefont {{Tauris}}, \citenamefont {{Langer}},
  \citenamefont {{Kramer}},\ and\ \citenamefont
  {{Izzard}}}]{2018MNRAS.481.1908K}\BibitemOpen
  \bibfield  {author} {\bibinfo {author} {\bibfnamefont {M.~U.}\ \bibnamefont
  {{Kruckow}}}, \bibinfo {author} {\bibfnamefont {T.~M.}\ \bibnamefont
  {{Tauris}}}, \bibinfo {author} {\bibfnamefont {N.}~\bibnamefont {{Langer}}},
  \bibinfo {author} {\bibfnamefont {M.}~\bibnamefont {{Kramer}}},\ and\
  \bibinfo {author} {\bibfnamefont {R.~G.}\ \bibnamefont {{Izzard}}},\
  }\bibfield  {title} {\bibinfo {title} {{Progenitors of gravitational wave
  mergers: binary evolution with the stellar grid-based code COMBINE}},\ }\href
  {https://doi.org/10.1093/mnras/sty2190} {\bibfield  {journal} {\bibinfo
  {journal} {\mnras}\ }\textbf {\bibinfo {volume} {481}},\ \bibinfo {pages}
  {1908} (\bibinfo {year} {2018})},\ \Eprint {https://arxiv.org/abs/1801.05433}
  {arXiv:1801.05433 [astro-ph.SR]} \BibitemShut {NoStop}\bibitem [{\citenamefont {Broekgaarden}\ \emph {et~al.}(2022)\citenamefont
  {Broekgaarden}, \citenamefont {Berger}, \citenamefont {Stevenson} \emph
  {et~al.}}]{Broekgaarden2022Rates}\BibitemOpen
  \bibfield  {author} {\bibinfo {author} {\bibfnamefont {F.~S.}\ \bibnamefont
  {Broekgaarden}}, \bibinfo {author} {\bibfnamefont {E.}~\bibnamefont
  {Berger}}, \bibinfo {author} {\bibfnamefont {S.}~\bibnamefont {Stevenson}},
  \emph {et~al.},\ }\bibfield  {title} {\bibinfo {title} {Impact of massive
  binary star and cosmic evolution on gravitational wave observations ii:
  Double compact object rates and properties},\ }\href
  {https://doi.org/10.1093/mnras/stac1677} {\bibfield  {journal} {\bibinfo
  {journal} {Monthly Notices of the Royal Astronomical Society}\ }\textbf
  {\bibinfo {volume} {516}},\ \bibinfo {pages} {5737} (\bibinfo {year}
  {2022})},\ \Eprint {https://arxiv.org/abs/2112.05763} {arXiv:2112.05763}
  \BibitemShut {NoStop}\bibitem [{\citenamefont {{Iorio}}\ \emph {et~al.}(2023)\citenamefont
  {{Iorio}}, \citenamefont {{Mapelli}}, \citenamefont {{Costa}}, \citenamefont
  {{Spera}}, \citenamefont {{Escobar}}, \citenamefont {{Sgalletta}},
  \citenamefont {{Trani}}, \citenamefont {{Korb}}, \citenamefont
  {{Santoliquido}}, \citenamefont {{Dall'Amico}}, \citenamefont {{Gaspari}},\
  and\ \citenamefont {{Bressan}}}]{2023MNRAS.524..426I}\BibitemOpen
  \bibfield  {author} {\bibinfo {author} {\bibfnamefont {G.}~\bibnamefont
  {{Iorio}}}, \bibinfo {author} {\bibfnamefont {M.}~\bibnamefont {{Mapelli}}},
  \bibinfo {author} {\bibfnamefont {G.}~\bibnamefont {{Costa}}}, \bibinfo
  {author} {\bibfnamefont {M.}~\bibnamefont {{Spera}}}, \bibinfo {author}
  {\bibfnamefont {G.~J.}\ \bibnamefont {{Escobar}}}, \bibinfo {author}
  {\bibfnamefont {C.}~\bibnamefont {{Sgalletta}}}, \bibinfo {author}
  {\bibfnamefont {A.~A.}\ \bibnamefont {{Trani}}}, \bibinfo {author}
  {\bibfnamefont {E.}~\bibnamefont {{Korb}}}, \bibinfo {author} {\bibfnamefont
  {F.}~\bibnamefont {{Santoliquido}}}, \bibinfo {author} {\bibfnamefont
  {M.}~\bibnamefont {{Dall'Amico}}}, \bibinfo {author} {\bibfnamefont
  {N.}~\bibnamefont {{Gaspari}}},\ and\ \bibinfo {author} {\bibfnamefont
  {A.}~\bibnamefont {{Bressan}}},\ }\bibfield  {title} {\bibinfo {title}
  {{Compact object mergers: exploring uncertainties from stellar and binary
  evolution with SEVN}},\ }\href {https://doi.org/10.1093/mnras/stad1630}
  {\bibfield  {journal} {\bibinfo  {journal} {\mnras}\ }\textbf {\bibinfo
  {volume} {524}},\ \bibinfo {pages} {426} (\bibinfo {year} {2023})},\ \Eprint
  {https://arxiv.org/abs/2211.11774} {arXiv:2211.11774 [astro-ph.HE]}
  \BibitemShut {NoStop}\bibitem [{\citenamefont {{Boesky}}\ \emph {et~al.}(2024)\citenamefont
  {{Boesky}}, \citenamefont {{Broekgaarden}},\ and\ \citenamefont
  {{Berger}}}]{2024ApJ...976...24B}\BibitemOpen
  \bibfield  {author} {\bibinfo {author} {\bibfnamefont {A.~P.}\ \bibnamefont
  {{Boesky}}}, \bibinfo {author} {\bibfnamefont {F.~S.}\ \bibnamefont
  {{Broekgaarden}}},\ and\ \bibinfo {author} {\bibfnamefont {E.}~\bibnamefont
  {{Berger}}},\ }\bibfield  {title} {\bibinfo {title} {{Investigating the
  Cosmological Rate of Compact Object Mergers from Isolated Massive Binary
  Stars}},\ }\href {https://doi.org/10.3847/1538-4357/ad7fe3} {\bibfield
  {journal} {\bibinfo  {journal} {\apj}\ }\textbf {\bibinfo {volume} {976}},\
  \bibinfo {eid} {24} (\bibinfo {year} {2024})},\ \Eprint
  {https://arxiv.org/abs/2405.01630} {arXiv:2405.01630 [astro-ph.HE]}
  \BibitemShut {NoStop}\bibitem [{\citenamefont {{Chattaraj}}\ \emph {et~al.}(2026)\citenamefont
  {{Chattaraj}}, \citenamefont {{Andrews}}, \citenamefont {{Bavera}},
  \citenamefont {{Briel}}, \citenamefont {{Chattopadhyay}}, \citenamefont
  {{Fragos}}, \citenamefont {{Gossage}}, \citenamefont {{Kalogera}},
  \citenamefont {{Kovlakas}}, \citenamefont {{Kruckow}}, \citenamefont
  {{Liotine}}, \citenamefont {{Rocha}}, \citenamefont {{Srivastava}},
  \citenamefont {{Sun}}, \citenamefont {{Teng}}, \citenamefont {{Xing}},\ and\
  \citenamefont {{Zapartas}}}]{2026ApJ...997...52C}\BibitemOpen
  \bibfield  {author} {\bibinfo {author} {\bibfnamefont {A.}~\bibnamefont
  {{Chattaraj}}}, \bibinfo {author} {\bibfnamefont {J.~J.}\ \bibnamefont
  {{Andrews}}}, \bibinfo {author} {\bibfnamefont {S.~S.}\ \bibnamefont
  {{Bavera}}}, \bibinfo {author} {\bibfnamefont {M.}~\bibnamefont {{Briel}}},
  \bibinfo {author} {\bibfnamefont {D.}~\bibnamefont {{Chattopadhyay}}},
  \bibinfo {author} {\bibfnamefont {T.}~\bibnamefont {{Fragos}}}, \bibinfo
  {author} {\bibfnamefont {S.}~\bibnamefont {{Gossage}}}, \bibinfo {author}
  {\bibfnamefont {V.}~\bibnamefont {{Kalogera}}}, \bibinfo {author}
  {\bibfnamefont {K.}~\bibnamefont {{Kovlakas}}}, \bibinfo {author}
  {\bibfnamefont {M.~U.}\ \bibnamefont {{Kruckow}}}, \bibinfo {author}
  {\bibfnamefont {C.}~\bibnamefont {{Liotine}}}, \bibinfo {author}
  {\bibfnamefont {K.~A.}\ \bibnamefont {{Rocha}}}, \bibinfo {author}
  {\bibfnamefont {P.~M.}\ \bibnamefont {{Srivastava}}}, \bibinfo {author}
  {\bibfnamefont {M.}~\bibnamefont {{Sun}}}, \bibinfo {author} {\bibfnamefont
  {E.}~\bibnamefont {{Teng}}}, \bibinfo {author} {\bibfnamefont
  {Z.}~\bibnamefont {{Xing}}},\ and\ \bibinfo {author} {\bibfnamefont
  {E.}~\bibnamefont {{Zapartas}}},\ }\bibfield  {title} {\bibinfo {title}
  {{Forming Double Neutron Stars Using Detailed Binary Evolution Models with
  POSYDON: Comparison to the Galactic Systems}},\ }\href
  {https://doi.org/10.3847/1538-4357/ae1f93} {\bibfield  {journal} {\bibinfo
  {journal} {\apj}\ }\textbf {\bibinfo {volume} {997}},\ \bibinfo {eid} {52}
  (\bibinfo {year} {2026})},\ \Eprint {https://arxiv.org/abs/2508.00186}
  {arXiv:2508.00186 [astro-ph.SR]} \BibitemShut {NoStop}\bibitem [{\citenamefont {Venet}\ and\ \citenamefont
  {Dvorkin}(2026)}]{Venet2026BNS}\BibitemOpen
  \bibfield  {author} {\bibinfo {author} {\bibfnamefont {M.}~\bibnamefont
  {Venet}}\ and\ \bibinfo {author} {\bibfnamefont {I.}~\bibnamefont
  {Dvorkin}},\ }\bibfield  {title} {\bibinfo {title} {Binary neutron star
  populations across cosmic time: the impact of binary stellar evolution
  uncertainties},\ }\href {https://doi.org/10.48550/arXiv.2607.27391}
  {\bibfield  {journal} {\bibinfo  {journal} {arXiv e-prints}\ ,\ \bibinfo
  {eid} {arXiv:2607.27391}} (\bibinfo {year} {2026})},\ \Eprint
  {https://arxiv.org/abs/2607.27391} {arXiv:2607.27391} \BibitemShut {NoStop}\bibitem [{\citenamefont {de~Mink}\ and\ \citenamefont
  {Belczynski}(2015)}]{deMink2015Initial}\BibitemOpen
  \bibfield  {author} {\bibinfo {author} {\bibfnamefont {S.~E.}\ \bibnamefont
  {de~Mink}}\ and\ \bibinfo {author} {\bibfnamefont {K.}~\bibnamefont
  {Belczynski}},\ }\bibfield  {title} {\bibinfo {title} {Merger rates of double
  neutron stars and stellar origin black holes: The impact of initial
  conditions on binary evolution predictions},\ }\href
  {https://doi.org/10.1088/0004-637X/814/1/58} {\bibfield  {journal} {\bibinfo
  {journal} {Astrophysical Journal}\ }\textbf {\bibinfo {volume} {814}},\
  \bibinfo {pages} {58} (\bibinfo {year} {2015})},\ \Eprint
  {https://arxiv.org/abs/1506.03573} {arXiv:1506.03573} \BibitemShut {NoStop}\bibitem [{\citenamefont {{Stevenson}}\ \emph {et~al.}(2015)\citenamefont
  {{Stevenson}}, \citenamefont {{Ohme}},\ and\ \citenamefont
  {{Fairhurst}}}]{2015ApJ...810...58S}\BibitemOpen
  \bibfield  {author} {\bibinfo {author} {\bibfnamefont {S.}~\bibnamefont
  {{Stevenson}}}, \bibinfo {author} {\bibfnamefont {F.}~\bibnamefont
  {{Ohme}}},\ and\ \bibinfo {author} {\bibfnamefont {S.}~\bibnamefont
  {{Fairhurst}}},\ }\bibfield  {title} {\bibinfo {title} {{Distinguishing
  Compact Binary Population Synthesis Models Using Gravitational Wave
  Observations of Coalescing Binary Black Holes}},\ }\href
  {https://doi.org/10.1088/0004-637X/810/1/58} {\bibfield  {journal} {\bibinfo
  {journal} {\apj}\ }\textbf {\bibinfo {volume} {810}},\ \bibinfo {eid} {58}
  (\bibinfo {year} {2015})},\ \Eprint {https://arxiv.org/abs/1504.07802}
  {arXiv:1504.07802 [astro-ph.HE]} \BibitemShut {NoStop}\bibitem [{\citenamefont {{Wysocki}}\ \emph {et~al.}(2018)\citenamefont
  {{Wysocki}}, \citenamefont {{Gerosa}}, \citenamefont {{O'Shaughnessy}},
  \citenamefont {{Belczynski}}, \citenamefont {{Gladysz}}, \citenamefont
  {{Berti}}, \citenamefont {{Kesden}},\ and\ \citenamefont
  {{Holz}}}]{gwastro-DanielW-PopsynKickPaper2017}\BibitemOpen
  \bibfield  {author} {\bibinfo {author} {\bibfnamefont {D.}~\bibnamefont
  {{Wysocki}}}, \bibinfo {author} {\bibfnamefont {D.}~\bibnamefont {{Gerosa}}},
  \bibinfo {author} {\bibfnamefont {R.}~\bibnamefont {{O'Shaughnessy}}},
  \bibinfo {author} {\bibfnamefont {K.}~\bibnamefont {{Belczynski}}}, \bibinfo
  {author} {\bibfnamefont {W.}~\bibnamefont {{Gladysz}}}, \bibinfo {author}
  {\bibfnamefont {E.}~\bibnamefont {{Berti}}}, \bibinfo {author} {\bibfnamefont
  {M.}~\bibnamefont {{Kesden}}},\ and\ \bibinfo {author} {\bibfnamefont
  {D.~E.}\ \bibnamefont {{Holz}}},\ }\bibfield  {title} {\bibinfo {title}
  {{Explaining LIGO's observations via isolated binary evolution with natal
  kicks}},\ }\href {https://doi.org/10.1103/PhysRevD.97.043014} {\bibfield
  {journal} {\bibinfo  {journal} {\prd}\ }\textbf {\bibinfo {volume} {97}},\
  \bibinfo {eid} {043014} (\bibinfo {year} {2018})},\ \Eprint
  {https://arxiv.org/abs/1709.01943} {arXiv:1709.01943 [astro-ph.HE]}
  \BibitemShut {NoStop}\bibitem [{\citenamefont {{Zevin}}\ \emph {et~al.}(2021)\citenamefont
  {{Zevin}}, \citenamefont {{Bavera}}, \citenamefont {{Berry}}, \citenamefont
  {{Kalogera}}, \citenamefont {{Fragos}}, \citenamefont {{Marchant}},
  \citenamefont {{Rodriguez}}, \citenamefont {{Antonini}}, \citenamefont
  {{Holz}},\ and\ \citenamefont {{Pankow}}}]{2021ApJ...910..152Z}\BibitemOpen
  \bibfield  {author} {\bibinfo {author} {\bibfnamefont {M.}~\bibnamefont
  {{Zevin}}}, \bibinfo {author} {\bibfnamefont {S.~S.}\ \bibnamefont
  {{Bavera}}}, \bibinfo {author} {\bibfnamefont {C.~P.~L.}\ \bibnamefont
  {{Berry}}}, \bibinfo {author} {\bibfnamefont {V.}~\bibnamefont {{Kalogera}}},
  \bibinfo {author} {\bibfnamefont {T.}~\bibnamefont {{Fragos}}}, \bibinfo
  {author} {\bibfnamefont {P.}~\bibnamefont {{Marchant}}}, \bibinfo {author}
  {\bibfnamefont {C.~L.}\ \bibnamefont {{Rodriguez}}}, \bibinfo {author}
  {\bibfnamefont {F.}~\bibnamefont {{Antonini}}}, \bibinfo {author}
  {\bibfnamefont {D.~E.}\ \bibnamefont {{Holz}}},\ and\ \bibinfo {author}
  {\bibfnamefont {C.}~\bibnamefont {{Pankow}}},\ }\bibfield  {title} {\bibinfo
  {title} {{One Channel to Rule Them All? Constraining the Origins of Binary
  Black Holes Using Multiple Formation Pathways}},\ }\href
  {https://doi.org/10.3847/1538-4357/abe40e} {\bibfield  {journal} {\bibinfo
  {journal} {\apj}\ }\textbf {\bibinfo {volume} {910}},\ \bibinfo {eid} {152}
  (\bibinfo {year} {2021})},\ \Eprint {https://arxiv.org/abs/2011.10057}
  {arXiv:2011.10057 [astro-ph.HE]} \BibitemShut {NoStop}\bibitem [{\citenamefont {{Bouffanais}}\ \emph {et~al.}(2021)\citenamefont
  {{Bouffanais}}, \citenamefont {{Mapelli}}, \citenamefont {{Santoliquido}},
  \citenamefont {{Giacobbo}}, \citenamefont {{Di Carlo}}, \citenamefont
  {{Rastello}}, \citenamefont {{Artale}},\ and\ \citenamefont
  {{Iorio}}}]{2021MNRAS.507.5224B}\BibitemOpen
  \bibfield  {author} {\bibinfo {author} {\bibfnamefont {Y.}~\bibnamefont
  {{Bouffanais}}}, \bibinfo {author} {\bibfnamefont {M.}~\bibnamefont
  {{Mapelli}}}, \bibinfo {author} {\bibfnamefont {F.}~\bibnamefont
  {{Santoliquido}}}, \bibinfo {author} {\bibfnamefont {N.}~\bibnamefont
  {{Giacobbo}}}, \bibinfo {author} {\bibfnamefont {U.~N.}\ \bibnamefont {{Di
  Carlo}}}, \bibinfo {author} {\bibfnamefont {S.}~\bibnamefont {{Rastello}}},
  \bibinfo {author} {\bibfnamefont {M.~C.}\ \bibnamefont {{Artale}}},\ and\
  \bibinfo {author} {\bibfnamefont {G.}~\bibnamefont {{Iorio}}},\ }\bibfield
  {title} {\bibinfo {title} {{New insights on binary black hole formation
  channels after GWTC-2: young star clusters versus isolated binaries}},\
  }\href {https://doi.org/10.1093/mnras/stab2438} {\bibfield  {journal}
  {\bibinfo  {journal} {\mnras}\ }\textbf {\bibinfo {volume} {507}},\ \bibinfo
  {pages} {5224} (\bibinfo {year} {2021})},\ \Eprint
  {https://arxiv.org/abs/2102.12495} {arXiv:2102.12495 [astro-ph.HE]}
  \BibitemShut {NoStop}\bibitem [{\citenamefont {Richards}\ \emph {et~al.}(2023)\citenamefont
  {Richards}, \citenamefont {Eldridge}, \citenamefont {Briel}, \citenamefont
  {Stevance},\ and\ \citenamefont {Willcox}}]{Richards2023Multiple}\BibitemOpen
  \bibfield  {author} {\bibinfo {author} {\bibfnamefont {S.~M.}\ \bibnamefont
  {Richards}}, \bibinfo {author} {\bibfnamefont {J.~J.}\ \bibnamefont
  {Eldridge}}, \bibinfo {author} {\bibfnamefont {M.~M.}\ \bibnamefont {Briel}},
  \bibinfo {author} {\bibfnamefont {H.~F.}\ \bibnamefont {Stevance}},\ and\
  \bibinfo {author} {\bibfnamefont {R.}~\bibnamefont {Willcox}},\ }\bibfield
  {title} {\bibinfo {title} {New constraints on the bray
  conservation-of-momentum natal kick model from multiple distinct
  observations},\ }\href {https://doi.org/10.1093/mnras/stad977} {\bibfield
  {journal} {\bibinfo  {journal} {MNRAS}\ }\textbf {\bibinfo {volume} {522}},\
  \bibinfo {pages} {3972} (\bibinfo {year} {2023})},\ \Eprint
  {https://arxiv.org/abs/2208.02407} {arXiv:2208.02407} \BibitemShut {NoStop}\bibitem [{\citenamefont {{Deng}}\ \emph {et~al.}(2024)\citenamefont {{Deng}},
  \citenamefont {{Li}}, \citenamefont {{Shao}},\ and\ \citenamefont
  {{Xu}}}]{2024ApJ...963...80D}\BibitemOpen
  \bibfield  {author} {\bibinfo {author} {\bibfnamefont {Z.-L.}\ \bibnamefont
  {{Deng}}}, \bibinfo {author} {\bibfnamefont {X.-D.}\ \bibnamefont {{Li}}},
  \bibinfo {author} {\bibfnamefont {Y.}~\bibnamefont {{Shao}}},\ and\ \bibinfo
  {author} {\bibfnamefont {K.}~\bibnamefont {{Xu}}},\ }\bibfield  {title}
  {\bibinfo {title} {{On the Formation of Double Neutron Stars in the Milky
  Way: Influence of Key Parameters}},\ }\href
  {https://doi.org/10.3847/1538-4357/ad2357} {\bibfield  {journal} {\bibinfo
  {journal} {\apj}\ }\textbf {\bibinfo {volume} {963}},\ \bibinfo {eid} {80}
  (\bibinfo {year} {2024})},\ \Eprint {https://arxiv.org/abs/2402.04658}
  {arXiv:2402.04658 [astro-ph.HE]} \BibitemShut {NoStop}\bibitem [{\citenamefont {{Chen}}\ \emph {et~al.}(2025)\citenamefont {{Chen}},
  \citenamefont {{Zhang}}, \citenamefont {{Lu}}, \citenamefont {{Liu}},\ and\
  \citenamefont {{Zeng}}}]{2025MNRAS.544L..89C}\BibitemOpen
  \bibfield  {author} {\bibinfo {author} {\bibfnamefont {Z.}~\bibnamefont
  {{Chen}}}, \bibinfo {author} {\bibfnamefont {J.}~\bibnamefont {{Zhang}}},
  \bibinfo {author} {\bibfnamefont {Y.}~\bibnamefont {{Lu}}}, \bibinfo {author}
  {\bibfnamefont {J.}~\bibnamefont {{Liu}}},\ and\ \bibinfo {author}
  {\bibfnamefont {C.}~\bibnamefont {{Zeng}}},\ }\bibfield  {title} {\bibinfo
  {title} {{Constraining common envelope evolution in binary neutron star
  formation with combined galactic and gravitational-wave observations}},\
  }\href {https://doi.org/10.1093/mnrasl/slaf105} {\bibfield  {journal}
  {\bibinfo  {journal} {\mnras}\ }\textbf {\bibinfo {volume} {544}},\ \bibinfo
  {pages} {L89} (\bibinfo {year} {2025})},\ \Eprint
  {https://arxiv.org/abs/2508.14397} {arXiv:2508.14397 [astro-ph.HE]}
  \BibitemShut {NoStop}\bibitem [{\citenamefont {{Taylor}}\ and\ \citenamefont
  {{Gerosa}}(2018)}]{2018MNRAS.481.4685T}\BibitemOpen
  \bibfield  {author} {\bibinfo {author} {\bibfnamefont {S.~R.}\ \bibnamefont
  {{Taylor}}}\ and\ \bibinfo {author} {\bibfnamefont {D.}~\bibnamefont
  {{Gerosa}}},\ }\bibfield  {title} {\bibinfo {title} {{Mining
  gravitational-wave catalogs to understand binary stellar evolution: A new
  hierarchical Bayesian framework}},\ }\href
  {https://doi.org/10.1103/PhysRevD.98.083017} {\bibfield  {journal} {\bibinfo
  {journal} {\prd}\ }\textbf {\bibinfo {volume} {98}},\ \bibinfo {eid} {083017}
  (\bibinfo {year} {2018})},\ \Eprint {https://arxiv.org/abs/1806.08365}
  {arXiv:1806.08365 [astro-ph.HE]} \BibitemShut {NoStop}\bibitem [{\citenamefont {{Wong}}\ and\ \citenamefont
  {{Gerosa}}(2019)}]{2019PhRvD.100h3015W}\BibitemOpen
  \bibfield  {author} {\bibinfo {author} {\bibfnamefont {K.~W.~K.}\
  \bibnamefont {{Wong}}}\ and\ \bibinfo {author} {\bibfnamefont
  {D.}~\bibnamefont {{Gerosa}}},\ }\bibfield  {title} {\bibinfo {title}
  {{Machine-learning interpolation of population-synthesis simulations to
  interpret gravitational-wave observations: A case study}},\ }\href
  {https://doi.org/10.1103/PhysRevD.100.083015} {\bibfield  {journal} {\bibinfo
   {journal} {\prd}\ }\textbf {\bibinfo {volume} {100}},\ \bibinfo {eid}
  {083015} (\bibinfo {year} {2019})},\ \Eprint
  {https://arxiv.org/abs/1909.06373} {arXiv:1909.06373 [astro-ph.HE]}
  \BibitemShut {NoStop}\bibitem [{\citenamefont {{Wong}}\ \emph {et~al.}(2021)\citenamefont {{Wong}},
  \citenamefont {{Breivik}}, \citenamefont {{Kremer}},\ and\ \citenamefont
  {{Callister}}}]{2021PhRvD.103h3021W}\BibitemOpen
  \bibfield  {author} {\bibinfo {author} {\bibfnamefont {K.~W.~K.}\
  \bibnamefont {{Wong}}}, \bibinfo {author} {\bibfnamefont {K.}~\bibnamefont
  {{Breivik}}}, \bibinfo {author} {\bibfnamefont {K.}~\bibnamefont
  {{Kremer}}},\ and\ \bibinfo {author} {\bibfnamefont {T.}~\bibnamefont
  {{Callister}}},\ }\bibfield  {title} {\bibinfo {title} {{Joint constraints on
  the field-cluster mixing fraction, common envelope efficiency, and globular
  cluster radii from a population of binary hole mergers via deep learning}},\
  }\href {https://doi.org/10.1103/PhysRevD.103.083021} {\bibfield  {journal}
  {\bibinfo  {journal} {\prd}\ }\textbf {\bibinfo {volume} {103}},\ \bibinfo
  {eid} {083021} (\bibinfo {year} {2021})},\ \Eprint
  {https://arxiv.org/abs/2011.03564} {arXiv:2011.03564 [astro-ph.HE]}
  \BibitemShut {NoStop}\bibitem [{\citenamefont {{Riley}}\ and\ \citenamefont
  {{Mandel}}(2023)}]{2023MNRAS.521.2201R}\BibitemOpen
  \bibfield  {author} {\bibinfo {author} {\bibfnamefont {J.}~\bibnamefont
  {{Riley}}}\ and\ \bibinfo {author} {\bibfnamefont {I.}~\bibnamefont
  {{Mandel}}},\ }\bibfield  {title} {\bibinfo {title} {{Surrogate Forward
  Models for Population Inference on Compact Binary Mergers}},\ }\href
  {https://doi.org/10.3847/1538-4357/accf90} {\bibfield  {journal} {\bibinfo
  {journal} {\apj}\ }\textbf {\bibinfo {volume} {950}},\ \bibinfo {eid} {80}
  (\bibinfo {year} {2023})},\ \Eprint {https://arxiv.org/abs/2303.00508}
  {arXiv:2303.00508 [astro-ph.SR]} \BibitemShut {NoStop}\bibitem [{\citenamefont {{Rauf}}\ \emph {et~al.}(2024)\citenamefont {{Rauf}},
  \citenamefont {{Howlett}}, \citenamefont {{Stevenson}}, \citenamefont
  {{Riley}},\ and\ \citenamefont {{Willcox}}}]{2024MNRAS.534.3506R}\BibitemOpen
  \bibfield  {author} {\bibinfo {author} {\bibfnamefont {L.}~\bibnamefont
  {{Rauf}}}, \bibinfo {author} {\bibfnamefont {C.}~\bibnamefont {{Howlett}}},
  \bibinfo {author} {\bibfnamefont {S.}~\bibnamefont {{Stevenson}}}, \bibinfo
  {author} {\bibfnamefont {J.}~\bibnamefont {{Riley}}},\ and\ \bibinfo {author}
  {\bibfnamefont {R.}~\bibnamefont {{Willcox}}},\ }\bibfield  {title} {\bibinfo
  {title} {{A trifecta of modelling tools: a Bayesian binary black hole model
  selection combining population synthesis and galaxy formation models}},\
  }\href {https://doi.org/10.1093/mnras/stae2288} {\bibfield  {journal}
  {\bibinfo  {journal} {\mnras}\ }\textbf {\bibinfo {volume} {534}},\ \bibinfo
  {pages} {3506} (\bibinfo {year} {2024})},\ \Eprint
  {https://arxiv.org/abs/2406.11885} {arXiv:2406.11885 [astro-ph.HE]}
  \BibitemShut {NoStop}\bibitem [{\citenamefont {{Colloms}}\ \emph {et~al.}(2025)\citenamefont
  {{Colloms}}, \citenamefont {{Berry}}, \citenamefont {{Veitch}},\ and\
  \citenamefont {{Zevin}}}]{2025ApJ...988..189C}\BibitemOpen
  \bibfield  {author} {\bibinfo {author} {\bibfnamefont {S.}~\bibnamefont
  {{Colloms}}}, \bibinfo {author} {\bibfnamefont {C.~P.~L.}\ \bibnamefont
  {{Berry}}}, \bibinfo {author} {\bibfnamefont {J.}~\bibnamefont {{Veitch}}},\
  and\ \bibinfo {author} {\bibfnamefont {M.}~\bibnamefont {{Zevin}}},\
  }\bibfield  {title} {\bibinfo {title} {{Exploring the Evolution of
  Gravitational-wave Emitters with Efficient Emulation: Constraining the
  Origins of Binary Black Holes Using Normalizing Flows}},\ }\href
  {https://doi.org/10.3847/1538-4357/ade546} {\bibfield  {journal} {\bibinfo
  {journal} {\apj}\ }\textbf {\bibinfo {volume} {988}},\ \bibinfo {eid} {189}
  (\bibinfo {year} {2025})},\ \Eprint {https://arxiv.org/abs/2503.03819}
  {arXiv:2503.03819 [astro-ph.HE]} \BibitemShut {NoStop}\bibitem [{\citenamefont {{Delfavero}}\ \emph {et~al.}(2023)\citenamefont
  {{Delfavero}}, \citenamefont {{O'Shaughnessy}}, \citenamefont {{Belczynski}},
  \citenamefont {{Drozda}},\ and\ \citenamefont
  {{Wysocki}}}]{popsyn-gwastro-STInterpFinal-Vera2023}\BibitemOpen
  \bibfield  {author} {\bibinfo {author} {\bibfnamefont {V.}~\bibnamefont
  {{Delfavero}}}, \bibinfo {author} {\bibfnamefont {R.}~\bibnamefont
  {{O'Shaughnessy}}}, \bibinfo {author} {\bibfnamefont {K.}~\bibnamefont
  {{Belczynski}}}, \bibinfo {author} {\bibfnamefont {P.}~\bibnamefont
  {{Drozda}}},\ and\ \bibinfo {author} {\bibfnamefont {D.}~\bibnamefont
  {{Wysocki}}},\ }\bibfield  {title} {\bibinfo {title} {{Iteratively comparing
  gravitational-wave observations to the evolution of massive stellar
  binaries}},\ }\href {https://doi.org/10.1103/PhysRevD.108.043023} {\bibfield
  {journal} {\bibinfo  {journal} {\prd}\ }\textbf {\bibinfo {volume} {108}},\
  \bibinfo {eid} {043023} (\bibinfo {year} {2023})},\ \Eprint
  {https://arxiv.org/abs/2303.05436} {arXiv:2303.05436 [gr-qc]} \BibitemShut
  {NoStop}\bibitem [{\citenamefont {Delfavero}\ \emph {et~al.}(2025)\citenamefont
  {Delfavero}, \citenamefont {Breivik}, \citenamefont {Thiele}, \citenamefont
  {O'Shaughnessy},\ and\ \citenamefont {Baker}}]{Delfavero2025DWD}\BibitemOpen
  \bibfield  {author} {\bibinfo {author} {\bibfnamefont {V.}~\bibnamefont
  {Delfavero}}, \bibinfo {author} {\bibfnamefont {K.}~\bibnamefont {Breivik}},
  \bibinfo {author} {\bibfnamefont {S.}~\bibnamefont {Thiele}}, \bibinfo
  {author} {\bibfnamefont {R.}~\bibnamefont {O'Shaughnessy}},\ and\ \bibinfo
  {author} {\bibfnamefont {J.~G.}\ \bibnamefont {Baker}},\ }\bibfield  {title}
  {\bibinfo {title} {Recovering injected astrophysics from the lisa double
  white dwarf binaries},\ }\href {https://doi.org/10.3847/1538-4357/ada9e2}
  {\bibfield  {journal} {\bibinfo  {journal} {Astrophysical Journal}\ }\textbf
  {\bibinfo {volume} {981}},\ \bibinfo {pages} {66} (\bibinfo {year} {2025})},\
  \Eprint {https://arxiv.org/abs/2409.15230} {arXiv:2409.15230} \BibitemShut
  {NoStop}\bibitem [{\citenamefont {{Andrews}}\ \emph {et~al.}(2018)\citenamefont
  {{Andrews}}, \citenamefont {{Zezas}},\ and\ \citenamefont
  {{Fragos}}}]{2018ApJS..237....1A}\BibitemOpen
  \bibfield  {author} {\bibinfo {author} {\bibfnamefont {J.~J.}\ \bibnamefont
  {{Andrews}}}, \bibinfo {author} {\bibfnamefont {A.}~\bibnamefont {{Zezas}}},\
  and\ \bibinfo {author} {\bibfnamefont {T.}~\bibnamefont {{Fragos}}},\
  }\bibfield  {title} {\bibinfo {title} {{dart\_board: Binary Population
  Synthesis with Markov Chain Monte Carlo}},\ }\href
  {https://doi.org/10.3847/1538-4365/aaca30} {\bibfield  {journal} {\bibinfo
  {journal} {\apjs}\ }\textbf {\bibinfo {volume} {237}},\ \bibinfo {eid} {1}
  (\bibinfo {year} {2018})},\ \Eprint {https://arxiv.org/abs/1710.11030}
  {arXiv:1710.11030 [astro-ph.HE]} \BibitemShut {NoStop}\bibitem [{\citenamefont {{Wong}}\ \emph {et~al.}(2023)\citenamefont {{Wong}},
  \citenamefont {{Breivik}}, \citenamefont {{Farr}},\ and\ \citenamefont
  {{Luger}}}]{2023ApJ...950..181W}\BibitemOpen
  \bibfield  {author} {\bibinfo {author} {\bibfnamefont {K.~W.~K.}\
  \bibnamefont {{Wong}}}, \bibinfo {author} {\bibfnamefont {K.}~\bibnamefont
  {{Breivik}}}, \bibinfo {author} {\bibfnamefont {W.~M.}\ \bibnamefont
  {{Farr}}},\ and\ \bibinfo {author} {\bibfnamefont {R.}~\bibnamefont
  {{Luger}}},\ }\bibfield  {title} {\bibinfo {title} {{Backward Population
  Synthesis: Mapping the Evolutionary History of Gravitational-wave
  Progenitors}},\ }\href {https://doi.org/10.3847/1538-4357/acc863} {\bibfield
  {journal} {\bibinfo  {journal} {\apj}\ }\textbf {\bibinfo {volume} {950}},\
  \bibinfo {eid} {181} (\bibinfo {year} {2023})},\ \Eprint
  {https://arxiv.org/abs/2206.04062} {arXiv:2206.04062 [astro-ph.HE]}
  \BibitemShut {NoStop}\bibitem [{\citenamefont {Gerosa}\ \emph {et~al.}(2018)\citenamefont {Gerosa}
  \emph {et~al.}}]{Gerosa2018Spin}\BibitemOpen
  \bibfield  {author} {\bibinfo {author} {\bibfnamefont {D.}~\bibnamefont
  {Gerosa}} \emph {et~al.},\ }\bibfield  {title} {\bibinfo {title} {Spin
  orientations of merging black holes formed from the evolution of stellar
  binaries},\ }\href {https://doi.org/10.1103/PhysRevD.98.084036} {\bibfield
  {journal} {\bibinfo  {journal} {Physical Review D}\ }\textbf {\bibinfo
  {volume} {98}},\ \bibinfo {pages} {084036} (\bibinfo {year} {2018})},\
  \Eprint {https://arxiv.org/abs/1808.02491} {arXiv:1808.02491} \BibitemShut
  {NoStop}\bibitem [{\citenamefont {Guerrero}\ \emph {et~al.}(2026)\citenamefont
  {Guerrero}, \citenamefont {Zevin}, \citenamefont {Maclean}, \citenamefont
  {Breivik}, \citenamefont {Rodriguez}, \citenamefont {Briel},\ and\
  \citenamefont {Holz}}]{Guerrero2026Codes}\BibitemOpen
  \bibfield  {author} {\bibinfo {author} {\bibfnamefont {A.~G.}\ \bibnamefont
  {Guerrero}}, \bibinfo {author} {\bibfnamefont {M.}~\bibnamefont {Zevin}},
  \bibinfo {author} {\bibfnamefont {D.~B.}\ \bibnamefont {Maclean}}, \bibinfo
  {author} {\bibfnamefont {K.}~\bibnamefont {Breivik}}, \bibinfo {author}
  {\bibfnamefont {C.~L.}\ \bibnamefont {Rodriguez}}, \bibinfo {author}
  {\bibfnamefont {M.~M.}\ \bibnamefont {Briel}},\ and\ \bibinfo {author}
  {\bibfnamefont {D.~E.}\ \bibnamefont {Holz}},\ }\bibfield  {title} {\bibinfo
  {title} {When the stars don't align: Investigating inconsistencies in binary
  black hole formation across population synthesis codes},\ }\href
  {https://doi.org/10.48550/arXiv.2608.21609} {\bibfield  {journal} {\bibinfo
  {journal} {arXiv e-prints}\ ,\ \bibinfo {eid} {arXiv:2608.21609}} (\bibinfo
  {year} {2026})},\ \Eprint {https://arxiv.org/abs/2608.21609}
  {arXiv:2608.21609} \BibitemShut {NoStop}\bibitem [{\citenamefont {{Hurley}}\ \emph {et~al.}(2002)\citenamefont
  {{Hurley}}, \citenamefont {{Tout}},\ and\ \citenamefont
  {{Pols}}}]{2002MNRAS.329..897H}\BibitemOpen
  \bibfield  {author} {\bibinfo {author} {\bibfnamefont {J.~R.}\ \bibnamefont
  {{Hurley}}}, \bibinfo {author} {\bibfnamefont {C.~A.}\ \bibnamefont
  {{Tout}}},\ and\ \bibinfo {author} {\bibfnamefont {O.~R.}\ \bibnamefont
  {{Pols}}},\ }\bibfield  {title} {\bibinfo {title} {{Evolution of binary stars
  and the effect of tides on binary populations}},\ }\href
  {https://doi.org/10.1046/j.1365-8711.2002.05038.x} {\bibfield  {journal}
  {\bibinfo  {journal} {\mnras}\ }\textbf {\bibinfo {volume} {329}},\ \bibinfo
  {pages} {897} (\bibinfo {year} {2002})},\ \Eprint
  {https://arxiv.org/abs/astro-ph/0201220} {arXiv:astro-ph/0201220}
  \BibitemShut {NoStop}\bibitem [{\citenamefont {{Hurley}}\ \emph {et~al.}(2000)\citenamefont
  {{Hurley}}, \citenamefont {{Pols}},\ and\ \citenamefont
  {{Tout}}}]{2000MNRAS.315..543H}\BibitemOpen
  \bibfield  {author} {\bibinfo {author} {\bibfnamefont {J.~R.}\ \bibnamefont
  {{Hurley}}}, \bibinfo {author} {\bibfnamefont {O.~R.}\ \bibnamefont
  {{Pols}}},\ and\ \bibinfo {author} {\bibfnamefont {C.~A.}\ \bibnamefont
  {{Tout}}},\ }\bibfield  {title} {\bibinfo {title} {{Comprehensive analytic
  formulae for stellar evolution as a function of mass and metallicity}},\
  }\href {https://doi.org/10.1046/j.1365-8711.2000.03426.x} {\bibfield
  {journal} {\bibinfo  {journal} {\mnras}\ }\textbf {\bibinfo {volume} {315}},\
  \bibinfo {pages} {543} (\bibinfo {year} {2000})},\ \Eprint
  {https://arxiv.org/abs/astro-ph/0001295} {arXiv:astro-ph/0001295 [astro-ph]}
  \BibitemShut {NoStop}\bibitem [{\citenamefont {Kim}\ \emph {et~al.}(2003)\citenamefont {Kim},
  \citenamefont {Kalogera},\ and\ \citenamefont {Lorimer}}]{Chunglee-nsns-1}\BibitemOpen
  \bibfield  {author} {\bibinfo {author} {\bibfnamefont {C.}~\bibnamefont
  {Kim}}, \bibinfo {author} {\bibfnamefont {V.}~\bibnamefont {Kalogera}},\ and\
  \bibinfo {author} {\bibfnamefont {D.~R.}\ \bibnamefont {Lorimer}},\
  }\bibfield  {title} {\bibinfo {title} {The probability distribution of binary
  pulsar coalescence rates. i. double neutron star systems in the galactic
  field},\ }\href {https://doi.org/10.1086/345740} {\bibfield  {journal}
  {\bibinfo  {journal} {ApJ}\ }\textbf {\bibinfo {volume} {584}},\ \bibinfo
  {pages} {985} (\bibinfo {year} {2003})},\ \Eprint
  {https://arxiv.org/abs/astro-ph/0207408} {arXiv:astro-ph/0207408}
  \BibitemShut {NoStop}\bibitem [{\citenamefont {{O'Shaughnessy}}\ and\ \citenamefont
  {{Kim}}(2010)}]{psr-SpinFb}\BibitemOpen
  \bibfield  {author} {\bibinfo {author} {\bibfnamefont {R.}~\bibnamefont
  {{O'Shaughnessy}}}\ and\ \bibinfo {author} {\bibfnamefont {C.}~\bibnamefont
  {{Kim}}},\ }\bibfield  {title} {\bibinfo {title} {Pulsar binary birthrates
  with spin-opening angle correlations},\ }\href
  {https://doi.org/10.1088/0004-637X/715/1/230} {\bibfield  {journal} {\bibinfo
   {journal} {The Astrophysical Journal}\ }\textbf {\bibinfo {volume} {715}},\
  \bibinfo {pages} {230} (\bibinfo {year} {2010})},\ \Eprint
  {https://arxiv.org/abs/0908.2049} {arXiv:0908.2049 [astro-ph.GA]}
  \BibitemShut {NoStop}\bibitem [{\citenamefont {Kim}\ \emph {et~al.}(2015)\citenamefont {Kim},
  \citenamefont {Perera},\ and\ \citenamefont
  {McLaughlin}}]{Kim2015DoublePulsar}\BibitemOpen
  \bibfield  {author} {\bibinfo {author} {\bibfnamefont {C.}~\bibnamefont
  {Kim}}, \bibinfo {author} {\bibfnamefont {B.~B.~P.}\ \bibnamefont {Perera}},\
  and\ \bibinfo {author} {\bibfnamefont {M.~A.}\ \bibnamefont {McLaughlin}},\
  }\bibfield  {title} {\bibinfo {title} {Implications of {PSR J0737-3039B} for
  the {Galactic NS--NS} binary merger rate},\ }\href
  {https://doi.org/10.1093/mnras/stu2729} {\bibfield  {journal} {\bibinfo
  {journal} {MNRAS}\ }\textbf {\bibinfo {volume} {448}},\ \bibinfo {pages}
  {928} (\bibinfo {year} {2015})},\ \Eprint {https://arxiv.org/abs/1308.4676}
  {arXiv:1308.4676 [astro-ph.SR]} \BibitemShut {NoStop}\bibitem [{\citenamefont {Pol}\ \emph {et~al.}(2019)\citenamefont {Pol},
  \citenamefont {McLaughlin},\ and\ \citenamefont {Lorimer}}]{Pol2019DNSRate}\BibitemOpen
  \bibfield  {author} {\bibinfo {author} {\bibfnamefont {N.}~\bibnamefont
  {Pol}}, \bibinfo {author} {\bibfnamefont {M.}~\bibnamefont {McLaughlin}},\
  and\ \bibinfo {author} {\bibfnamefont {D.~R.}\ \bibnamefont {Lorimer}},\
  }\bibfield  {title} {\bibinfo {title} {Future prospects for ground-based
  gravitational wave detectors---the galactic double neutron star merger rate
  revisited},\ }\href {https://doi.org/10.3847/1538-4357/aaf006} {\bibfield
  {journal} {\bibinfo  {journal} {The Astrophysical Journal}\ }\textbf
  {\bibinfo {volume} {870}},\ \bibinfo {eid} {71} (\bibinfo {year} {2019})},\
  \Eprint {https://arxiv.org/abs/1811.04086} {arXiv:1811.04086 [astro-ph.HE]}
  \BibitemShut {NoStop}\bibitem [{\citenamefont {Cutler}\ and\ \citenamefont
  {Flanagan}(1994)}]{CutlerFlanagan1994}\BibitemOpen
  \bibfield  {author} {\bibinfo {author} {\bibfnamefont {C.}~\bibnamefont
  {Cutler}}\ and\ \bibinfo {author} {\bibfnamefont {E.~E.}\ \bibnamefont
  {Flanagan}},\ }\bibfield  {title} {\bibinfo {title} {Gravitational waves from
  merging compact binaries: How accurately can one extract the binary's
  parameters from the inspiral waveform?},\ }\href
  {https://doi.org/10.1103/PhysRevD.49.2658} {\bibfield  {journal} {\bibinfo
  {journal} {Phys. Rev. D}\ }\textbf {\bibinfo {volume} {49}},\ \bibinfo
  {pages} {2658} (\bibinfo {year} {1994})},\ \Eprint
  {https://arxiv.org/abs/gr-qc/9402014} {arXiv:gr-qc/9402014} \BibitemShut
  {NoStop}\bibitem [{\citenamefont {Vallisneri}(2008)}]{Vallisneri2008}\BibitemOpen
  \bibfield  {author} {\bibinfo {author} {\bibfnamefont {M.}~\bibnamefont
  {Vallisneri}},\ }\bibfield  {title} {\bibinfo {title} {Use and abuse of the
  {Fisher} information matrix in the assessment of gravitational-wave
  parameter-estimation prospects},\ }\href
  {https://doi.org/10.1103/PhysRevD.77.042001} {\bibfield  {journal} {\bibinfo
  {journal} {Phys. Rev. D}\ }\textbf {\bibinfo {volume} {77}},\ \bibinfo
  {pages} {042001} (\bibinfo {year} {2008})},\ \Eprint
  {https://arxiv.org/abs/gr-qc/0703086} {arXiv:gr-qc/0703086} \BibitemShut
  {NoStop}\bibitem [{\citenamefont {Gair}\ \emph {et~al.}(2023)\citenamefont {Gair},
  \citenamefont {Antonelli},\ and\ \citenamefont
  {Barbieri}}]{GairAntonelliBarbieri2023}\BibitemOpen
  \bibfield  {author} {\bibinfo {author} {\bibfnamefont {J.~R.}\ \bibnamefont
  {Gair}}, \bibinfo {author} {\bibfnamefont {A.}~\bibnamefont {Antonelli}},\
  and\ \bibinfo {author} {\bibfnamefont {R.}~\bibnamefont {Barbieri}},\
  }\bibfield  {title} {\bibinfo {title} {A {Fisher} matrix for
  gravitational-wave population inference},\ }\href
  {https://doi.org/10.1093/mnras/stac3560} {\bibfield  {journal} {\bibinfo
  {journal} {MNRAS}\ }\textbf {\bibinfo {volume} {519}},\ \bibinfo {pages}
  {2736} (\bibinfo {year} {2023})},\ \Eprint {https://arxiv.org/abs/2205.07893}
  {arXiv:2205.07893 [gr-qc]} \BibitemShut {NoStop}\bibitem [{\citenamefont {Cutler}\ and\ \citenamefont
  {Vallisneri}(2007)}]{CutlerVallisneri2007}\BibitemOpen
  \bibfield  {author} {\bibinfo {author} {\bibfnamefont {C.}~\bibnamefont
  {Cutler}}\ and\ \bibinfo {author} {\bibfnamefont {M.}~\bibnamefont
  {Vallisneri}},\ }\bibfield  {title} {\bibinfo {title} {{LISA} detections of
  massive black hole inspirals: Parameter extraction errors due to inaccurate
  template waveforms},\ }\href {https://doi.org/10.1103/PhysRevD.76.104018}
  {\bibfield  {journal} {\bibinfo  {journal} {Phys. Rev. D}\ }\textbf {\bibinfo
  {volume} {76}},\ \bibinfo {pages} {104018} (\bibinfo {year} {2007})},\
  \Eprint {https://arxiv.org/abs/0707.2982} {arXiv:0707.2982 [gr-qc]}
  \BibitemShut {NoStop}\bibitem [{\citenamefont {Vallisneri}\ and\ \citenamefont
  {Yunes}(2013)}]{VallisneriYunes2013}\BibitemOpen
  \bibfield  {author} {\bibinfo {author} {\bibfnamefont {M.}~\bibnamefont
  {Vallisneri}}\ and\ \bibinfo {author} {\bibfnamefont {N.}~\bibnamefont
  {Yunes}},\ }\bibfield  {title} {\bibinfo {title} {Stealth bias in
  gravitational-wave parameter estimation},\ }\href
  {https://doi.org/10.1103/PhysRevD.87.102002} {\bibfield  {journal} {\bibinfo
  {journal} {Phys. Rev. D}\ }\textbf {\bibinfo {volume} {87}},\ \bibinfo
  {pages} {102002} (\bibinfo {year} {2013})},\ \Eprint
  {https://arxiv.org/abs/1301.2627} {arXiv:1301.2627 [gr-qc]} \BibitemShut
  {NoStop}\bibitem [{\citenamefont {Qazalbash}\ \emph {et~al.}(2026)\citenamefont
  {Qazalbash}, \citenamefont {Zeeshan},\ and\ \citenamefont
  {O'Shaughnessy}}]{gwastro-mergers-zeeshan-gwkokab}\BibitemOpen
  \bibfield  {author} {\bibinfo {author} {\bibfnamefont {M.}~\bibnamefont
  {Qazalbash}}, \bibinfo {author} {\bibfnamefont {M.}~\bibnamefont {Zeeshan}},\
  and\ \bibinfo {author} {\bibfnamefont {R.}~\bibnamefont {O'Shaughnessy}},\
  }\bibfield  {title} {\bibinfo {title} {Implementation to identify the
  properties of multiple populations of gravitational wave sources},\ }\href
  {https://doi.org/10.1103/krnm-3vrf} {\bibfield  {journal} {\bibinfo
  {journal} {\prd}\ }\textbf {\bibinfo {volume} {113}},\ \bibinfo {pages}
  {103003} (\bibinfo {year} {2026})},\ \Eprint
  {https://arxiv.org/abs/2509.13638} {arXiv:2509.13638 [gr-qc]} \BibitemShut
  {NoStop}\bibitem [{\citenamefont {O'Shaughnessy}\ \emph {et~al.}(2010)\citenamefont
  {O'Shaughnessy}, \citenamefont {Kalogera},\ and\ \citenamefont
  {Belczynski}}]{PSellipticals}\BibitemOpen
  \bibfield  {author} {\bibinfo {author} {\bibfnamefont {R.}~\bibnamefont
  {O'Shaughnessy}}, \bibinfo {author} {\bibfnamefont {V.}~\bibnamefont
  {Kalogera}},\ and\ \bibinfo {author} {\bibfnamefont {K.}~\bibnamefont
  {Belczynski}},\ }\bibfield  {title} {\bibinfo {title} {Binary compact object
  coalescence rates: The role of elliptical galaxies},\ }\href
  {https://doi.org/10.1088/0004-637X/716/1/615} {\bibfield  {journal} {\bibinfo
   {journal} {\apj}\ }\textbf {\bibinfo {volume} {716}},\ \bibinfo {pages}
  {615} (\bibinfo {year} {2010})},\ \Eprint {https://arxiv.org/abs/0908.3635}
  {arXiv:0908.3635} \BibitemShut {NoStop}\bibitem [{\citenamefont {O'Shaughnessy}\ \emph {et~al.}(2026)\citenamefont
  {O'Shaughnessy}, \citenamefont {Mechum}, \citenamefont {Qazalbash},
  \citenamefont {Rosenberg},\ and\ \citenamefont
  {Zeeshan}}]{gwastro-PopulationReconstruct-Hierarchical-PhysicalParam1-Clusters}\BibitemOpen
  \bibfield  {author} {\bibinfo {author} {\bibfnamefont {R.}~\bibnamefont
  {O'Shaughnessy}}, \bibinfo {author} {\bibfnamefont {R.}~\bibnamefont
  {Mechum}}, \bibinfo {author} {\bibfnamefont {M.}~\bibnamefont {Qazalbash}},
  \bibinfo {author} {\bibfnamefont {Z.}~\bibnamefont {Rosenberg}},\ and\
  \bibinfo {author} {\bibfnamefont {M.}~\bibnamefont {Zeeshan}},\ }\bibfield
  {title} {\bibinfo {title} {Physics-based phenomenological modeling of binary
  black hole hierarchical formation 1: Synthetic universes from globular
  cluster simulations for {GWTC}},\ }\href
  {https://doi.org/10.48550/arXiv.2609.05996} {\bibfield  {journal} {\bibinfo
  {journal} {arXiv e-prints}\ ,\ \bibinfo {eid} {arXiv:2609.05996}} (\bibinfo
  {year} {2026})},\ \Eprint {https://arxiv.org/abs/2609.05996}
  {arXiv:2609.05996 [gr-qc]} \BibitemShut {NoStop}\bibitem [{\citenamefont {{Bhattacharya}}\ \emph {et~al.}(1992)\citenamefont
  {{Bhattacharya}}, \citenamefont {{Wijers}}, \citenamefont {{Hartman}},\ and\
  \citenamefont {{Verbunt}}}]{gwastro-BhattacharyaDecayMagneticFields1992}\BibitemOpen
  \bibfield  {author} {\bibinfo {author} {\bibfnamefont {D.}~\bibnamefont
  {{Bhattacharya}}}, \bibinfo {author} {\bibfnamefont {R.~A.~M.~J.}\
  \bibnamefont {{Wijers}}}, \bibinfo {author} {\bibfnamefont {J.~W.}\
  \bibnamefont {{Hartman}}},\ and\ \bibinfo {author} {\bibfnamefont
  {F.}~\bibnamefont {{Verbunt}}},\ }\bibfield  {title} {\bibinfo {title} {{On
  the decay of the magnetic fields of single radio pulsars.}},\ }\href@noop {}
  {\bibfield  {journal} {\bibinfo  {journal} {\aap}\ }\textbf {\bibinfo
  {volume} {254}},\ \bibinfo {pages} {198} (\bibinfo {year}
  {1992})}\BibitemShut {NoStop}\bibitem [{\citenamefont {{Peters}}(1964)}]{Peters:1964}\BibitemOpen
  \bibfield  {author} {\bibinfo {author} {\bibfnamefont {P.~C.}\ \bibnamefont
  {{Peters}}},\ }\bibfield  {title} {\bibinfo {title} {{Gravitational Radiation
  and the Motion of Two Point Masses}},\ }\href
  {https://doi.org/10.1103/PhysRev.136.B1224} {\bibfield  {journal} {\bibinfo
  {journal} {Physical Review}\ }\textbf {\bibinfo {volume} {136}},\ \bibinfo
  {pages} {1224} (\bibinfo {year} {1964})}\BibitemShut {NoStop}\bibitem [{\citenamefont {{Sobol'}}(1967)}]{Sobol1967DistributionPointsCube}\BibitemOpen
  \bibfield  {author} {\bibinfo {author} {\bibfnamefont {I.~M.}\ \bibnamefont
  {{Sobol'}}},\ }\bibfield  {title} {\bibinfo {title} {{On the distribution of
  points in a cube and the approximate evaluation of integrals}},\ }\href
  {https://doi.org/10.1016/0041-5553(67)90144-9} {\bibfield  {journal}
  {\bibinfo  {journal} {USSR Computational Mathematics and Mathematical
  Physics}\ }\textbf {\bibinfo {volume} {7}},\ \bibinfo {pages} {86} (\bibinfo
  {year} {1967})}\BibitemShut {NoStop}\bibitem [{\citenamefont {{Owen}}(1995)}]{Owen1995RandomlyPermutedNets}\BibitemOpen
  \bibfield  {author} {\bibinfo {author} {\bibfnamefont {A.~B.}\ \bibnamefont
  {{Owen}}},\ }\bibfield  {title} {\bibinfo {title} {{Randomly Permuted
  (t,m,s)-Nets and (t,s)-Sequences}},\ }in\ \href
  {https://doi.org/10.1007/978-1-4612-2552-2_19} {\emph {\bibinfo {booktitle}
  {Monte Carlo and Quasi-Monte Carlo Methods in Scientific Computing}}}\
  (\bibinfo  {publisher} {Springer New York},\ \bibinfo {year} {1995})\ pp.\
  \bibinfo {pages} {299--317}\BibitemShut {NoStop}\bibitem [{\citenamefont
  {{Matou{\v{s}}ek}}(1998)}]{Matousek1998L2DiscrepancyAnchoredBoxes}\BibitemOpen
  \bibfield  {author} {\bibinfo {author} {\bibfnamefont {J.}~\bibnamefont
  {{Matou{\v{s}}ek}}},\ }\bibfield  {title} {\bibinfo {title} {{On the
  L$_2$-Discrepancy for Anchored Boxes}},\ }\href
  {https://doi.org/10.1006/jcom.1998.0489} {\bibfield  {journal} {\bibinfo
  {journal} {Journal of Complexity}\ }\textbf {\bibinfo {volume} {14}},\
  \bibinfo {pages} {527} (\bibinfo {year} {1998})}\BibitemShut {NoStop}\bibitem [{\citenamefont {{Virtanen}}\ \emph {et~al.}(2020)\citenamefont
  {{Virtanen}}, \citenamefont {{Gommers}}, \citenamefont {{Oliphant}},
  \citenamefont {{Haberland}}, \citenamefont {{Reddy}}, \citenamefont
  {{Cournapeau}}, \citenamefont {{Burovski}}, \citenamefont {{Peterson}},
  \citenamefont {{Weckesser}}, \citenamefont {{Bright}}, \citenamefont {{van
  der Walt}}, \citenamefont {{Brett}}, \citenamefont {{Wilson}}, \citenamefont
  {{Millman}}, \citenamefont {{Mayorov}}, \citenamefont {{Nelson}},
  \citenamefont {{Jones}}, \citenamefont {{Kern}}, \citenamefont {{Larson}},
  \citenamefont {{Carey}}, \citenamefont {{Polat}}, \citenamefont {{Feng}},
  \citenamefont {{Moore}}, \citenamefont {{VanderPlas}}, \citenamefont
  {{Laxalde}}, \citenamefont {{Perktold}}, \citenamefont {{Cimrman}},
  \citenamefont {{Henriksen}}, \citenamefont {{Quintero}}, \citenamefont
  {{Harris}}, \citenamefont {{Archibald}}, \citenamefont {{Ribeiro}},
  \citenamefont {{Pedregosa}}, \citenamefont {{van Mulbregt}},\ and\
  \citenamefont {{SciPy 1.0
  Contributors}}}]{gwastro-VirtanenScipyFundamentalAlgorithms2020}\BibitemOpen
  \bibfield  {author} {\bibinfo {author} {\bibfnamefont {P.}~\bibnamefont
  {{Virtanen}}}, \bibinfo {author} {\bibfnamefont {R.}~\bibnamefont
  {{Gommers}}}, \bibinfo {author} {\bibfnamefont {T.~E.}\ \bibnamefont
  {{Oliphant}}}, \bibinfo {author} {\bibfnamefont {M.}~\bibnamefont
  {{Haberland}}}, \bibinfo {author} {\bibfnamefont {T.}~\bibnamefont
  {{Reddy}}}, \bibinfo {author} {\bibfnamefont {D.}~\bibnamefont
  {{Cournapeau}}}, \bibinfo {author} {\bibfnamefont {E.}~\bibnamefont
  {{Burovski}}}, \bibinfo {author} {\bibfnamefont {P.}~\bibnamefont
  {{Peterson}}}, \bibinfo {author} {\bibfnamefont {W.}~\bibnamefont
  {{Weckesser}}}, \bibinfo {author} {\bibfnamefont {J.}~\bibnamefont
  {{Bright}}}, \bibinfo {author} {\bibfnamefont {S.~J.}\ \bibnamefont {{van der
  Walt}}}, \bibinfo {author} {\bibfnamefont {M.}~\bibnamefont {{Brett}}},
  \bibinfo {author} {\bibfnamefont {J.}~\bibnamefont {{Wilson}}}, \bibinfo
  {author} {\bibfnamefont {K.~J.}\ \bibnamefont {{Millman}}}, \bibinfo {author}
  {\bibfnamefont {N.}~\bibnamefont {{Mayorov}}}, \bibinfo {author}
  {\bibfnamefont {A.~R.~J.}\ \bibnamefont {{Nelson}}}, \bibinfo {author}
  {\bibfnamefont {E.}~\bibnamefont {{Jones}}}, \bibinfo {author} {\bibfnamefont
  {R.}~\bibnamefont {{Kern}}}, \bibinfo {author} {\bibfnamefont
  {E.}~\bibnamefont {{Larson}}}, \bibinfo {author} {\bibfnamefont {C.~J.}\
  \bibnamefont {{Carey}}}, \bibinfo {author} {\bibfnamefont
  {{\.I}.}~\bibnamefont {{Polat}}}, \bibinfo {author} {\bibfnamefont
  {Y.}~\bibnamefont {{Feng}}}, \bibinfo {author} {\bibfnamefont {E.~W.}\
  \bibnamefont {{Moore}}}, \bibinfo {author} {\bibfnamefont {J.}~\bibnamefont
  {{VanderPlas}}}, \bibinfo {author} {\bibfnamefont {D.}~\bibnamefont
  {{Laxalde}}}, \bibinfo {author} {\bibfnamefont {J.}~\bibnamefont
  {{Perktold}}}, \bibinfo {author} {\bibfnamefont {R.}~\bibnamefont
  {{Cimrman}}}, \bibinfo {author} {\bibfnamefont {I.}~\bibnamefont
  {{Henriksen}}}, \bibinfo {author} {\bibfnamefont {E.~A.}\ \bibnamefont
  {{Quintero}}}, \bibinfo {author} {\bibfnamefont {C.~R.}\ \bibnamefont
  {{Harris}}}, \bibinfo {author} {\bibfnamefont {A.~M.}\ \bibnamefont
  {{Archibald}}}, \bibinfo {author} {\bibfnamefont {A.~H.}\ \bibnamefont
  {{Ribeiro}}}, \bibinfo {author} {\bibfnamefont {F.}~\bibnamefont
  {{Pedregosa}}}, \bibinfo {author} {\bibfnamefont {P.}~\bibnamefont {{van
  Mulbregt}}},\ and\ \bibinfo {author} {\bibnamefont {{SciPy 1.0
  Contributors}}},\ }\bibfield  {title} {\bibinfo {title} {{SciPy 1.0:
  fundamental algorithms for scientific computing in Python}},\ }\href
  {https://doi.org/10.1038/s41592-019-0686-2} {\bibfield  {journal} {\bibinfo
  {journal} {Nature Methods}\ }\textbf {\bibinfo {volume} {17}},\ \bibinfo
  {pages} {261} (\bibinfo {year} {2020})},\ \Eprint
  {https://arxiv.org/abs/1907.10121} {arXiv:1907.10121 [cs.MS]} \BibitemShut
  {NoStop}\bibitem [{\citenamefont {Belczynski}\ \emph {et~al.}(2010)\citenamefont
  {Belczynski}, \citenamefont {Lorimer}, \citenamefont {Ridley},\ and\
  \citenamefont {Curran}}]{Belczynski2010DRP}\BibitemOpen
  \bibfield  {author} {\bibinfo {author} {\bibfnamefont {K.}~\bibnamefont
  {Belczynski}}, \bibinfo {author} {\bibfnamefont {D.~R.}\ \bibnamefont
  {Lorimer}}, \bibinfo {author} {\bibfnamefont {J.~P.}\ \bibnamefont
  {Ridley}},\ and\ \bibinfo {author} {\bibfnamefont {S.~J.}\ \bibnamefont
  {Curran}},\ }\bibfield  {title} {\bibinfo {title} {Double and single recycled
  pulsars: an evolutionary puzzle?},\ }\href
  {https://doi.org/10.1111/j.1365-2966.2010.16970.x} {\bibfield  {journal}
  {\bibinfo  {journal} {MNRAS}\ }\textbf {\bibinfo {volume} {407}},\ \bibinfo
  {pages} {1245} (\bibinfo {year} {2010})},\ \Eprint
  {https://arxiv.org/abs/0907.3486} {arXiv:0907.3486} \BibitemShut {NoStop}\bibitem [{\citenamefont {{Graber}}\ \emph {et~al.}(2024)\citenamefont
  {{Graber}}, \citenamefont {{Ronchi}}, \citenamefont {{Pardo-Araujo}},\ and\
  \citenamefont {{Rea}}}]{2024arXiv231214848G}\BibitemOpen
  \bibfield  {author} {\bibinfo {author} {\bibfnamefont {V.}~\bibnamefont
  {{Graber}}}, \bibinfo {author} {\bibfnamefont {M.}~\bibnamefont {{Ronchi}}},
  \bibinfo {author} {\bibfnamefont {C.}~\bibnamefont {{Pardo-Araujo}}},\ and\
  \bibinfo {author} {\bibfnamefont {N.}~\bibnamefont {{Rea}}},\ }\bibfield
  {title} {\bibinfo {title} {{Isolated Pulsar Population Synthesis with
  Simulation-based Inference}},\ }\href
  {https://doi.org/10.3847/1538-4357/ad3e78} {\bibfield  {journal} {\bibinfo
  {journal} {\apj}\ }\textbf {\bibinfo {volume} {968}},\ \bibinfo {eid} {16}
  (\bibinfo {year} {2024})},\ \Eprint {https://arxiv.org/abs/2312.14848}
  {arXiv:2312.14848 [astro-ph.HE]} \BibitemShut {NoStop}\bibitem [{\citenamefont {Pardo~Araujo}\ \emph {et~al.}(2025)\citenamefont
  {Pardo~Araujo}, \citenamefont {Ronchi}, \citenamefont {Graber},\ and\
  \citenamefont {Rea}}]{PardoAraujo2025FluxSBI}\BibitemOpen
  \bibfield  {author} {\bibinfo {author} {\bibfnamefont {C.}~\bibnamefont
  {Pardo~Araujo}}, \bibinfo {author} {\bibfnamefont {M.}~\bibnamefont
  {Ronchi}}, \bibinfo {author} {\bibfnamefont {V.}~\bibnamefont {Graber}},\
  and\ \bibinfo {author} {\bibfnamefont {N.}~\bibnamefont {Rea}},\ }\bibfield
  {title} {\bibinfo {title} {Radio pulsar population synthesis with consistent
  flux measurements using simulation-based inference},\ }\href
  {https://doi.org/10.1051/0004-6361/202453314} {\bibfield  {journal} {\bibinfo
   {journal} {Astronomy \& Astrophysics}\ }\textbf {\bibinfo {volume} {696}},\
  \bibinfo {pages} {A114} (\bibinfo {year} {2025})},\ \Eprint
  {https://arxiv.org/abs/2412.04070} {arXiv:2412.04070} \BibitemShut {NoStop}\bibitem [{\citenamefont
  {{O'Shaughnessy}}(2026{\natexlab{a}})}]{ScorpioInPrep}\BibitemOpen
  \bibfield  {author} {\bibinfo {author} {\bibfnamefont {R.}~\bibnamefont
  {{O'Shaughnessy}}},\ }\bibfield  {title} {\bibinfo {title} {A probabilistic,
  stripped-native remnant mass--kick model calibrated jointly to
  gravitational-wave and resolved-binary observations}} (\bibinfo {year}
  {2026}{\natexlab{a}}),\ \bibinfo {note} {in preparation}\BibitemShut
  {NoStop}\bibitem [{\citenamefont {{Blaauw}}(1961)}]{1961BAN....15..265B}\BibitemOpen
  \bibfield  {author} {\bibinfo {author} {\bibfnamefont {A.}~\bibnamefont
  {{Blaauw}}},\ }\bibfield  {title} {\bibinfo {title} {{On the origin of the O-
  and B-type stars with high velocities (the ``run-away'' stars), and some
  related problems}},\ }\href@noop {} {\bibfield  {journal} {\bibinfo
  {journal} {Bulletin of the Astronomical Institutes of the Netherlands}\
  }\textbf {\bibinfo {volume} {15}},\ \bibinfo {pages} {265} (\bibinfo {year}
  {1961})}\BibitemShut {NoStop}\bibitem [{\citenamefont {{Tauris}}\ and\ \citenamefont {{van den
  Heuvel}}(2006)}]{2006csxs.book..623T}\BibitemOpen
  \bibfield  {author} {\bibinfo {author} {\bibfnamefont {T.~M.}\ \bibnamefont
  {{Tauris}}}\ and\ \bibinfo {author} {\bibfnamefont {E.~P.~J.}\ \bibnamefont
  {{van den Heuvel}}},\ }\bibfield  {title} {\bibinfo {title} {{Formation and
  evolution of compact stellar X-ray sources}},\ }in\ \href
  {https://doi.org/10.48550/arXiv.astro-ph/0303456} {\emph {\bibinfo
  {booktitle} {Compact stellar X-ray sources}}},\ Vol.~\bibinfo {volume} {39},\
  \bibinfo {editor} {edited by\ \bibinfo {editor} {\bibfnamefont {W.~H.~G.}\
  \bibnamefont {{Lewin}}}\ and\ \bibinfo {editor} {\bibfnamefont
  {M.}~\bibnamefont {{van der Klis}}}}\ (\bibinfo  {publisher} {Cambridge
  University Press},\ \bibinfo {year} {2006})\ pp.\ \bibinfo {pages}
  {623--665}\BibitemShut {NoStop}\bibitem [{\citenamefont {{Postnov}}\ and\ \citenamefont
  {{Yungelson}}(2014)}]{2014LRR....17....3P}\BibitemOpen
  \bibfield  {author} {\bibinfo {author} {\bibfnamefont {K.~A.}\ \bibnamefont
  {{Postnov}}}\ and\ \bibinfo {author} {\bibfnamefont {L.~R.}\ \bibnamefont
  {{Yungelson}}},\ }\bibfield  {title} {\bibinfo {title} {{The Evolution of
  Compact Binary Star Systems}},\ }\href {https://doi.org/10.12942/lrr-2014-3}
  {\bibfield  {journal} {\bibinfo  {journal} {Living Reviews in Relativity}\
  }\textbf {\bibinfo {volume} {17}},\ \bibinfo {eid} {3} (\bibinfo {year}
  {2014})},\ \Eprint {https://arxiv.org/abs/1403.4754} {arXiv:1403.4754
  [astro-ph.HE]} \BibitemShut {NoStop}\bibitem [{\citenamefont {{Tauris}}\ and\ \citenamefont {{van den
  Heuvel}}(2023)}]{2023pbse.book.....T}\BibitemOpen
  \bibfield  {author} {\bibinfo {author} {\bibfnamefont {T.~M.}\ \bibnamefont
  {{Tauris}}}\ and\ \bibinfo {author} {\bibfnamefont {E.~P.~J.}\ \bibnamefont
  {{van den Heuvel}}},\ }\href {https://doi.org/10.48550/arXiv.2305.09388}
  {\emph {\bibinfo {title} {{Physics of Binary Star Evolution. From Stars to
  X-ray Binaries and Gravitational Wave Sources}}}}\ (\bibinfo  {publisher}
  {Princeton University Press},\ \bibinfo {year} {2023})\BibitemShut {NoStop}\bibitem [{\citenamefont {{Pfahl}}\ \emph {et~al.}(2002)\citenamefont
  {{Pfahl}}, \citenamefont {{Rappaport}},\ and\ \citenamefont
  {{Podsiadlowski}}}]{2002ApJ...573..283P}\BibitemOpen
  \bibfield  {author} {\bibinfo {author} {\bibfnamefont {E.}~\bibnamefont
  {{Pfahl}}}, \bibinfo {author} {\bibfnamefont {S.}~\bibnamefont
  {{Rappaport}}},\ and\ \bibinfo {author} {\bibfnamefont {P.}~\bibnamefont
  {{Podsiadlowski}}},\ }\bibfield  {title} {\bibinfo {title} {{A Comprehensive
  Study of Neutron Star Retention in Globular Clusters}},\ }\href
  {https://doi.org/10.1086/340494} {\bibfield  {journal} {\bibinfo  {journal}
  {\apj}\ }\textbf {\bibinfo {volume} {573}},\ \bibinfo {pages} {283} (\bibinfo
  {year} {2002})},\ \Eprint {https://arxiv.org/abs/astro-ph/0106141}
  {arXiv:astro-ph/0106141} \BibitemShut {NoStop}\bibitem [{\citenamefont {Dossena}\ \emph {et~al.}(2026)\citenamefont
  {Dossena}, \citenamefont {Gerosa},\ and\ \citenamefont
  {Bruel}}]{Dossena2026Tilts}\BibitemOpen
  \bibfield  {author} {\bibinfo {author} {\bibfnamefont {S.}~\bibnamefont
  {Dossena}}, \bibinfo {author} {\bibfnamefont {D.}~\bibnamefont {Gerosa}},\
  and\ \bibinfo {author} {\bibfnamefont {T.}~\bibnamefont {Bruel}},\ }\bibfield
   {title} {\bibinfo {title} {Not all spins of black-hole binaries formed in
  clusters are isotropic and not all spins of black-hole binaries formed in
  isolation are aligned},\ }\href {https://doi.org/10.48550/arXiv.2609.31879}
  {\bibfield  {journal} {\bibinfo  {journal} {arXiv e-prints}\ ,\ \bibinfo
  {eid} {arXiv:2609.31879}} (\bibinfo {year} {2026})},\ \Eprint
  {https://arxiv.org/abs/2609.31879} {arXiv:2609.31879 [astro-ph.HE]}
  \BibitemShut {NoStop}\bibitem [{\citenamefont {{Fryer}}\ \emph {et~al.}(2012)\citenamefont
  {{Fryer}}, \citenamefont {{Belczynski}}, \citenamefont {{Wiktorowicz}},
  \citenamefont {{Dominik}}, \citenamefont {{Kalogera}},\ and\ \citenamefont
  {{Holz}}}]{2012ApJ...749...91F}\BibitemOpen
  \bibfield  {author} {\bibinfo {author} {\bibfnamefont {C.~L.}\ \bibnamefont
  {{Fryer}}}, \bibinfo {author} {\bibfnamefont {K.}~\bibnamefont
  {{Belczynski}}}, \bibinfo {author} {\bibfnamefont {G.}~\bibnamefont
  {{Wiktorowicz}}}, \bibinfo {author} {\bibfnamefont {M.}~\bibnamefont
  {{Dominik}}}, \bibinfo {author} {\bibfnamefont {V.}~\bibnamefont
  {{Kalogera}}},\ and\ \bibinfo {author} {\bibfnamefont {D.~E.}\ \bibnamefont
  {{Holz}}},\ }\bibfield  {title} {\bibinfo {title} {{Compact Remnant Mass
  Function: Dependence on the Explosion Mechanism and Metallicity}},\ }\href
  {https://doi.org/10.1088/0004-637X/749/1/91} {\bibfield  {journal} {\bibinfo
  {journal} {\apj}\ }\textbf {\bibinfo {volume} {749}},\ \bibinfo {eid} {91}
  (\bibinfo {year} {2012})},\ \Eprint {https://arxiv.org/abs/1110.1726}
  {arXiv:1110.1726 [astro-ph.SR]} \BibitemShut {NoStop}\bibitem [{\citenamefont {{Anik}}\ \emph {et~al.}(2026)\citenamefont {{Anik}},
  \citenamefont {{Steiner}},\ and\ \citenamefont
  {{O'Shaughnessy}}}]{2025arXiv251212130A}\BibitemOpen
  \bibfield  {author} {\bibinfo {author} {\bibfnamefont {M.~H.}\ \bibnamefont
  {{Anik}}}, \bibinfo {author} {\bibfnamefont {A.~W.}\ \bibnamefont
  {{Steiner}}},\ and\ \bibinfo {author} {\bibfnamefont {R.}~\bibnamefont
  {{O'Shaughnessy}}},\ }\bibfield  {title} {\bibinfo {title} {Inference of
  neutron star mass distributions and the dense matter equation of state from
  multi-messenger observations},\ }\href
  {https://doi.org/10.3847/1538-4357/ae75e4} {\bibfield  {journal} {\bibinfo
  {journal} {\apj}\ }\textbf {\bibinfo {volume} {1005}},\ \bibinfo {eid} {148}
  (\bibinfo {year} {2026})},\ \Eprint {https://arxiv.org/abs/2512.12130}
  {arXiv:2512.12130 [astro-ph.HE]} \BibitemShut {NoStop}\bibitem [{\citenamefont {{Mandel}}\ and\ \citenamefont
  {{M{\"u}ller}}(2020)}]{2020MNRAS.499.3214M}\BibitemOpen
  \bibfield  {author} {\bibinfo {author} {\bibfnamefont {I.}~\bibnamefont
  {{Mandel}}}\ and\ \bibinfo {author} {\bibfnamefont {B.}~\bibnamefont
  {{M{\"u}ller}}},\ }\bibfield  {title} {\bibinfo {title} {{Simple recipes for
  compact remnant masses and natal kicks}},\ }\href
  {https://doi.org/10.1093/mnras/staa3043} {\bibfield  {journal} {\bibinfo
  {journal} {\mnras}\ }\textbf {\bibinfo {volume} {499}},\ \bibinfo {pages}
  {3214} (\bibinfo {year} {2020})},\ \Eprint {https://arxiv.org/abs/2006.08360}
  {arXiv:2006.08360 [astro-ph.HE]} \BibitemShut {NoStop}\bibitem [{\citenamefont {{Abbott}}\ \emph {et~al.}(2020)\citenamefont
  {{Abbott}}, \citenamefont {{Abbott}}, \citenamefont {{Abbott}}, \citenamefont
  {{Abraham}}, \citenamefont {{Acernese}}, \citenamefont {{Ackley}},
  \citenamefont {{Adams}}, \citenamefont {{Adhikari}} \emph
  {et~al.}}]{2020ApJ...892L...3A}\BibitemOpen
  \bibfield  {author} {\bibinfo {author} {\bibfnamefont {B.~P.}\ \bibnamefont
  {{Abbott}}}, \bibinfo {author} {\bibfnamefont {R.}~\bibnamefont {{Abbott}}},
  \bibinfo {author} {\bibfnamefont {T.~D.}\ \bibnamefont {{Abbott}}}, \bibinfo
  {author} {\bibfnamefont {S.}~\bibnamefont {{Abraham}}}, \bibinfo {author}
  {\bibfnamefont {F.}~\bibnamefont {{Acernese}}}, \bibinfo {author}
  {\bibfnamefont {K.}~\bibnamefont {{Ackley}}}, \bibinfo {author}
  {\bibfnamefont {C.}~\bibnamefont {{Adams}}}, \bibinfo {author} {\bibfnamefont
  {R.~X.}\ \bibnamefont {{Adhikari}}}, \emph {et~al.},\ }\bibfield  {title}
  {\bibinfo {title} {{GW190425: Observation of a Compact Binary Coalescence
  with Total Mass {\ensuremath{\sim}} 3.4 M$_{\odot}$}},\ }\href
  {https://doi.org/10.3847/2041-8213/ab75f5} {\bibfield  {journal} {\bibinfo
  {journal} {\apjl}\ }\textbf {\bibinfo {volume} {892}},\ \bibinfo {eid} {L3}
  (\bibinfo {year} {2020})},\ \Eprint {https://arxiv.org/abs/2001.01761}
  {arXiv:2001.01761 [astro-ph.HE]} \BibitemShut {NoStop}\bibitem [{\citenamefont
  {{O'Shaughnessy}}(2026{\natexlab{b}})}]{AdSnPopsynP3InPrep}\BibitemOpen
  \bibfield  {author} {\bibinfo {author} {\bibfnamefont {R.}~\bibnamefont
  {{O'Shaughnessy}}},\ }\bibfield  {title} {\bibinfo {title} {Channels as
  regions: an automatically differentiable rapid binary population synthesis}}
  (\bibinfo {year} {2026}{\natexlab{b}}),\ \bibinfo {note} {in
  preparation}\BibitemShut {NoStop}\bibitem [{\citenamefont {{Nieuwenhuijzen}}\ and\ \citenamefont {{de
  Jager}}(1990)}]{1990A&A...231..134N}\BibitemOpen
  \bibfield  {author} {\bibinfo {author} {\bibfnamefont {H.}~\bibnamefont
  {{Nieuwenhuijzen}}}\ and\ \bibinfo {author} {\bibfnamefont {C.}~\bibnamefont
  {{de Jager}}},\ }\bibfield  {title} {\bibinfo {title} {{Parametrization of
  stellar rates of mass loss as functions of the fundamental stellar parameters
  M, L, and R.}},\ }\href@noop {} {\bibfield  {journal} {\bibinfo  {journal}
  {\aap}\ }\textbf {\bibinfo {volume} {231}},\ \bibinfo {pages} {134} (\bibinfo
  {year} {1990})}\BibitemShut {NoStop}\bibitem [{\citenamefont {{Vink}}\ \emph {et~al.}(2001)\citenamefont {{Vink}},
  \citenamefont {{de Koter}},\ and\ \citenamefont
  {{Lamers}}}]{2001AA...369..574V}\BibitemOpen
  \bibfield  {author} {\bibinfo {author} {\bibfnamefont {J.~S.}\ \bibnamefont
  {{Vink}}}, \bibinfo {author} {\bibfnamefont {A.}~\bibnamefont {{de Koter}}},\
  and\ \bibinfo {author} {\bibfnamefont {H.~J.~G.~L.~M.}\ \bibnamefont
  {{Lamers}}},\ }\bibfield  {title} {\bibinfo {title} {{Mass-loss predictions
  for O and B stars as a function of metallicity}},\ }\href
  {https://doi.org/10.1051/0004-6361:20010127} {\bibfield  {journal} {\bibinfo
  {journal} {\aap}\ }\textbf {\bibinfo {volume} {369}},\ \bibinfo {pages} {574}
  (\bibinfo {year} {2001})},\ \Eprint {https://arxiv.org/abs/astro-ph/0101509}
  {arXiv:astro-ph/0101509 [astro-ph]} \BibitemShut {NoStop}\bibitem [{\citenamefont {{Claeys}}\ \emph {et~al.}(2014)\citenamefont
  {{Claeys}}, \citenamefont {{Pols}}, \citenamefont {{Izzard}}, \citenamefont
  {{Vink}},\ and\ \citenamefont {{Verbunt}}}]{2014A&A...563A..83C}\BibitemOpen
  \bibfield  {author} {\bibinfo {author} {\bibfnamefont {J.~S.~W.}\
  \bibnamefont {{Claeys}}}, \bibinfo {author} {\bibfnamefont {O.~R.}\
  \bibnamefont {{Pols}}}, \bibinfo {author} {\bibfnamefont {R.~G.}\
  \bibnamefont {{Izzard}}}, \bibinfo {author} {\bibfnamefont {J.}~\bibnamefont
  {{Vink}}},\ and\ \bibinfo {author} {\bibfnamefont {F.~W.~M.}\ \bibnamefont
  {{Verbunt}}},\ }\bibfield  {title} {\bibinfo {title} {{Theoretical
  uncertainties of the Type Ia supernova rate}},\ }\href
  {https://doi.org/10.1051/0004-6361/201322714} {\bibfield  {journal} {\bibinfo
   {journal} {\aap}\ }\textbf {\bibinfo {volume} {563}},\ \bibinfo {eid} {A83}
  (\bibinfo {year} {2014})},\ \Eprint {https://arxiv.org/abs/1401.2895}
  {arXiv:1401.2895 [astro-ph.SR]} \BibitemShut {NoStop}\bibitem [{\citenamefont {{Eggleton}}(1983)}]{1983ApJ...268..368E}\BibitemOpen
  \bibfield  {author} {\bibinfo {author} {\bibfnamefont {P.~P.}\ \bibnamefont
  {{Eggleton}}},\ }\bibfield  {title} {\bibinfo {title} {{Aproximations to the
  radii of Roche lobes.}},\ }\href {https://doi.org/10.1086/160960} {\bibfield
  {journal} {\bibinfo  {journal} {\apj}\ }\textbf {\bibinfo {volume} {268}},\
  \bibinfo {pages} {368} (\bibinfo {year} {1983})}\BibitemShut {NoStop}\bibitem [{\citenamefont {{Bondi}}\ and\ \citenamefont
  {{Hoyle}}(1944)}]{1944MNRAS.104..273B}\BibitemOpen
  \bibfield  {author} {\bibinfo {author} {\bibfnamefont {H.}~\bibnamefont
  {{Bondi}}}\ and\ \bibinfo {author} {\bibfnamefont {F.}~\bibnamefont
  {{Hoyle}}},\ }\bibfield  {title} {\bibinfo {title} {{On the mechanism of
  accretion by stars}},\ }\href {https://doi.org/10.1093/mnras/104.5.273}
  {\bibfield  {journal} {\bibinfo  {journal} {\mnras}\ }\textbf {\bibinfo
  {volume} {104}},\ \bibinfo {pages} {273} (\bibinfo {year}
  {1944})}\BibitemShut {NoStop}\bibitem [{\citenamefont {{Hjellming}}\ and\ \citenamefont
  {{Webbink}}(1987)}]{1987ApJ...318..794H}\BibitemOpen
  \bibfield  {author} {\bibinfo {author} {\bibfnamefont {M.~S.}\ \bibnamefont
  {{Hjellming}}}\ and\ \bibinfo {author} {\bibfnamefont {R.~F.}\ \bibnamefont
  {{Webbink}}},\ }\bibfield  {title} {\bibinfo {title} {{Thresholds for Rapid
  Mass Transfer in Binary System. I. Polytropic Models}},\ }\href
  {https://doi.org/10.1086/165412} {\bibfield  {journal} {\bibinfo  {journal}
  {\apj}\ }\textbf {\bibinfo {volume} {318}},\ \bibinfo {pages} {794} (\bibinfo
  {year} {1987})}\BibitemShut {NoStop}\bibitem [{\citenamefont {{Webbink}}(1984)}]{1984ApJ...277..355W}\BibitemOpen
  \bibfield  {author} {\bibinfo {author} {\bibfnamefont {R.~F.}\ \bibnamefont
  {{Webbink}}},\ }\bibfield  {title} {\bibinfo {title} {{Double white dwarfs as
  progenitors of R Coronae Borealis stars and type I supernovae.}},\ }\href
  {https://doi.org/10.1086/161701} {\bibfield  {journal} {\bibinfo  {journal}
  {\apj}\ }\textbf {\bibinfo {volume} {277}},\ \bibinfo {pages} {355} (\bibinfo
  {year} {1984})}\BibitemShut {NoStop}\bibitem [{\citenamefont {{Rappaport}}\ \emph {et~al.}(1983)\citenamefont
  {{Rappaport}}, \citenamefont {{Verbunt}},\ and\ \citenamefont
  {{Joss}}}]{1983ApJ...275..713R}\BibitemOpen
  \bibfield  {author} {\bibinfo {author} {\bibfnamefont {S.}~\bibnamefont
  {{Rappaport}}}, \bibinfo {author} {\bibfnamefont {F.}~\bibnamefont
  {{Verbunt}}},\ and\ \bibinfo {author} {\bibfnamefont {P.~C.}\ \bibnamefont
  {{Joss}}},\ }\bibfield  {title} {\bibinfo {title} {{A new technique for
  calculations of binary stellar evolution application to magnetic braking.}},\
  }\href {https://doi.org/10.1086/161569} {\bibfield  {journal} {\bibinfo
  {journal} {\apj}\ }\textbf {\bibinfo {volume} {275}},\ \bibinfo {pages} {713}
  (\bibinfo {year} {1983})}\BibitemShut {NoStop}\bibitem [{\citenamefont {{Kiel}}\ \emph {et~al.}(2008)\citenamefont {{Kiel}},
  \citenamefont {{Hurley}}, \citenamefont {{Bailes}},\ and\ \citenamefont
  {{Murray}}}]{psr-popsyn-AustralianDetailedPopsyn2008}\BibitemOpen
  \bibfield  {author} {\bibinfo {author} {\bibfnamefont {P.~D.}\ \bibnamefont
  {{Kiel}}}, \bibinfo {author} {\bibfnamefont {J.~R.}\ \bibnamefont
  {{Hurley}}}, \bibinfo {author} {\bibfnamefont {M.}~\bibnamefont {{Bailes}}},\
  and\ \bibinfo {author} {\bibfnamefont {J.~R.}\ \bibnamefont {{Murray}}},\
  }\bibfield  {title} {\bibinfo {title} {{Populating the Galaxy with pulsars -
  I. Stellar and binary evolution}},\ }\href
  {https://doi.org/10.1111/j.1365-2966.2008.13402.x} {\bibfield  {journal}
  {\bibinfo  {journal} {\mnras}\ }\textbf {\bibinfo {volume} {388}},\ \bibinfo
  {pages} {393} (\bibinfo {year} {2008})},\ \Eprint
  {https://arxiv.org/abs/0805.0059} {arXiv:0805.0059 [astro-ph]} \BibitemShut
  {NoStop}\bibitem [{\citenamefont {{Tauris}}\ \emph {et~al.}(2012)\citenamefont
  {{Tauris}}, \citenamefont {{Langer}},\ and\ \citenamefont
  {{Kramer}}}]{gwastro-TaurisFormationMillisecondPulsars2012}\BibitemOpen
  \bibfield  {author} {\bibinfo {author} {\bibfnamefont {T.~M.}\ \bibnamefont
  {{Tauris}}}, \bibinfo {author} {\bibfnamefont {N.}~\bibnamefont {{Langer}}},\
  and\ \bibinfo {author} {\bibfnamefont {M.}~\bibnamefont {{Kramer}}},\
  }\bibfield  {title} {\bibinfo {title} {{Formation of millisecond pulsars with
  CO white dwarf companions - II. Accretion, spin-up, true ages and comparison
  to MSPs with He white dwarf companions}},\ }\href
  {https://doi.org/10.1111/j.1365-2966.2012.21446.x} {\bibfield  {journal}
  {\bibinfo  {journal} {\mnras}\ }\textbf {\bibinfo {volume} {425}},\ \bibinfo
  {pages} {1601} (\bibinfo {year} {2012})},\ \Eprint
  {https://arxiv.org/abs/1206.1862} {arXiv:1206.1862 [astro-ph.SR]}
  \BibitemShut {NoStop}\bibitem [{\citenamefont {{Wong}}\ \emph {et~al.}(2020)\citenamefont {{Wong}},
  \citenamefont {{Contardo}},\ and\ \citenamefont
  {{Ho}}}]{2020arXiv200209491W}\BibitemOpen
  \bibfield  {author} {\bibinfo {author} {\bibfnamefont {K.~W.~K.}\
  \bibnamefont {{Wong}}}, \bibinfo {author} {\bibfnamefont {G.}~\bibnamefont
  {{Contardo}}},\ and\ \bibinfo {author} {\bibfnamefont {S.}~\bibnamefont
  {{Ho}}},\ }\bibfield  {title} {\bibinfo {title} {{Gravitational-wave
  population inference with deep flow-based generative network}},\ }\href
  {https://doi.org/10.1103/PhysRevD.101.123005} {\bibfield  {journal} {\bibinfo
   {journal} {\prd}\ }\textbf {\bibinfo {volume} {101}},\ \bibinfo {eid}
  {123005} (\bibinfo {year} {2020})},\ \Eprint
  {https://arxiv.org/abs/2002.09491} {arXiv:2002.09491 [astro-ph.IM]}
  \BibitemShut {NoStop}\bibitem [{\citenamefont {Bates}\ \emph {et~al.}(2014)\citenamefont {Bates},
  \citenamefont {Lorimer}, \citenamefont {Rane},\ and\ \citenamefont
  {Swiggum}}]{Bates2014PsrPopPy}\BibitemOpen
  \bibfield  {author} {\bibinfo {author} {\bibfnamefont {S.~D.}\ \bibnamefont
  {Bates}}, \bibinfo {author} {\bibfnamefont {D.~R.}\ \bibnamefont {Lorimer}},
  \bibinfo {author} {\bibfnamefont {A.}~\bibnamefont {Rane}},\ and\ \bibinfo
  {author} {\bibfnamefont {J.}~\bibnamefont {Swiggum}},\ }\bibfield  {title}
  {\bibinfo {title} {{PsrPopPy}: an open-source package for pulsar population
  simulations},\ }\href {https://doi.org/10.1093/mnras/stu157} {\bibfield
  {journal} {\bibinfo  {journal} {MNRAS}\ }\textbf {\bibinfo {volume} {439}},\
  \bibinfo {pages} {2893} (\bibinfo {year} {2014})},\ \Eprint
  {https://arxiv.org/abs/1311.3427} {arXiv:1311.3427} \BibitemShut {NoStop}\bibitem [{\citenamefont {Cordes}\ and\ \citenamefont
  {Lazio}(2002)}]{psr-ElectronDensityModel-NE2001}\BibitemOpen
  \bibfield  {author} {\bibinfo {author} {\bibfnamefont {J.~M.}\ \bibnamefont
  {Cordes}}\ and\ \bibinfo {author} {\bibfnamefont {T.~J.~W.}\ \bibnamefont
  {Lazio}},\ }\bibfield  {title} {\bibinfo {title} {{NE2001.I.} a new model for
  the galactic distribution of free electrons and its fluctuations},\ }\href
  {https://doi.org/10.48550/arXiv.astro-ph/0207156} {\bibfield  {journal}
  {\bibinfo  {journal} {arXiv e-prints}\ ,\ \bibinfo {eid} {astro-ph/0207156}}
  (\bibinfo {year} {2002})},\ \Eprint {https://arxiv.org/abs/astro-ph/0207156}
  {arXiv:astro-ph/0207156} \BibitemShut {NoStop}\bibitem [{\citenamefont {Haslam}\ \emph {et~al.}(1982)\citenamefont {Haslam},
  \citenamefont {Salter}, \citenamefont {Stoffel},\ and\ \citenamefont
  {Wilson}}]{1982AAS...47....1H}\BibitemOpen
  \bibfield  {author} {\bibinfo {author} {\bibfnamefont {C.~G.~T.}\
  \bibnamefont {Haslam}}, \bibinfo {author} {\bibfnamefont {C.~J.}\
  \bibnamefont {Salter}}, \bibinfo {author} {\bibfnamefont {H.}~\bibnamefont
  {Stoffel}},\ and\ \bibinfo {author} {\bibfnamefont {W.~E.}\ \bibnamefont
  {Wilson}},\ }\bibfield  {title} {\bibinfo {title} {A 408 mhz all-sky
  continuum survey. ii. the atlas of contour maps},\ }\href@noop {} {\bibfield
  {journal} {\bibinfo  {journal} {Astronomy and Astrophysics Supplement
  Series}\ }\textbf {\bibinfo {volume} {47}},\ \bibinfo {pages} {1} (\bibinfo
  {year} {1982})}\BibitemShut {NoStop}\bibitem [{\citenamefont {Drummond}\ \emph {et~al.}(2026)\citenamefont
  {Drummond}, \citenamefont {Chatziioannou},\ and\ \citenamefont
  {Fonseca}}]{Drummond2026RadioSelection}\BibitemOpen
  \bibfield  {author} {\bibinfo {author} {\bibfnamefont {L.~V.}\ \bibnamefont
  {Drummond}}, \bibinfo {author} {\bibfnamefont {K.}~\bibnamefont
  {Chatziioannou}},\ and\ \bibinfo {author} {\bibfnamefont {E.}~\bibnamefont
  {Fonseca}},\ }\bibfield  {title} {\bibinfo {title} {Observational selection
  effects on radio pulsars are minimal for masses, but significant for orbits
  and spins},\ }\href {https://doi.org/10.48550/arXiv.2609.03157} {\bibfield
  {journal} {\bibinfo  {journal} {arXiv e-prints}\ ,\ \bibinfo {eid}
  {arXiv:2609.03157}} (\bibinfo {year} {2026})},\ \Eprint
  {https://arxiv.org/abs/2609.03157} {arXiv:2609.03157 [astro-ph.HE]}
  \BibitemShut {NoStop}\bibitem [{\citenamefont {{Posselt}}\ \emph {et~al.}(2021)\citenamefont
  {{Posselt}}, \citenamefont {{Karastergiou}}, \citenamefont {{Johnston}},
  \citenamefont {{Parthasarathy}}, \citenamefont {{Keith}}, \citenamefont
  {{Oswald}}, \citenamefont {{Song}}, \citenamefont {{Weltevrede}},
  \citenamefont {{Barr}}, \citenamefont {{Buchner}}, \citenamefont {{Geyer}},
  \citenamefont {{Kramer}}, \citenamefont {{Reardon}}, \citenamefont
  {{Serylak}}, \citenamefont {{Shannon}}, \citenamefont {{Spiewak}},\ and\
  \citenamefont {{Venkatraman Krishnan}}}]{Posselt2021Widths}\BibitemOpen
  \bibfield  {author} {\bibinfo {author} {\bibfnamefont {B.}~\bibnamefont
  {{Posselt}}}, \bibinfo {author} {\bibfnamefont {A.}~\bibnamefont
  {{Karastergiou}}}, \bibinfo {author} {\bibfnamefont {S.}~\bibnamefont
  {{Johnston}}}, \bibinfo {author} {\bibfnamefont {A.}~\bibnamefont
  {{Parthasarathy}}}, \bibinfo {author} {\bibfnamefont {M.~J.}\ \bibnamefont
  {{Keith}}}, \bibinfo {author} {\bibfnamefont {L.~S.}\ \bibnamefont
  {{Oswald}}}, \bibinfo {author} {\bibfnamefont {X.}~\bibnamefont {{Song}}},
  \bibinfo {author} {\bibfnamefont {P.}~\bibnamefont {{Weltevrede}}}, \bibinfo
  {author} {\bibfnamefont {E.~D.}\ \bibnamefont {{Barr}}}, \bibinfo {author}
  {\bibfnamefont {S.}~\bibnamefont {{Buchner}}}, \bibinfo {author}
  {\bibfnamefont {M.}~\bibnamefont {{Geyer}}}, \bibinfo {author} {\bibfnamefont
  {M.}~\bibnamefont {{Kramer}}}, \bibinfo {author} {\bibfnamefont {D.~J.}\
  \bibnamefont {{Reardon}}}, \bibinfo {author} {\bibfnamefont {M.}~\bibnamefont
  {{Serylak}}}, \bibinfo {author} {\bibfnamefont {R.~M.}\ \bibnamefont
  {{Shannon}}}, \bibinfo {author} {\bibfnamefont {R.}~\bibnamefont
  {{Spiewak}}},\ and\ \bibinfo {author} {\bibfnamefont {V.}~\bibnamefont
  {{Venkatraman Krishnan}}},\ }\bibfield  {title} {\bibinfo {title} {{The
  Thousand-Pulsar-Array programme on MeerKAT -- VI. Pulse widths of a large and
  diverse sample of radio pulsars}},\ }\href
  {https://doi.org/10.1093/mnras/stab2775} {\bibfield  {journal} {\bibinfo
  {journal} {\mnras}\ }\textbf {\bibinfo {volume} {508}},\ \bibinfo {pages}
  {4249} (\bibinfo {year} {2021})},\ \Eprint {https://arxiv.org/abs/2109.12007}
  {arXiv:2109.12007 [astro-ph.HE]} \BibitemShut {NoStop}\bibitem [{\citenamefont {{Wang}}\ \emph {et~al.}(2023)\citenamefont {{Wang}},
  \citenamefont {{Han}}, \citenamefont {{Xu}}, \citenamefont {{Wang}},
  \citenamefont {{Yan}}, \citenamefont {{Jing}}, \citenamefont {{Su}},
  \citenamefont {{Zhou}},\ and\ \citenamefont {{Wang}}}]{Wang2023FASTProfiles}\BibitemOpen
  \bibfield  {author} {\bibinfo {author} {\bibfnamefont {P.~F.}\ \bibnamefont
  {{Wang}}}, \bibinfo {author} {\bibfnamefont {J.~L.}\ \bibnamefont {{Han}}},
  \bibinfo {author} {\bibfnamefont {J.}~\bibnamefont {{Xu}}}, \bibinfo {author}
  {\bibfnamefont {C.}~\bibnamefont {{Wang}}}, \bibinfo {author} {\bibfnamefont
  {Y.}~\bibnamefont {{Yan}}}, \bibinfo {author} {\bibfnamefont {W.~C.}\
  \bibnamefont {{Jing}}}, \bibinfo {author} {\bibfnamefont {W.~Q.}\
  \bibnamefont {{Su}}}, \bibinfo {author} {\bibfnamefont {D.~J.}\ \bibnamefont
  {{Zhou}}},\ and\ \bibinfo {author} {\bibfnamefont {T.}~\bibnamefont
  {{Wang}}},\ }\bibfield  {title} {\bibinfo {title} {{FAST Pulsar Database. I.
  Polarization Profiles of 682 Pulsars}},\ }\href
  {https://doi.org/10.1088/1674-4527/acea1f} {\bibfield  {journal} {\bibinfo
  {journal} {Research in Astronomy and Astrophysics}\ }\textbf {\bibinfo
  {volume} {23}},\ \bibinfo {eid} {104002} (\bibinfo {year} {2023})},\ \Eprint
  {https://arxiv.org/abs/2307.10340} {arXiv:2307.10340 [astro-ph.HE]}
  \BibitemShut {NoStop}\bibitem [{\citenamefont {Philippov}\ \emph {et~al.}(2014)\citenamefont
  {Philippov}, \citenamefont {Tchekhovskoy},\ and\ \citenamefont
  {Li}}]{Philippov2014Obliquity}\BibitemOpen
  \bibfield  {author} {\bibinfo {author} {\bibfnamefont {A.}~\bibnamefont
  {Philippov}}, \bibinfo {author} {\bibfnamefont {A.}~\bibnamefont
  {Tchekhovskoy}},\ and\ \bibinfo {author} {\bibfnamefont {J.~G.}\ \bibnamefont
  {Li}},\ }\bibfield  {title} {\bibinfo {title} {Time evolution of pulsar
  obliquity angle from 3d simulations of magnetospheres},\ }\href
  {https://doi.org/10.1093/mnras/stu591} {\bibfield  {journal} {\bibinfo
  {journal} {MNRAS}\ }\textbf {\bibinfo {volume} {441}},\ \bibinfo {pages}
  {1879} (\bibinfo {year} {2014})},\ \Eprint {https://arxiv.org/abs/1311.1513}
  {arXiv:1311.1513} \BibitemShut {NoStop}\bibitem [{\citenamefont {{Lyne}}\ and\ \citenamefont
  {{Manchester}}(1988)}]{Lyne1988Shape}\BibitemOpen
  \bibfield  {author} {\bibinfo {author} {\bibfnamefont {A.~G.}\ \bibnamefont
  {{Lyne}}}\ and\ \bibinfo {author} {\bibfnamefont {R.~N.}\ \bibnamefont
  {{Manchester}}},\ }\bibfield  {title} {\bibinfo {title} {{The shape of pulsar
  radio beams}},\ }\href@noop {} {\bibfield  {journal} {\bibinfo  {journal}
  {\mnras}\ }\textbf {\bibinfo {volume} {234}},\ \bibinfo {pages} {477}
  (\bibinfo {year} {1988})}\BibitemShut {NoStop}\bibitem [{\citenamefont {{Tauris}}\ and\ \citenamefont
  {{Manchester}}(1998)}]{Tauris1998Beaming}\BibitemOpen
  \bibfield  {author} {\bibinfo {author} {\bibfnamefont {T.~M.}\ \bibnamefont
  {{Tauris}}}\ and\ \bibinfo {author} {\bibfnamefont {R.~N.}\ \bibnamefont
  {{Manchester}}},\ }\bibfield  {title} {\bibinfo {title} {{On the Evolution of
  Pulsar Beams}},\ }\href {https://doi.org/10.1046/j.1365-8711.1998.01369.x}
  {\bibfield  {journal} {\bibinfo  {journal} {\mnras}\ }\textbf {\bibinfo
  {volume} {298}},\ \bibinfo {pages} {625} (\bibinfo {year}
  {1998})}\BibitemShut {NoStop}\bibitem [{Transparent Agentic
  Science()}]{whitepaper-transparent-agentic-science}\BibitemOpen
  Transparent Agentic Science,\ \href
  {https://github.com/oshaughnessy-junior/whitepaper-transparent-agentic-science}
  {\bibinfo {title} {{Transparent Agentic Science}: Reproducible computation,
  adversarial review, and human-directed writing}},\ \bibinfo {howpublished}
  {\url{https://github.com/oshaughnessy-junior/whitepaper-transparent-agentic-science}}
  (\bibinfo {year} {2026}),\ \bibinfo {note} {workflow documentation and
  reusable skills}\BibitemShut {NoStop}\bibitem [{\citenamefont {{Grichener}}\ \emph {et~al.}(2026)\citenamefont
  {{Grichener}}, \citenamefont {{Disberg}}, \citenamefont {{Hirai}},\ and\
  \citenamefont {{Mandel}}}]{Grichener2026Eccentricity}\BibitemOpen
  \bibfield  {author} {\bibinfo {author} {\bibfnamefont {A.}~\bibnamefont
  {{Grichener}}}, \bibinfo {author} {\bibfnamefont {P.}~\bibnamefont
  {{Disberg}}}, \bibinfo {author} {\bibfnamefont {R.}~\bibnamefont {{Hirai}}},\
  and\ \bibinfo {author} {\bibfnamefont {I.}~\bibnamefont {{Mandel}}},\
  }\bibfield  {title} {\bibinfo {title} {{On Bimodality in the Eccentricity
  Distribution of Galactic Double Neutron Stars}},\ }\href
  {https://doi.org/10.48550/arXiv.2608.00174} {\bibfield  {journal} {\bibinfo
  {journal} {arXiv e-prints}\ ,\ \bibinfo {eid} {arXiv:2608.00174}} (\bibinfo
  {year} {2026})},\ \Eprint {https://arxiv.org/abs/2608.00174}
  {arXiv:2608.00174 [astro-ph.SR]} \BibitemShut {NoStop}\end{thebibliography}
\end{document}

\title{Supplemental Material for Selection-aware inference of supernova kicks}
\author{R.~O'Shaughnessy}
\affiliation{Center for Computational Relativity and Gravitation, Rochester Institute of Technology}
\date{2026-11-26}
\maketitle

\section{Checks on the radio-selected likelihood}
Figures 3 and 4 of the paper share 18 fixed progenitors, the five-survey
response, and the same DNS measurement bins. Figure 3 uses a frozen
random catalog of 640 DNS detections. Figure 4 uses expected cell counts
for 640 DNS and 3,840 young pulsar detections. Each class is normalized
separately after pooling absolute cell integrals over progenitors,
integration scrambles, and all origins within that class. The young
sample includes its missing-astrometry cell. The relative cohort sizes
provide no rate constraint.

The $13\times19$ grid varies the logarithms of the shared kick multiplier
and inter-supernova common-envelope (CE) efficiency. The horizontal
coordinate in the figures is the median ordinary core-collapse kick,
$e^{5.60}s_k\,{\rm km\,s^{-1}}$. Bicubic interpolation reproduces the
evaluated likelihoods at the grid nodes. Three independent,
higher-resolution checks test the expected-count DNS surface in Fig.~4.
For Fig.~3, the negative Hessian of the random-catalog interpolant at
its maximum gives local log-parameter widths $(0.060,0.137)$. Replacing
the spline curvature with local quadratic fits on $3\times3$ or
$5\times5$ patches changes the kick and CE widths by up to 7.3\% and 15.8\%,
respectively. The replicated contours rescale this
local curvature. The evaluated grid sets the resolution of the likelihood.

The denominator-only error adds $640\ln S_1$ to the log likelihood,
where $S_1$ is the probability per starting binary of forming a bound
neutron-star system at the first explosion. Integrating the analytic
bound fraction over the continuous kick distributions reproduces
the forward-code $S_1$ totals to within $6\times10^{-8}$ relative at
two checked points. At the correctly normalized maximum
$\partial\ln S_1/\partial\ln s_k=-0.867$. Eight direct checks with
four scrambles use $2^{14}$ SN1 and $2^{16}$ SN2 draws per integration
bank. At the best checked point $(37.8\,{\rm km\,s^{-1}},0.322)$ the
correct log likelihood falls by 213 and the extra survival term rises
by 712. The net gain is 499.3, with a paired integration standard
error of 1.2. The evaluated surfaces and check results are recorded in
\path{data/paper1_joint_radio/} and \path{data/paper1_radio_dns_recovery/}.

The paper's four-coordinate DNS endpoint calculation has a separate
integration check. Refining the explosion scatter and direction rules
on the same Sobol progenitors shifts the 64-endpoint local estimate by
less than one statistical width. Forward automatic derivatives were
compared with centered finite differences and analytic SN1 survival
with independent vector impulses. A binary-by-binary comparison with
\cosmic{}~4.1's \texttt{kick.f} imposed the same kicks and phases.
All 203 cases agreed on bound versus unbound. For the 110 bound cases
the median fractional separation and absolute eccentricity residuals
were $1.2\times10^{-4}$ and $7\times10^{-5}$, respectively.

\section{COSMIC progenitor calibration}
The archive uses the paper's baseline birth density $p_b$ and fixed
evolution physics. It contains \ArchiveBaselineBirths{} unweighted
births, \ArchivePartitionEvolved{} evolved births from
\ArchivePartitionDraws{} partition-proposal draws, and a separate
\ArchiveJointBirths{}-birth cohort targeted at pre-SN2 ancestors.
The partition proposal evolves a draw $x$ with its recorded inclusion
probability $r(x)$, floored at \PartitionFloor{}. For $N_b$ baseline and
$N_p$ partition draws, the fraction in class $A$ is estimated by summing
$1/[N_b+N_p r(x)]$ over its retained rows in both samples. The denominator
accounts for all original draws. The targeted cohort uses its own
proposal weights and agrees with the pooled yields within
\ArchiveJointMaxPull{} sampling standard errors.

\begin{table}[htbp]
\caption{Weighted yields per baseline birth. Effective sizes are
$(\sum_i w_i)^2/\sum_i w_i^2$ within each class. The AD-image neighborhood
surrounds the propagated image of the paper's pre-SN2 input figure.}
\label{tab:critical_yields}
\begin{tabular}{lrr}
\toprule
Class & Yield & Effective size\\
\midrule
Pre-SN1 stripped-star family & $5.39\times10^{-1}$ & 2,102,026\\
Prescribed starting boxes & $3.96\times10^{-3}$ & 13,355\\
Circular He-star--NS arrivals & $2.57\times10^{-3}$ & 27,326\\
Wide-orbit CE arrivals & $1.08\times10^{-3}$ & 12,869\\
Double-core arrivals & $1.31\times10^{-3}$ & 13,670\\
AD-image neighborhood & $3.06\times10^{-5}$ & 644\\
Bound DNS & $1.49\times10^{-3}$ & 15,536\\
\bottomrule
\end{tabular}
\end{table}

\paragraph{Pre-SN1 fit.}
The bounded Gaussian-mixture transport uses \FitTrainRows{} training
prefixes and \FitTestRows{} held-out prefixes from separate births.
Across the six fitted coordinates, the largest marginal CDF differences
are \FitMaxCdfNoCe{} and \FitMaxCdfCe{} for histories without and with
earlier CE, below familywise envelopes of \FitCdfLimitNoCe{} and
\FitCdfLimitCe{}. The largest correlation differences are
\FitCorrNoCe{} and \FitCorrCe{}.

\paragraph{Pre-SN2 arrivals.}
Of the 1,080,000 training births, 900,000 use a defensive mixture
with 25\% of the base density and 75\% in fitted proposals for the
two inter-SN CE histories. Each arrival receives its full $p_b/q$
ratio before normalization. The 28,281 training arrivals have
effective size 5,105 and weighted inter-SN CE fraction 49.2\%.
The independent base-proposal sample yields 452 arrivals in
160,000 births. The extractor stops at the pre-SN2 state, and the
cuts define a circular He-star--NS family. The weighted empirical
rows retain their joint masses and orbit. The period-tilted
orientation response bank samples the nominal and two endpoint-tilted
priors in proportions $(1/2,1/4,1/4)$ with exact importance correction.

\paragraph{Evolution from the same starting states.}
For \AdWideStates{} map arrivals from wide-orbit \cosmic{} prefixes,
median exploding-mass and log-period offsets from the corresponding
\cosmic{} states are $\AdWideDeltaMass\,\Msun$ and
\AdWideDeltaLogP{} dex, with period rank correlation
\AdWidePeriodRank{}. Circularization at fixed angular momentum
changes the period offset to \AdWideDeltaLogPJcirc{} dex. These
prefixes are selected on \cosmic{} arrival. A separate calculation
evaluates 256 starting states per class box for the differentiable
map. The \cosmic{} fractions use all archive states in each box.
Table~\ref{tab:critical_archive_boxes} gives the
survival and arrival comparisons. The map uses Mandel--M\"uller
remnants and the archive uses \cosmic{}'s delayed remnants. The
comparison tests where the starting boxes lie on the archive
and how the two models propagate their states.

\begin{table}[htbp]
\caption{Evolution from the pre-SN1 class boxes. Rates are percentages
except for the mean SN1 kick, which averages neutron-star-forming events.}
\label{tab:critical_archive_boxes}
\begin{tabular}{lrrrr}
\toprule
 & \multicolumn{2}{c}{W} & \multicolumn{2}{c}{C}\\
 & Map & \cosmic{} & Map & \cosmic{}\\
\midrule
Bound after SN1 & \AdClassWBoundPercent & \ClassWBoundPercent & \AdClassCBoundPercent & \ClassCBoundPercent\\
Reach SN2 if bound & \AdClassWArrivalGivenBoundPercent & \ClassWArrivalGivenBoundPercent & \AdClassCArrivalGivenBoundPercent & \ClassCArrivalGivenBoundPercent\\
Reach SN2 & \AdClassWArrivalPercent & \ClassWArrivalPercent & \AdClassCArrivalPercent & \ClassCArrivalPercent\\
Bound DNS & \AdClassWDnsPercent & \ClassWDnsPercent & \AdClassCDnsPercent & \ClassCDnsPercent\\
Mean SN1 kick ($\mathrm{km\,s^{-1}}$) & \AdClassWKick & \ClassWKick & \AdClassCKick & \ClassCKick\\
\bottomrule
\end{tabular}
\end{table}

\section{Sensitivity to initial populations and native COMPAS evolution}
For Fig.~9, we multiply each parent's absolute DNS cell response by its
new-to-old birth-weight ratio before pooling and normalizing. The
frozen catalog, stellar trajectories and five-survey response are
shared across fits. \compas{} 03.31.00 pre-SN1 states
\cite{2025arXiv250602316C} within the paper's starting support give
W fractions of 0.319 with default core retention and 0.255 with it
disabled, versus 0.724 in the original family. The second test
applies period weights proportional to $P_1^{\pm0.5}$ within the
fixed family. The mixture changes preserve the corrected peak because
selected DNSs come almost entirely from W. The period tilts move the
corrected kick peak to 245--322\,km\,s$^{-1}$. Direct gains under the
inconsistent objective are 499, 529 and 534 for the original and
alternative mixtures and 324 and 939 for the negative and positive
tilts. Each gain is relative to the correctly normalized peak of its
own measure, with paired integration standard errors of 1.1--1.5.

Figure 10 substitutes native \compas{} tracks from the initial stellar
population through both explosions while retaining the catalog and
radio response. The starting population also changes, and the native
CE parameter acts throughout evolution. Two independent evaluation
birth sets supply 65,536 births per CE value in the pooled nine-value
scan. Importance ratios restore the intended birth density and
native remnant-dependent kick laws, retaining the zero-kick
accretion-induced-collapse atom. The omitted factor is the probability
per initial binary of forming a bound NS system in the first collapse
episode by 100 Myr. An independent direct-default run agrees with the
weighted first-NS-bound, DNS-birth and radio-selected yields within
0.57 combined sampling standard errors. The 384-to-768-node survey-age
refinement changes pooled log likelihood by at most 0.021 at checked
points. Independent births, CE-grid refinement and whole-birth omissions
shift the native peak by more than its local widths. The downward
kick shift under inconsistent normalization persists across these
checks, including fits that reach the CE grid edge.

\section{Radio calibration and rate inputs}
\paragraph{Five-survey benchmark.}
The archived snapshot calculation of Ref.~\cite{Pol2019DNSRate} counts
detections in each survey, including rediscoveries. The five-class rate
counts unique detections using the union of the survey responses. Both
coefficients condition on active sources whose beams cross Earth. The
fresh run of 500,000 sources per class reproduces the five coefficients
extracted from the authors' archived simulations within 0.84 combined
Monte Carlo standard errors. This checks the simulation outputs at
matched survey assumptions. Empirical survey completeness requires
separate validation.

The class inputs use the historical beam and lifetime choices of
Ref.~\cite{psr-SpinFb}. A single count per class with a flat abundance
prior gives a rate density proportional to
$\mathcal R_i e^{-\alpha_i\tau_i\mathcal R_i/f_{b,i}}$. We normalize
before averaging over nuisance parameters and convolve the independent
class densities. Removing J1906$+$0746 gives a median of
$18.8\,{\rm Myr}^{-1}$ and a 90\% interval $[7.2,45.5]\,{\rm Myr}^{-1}$
with other inputs fixed.
\begin{table}[htbp]
\caption{Five-class rate inputs. $\alpha_i$ is the unique-source detection
fraction, $f_{b,i}$ the beaming correction, and $\tau_i$ the observable
lifetime. Bracketed beaming ranges denote log-uniform priors.
Class rates are in $\mathrm{Myr}^{-1}$.}
\begin{tabular}{lrrrrr}
\hline
Pulsar & $10^3\alpha_i$ & $f_{b,i}$ & $\tau_i$ (Myr) & Median & 90\% interval \\
\hline
J0737$-$3039A & 2.152 & [1.55,6] & 228 & 10.20 & [1.96,35.83] \\
J1756$-$2251 & 6.462 & [1.68,6] & 2032 & 0.40 & [0.08,1.37] \\
J1906$+$0746 & 13.040 & 3.37 & 82.1 & 5.28 & [1.12,14.93] \\
B1913$+$16 & 4.416 & 5.72 & 366 & 5.93 & [1.26,16.77] \\
B1534$+$12 & 6.996 & 6.04 & 2930 & 0.49 & [0.10,1.40] \\
\hline
\end{tabular}

\end{table}

\paragraph{Pulse-width likelihood.}
We use ATNF v2.9.0 $W_{10}$ values measured by MeerKAT at 1284\,MHz
or, where unavailable, FAST at 1250\,MHz
\cite{Manchester2005ATNF,Posselt2021Widths,Wang2023FASTProfiles}.
We exclude extragalactic and cluster pulsars and objects with a
predicted scattering tail above $1^\circ$ of longitude. The normal
sample requires isolation, $P>32\,{\rm ms}$ and a dipole field above
$3.1\times10^{10}\,{\rm G}$ (\PbNNormal{} objects). The mildly recycled
sample uses the lower field range and $15<P<200\,{\rm ms}$
(\PbNRecycled{} objects).
For magnetic inclination $\alpha$, sight-line angle $\zeta$ and cone
opening $\rho$, the pulse width obeys
$\cos\rho=\cos\alpha\cos\zeta+\sin\alpha\sin\zeta\cos(W/2)$
\cite{Gil1984Geometry}. The width distribution integrates isotropic
sight lines conditioned on a pulse with $0<W<2\pi$, and averages
over inclinations and lognormal scatter in
$\rho=\rho_1(P/1\,{\rm s})^{-\epsilon}$, capped at $90^\circ$.
The likelihood includes Gaussian error in $\log_{10}W$ and a
log-uniform outlier component over $1^\circ$--$360^\circ$.
We compare isotropic inclinations with the plasma-torque model of
Ref.~\cite{Philippov2014Obliquity}, integrating over a lognormal birth
period truncated at the present period. The nominal birth-period median
and log-period dispersion are \PbPZeroMedianMs{}\,ms and
\PbPZeroSigma{}\,dex. The plasma fit gives $\rho_1=\PbRhoOnePlasma^\circ$
and $\epsilon=\PbRhoExpPlasma\pm\PbRhoExpPlasmaErr$, with exponent
uncertainty from mock fits. The log likelihood rises by
\PbDlnLPlasma{} relative to isotropic inclinations. A second birth-period
prior gives $\epsilon=\PbRhoExpPlasmaCensus$. The five-class rate retains
the published beam corrections.

\section{Conditional low-kick-fraction recovery}
The one-explosion controls start from the pre-SN2 empirical prior above
and set the new neutron-star mass to $1.33\,\Msun$. They compare a
birth-mixture likelihood with survival, a fit that interprets the
selected mixture as natal, and an unnormalized product. For an injected
birth fraction $f_{\rm low}=0.5$, the bound sample is enriched in low
kicks because the low-kick component enters the orbital sample more
often. At 1,100 detections, the offset from reading the selected
mixture as natal is 1.29 catalog standard deviations, rising to 3.71
at 11,000. In the 11-object mocks the three fitted fractions have
medians \DnsMockRight, \DnsMockNatal, and \DnsMockProduct, respectively.

The field illustration uses the 11 Galactic field DNSs with total
mass in Ref.~\cite{2023A&A...678A.187C}, excluding cluster systems.
We integrate the Peters equations backward with measured component
masses, or equal masses if only the total is available, over ages
from zero to the smaller of characteristic age and 12 Gyr
\cite{Peters:1964,Grichener2026Eccentricity}. Each birth-orbit Gaussian
is centered at its range midpoint. The variance adds the squared
half-range to a floor of $0.05^2$ in $\log_{10}(P/{\rm day})$ or
$0.02^2$ in eccentricity. The correctly normalized, natal-reading,
and unnormalized fits give \DnsRealRight, \DnsRealNatal, and
\DnsRealProduct, respectively. Substituting the independent
\cosmic{} arrival prior gives \DnsRealAltRight{} for the correctly
normalized fit. The illustration depends on the assumed progenitor
and orbital measurement models. A Galactic inference would require
the additional selection effects discussed in the paper.
The archived catalog inputs and recovery results are in
\path{data/presn2_controls/dns_kick_bias.json}.

\section{Code checks and AI use}
The author used Claude and Codex for research, software, numerical
checks, and manuscript preparation, as detailed in the main article.
The analysis included independent vector-impulse tests, binary-by-binary
\cosmic{} parity, integration refinements, and adversarial review of
the implementation and interpretation. The native \compas{} review
checked first-collapse classification and the zero-kick
accretion-induced-collapse event against the C++ implementation and
stored events. Follow-up tests verified the corrections and
characterized the remaining sampling sensitivity. The author takes
responsibility for the code, results, and manuscript.